\documentclass[11pt,a4paper]{article}
\pdfoutput=1 

\usepackage{jheppub}
\usepackage{latexsym}
\usepackage{revsymb}
\usepackage{bm}
\usepackage{subfigure}
\usepackage{color}
\usepackage[usenames,dvipsnames,svgnames,table]{xcolor}

\renewcommand{\tilde}[1]{\overset{\lower1pt\hbox{$\scriptstyle{\sim}$}}{#1}}
\newcommand{\backtilde}[1]{\overset{\lower1pt\hbox{$\scriptstyle{\backsim}$}}{#1}}
\newcommand{\edlit}[1]{\overset{\lower1pt\hbox{$\scriptstyle{\backsim}$}}{#1}}

\newcommand {\bea} {\begin{eqnarray}}
\newcommand {\eea} {\end{eqnarray}}
\newcommand {\ba} {\begin{array}}
\newcommand {\ea} {\end{array}}

\preprint{IP/BBSR/2015-2}

\title{Running of Oscillation Parameters in Matter with 
Flavor-Diagonal Non-Standard Interactions of the Neutrino}
 
\author[a]{Sanjib Kumar Agarwalla,}
\author[b]{Yee Kao,}
\author[a]{Debashis Saha,}
\author[c]{and Tatsu Takeuchi$\,$}

\affiliation[a]{Institute of Physics, Sachivalaya Marg, Sainik School Post, Bhubaneswar 751005, Orissa, India}
\affiliation[b]{Department of Chemistry and Physics, Western Carolina University, Cullowhee, NC 28723, USA}  
\affiliation[c]{Center for Neutrino Physics, Physics Department, Virginia Tech, Blacksburg, VA 24061, USA}

\emailAdd{sanjib@iopb.res.in}
\emailAdd{ykao@email.wcu.edu}
\emailAdd{debasaha@iopb.res.in}
\emailAdd{takeuchi@vt.edu}

\abstract{
In this article we unravel the role of matter effect in neutrino oscillation in the presence 
of lepton-flavor-conserving, non-universal non-standard interactions (NSI's) of the neutrino.
Employing the Jacobi method,
we derive approximate analytical expressions for the effective mass-squared differences and
mixing angles in matter.
It is shown that, within the effective mixing matrix,
the Standard Model (SM) $W$-exchange interaction only affects $\theta_{12}$ and $\theta_{13}$, 
while the flavor-diagonal NSI's only affect $\theta_{23}$.  
The CP-violating phase $\delta$ remains unaffected.
Using our simple and compact analytical approximation, 
we study the impact of the flavor-diagonal NSI's 
on the neutrino oscillation probabilities for various appearance and disappearance channels. 
At higher energies and longer baselines, it is found that
the impact of the NSI's can be significant in the $\nu_\mu\rightarrow\nu_\mu$ channel,
which can probed in future atmospheric neutrino experiments, 
if the NSI's are of the order of their current upper bounds.
Our analysis also enables us to explore the 
possible degeneracy between the octant of $\theta_{23}$ and the sign of the NSI parameter 
for a given choice of mass hierarchy in a simple manner.
}

\arxivnumber{1506.08464}

\begin{document}
\maketitle
\flushbottom

\section{Introduction and Motivation}
\label{introduction}

The recent measurement of the moderately large value of the 1-3 mixing angle 
\cite{An:2015rpe,An:2013zwz,An:2012bu,Ahn:2012nd,Abe:2011fz,Abe:2012tg,Abe:2013sxa,Adamson:2013ue,Abe:2013hdq,Abe:2013xua,Abe:2015awa}, quite close to its previous upper limit \cite{Apollonio:1999ae,Piepke:2002ju}, 
strongly validates the standard three-flavor oscillation model 
of neutrinos \cite{Agashe:2014kda,Hewett:2012ns}, which has been quite successful 
in explaining all the neutrino oscillation data available so far \cite{Gonzalez-Garcia:2014bfa,Capozzi:2013csa,Forero:2014bxa},
except for a few anomalies observed at very-short-baseline experiments \cite{Abazajian:2012ys}. 
This fairly large value of $\theta_{13}$ greatly enhances 
the role of matter effects\footnote{For a recent review, see Ref.~\cite{Blennow:2013rca}.} 
\cite{Wolfenstein:1977ue,Mikheev:1986gs,Mikheev:1986wj} in currently running 
and upcoming long-baseline \cite{Agarwalla:2014fva,Pascoli:2013wca,Feldman:2013vca}
and atmospheric \cite{Wendell:2015ona,Ahmed:2015jtv,Devi:2014yaa,Aartsen:2014oha} 
neutrino oscillation experiments aimed at determining the remaining fundamental unknowns, 
in particular, the neutrino mass hierarchy,\footnote{There are two possibilities: it can be either `normal' if $\delta m^2_{31} \equiv m^2_3 -m^2_1 > 0$, or `inverted' if $\delta m^2_{31} < 0$.}
possible presence of a CP-violating phase $\delta$, 
and the octant ambiguity of $\theta_{23}$ \cite{Fogli:1996pv} if the 2-3 mixing angle is non-maximal. 
A clear understanding of the sub-leading three-flavor effects \cite{Agarwalla:2013hma,Minakata:2012ue} 
in the neutrino oscillation probabilities 
in matter is mandatory to achieve the above goals.

Furthermore, in addition to the Standard Model (SM) $W$-exchange interaction, 
various models of physics beyond the SM predict non-standard interactions (NSI) of the neutrino \cite{Wolfenstein:1977ue,Valle:1987gv,Guzzo:1991hi,Roulet:1991sm,Grossman:1995wx}, 
which could affect neutrino propagation through matter.
Indeed, such NSI's\footnote{Present status and future prospects of NSI's 
are discussed in recent reviews \cite{Ohlsson:2012kf,Miranda:2015dra}.} 
arise naturally in many neutrino-mass models
\cite{Minkowski:1977sc,Yanagida:1979as,Mohapatra:1979ia,GellMann:1980vs,Schechter:1980gr,Lazarides:1980nt,Mohapatra:1986bd,Akhmedov:1995ip,Akhmedov:1995vm,Dev:2009aw,Boucenna:2014zba,Cheng:1980qt,Zee:1980ai,Babu:1988ki,Diaz:1997xc,Hirsch:2000ef}
which attempt to explain the small neutrino masses and the relatively large neutrino mixing angles, 
as suggested by current oscillation data \cite{Gonzalez-Garcia:2014bfa,Capozzi:2013csa,Forero:2014bxa},
as well as in many other models to be discussed in a later section.
Thus, understanding how the presence of NSI's would affect the three-flavor neutrino oscillation probabilities in matter
is crucial in extracting the fundamental unknowns listed above,
and also in searching for new-physics signatures in neutrino oscillation data.

While the three-flavor oscillation probabilities in matter can be calculated numerically on a computer, 
with or without NSI's, and properly taking into account the changing mass-density along the baseline, the computer program is a blackbox which does not offer any deep understanding as to why the probabilities depend on the input parameters in a particular way.
If the mass-density along the baseline is approximated by an average constant value, then
exact analytical expressions for the three-flavor oscillation probabilities can, in principle, be derived
as was done for the SM case in Refs.~\cite{Barger:1980tf,Zaglauer:1988gz,Kimura:2002hb,Kimura:2002wd}.
But even for the SM case, the expressions are extremely lengthy and too complicated 
to yield much physical insight.
Thus, for the SM case,
further approximations which simplify the analytic expressions while maintaining the essential physics
have been developed by various authors 
\cite{Petcov:1986qg,Kim:1986vg,Arafune:1996bt,Arafune:1997hd,Ohlsson:1999xb,Freund:2001pn,Cervera:2000kp,Akhmedov:2004ny,Honda:2006hp,Asano:2011nj,Agarwalla:2013tza,Minakata:2015gra}
to help us in this direction.
Indeed, approximate analytical expressions 
of the SM neutrino oscillation probabilities in constant-density matter have played important roles in
understanding the nature of the flavor transitions as functions 
of baseline $L$ and/or neutrino energy  $E$
\cite{Freund:2001pn,Cervera:2000kp,Akhmedov:2004ny,Asano:2011nj}. 
To obtain similar insights for the NSI case,
approximate analytical expressions for the three-flavor oscillation probabilities 
in constant-density matter in the presence of NSI's
are called for. 
Once they have provided us with the intuition we seek,
on how and why the oscillation probabilities behave in a particular way, 
we can then resort to numerical techniques to further refine the analysis, e.g. taking into
account the non-constant mass-density, if the need arises.

In previous works \cite{Agarwalla:2013tza,Honda:2006hp}, we looked at the matter effect
on neutrino oscillation due to the SM $W$-exchange 
interaction between the matter electrons and the propagating electron 
neutrinos. 
Employing the Jacobi method \cite{Jacobi:1846}, we showed 
that the said matter effects could be absorbed into the `running' of 
the effective mass-squared-differences, and the 
effective mixing angles $\theta_{12}$ and $\theta_{13}$ in matter 
as functions of the parameter $a=2\sqrt{2}G_F N_e E$, while the 
effective values of $\theta_{23}$ and the CP-violating phase $\delta$ 
remained unaffected. Here, $G_F$ is the Fermi muon decay constant,
$N_e$ is the electron density, and $E$ is the energy of the neutrino.
The approximate neutrino oscillation probabilities were obtained by simply
replacing the oscillation parameters in the vacuum expressions for the probabilities with
their running in-matter counterparts.

This running-effective-parameter approach has several advantages over other 
approaches which approximate the neutrino oscillation probabilities directly.  
First, the resulting expressions for the probabilities are strictly positive, which is not always the 
case when the probabilities are directly expanded in some small parameter, and the
series truncated after a few terms.
Second, the behavior of the oscillation probabilities as functions of $L$ and $E$ can
be understood easily as due to the running of the oscillation parameters with $a$,
as was shown in several examples in Refs.~\cite{Agarwalla:2013tza,Honda:2006hp}.
Third, as will be shown later, it is very convenient in exploring possible 
correlations and degeneracies among the mass-mixing parameters that 
may appear in matter in a non-trivial fashion.

In this paper, we extend our previous analysis and investigate 
how our conclusions are modified in the presence of neutrino
NSI's of the form
\begin{equation}
\mathcal{L}_{\mathrm{NC-NSI}} \;=\;
-\sum_{\alpha\beta f}
2\sqrt{2}G_F \varepsilon_{\alpha\beta}^{fC}
\bigl(\overline{\nu_\alpha}\gamma^\mu P_L \nu_\beta\bigr)
\bigl(\overline{f}\gamma_\mu P_C f\bigr)
\;,
\label{H_NC-NSI}
\end{equation}
where subscripts $\alpha, \beta = e,\mu,\tau$ label the 
neutrino flavor, $f = e,u,d$ mark the matter fermions, 
$C=L, R$ denotes the chirality of the $ff$ current,
and $\varepsilon_{\alpha\beta}^{fC}$ are dimensionless
quantities which parametrize the strengths of the interactions.
The hermiticity of the interaction demands
\begin{equation}
\varepsilon_{\beta\alpha}^{fC} \;=\; (\varepsilon_{\alpha\beta}^{fC})^*
\;.
\end{equation}
For neutrino propagation through matter, the relevant combinations are
\begin{equation}
\varepsilon_{\alpha\beta}
\;\equiv\; 
\sum_{f=e,u,d}
\varepsilon_{\alpha\beta}^{f}
\dfrac{N_f}{N_e}
\;\equiv\;
\sum_{f=e,u,d}
\left(
\varepsilon_{\alpha\beta}^{fL}+
\varepsilon_{\alpha\beta}^{fR}
\right)\dfrac{N_f}{N_e}
\;,
\label{epsilondef}
\end{equation}
where $N_f$ denotes the density of fermion $f$.
In this current work, we limit our investigation to flavor-diagonal NSI's, that is, we only
allow the $\varepsilon_{\alpha\beta}$'s with $\alpha=\beta$ to be non-zero.
The case of flavor non-diagonal NSI's will be considered in 
a separate work \cite{AgarwallaKaoSunTakeuchi:2015}.

In the following, we will show how the presence of such flavor-diagonal NSI's 
affect the running of the effective neutrino oscillation parameters (the mass-squared differences,
mixing angles, and CP-violating phase), and ultimately how they alter the oscillation probabilities.  
We find that due to the expected smallness of the $\varepsilon_{\alpha\alpha}$'s as compared 
to the SM $W$-exchange interaction, there is a clear separation 
in the ranges of $a$ at which the NSI's and the SM interaction are respectively relevant.
Furthermore, within the effective neutrino mixing matrix,
the SM interaction only affects the running of $\theta_{12}$ and $\theta_{13}$, while the flavor-diagonal NSI's only affect the running of 
$\theta_{23}$.  
The CP-violating phase $\delta$ remains unaffected and maintains its vacuum value.

We note that similar studies have been performed 
in the past by many authors e.g. in Refs.~\cite{GonzalezGarcia:2001mp,Ota:2001pw,Yasuda:2007jp,Kopp:2007ne,Ribeiro:2007ud,Blennow:2008eb,Kikuchi:2008vq,Meloni:2009ia,Asano:2011nj}.
This work differs from these existing works in the use of 
the Jacobi method \cite{Jacobi:1846} to derive compact analytical formulae for the running
effective mass-squared differences and effective mixing angles,  
which provide a clear and simple picture of how neutrino 
NSI's affect neutrino oscillation.
We also note that we have addressed the same problem with a similar approach previously in Refs.~\cite{Honda:2006gv,Honda:2006di,Honda:2007yd,Honda:2007wv}.
The current paper updates these works by allowing for a non-maximal value of $\theta_{23}$,
a generic value of the CP-violating phase $\delta$, a larger range of $a$, and refinements on how the matter effect is absorbed into the running parameters.

This paper is organized as follows.
We start section~\ref{neutrino-NSI} with a discussion on neutrino NSI's:
how and where they arise, how they affect the propagation of the neutrinos in matter, 
and show that the linear combinations relevant for neutrino oscillation are
$\eta = (\varepsilon_{\mu\mu}-\varepsilon_{\tau\tau})/2$ and $\zeta = \varepsilon_{ee}-(\varepsilon_{\mu\mu}+\varepsilon_{\tau\tau})/2$.
This is followed by a brief discussion on the theoretical expectation for the
sizes of these parameters, and their current experimental bounds.
In section~\ref{neutrino}, we use the Jacobi method 
to calculate how the NSI parameter $\eta$ affects 
the running of the effective mass-squared 
differences, effective mixing angles, and the effective CP-violating phase 
as functions of $\hat{a}=a(1+\zeta)$ for the neutrinos. 
To check the accuracy of our method,
we also present a comparison between our approximate 
analytical probability expressions and exact numerical 
calculations (for constant matter density) towards the end of this section.
Section~\ref{sec:applications} describes the
advantages of our analytical probability expressions 
to figure out the suitable testbeds to probe these 
NSI's of the neutrino. In this section, we also present
simple and compact analytical expressions exposing 
the possible correlations and degeneracies between 
$\theta_{23}$ and the NSI parameter $\eta$ under 
such situations. Finally, we summarize and draw our 
conclusions in section~\ref{summary-conclusions}.
The derivation of the running oscillation parameters for 
the anti-neutrino case is relegated to appendix~\ref{anti-neutrino} 
where we also compare our analytical results with exact numerical 
probabilities. In appendix~\ref{Constant-vs-varying-density}, we 
examine the differences in the exact numerical probabilities with 
line-averaged constant Earth density and varying Earth density 
profile for 8770 km and 10000 km baselines.

\section{A Brief Tour of Non-Standard Interactions of the Neutrino}
\label{neutrino-NSI}

In this section, we first briefly discuss the various categories 
of models that give rise to NSI's of neutrino.

\subsection{Models that Predict NSI's of the Neutrino}
\label{NSI-models}

NSI's of the neutrino arise in a variety of beyond 
the Standard Model (BSM) scenarios 
\cite{Honda:2006gv,Honda:2006di,Honda:2007yd,Honda:2007wv,Antusch:2008tz,Ohlsson:2012kf}
via the direct tree-level exchange of new particles, 
or via flavor distinguishing radiative corrections to the $Z\nu\nu$ vertex 
\cite{Loinaz:1998jg,Lebedev:1999vc,Chang:2000xy},
or indirectly via the non-unitarity of the 
lepton mixing matrix \cite{Antusch:2006vwa}.
BSM models which predict neutrino NSI's include:
\begin{enumerate}

\item 
Models with a generation distinguishing $Z'$ boson.
This class includes 
gauged $L_e-L_\mu$ and
gauged $L_e-L_\tau$ \cite{He:1990pn,He:1991qd}, 
gauged $B-\alpha L_e-\beta L_\mu-\gamma L_\tau$ (with $\alpha+\beta+\gamma=3$) 
\cite{Ma:1997nq,Ma:1998dp,Ma:1998dr,Ma:1998zg,Ma:1998xd,Chang:2000xy}, and
topcolor assisted technicolor \cite{Hill:1994hp,Loinaz:1998jg}.

\item 
Models with leptoquarks 
\cite{Buchmuller:1986iq,Buchmuller:1986zs,Leurer:1993em,Davidson:1993qk,Hewett:1997ce} 
and/or bileptons \cite{Cuypers:1996ia}.
These can be either scalar or vector particles.
This class includes various Grand Unification Theory (GUT) models 
and extended technicolor (ETC) \cite{Dimopoulos:1979es,Eichten:1979ah,Dimopoulos:1980fj}.

\item 
The Supersymmetric Standard Model with $R$-parity violation 
\cite{Barger:1989rk,Dreiner:1997uz,Barbier:2004ez,Lebedev:1999vc,Lebedev:1999ze}. 
The super-partners of the SM particles play the role of the leptoquarks and bileptons
of class 2.

\item 
Extended Higgs models.
This class includes the Zee model \cite{Zee:1980ai}, the Zee-Babu model \cite{Zee:1985id,Babu:1988ki,Ohlsson:2009vk}, and
various models with $SU(2)$ singlet \cite{Chikashige:1980ui,Schechter:1981cv,Fukugita:1988qe,Bilenky:1993bt,Antusch:2008tz} 
and $SU(2)$ triplet Higgses \cite{Gelmini:1980re,Schechter:1981cv,Malinsky:2008qn}, as well as the generation
distinguishing $Z'$ models listed under class 1.

\item 
Models with non-unitary neutrino mixing matrices 
\cite{Langacker:1988up,Bilenky:1992wv,Nardi:1994iv,Tommasini:1995ii,Bergmann:1998rg,Czakon:2001em,Fornengo:2001pm,Bekman:2002zk,Loinaz:2002ep,Loinaz:2003gc,Loinaz:2004qc,Barger:2004db,Antusch:2006vwa,Malinsky:2009df,Malinsky:2009gw,Abazajian:2012ys,Escrihuela:2015wra}.
Apparent non-unitarity of the mixing matrix 
for the three light neutrino flavors would 
result when it is part of a larger mixing matrix 
involving heavier and/or sterile fields.

\end{enumerate}
Systematic studies of how NSI's can arise in these, 
and other BSM theories can be found, for instance, 
in Refs.~\cite{Honda:2007wv,Antusch:2008tz,Gavela:2008ra,Ohlsson:2012kf}.
Thus, the ability to detect NSI's in neutrino experiments would complement the
direct searches for new particles at the LHC for a variety of BSM models.
Next, we focus our attention to see the role of lepton-flavor-conserving
NSI's when neutrinos travel through the matter.

\subsection{Lepton-Flavor-Conserving NSI's}
\label{LFC-NSI's}

The NSI's of Eq.~(\ref{epsilondef}) modify 
the effective Hamiltonian for neutrino propagation 
in matter in the flavor basis to
\begin{equation}
H \;=\; \dfrac{1}{2E}
\left(
U
\begin{bmatrix} 
m_1^2 & 0 & 0 \\ 
0 & m_2^2 & 0 \\ 
0 & 0 & m_3^2 
\end{bmatrix}
U^\dagger
+
a
\begin{bmatrix}
1 + \varepsilon_{ee}  & \varepsilon_{e\mu}      & \varepsilon_{e\tau}   \\
\varepsilon_{e\mu}^*  & \varepsilon_{\mu\mu}    & \varepsilon_{\mu\tau} \\
\varepsilon_{e\tau}^* & \varepsilon_{\mu\tau}^* & \varepsilon_{\tau\tau}
\end{bmatrix}
\right)
\end{equation}
where $U$ is the vacuum Pontecorvo-Maki-Nakagawa-Sakata (PMNS) matrix 
\cite{Pontecorvo:1957vz,Maki:1962mu,Pontecorvo:1967fh}, $E$ is the neutrino 
energy, the matter-effect parameter $a$ is given by 
\begin{equation}
a 
\;=\; 2\sqrt{2}G_F N_e E
\;=\; 7.6324\times 10^{-5}(\mathrm{eV}^2)
\left(\dfrac{\rho}{\mathrm{g/cm^3}}\right)
\left(\dfrac{E}{\mathrm{GeV}}\right)
\;.
\label{matter-term}
\end{equation}
For Earth matter, we can assume $N_n\approx N_p = N_e$, 
in which case $N_u \approx N_d \approx 3N_e$.
Therefore,
\begin{equation}
\varepsilon_{\alpha\beta} \;=\;
\varepsilon_{\alpha\beta}^{\oplus} \;\approx\; 
\varepsilon_{\alpha\beta}^{e}
+3\,\varepsilon_{\alpha\beta}^{u}
+3\,\varepsilon_{\alpha\beta}^{d}
\;.
\end{equation}

Since we restrict our attention to 
lepton-flavor-conserving NSI's in this paper,
the effective Hamiltonian in the 
flavor basis takes the form
\begin{equation}
H \;=\; \dfrac{1}{2E}
\left(
U
\begin{bmatrix} 
m_1^2 & 0 & 0 \\ 
0 & m_2^2 & 0 \\ 
0 & 0 & m_3^2 
\end{bmatrix}
U^\dagger
+
a
\begin{bmatrix}
1 + \varepsilon_{ee} & 0 & 0 \\
0 & \varepsilon_{\mu\mu} & 0 \\
0 & 0 & \varepsilon_{\tau\tau}
\end{bmatrix}
\right)
\;.
\label{eq:epsilons}
\end{equation}
In the absence of off-diagonal terms, 
we can rewrite the matter effect matrix 
as follows:
\begin{eqnarray}
\lefteqn{\begin{bmatrix}
1 + \varepsilon_{ee} & 0 & 0   \\
0 & \varepsilon_{\mu\mu} & 0 \\
0 & 0 & \varepsilon_{\tau\tau}
\end{bmatrix}
}
\cr
& = &
\begin{bmatrix}
1 + \varepsilon_{ee} - \dfrac{\varepsilon_{\mu\mu}+\varepsilon_{\tau\tau}}{2} & 0 & 0 \\
0 & \left(\dfrac{\varepsilon_{\mu\mu}-\varepsilon_{\tau\tau}}{2}\right) & 0 \\
0 & 0 & -\left(\dfrac{\varepsilon_{\mu\mu}-\varepsilon_{\tau\tau}}{2}\right) 
\end{bmatrix}
+ \left(\dfrac{\varepsilon_{\mu\mu}+\varepsilon_{\tau\tau}}{2}\right)
\begin{bmatrix}
1 & 0 & 0 \\ 0 & 1 & 0 \\ 0 & 0 & 1
\end{bmatrix}
\cr
& = &
a\left(
1 + \varepsilon_{ee} - \dfrac{\varepsilon_{\mu\mu}+\varepsilon_{\tau\tau}}{2}
\right)
\begin{bmatrix}
1 & 0 & 0 \\
0 & \left(\dfrac{\varepsilon_{\mu\mu}-\varepsilon_{\tau\tau}}{2}\right) & 0 \\
0 & 0 & -\left(\dfrac{\varepsilon_{\mu\mu}-\varepsilon_{\tau\tau}}{2}\right) 
\end{bmatrix}
\cr & &
+ \mbox{ $O(\varepsilon^2)$ terms }
+ \mbox{ unit matrix term } 
\;,
\end{eqnarray}
where we have assumed that 
the $\varepsilon$'s are small compared to 1.
Let
\begin{equation}
\eta
\;\equiv\;
\dfrac{\varepsilon_{\mu\mu}-\varepsilon_{\tau\tau}}{2}
\;,\qquad
\zeta
\;\equiv\;
\varepsilon_{ee} - \dfrac{\varepsilon_{\mu\mu}+\varepsilon_{\tau\tau}}{2}
\;,
\end{equation}
and
\begin{equation}
\hat{a} \;\equiv\; a(1+\zeta)\;.
\label{eq:a-and-ahat}
\end{equation}
Note that the NSI parameter $\eta$ introduced here 
is related to the parameter $\xi$ which was used in
Ref.~\cite{Honda:2006gv} by
\begin{equation}
\eta \;=\; -\dfrac{\xi}{2}\;.
\end{equation}
Thus, the effective Hamiltonian can be taken to be
\begin{equation}
H \;=\; \dfrac{1}{2E}
\left(
U
\begin{bmatrix} 
m_1^2 & 0 & 0 \\ 
0 & m_2^2 & 0 \\ 
0 & 0 & m_3^2 
\end{bmatrix}
U^\dagger
+
\hat{a}
\begin{bmatrix}
1 & 0 & 0   \\
0 & \eta & 0 \\
0 & 0 & -\eta
\end{bmatrix}
\right)
\;.
\label{eq:eta-zeta}
\end{equation}
Note that the $\eta=0$ case simply replaces 
$a$ with $\hat{a}=a(1+\zeta)$ in the 
SM Hamiltonian. Thus, the 
effective mass-squared differences,
mixing angles, and CP-violating phase have the same
functional dependence on $\hat{a}$ as 
they had on $a$ in the SM case.
In other words, a non-zero $\zeta$ simply rescales 
the value of $a$ by a constant factor. 
The presence of a non-zero $\eta$, 
on the other hand, requires us to diagonalize 
$H$ with a different unitary matrix from 
the $\eta=0$ case, and this will introduce 
corrections to the effective oscillation parameters 
beyond a simple rescaling of $a$.
Next, we discuss the constraints that 
we have at present on these NSI 
parameters.

\subsection{Existing Bounds on the NSI parameters}
\label{existing-bounds}

Before looking at the matter effect due to the neutrino NSI's,
let us first look at what is currently known about the sizes of the 
$\varepsilon_{\alpha\beta}^{fC}$'s and the combinations
$\eta = (\varepsilon_{\mu\mu}-\varepsilon_{\tau\tau})/2$ and
$\zeta = \varepsilon_{ee} - (\varepsilon_{\mu\mu}+\varepsilon_{\tau\tau})/2$.

\subsubsection{Theoretical Expectation}
\label{theoretical-expectations}

Theoretically, the sizes of the NSI's are generically expected to be small
since they are putatively due to BSM physics at a much higher scale than 
the electroweak scale, or to loop effects. If they arise from the tree level 
exchange of new particles of mass $\Lambda$, which would be described 
by dimension six operators, we can expect the 
$\varepsilon_{\alpha\beta}^{fC}$'s to be of order $O(M_W^2/\Lambda^2)$.
Loop effects involving a heavy particle of mass $\Lambda$ would be further 
suppressed by a factor of $O(1/4\pi)$ or more. Processes that lead to 
dimension eight operators would be of order $O(M_W^4/\Lambda^4)$.
Thus, if we assume $\Lambda=O(1\,\mathrm{TeV})$, we expect the
$\varepsilon_{\alpha\beta}^{fC}$'s, and consequently $\eta$ and $\zeta$, 
to be $O(10^{-3})$ or smaller.

\subsubsection{Direct Experimental Bounds}
\label{direct-bounds}

Direct experimental bounds on the flavor-diagonal NSI parameters 
$\varepsilon_{\alpha\alpha}^{fC}$'s are available from 
a variety of sources. 
These include 
$\nu_e e$ scattering data from 
LAMPF \cite{Allen:1992qe} and LSND \cite{Auerbach:2001wg},
$\overline{\nu_e} e$ scattering data from the reactor experiments
Irvine \cite{Reines:1976pv},
Krasnoyarsk \cite{Vidyakin:1992nf}, 
Rovno \cite{Derbin:1993wy},
MUNU \cite{Daraktchieva:2003dr},
and Texono \cite{Deniz:2009mu},
$\nu_e q$ scattering data from CHARM \cite{Dorenbosch:1986tb},
$\nu_\mu e$ scattering data from CHARM~II \cite{Vilain:1993xb,Vilain:1994qy},
$\nu_\mu q$ scattering data from NuTeV \cite{Zeller:2001hh,Ball:2009mk,Bentz:2009yy},
$e^+e^-\rightarrow \nu\overline{\nu}\gamma$ data from the LEP experiments 
ALEPH \cite{Barate:1997ue,Barate:1998ci,Heister:2002ut}, 
L3 \cite{Acciarri:1997dq,Acciarri:1998hb,Acciarri:1999kp}
OPAL \cite{Ackerstaff:1996sd,Ackerstaff:1997ze,Abbiendi:1998yu,Abbiendi:2000hh},
and DELPHI \cite{Abdallah:2003np},
and neutrino oscillation data from 
Super-Kamiokande \cite{Mitsuka:2011ty}, 
IceCube and DeepCore \cite{Gross:2013iq,Collaboration:2011ym},
KamLAND \cite{Abe:2008aa}, 
SNO \cite{Aharmim:2008kc}, and
Borexino \cite{Arpesella:2008mt,Bellini:2011rx}.
These have been analyzed by various authors 
in Refs.~\cite{Berezhiani:2001rs,Berezhiani:2001rt,Hirsch:2002uv,Davidson:2003ha,Maltoni:2002kq,Friedland:2004pp,Friedland:2004ah,Friedland:2005vy,Barranco:2005ps,Barranco:2007ej,Bolanos:2008km,Escrihuela:2009up,Biggio:2009kv,Biggio:2009nt,Forero:2011zz,Escrihuela:2011cf,Agarwalla:2012wf,Esmaili:2013fva,Khan:2013hva,Khan:2014zwa,Girardi:2014gna,Girardi:2014kca,Agarwalla:2014bsa}, and collecting the results of the most recent analyses, 
we place the following 90\% C.L. bounds on the 
flavor-diagonal vectorial NSI couplings:
\begin{equation}
\begin{array}{lll}
|\varepsilon_{ee}^{e}| \;<\; 0.1\;, \qquad
& |\varepsilon_{ee}^{u}| \;<\; 1\;, \qquad
& |\varepsilon_{ee}^{d}| \;<\; 1\;,
\\
|\varepsilon_{\mu\mu}^{e}| \;<\; 0.04\;, \qquad\qquad
& |\varepsilon_{\mu\mu}^{u}| \;<\; 0.04\;, \qquad\qquad
& |\varepsilon_{\mu\mu}^{d}| \;<\; 0.04\;,
\\
|\varepsilon_{\tau\tau}^{e}| \;<\; 0.6\;, \qquad
& |\varepsilon_{\tau\tau}^{u}| \;<\; 0.05\;, \qquad
&|\varepsilon_{\tau\tau}^{d}| \;<\; 0.05\;.
\end{array}
\end{equation}
Note that the current experimental bounds on the NSI couplings $\varepsilon_{\alpha\beta}^{fC}$ are weak compared to the theoretical expectation
of $O(10^{-3})$, the 90\% C.L. upper bound on their
absolute values ranging from $O(10^{-2})$ to $O(1)$.
To combine these bounds into bounds on the $\varepsilon_{\alpha\beta}$'s,
we follow the procedure of Ref.~\cite{Biggio:2009nt},
\begin{equation}
|\varepsilon_{\alpha\beta}|
\;\alt\;
\sqrt{ |\varepsilon_{\alpha\beta}^{e}|^2 +
|3\varepsilon_{\alpha\beta}^{u}|^2 +
|3\varepsilon_{\alpha\beta}^{d}|^2
}
\;,
\end{equation}
and find
\begin{equation}
|\varepsilon_{ee}| \;<\; 4\;,\qquad
|\varepsilon_{\mu\mu}| \;<\; 0.2\;,\qquad
|\varepsilon_{\tau\tau}| \;<\; 0.6\;.
\end{equation}
Neglecting possible correlations among these parameters, these bounds can again be combined to
yield
\begin{eqnarray}
|\eta| 
& < & \sqrt{\dfrac{1}{4}|\varepsilon_{\mu\mu}|^2 + \dfrac{1}{4}|\varepsilon_{\tau\tau}|^2}
\;=\; 0.3\;,\cr
|\zeta| & < & \sqrt{|\varepsilon_{ee}|^2 + \dfrac{1}{4}|\varepsilon_{\mu\mu}|^2 + \dfrac{1}{4}|\varepsilon_{\tau\tau}|^2}
\;=\; 4\;.
\end{eqnarray}

A tighter bound exists for $\eta$ which has been obtained 
directly using solar and atmospheric neutrino data 
in Ref.~\cite{Maltoni:2002kq}, and more recently 
in Ref.~\cite{GonzalezGarcia:2011my} using atmospheric 
and MINOS data.
In Ref.~\cite{Maltoni:2002kq}, only NSI's with the $d$-quarks 
were considered and the following $3\sigma$ (99.7\% C.L.) 
bounds were obtained\footnote{The notation used in 
Ref.~\cite{Maltoni:2002kq} is $\varepsilon=\varepsilon^{d}_{\mu\tau}$
and $\varepsilon'=\varepsilon^{d}_{\tau\tau}-\varepsilon^{d}_{\mu\mu}$.}:
\begin{eqnarray}
-0.03\;<\; \varepsilon^{d}_{\mu\tau} & < & 0.02\;,\cr
|\varepsilon^{d}_{\tau\tau}-\varepsilon^{d}_{\mu\mu}| & < & 0.05\;.
\label{ATMOSd2}
\end{eqnarray}
Since $N_d = N_u = 3N_e$, Ref.~\cite{Maltoni:2002kq} is actually constraining
$\varepsilon_{\alpha\beta}/3$, so this result can be interpreted as
\begin{eqnarray}
-0.09 \;<\; \varepsilon_{\mu\tau} & < & 0.06\;,\cr
|\varepsilon_{\mu\mu}-\varepsilon_{\tau\tau}| & < & 0.15\;.
\label{ATMOS3sigma1}
\end{eqnarray}
Rescaling to $1.64\sigma$ (90\% C.L.), we find 
\begin{equation}
|\eta| 
\;=\; \left|\dfrac{\varepsilon_{\mu\mu}-\varepsilon_{\tau\tau}}{2}\right|
\;<\; 0.04
\;. 
\label{etabound1}
\end{equation}
Ref.~\cite{GonzalezGarcia:2011my} gives slightly 
different 90\% C.L. bounds of
\begin{eqnarray}
|\varepsilon_{\mu\tau}| & < & 0.035\;,\cr
|\varepsilon_{\mu\mu}-\varepsilon_{\tau\tau}| & < & 0.11\;,
\label{ATMOS3sigma2}
\end{eqnarray}
which translates to
\begin{equation}
|\eta| 
\;=\; \left|\dfrac{\varepsilon_{\mu\mu}-\varepsilon_{\tau\tau}}{2}\right|
\;<\; 0.055
\;. 
\label{etabound2}
\end{equation}
Thus, though $\eta$ and $\zeta$ are expected 
theoretically to be $O(10^{-3})$, their current 
90\% C.L. experimental bounds are respectively 
$\sim\!0.05$ and $O(1)$.

\section{Effective Mixing Angles and Effective Mass-Squared Differences \\ -- Neutrino Case}
\label{neutrino}

\subsection{Setup of the Problem}

\begin{table}[t]
\begin{center}
\begin{tabular}{|c|c|c|} 
\hline
Parameter & Best-fit Value \& 1$\sigma$ Range & Benchmark Value
\\
\hline
\hline
$\quad\delta m^2_{21}\quad$ 
& $\quad (7.50\pm 0.185) \times 10^{-5} \ {\rm eV}^2\quad$  
& $\quad 7.50\times 10^{-5} \ {\rm eV}^2\quad$
\\
\hline
$\delta m^2_{31}$ 
& $(2.47^{+0.069}_{-0.067}) \times 10^{-3} \ {\rm eV}^2$ 
& $2.47\times 10^{-3} \ {\rm eV}^2$
\\
\hline
$\sin^2\theta_{23}$ & $0.41^{+0.037}_{-0.025}\oplus 0.59^{+0.021}_{-0.022}$ &
$0.41$
\\
$\theta_{23}/^{\circ}$ & $40.0^{+2.1}_{-1.5}\oplus 50.4^{+1.2}_{-1.3}$ & \\
$\theta_{23}/\mathrm{rad}$ & $0.698^{+0.037}_{-0.026}\oplus 0.880^{+0.021}_{-0.023}$ & \\
\hline
$\sin^2\theta_{12}$ & $0.30\pm 0.013$ & $0.30$ \\
$\theta_{12}/^{\circ}$ & $33.3\pm 0.8$ & \\
$\theta_{12}/\mathrm{rad}$ & $0.580\pm 0.014$ & \\
\hline
$\sin^2\theta_{13}$ & $0.023\pm 0.0023$ & $0.023$ \\
$\theta_{13}/^{\circ}$ & $8.6^{+0.44}_{-0.46}$ & \\
$\theta_{13}/\mathrm{rad}$ & $0.15\pm 0.01$ & \\
\hline
$\delta/^{\circ}$ & $300^{+66}_{-138}$ & $0$ \\
$\delta/\pi$ & $1.67^{+0.37}_{-0.77}$ & \\
\hline
\end{tabular}
\caption{Second column shows the best-fit values and 1$\sigma$
uncertainties on the oscillation parameters taken from 
Ref.~\cite{GonzalezGarcia:2012sz}.
We use the values listed in the third column as benchmark values
for which we calculate our oscillation probabilities in this work.}
\label{tab:bench}
\end{center}
\end{table}

As we have seen, in the presence of non-zero $\eta$ and $\zeta$,
the effective Hamiltonian (times $2E$) for neutrino propagation 
in Earth matter is given by
\begin{equation}
H_\eta \;=\;
\tilde{U}
\begin{bmatrix} \lambda_1 & 0 & 0 \\
                          0 & \lambda_2 & 0 \\
                          0 & 0 & \lambda_3
\end{bmatrix}
\tilde{U}^\dagger
\;=\; 
\underbrace{
U
\begin{bmatrix} 
0 & 0 & 0 \\ 
0 & \delta m_{21}^2 & 0 \\ 
0 & 0 & \delta m_{31}^2 
\end{bmatrix}
U^\dagger
+
\hat{a}
\underbrace{
\begin{bmatrix}
1 & 0 & 0   \\
0 & 0 & 0 \\
0 & 0 & 0
\end{bmatrix}
}_{\displaystyle \equiv M_a}
}_{\displaystyle \equiv H_0}
+
\hat{a}\eta
\underbrace{
\begin{bmatrix}
0 & 0 & 0   \\
0 & 1 & 0 \\
0 & 0 & -1
\end{bmatrix}
}_{\displaystyle \equiv M_\eta}
\;,
\label{Hdef}
\end{equation}
where $\hat{a}=a(1+\zeta)$.
The problem is to diagonalize $H_\eta = H_0 + \hat{a}\eta M_\eta$ 
and find the eigenvalues $\lambda_i$ ($i=1,2,3$) and
the diagonalization matrix $\tilde{U}$ as functions of $\hat{a}$ and $\eta$.

To this end, we utilize the method used in 
Refs.~\cite{Agarwalla:2013tza,Honda:2006hp} 
where approximate expressions for the 
$\lambda_i$'s and $\tilde{U}$ were derived 
for $H_0$, the $\eta=0$ case, using the 
Jacobi method \cite{Jacobi:1846}. 
The Jacobi method entails diagonalizing 
$2\times 2$ submatrices of a matrix in the order which
requires the the largest rotation angles until 
the off-diagonal elements are negligibly small. 
In the case of $H_0$, it was discovered in 
Refs.~\cite{Agarwalla:2013tza,Honda:2006hp} 
that two $2\times 2$ rotations were sufficient 
to render it approximately diagonal, and that these
two rotation angles could be absorbed into `running' 
values of $\theta_{12}$ and $\theta_{13}$.
The procedure that we use in the following 
for $H_\eta$ is to add on the $\hat{a}\eta M_\eta$
term to $H_0$ \textit{after} it is approximately 
diagonalized, and then proceed with a third 
$2\times 2$ rotation to rotate away the 
additional off-diagonal terms.

As the order parameter to evaluate the sizes 
of the off-diagonal elements, we use
\begin{equation}
\epsilon 
\;\equiv\; \sqrt{\dfrac{\delta m^2_{21}}{|\delta m^2_{31}|}}
\;\approx\; 0.17
\;,
\end{equation}
and consider $H_0$ and $H_\eta$ to be approximately 
diagonalized when the rotation angles required for
further diagonalization are of order $\epsilon^3=0.005$ or smaller.
Note that we are using a slightly different epsilon ($\epsilon$) here to
distinguish this quantity from the NSI's ($\varepsilon_{\alpha\beta}$). 

The eigenvalues $\lambda_i$ ($i=1,2,3$) 
and the diagonalization matrix 
$\tilde{U}$ of $H_\eta$ are necessarily functions 
of $\hat{a}=2\sqrt{2}G_F N_e E(1+\zeta)$.
In order to parametrize the size of $\hat{a}$, 
we find it convenient to introduce the 
log-scale variable \cite{Agarwalla:2013tza,Honda:2006hp}
\begin{equation}
\beta \equiv -\log_{\epsilon}\dfrac{\hat{a}}{|\delta m^2_{31}|}\;,
\end{equation}
so that 
$\hat{a}=\delta m^2_{21}$ corresponds to $\beta=-2$, and 
$\hat{a}=|\delta m^2_{31}|$ corresponds to $\beta=0$.
In the following, various quantities will be plotted 
as functions of $\beta$.

Unless otherwise stated, we use the benchmark 
values of the various oscillation parameters as given 
in the third column of Table~\ref{tab:bench} 
to draw our plots. These values are taken from 
Ref.~\cite{GonzalezGarcia:2012sz} and correspond to 
the case in which reactor fluxes have been left free 
in the fit and short-baseline reactor data 
with $L \leq 100$ m are included. 
For $\sin^2\theta_{23}$, we consider the 
benchmark value which lies in the lower octant 
and CP-violating phase $\delta$ is assumed 
to be zero. These choices of the oscillation parameters 
are well within their 3$\sigma$ allowed ranges 
which are obtained in recent global fit analyses 
\cite{Forero:2014bxa,Capozzi:2013csa,Gonzalez-Garcia:2014bfa}.
We also present results considering other allowed values 
of $\sin^2\theta_{23}$ and $\delta$ which we discuss 
in section~\ref{comparison}.
In the evaluation of the sizes of the elements of 
the effective Hamiltonian, we will assume
$\theta_{13} \approx 0.15 = O(\epsilon)$, 
$\cos(2\theta_{12})/2 \approx 0.2 = O(\epsilon)$, 
and $|\cos(2\theta_{23})| \approx 0.18 = O(\epsilon)$.
We also assume that the NSI parameter $\eta$ 
is of order $\epsilon^2 = 0.03$ (or smaller), 
since the current 90\% C.L. ($1.64\sigma$) 
upper bound on $|\eta|$ was $\sim 0.05$, 
cf. Eqs.~(\ref{etabound1}) and (\ref{etabound2}),
though we allow it to be as large as $0.1$ 
in our plots to enhance and make visible 
the effect of a non-zero $\eta$.

\subsection{Diagonalization of the Effective Hamiltonian}

\subsubsection{Change to the Mass Eigenbasis in Vacuum}

Introducing the matrix
\begin{equation}
Q_3 = \mathrm{diag}(1,1,e^{i\delta})\;,
\label{Qdef}
\end{equation}
we begin by partially diagonalizing the Hamiltonian $H_\eta$ as
\begin{eqnarray}
H_\eta' & = & Q_3^\dagger U^\dagger H_\eta U Q_3\cr
& = &
\underbrace{%
\left[ \begin{array}{ccc} 0 & 0 & 0 \\
                          0 & \delta m^2_{21} & 0 \\
                          0 & 0 & \delta m^2_{31}
       \end{array}
\right] +
\hat{a}
\underbrace{%
Q_3^\dagger U^\dagger
\underbrace{%
\left[ \begin{array}{ccc} 1 & 0 & 0 \\
                          0 & 0 & 0 \\
                          0 & 0 & 0 
       \end{array}
\right] 
}_{\displaystyle M_a}
U Q_3
}_{\displaystyle \equiv M'_a(\theta_{12},\theta_{13},\theta_{23})} 
}_{\displaystyle \equiv H_0'}
+ 
\hat{a}\eta\,
\underbrace{%
Q_3^\dagger U^\dagger
\underbrace{%
\left[ \begin{array}{ccc} 0 & 0 & 0 \\
                          0 & 1 & 0 \\
                          0 & 0 & -1 
       \end{array}
\right] 
}_{\displaystyle M_\eta}
U Q_3
}_{\displaystyle \equiv M'_\eta(\theta_{12},\theta_{13},\theta_{23},\delta)}  
\;,
\end{eqnarray}
where
\begin{equation}
M'_{a}(\theta_{12},\theta_{13},\theta_{23}) \;=\; 
Q_3^\dagger
\left[ \begin{array}{ccc} U^*_{e1}U_{e1} & U^*_{e1}U_{e2} & U^*_{e1}U_{e3} \\
                          U^*_{e2}U_{e1} & U^*_{e2}U_{e2} & U^*_{e2}U_{e3} \\
                          U^*_{e3}U_{e1} & U^*_{e3}U_{e2} & U^*_{e3}U_{e3} 
       \end{array}
\right] Q_3
\;=\;
\left[ \begin{array}{ccc}
       c_{12}^2 c_{13}^2    & c_{12}s_{12}c_{13}^2 & c_{12}c_{13}s_{13} \\
       c_{12}s_{12}c_{13}^2 & s_{12}^2 c_{13}^2    & s_{12}c_{13}s_{13} \\
       c_{12}c_{13}s_{13}   & s_{12}c_{13}s_{13}   & s_{13}^2
       \end{array}
\right]\;,
\end{equation}
and
\begin{eqnarray}
\lefteqn{M'_{\eta}(\theta_{12},\theta_{13},\theta_{23},\delta)} \cr
& = &
Q_3^\dagger\left\{
\left[ \begin{array}{ccc} 
U^*_{\mu 1}U_{\mu 1} & U^*_{\mu 1}U_{\mu 2} & U^*_{\mu 1}U_{\mu 3} \\
U^*_{\mu 2}U_{\mu 1} & U^*_{\mu 2}U_{\mu 2} & U^*_{\mu 2}U_{\mu 3} \\
U^*_{\mu 3}U_{\mu 1} & U^*_{\mu 3}U_{\mu 2} & U^*_{\mu 3}U_{\mu 3} 
\end{array}
\right] - 
\left[ \begin{array}{ccc} 
U^*_{\tau 1}U_{\tau 1} & U^*_{\tau 1}U_{\tau 2} & U^*_{\tau 1}U_{\tau 3} \\
U^*_{\tau 2}U_{\tau 1} & U^*_{\tau 2}U_{\tau 2} & U^*_{\tau 2}U_{\tau 3} \\
U^*_{\tau 3}U_{\tau 1} & U^*_{\tau 3}U_{\tau 2} & U^*_{\tau 3}U_{\tau 3} 
\end{array}
\right] \right\} Q_3 \cr
& = &
\left[ \begin{array}{l}
\sin(2\theta_{12})\sin(2\theta_{23})s_{13}\cos\delta 
+ (s_{12}^2-c_{12}^2 s_{13}^2)\cos(2\theta_{23}) 
\\
(s_{12}^2 e^{-i\delta} - c_{12}^2 e^{i\delta})s_{13}\sin(2\theta_{23}) 
-(1+s_{13}^2)s_{12}c_{12}\cos(2\theta_{23}) 
\\
-s_{12}c_{13}\sin(2\theta_{23})e^{-i\delta} + c_{12}s_{13}c_{13}\cos(2\theta_{23}) 
\end{array} \right. \cr
&   & \qquad\qquad
\begin{array}{l}
(s_{12}^2 e^{i\delta} - c_{12}^2 e^{-i\delta})s_{13}\sin(2\theta_{23}) 
-(1+s_{13}^2)s_{12}c_{12}\cos(2\theta_{23}) 
\\
- \sin(2\theta_{12})\sin(2\theta_{23})s_{13}\cos\delta 
+ (c_{12}^2-s_{12}^2 s_{13}^2)\cos(2\theta_{23}) 
\\
\phantom{-}c_{12}c_{13}\sin(2\theta_{23})e^{-i\delta} + s_{12}s_{13}c_{13}\cos(2\theta_{23}) 
\end{array} \cr
&   & \qquad\qquad\qquad\qquad
\left. \begin{array}{l}
-s_{12}c_{13}\sin(2\theta_{23})e^{+i\delta} + c_{12}s_{13}c_{13}\cos(2\theta_{23}) \\
\phantom{-}c_{12}c_{13}\sin(2\theta_{23})e^{+i\delta} + s_{12}s_{13}c_{13}\cos(2\theta_{23}) \\
 -c_{13}^2\cos(2\theta_{23})
\end{array} \right] \;.
\end{eqnarray}
Using $\cos(2\theta_{23}) = O(\epsilon)$ and
$\theta_{13} = O(\epsilon)$, we estimate the sizes of the elements of 
$M_a$ to be:
\begin{equation}
M_{a} \;=\;
\left[ \begin{array}{ccc}
       O(1) & O(1) & O(\epsilon) \\
       O(1) & O(1) & O(\epsilon) \\
       O(\epsilon) & O(\epsilon) & O(\epsilon^2)
       \end{array}
\right]\;,
\end{equation}
and those of $M_\eta$ to be
\begin{equation}
M_\eta \;=\; 
\left[ \begin{array}{ccc} O(\epsilon) & O(\epsilon) & O(1) \\
                          O(\epsilon) & O(\epsilon) & O(1) \\
                          O(1) & O(1) & O(\epsilon)
       \end{array}
\right]\;.
\end{equation}
Given that we have assumed $\eta=O(\epsilon^2)$ or smaller,
the off-diagonal elements of $\eta M_\eta$ are suppressed
compared to those of $M_a$, and only become important for 
$\hat{a} \agt |\delta m^2_{31}|$, or equivalently,
$\beta \agt 0$.

\subsubsection{Diagonalization of a $2\times 2$ hermitian matrix}

The Jacobi method entails diagonalizing $2\times 2$ submatrices repeatedly.
For this, it is convenient to note that given a $2\times 2$ hermitian matrix,
\begin{equation}
\begin{bmatrix}
A & B\, e^{iD} \\ B\, e^{-iD} & C
\end{bmatrix}
\;=\;
\begin{bmatrix}
1 & 0 \\ 0 & e^{-iD}
\end{bmatrix}
\begin{bmatrix}
A & B \\ B & C
\end{bmatrix}
\begin{bmatrix}
1 & 0 \\ 0 & e^{iD}
\end{bmatrix}
\;,\qquad
A, B, C, D \in \mathbb{R}\;,
\end{equation}
the unitary matrix which diagonalizes this is given by
\begin{equation}
U \;=\; 
\begin{bmatrix}
 c_\omega         & s_\omega e^{iD} \\
-s_\omega e^{-iD} & c_\omega 
\end{bmatrix}
\;=\;
\begin{bmatrix}
1 & 0 \\ 0 & e^{-iD}
\end{bmatrix}
\begin{bmatrix}
 c_\omega & s_\omega \\
-s_\omega & c_\omega 
\end{bmatrix}
\begin{bmatrix}
1 & 0 \\ 0 & e^{iD}
\end{bmatrix}
\;,
\end{equation}
where
\begin{equation}
c_\omega\;=\;\cos\omega\;,\qquad
s_\omega\;=\;\sin\omega\;,\qquad
\tan 2\omega \;=\; \dfrac{2B}{C-A}\;.
\label{tan2omega}
\end{equation}
That is,
\begin{equation}
U^\dagger 
\begin{bmatrix}
A & B\, e^{iD} \\ B\, e^{-iD} & C
\end{bmatrix}
U
\;=\;
\begin{bmatrix}
\Lambda_1 & 0 \\ 0 & \Lambda_2
\end{bmatrix}
\;,
\end{equation}
where
\begin{eqnarray}
\Lambda_1 
& = & \dfrac{A c^2_\omega - C s^2_\omega}{c^2_\omega-s^2_\omega}
\;=\; \dfrac{(A+C)\mp\sqrt{(A-C)^2+4B^2}}{2}\;,\cr
\Lambda_2 
& = & \dfrac{C c^2_\omega - A s^2_\omega}{c^2_\omega-s^2_\omega}
\;=\; \dfrac{(A+C)\pm\sqrt{(A-C)^2+4B^2}}{2}
\;,
\end{eqnarray}
the double signs corresponding to the two possible quadrants for 
$2\omega$ that satisfy Eq.~(\ref{tan2omega}).
In applying the above formula to our problem, care needs to be taken to choose
the correct quadrant and sign combination so that the resulting effective mixing angles
and mass-squared eigenvalues
run smoothly from their vacuum values.

\subsubsection{$\eta=0$ Case, First and Second Rotations}

As mentioned above, we will first approximately diagonalize $H'_0$, and
add on the $\hat{a}\eta M'_\eta$ term afterwards.
Here, we reproduce how the Jacobi method was applied to $H_0'$ 
in Refs.~\cite{Agarwalla:2013tza,Honda:2006hp}.
There, a $(1,2)$ rotation was applied to $H_0'$, followed by a $(2,3)$ rotation,
which was sufficient to approximately diagonalize $H_0'$.

\begin{enumerate}
\item{First Rotation}

\begin{figure}[t]
\subfigure[]{\includegraphics[height=5.1cm]{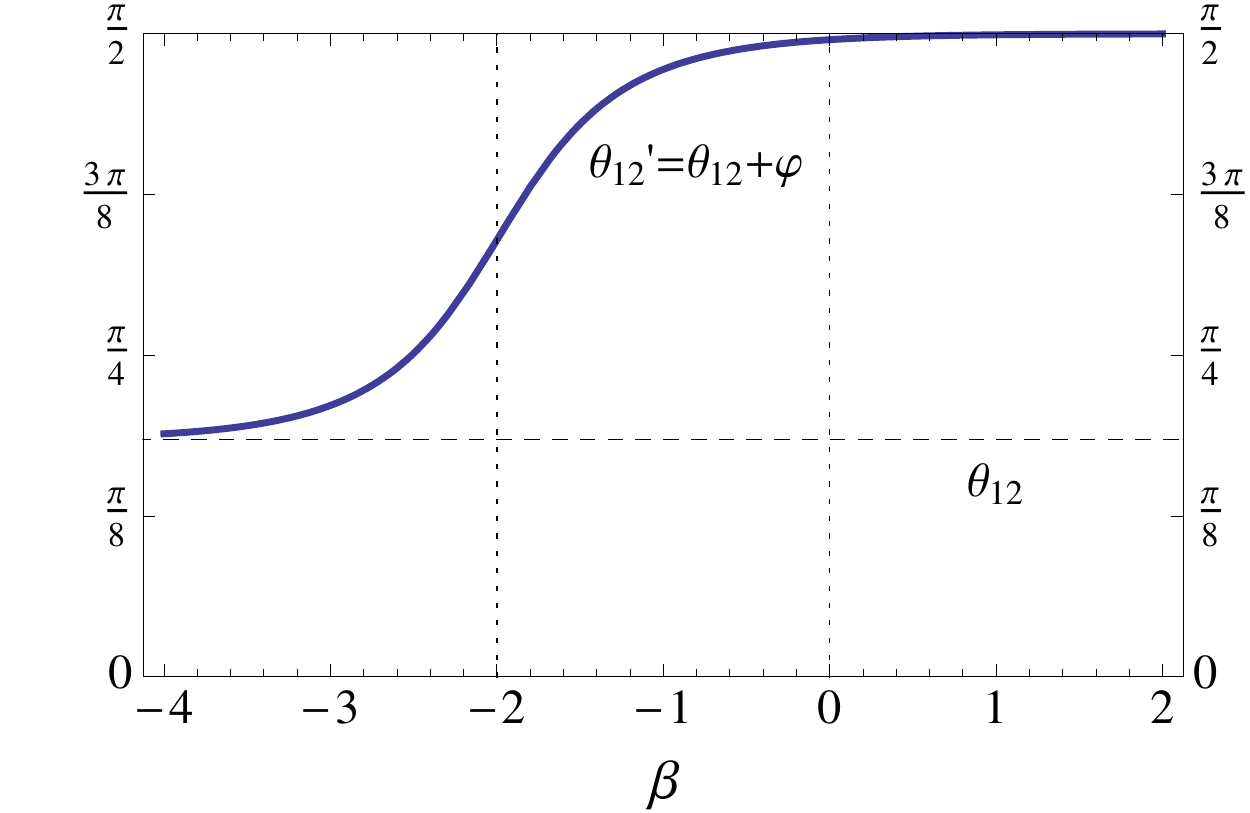}}
\subfigure[]{\includegraphics[height=5cm]{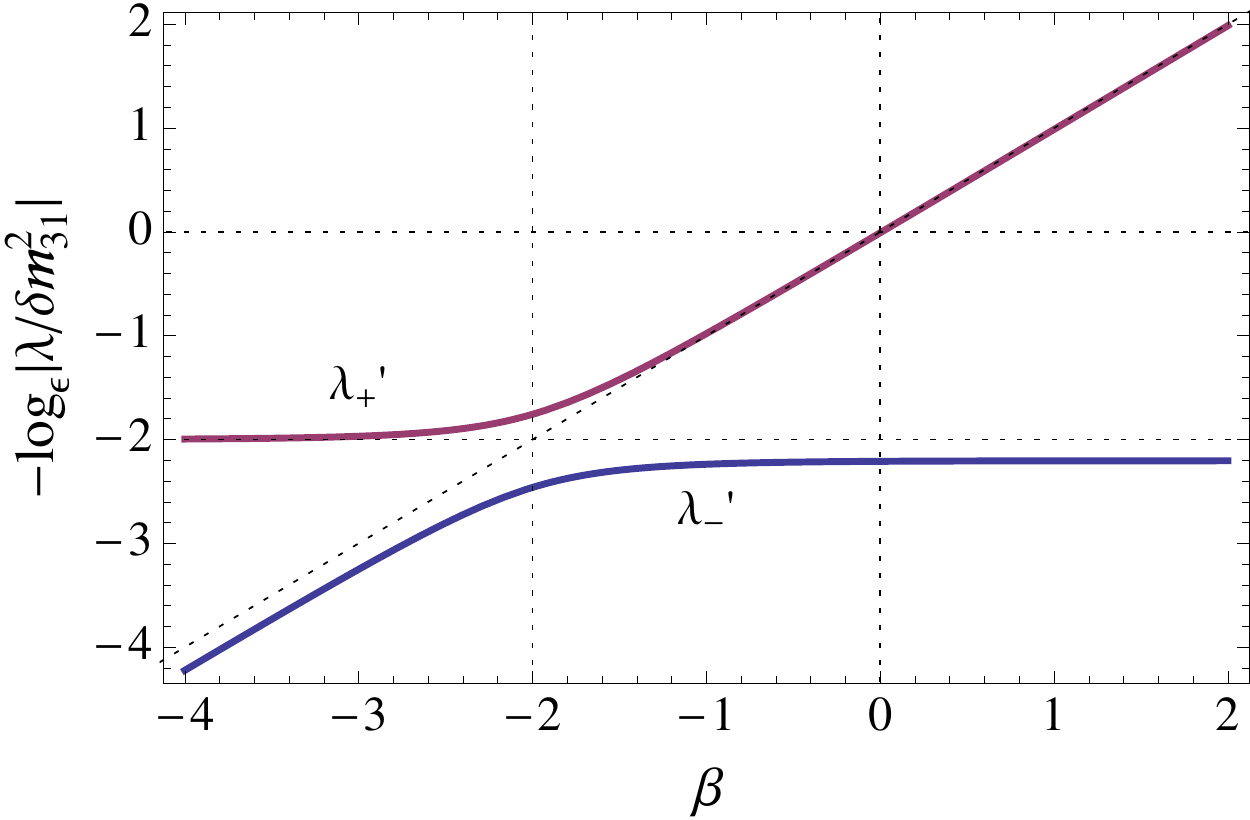}}
\caption{(a) The dependence of $\theta'_{12}$ on 
$\beta=-\log_{\,\epsilon}\left(\hat{a}/|\delta m^2_{31}|\right)$.
(b) The $\beta$-dependence of $\lambda'_{\pm}$.
}
\label{theta12primeplot}
\end{figure}

\begin{figure}[t]
\subfigure[]{\includegraphics[height=5cm]{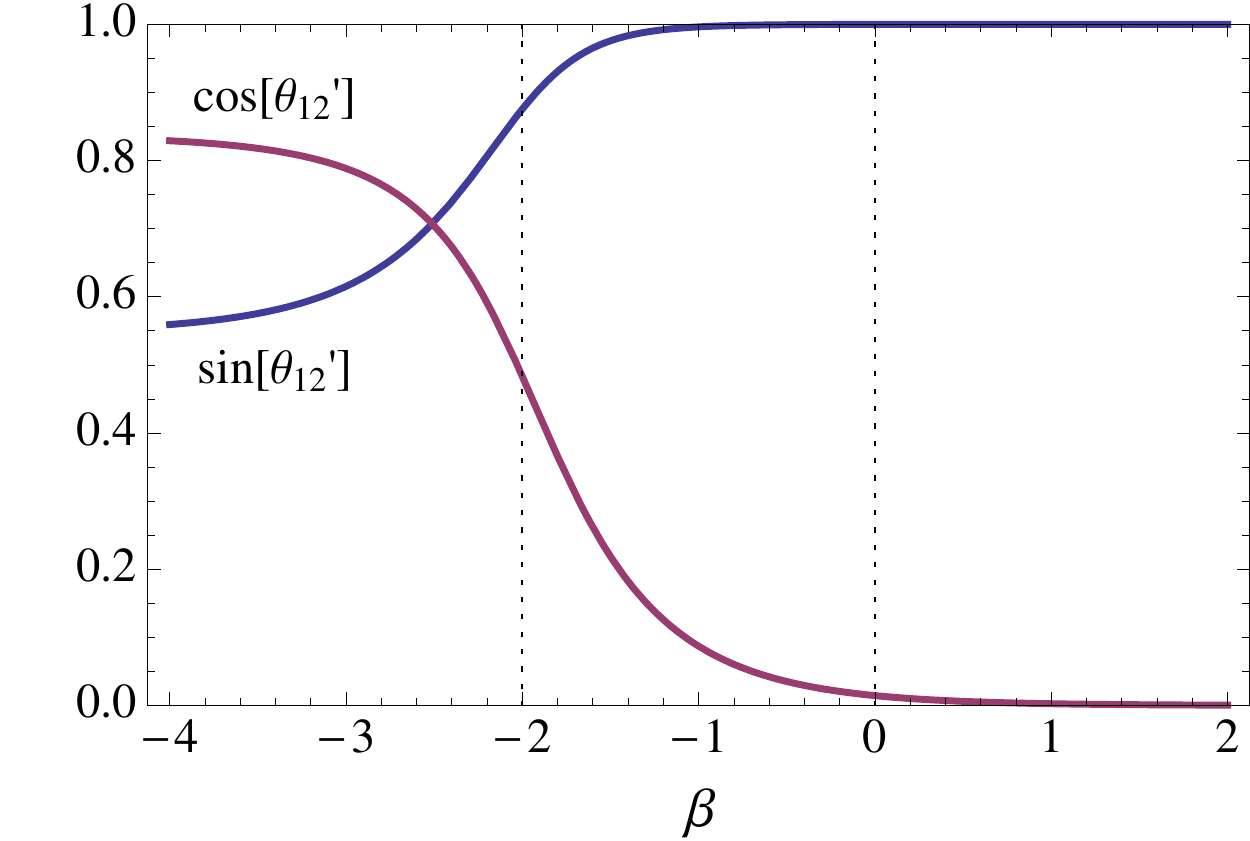}}
\subfigure[]{\includegraphics[height=5cm]{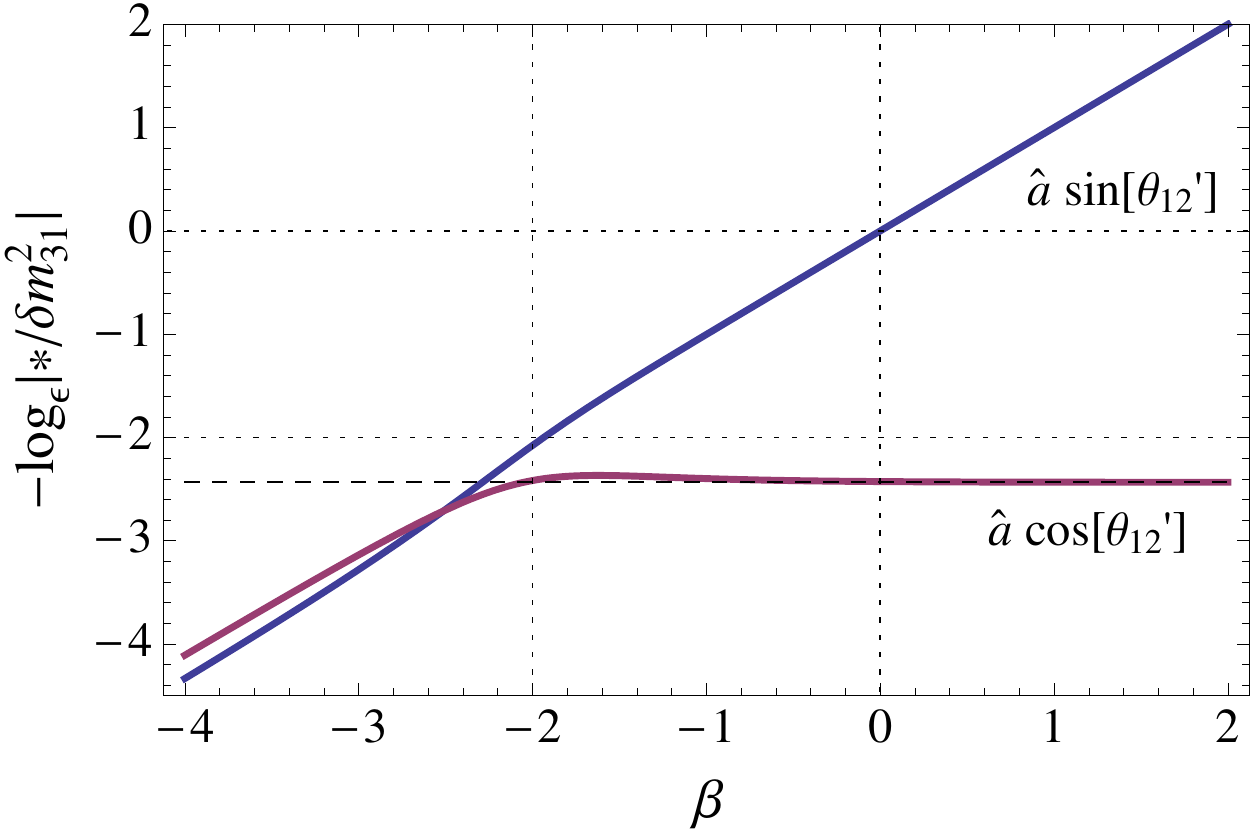}}
\caption{(a) The dependence of $s'_{12}=\sin\theta'_{12}$ and
$c'_{12}=\cos\theta'_{12}$ on 
$\beta=-\log_{\,\epsilon}\left(\hat{a}/|\delta m^2_{31}|\right)$.
(b) The dependence of $\hat{a} s'_{12}$ and $\hat{a} c'_{12}$ on $\beta$.
The values are given in units of $|\delta m^2_{31}|$.
The asymptotic value of $\hat{a} c'_{12}$ is $\delta m^2_{21}s_{12}c_{12}/c_{13}^2
\approx 0.014\;|\delta m^2_{31}| = O(\varepsilon^2|\delta m^2_{31}|)$.
}
\label{s12primec12primeplot}
\end{figure}

Define the matrix $V$ as:
\begin{equation}
V \;=\; 
\begin{bmatrix}
 c_{\varphi} &  s_{\varphi} & 0 \\
	                     -s_{\varphi} &  c_{\varphi} & 0 \\
	                      0 & 0 & 1
\end{bmatrix}
\;,
\label{Vdef}
\end{equation}
where
\begin{equation}
c_{\varphi} \,=\,\cos\varphi\;,\quad
s_{\varphi} \,=\,\sin\varphi\;,\quad
\tan 2\varphi \,\equiv\, 
\dfrac{\hat{a} c_{13}^2\sin2\theta_{12}}{\delta m^2_{21}-\hat{a} c_{13}^2\cos2\theta_{12}}\;,\quad
\left(0\le\varphi<\frac{\pi}{2}-\theta_{12}\right)\;.
\label{phi1def}
\end{equation}
Then,
\begin{equation}
H_0'' 
= V^\dagger H_0' V 
=
\left[ \begin{array}{ccc}
       \lambda'_- & 0 & \hat{a} c'_{12} c_{13}s_{13} \\
       0 & \lambda'_+ & \hat{a} s'_{12} c_{13}s_{13} \\
       \hat{a} c'_{12} c_{13}s_{13} & 
       \hat{a} s'_{12} c_{13}s_{13} &
       \delta m^2_{31} + \hat{a}\, s_{13}^2
       \end{array}
\right] \;,
\label{Hdoubleprimedef}
\end{equation}
where
\begin{equation}
{c}_{12}' = \cos\theta_{12}' \;,\quad
{s}_{12}' = \sin\theta_{12}' \;,\quad
{\theta}_{12}' = \theta_{12} + \varphi \;,
\label{theta12primedef}
\end{equation}
and
\begin{equation}
\lambda'_{\pm}
\;=\; \dfrac{ (\hat{a} c_{13}^2+\delta m^2_{21})
          \pm\sqrt{ (\hat{a} c_{13}^2-\delta m^2_{21})^2 + 4 \hat{a} c_{13}^2 s_{12}^2 \delta m^2_{21} }
        }
        { 2 }\;.
\label{lambdaprimeplusminusdef}
\end{equation}
The angle $\theta_{12}'=\theta_{12}+\varphi$ can be calculated directly
without calculating $\varphi$ via
\begin{equation}
\tan 2\theta'_{12} \;=\; 
\dfrac{\delta m^2_{21} \sin 2\theta_{12}}{\delta m^2_{21}\cos 2\theta_{12} - \hat{a}c^2_{13}}
\;,\qquad
\left(\theta_{12}\le\theta'_{12}\le\dfrac{\pi}{2}\right)\;.
\end{equation}
As $\beta$ is increased beyond $-2$, the $\lambda'_\pm$ asymptote to
\begin{eqnarray}
\lambda'_+ & \rightarrow & \hat{a} c_{13}^2 + \delta m^2_{21} s_{12}^2\;,\cr
\lambda'_- & \rightarrow & \delta m^2_{21}c_{12}^2\;.
\end{eqnarray}
The dependences of $\theta'_{12}$ and $\lambda'_{\pm}$ 
on $\beta$ are plotted in Fig.~\ref{theta12primeplot}.
Note that $\theta'_{12}$ increases monotonically from $\theta_{12}$ to $\pi/2$ with increasing $\beta$.
The $\beta$-dependence of $s'_{12}=\sin\theta'_{12}$ and $c'_{12}=\cos\theta'_{12}$
are shown in Fig.~\ref{s12primec12primeplot}(a).
As $\beta$ is increased beyond $-2$, that is $\hat{a}=\delta m^2_{21}$, $s'_{12}$ grows rapidly to one while
$c'_{12}$ damps quickly to zero.
In fact, the product $\hat{a}c'_{12}$ stops increasing at around $\beta=-2$ and plateaus to
the asymptotic value of $\delta m^2_{21}s_{12}c_{12}/c_{13}^2
\approx 0.014\;|\delta m^2_{31}| = O(\varepsilon^2)|\delta m^2_{31}|$
as shown in Fig.~\ref{s12primec12primeplot}(b).
That is: 
\begin{eqnarray}
\hat{a}s_{12}' & = & |\delta m^2_{31}|\, O(\epsilon^{-\beta})\;,\cr
\hat{a}c_{12}' & = & |\delta m^2_{31}|\, O(\epsilon^{-\min(\beta,-2)})
\;\le\; |\delta m^2_{31}|\,O(\epsilon^2)\;.
\end{eqnarray}
Note also that the scales of $\lambda'_\pm$ are given by
\begin{eqnarray}
\lambda'_+ 
& = & O(\max(\delta m^2_{21},\hat{a}))
\;=\; |\delta m^2_{31}|\, O(\epsilon^{-\max(\beta,-2)})
\;,\cr
\lambda'_- 
& = & O(\min(\delta m^2_{21},\hat{a}))
\;=\; |\delta m^2_{31}|\, O(\epsilon^{-\min(\beta,-2)})
\;.
\end{eqnarray}
%

\item{Second Rotation}

Given that the $(1,3)$ element of $H''_0$, namely
$\hat{a}c'_{12}c_{13}s_{13}$, is at most of order 
$|\delta m^2_{31}|O(\epsilon^3)$ for all $\hat{a}$, 
whereas the $(2,3)$ element $\hat{a}s'_{12}c_{13}s_{13}$ 
will continue to increase with $\hat{a}$, it is
the $(2,3)$ submatrix that needs to be diagonalized next.
The matrix $W$ which diagonalizes 
the $(2,3)$ submatrix of $H_0''$ is
\begin{equation}
W = 
\begin{bmatrix} 1 & 0 & 0 \\
                          0 &  c_{\phi} &  s_{\phi} \\
	                      0 & -s_{\phi} &  c_{\phi} 
\end{bmatrix}
\;,
\label{Wdef}
\end{equation}
where $c_{\phi}=\cos\phi$, $s_{\phi}=\sin\phi$, and
\begin{equation}
\tan 2\phi \;\equiv\; 
\dfrac{2\hat{a} s'_{12}s_{13}c_{13}}
      {\delta m^2_{31}+\hat{a} s_{13}^2 - \lambda'_+} 
\;\approx\;
\dfrac{\hat{a} \sin 2\theta_{13}}
      {(\delta m^2_{31}-\delta m^2_{21} s^2_{12}) -\hat{a}\cos 2\theta_{13}}
\;.
\label{phi2def}
\end{equation}
The angle $\phi$ is in the first quadrant 
when $\delta m^2_{31} > 0$ (normal hierarchy)
and increases from zero to 
$\frac{\pi}{2}-\theta_{13}$ as $\beta$ is increased.
$\phi$ is in the fourth quadrant when $\delta m^2_{31} < 0$ (inverted hierarchy)
and decreases from zero to $-\theta_{13}$ as $\beta$ is increased.
This $\beta$-dependence of $\phi$ is shown in Fig.~\ref{phiplot}(a) for both mass hierarchies.

\begin{figure}[t]
\subfigure[]{\includegraphics[height=5cm]{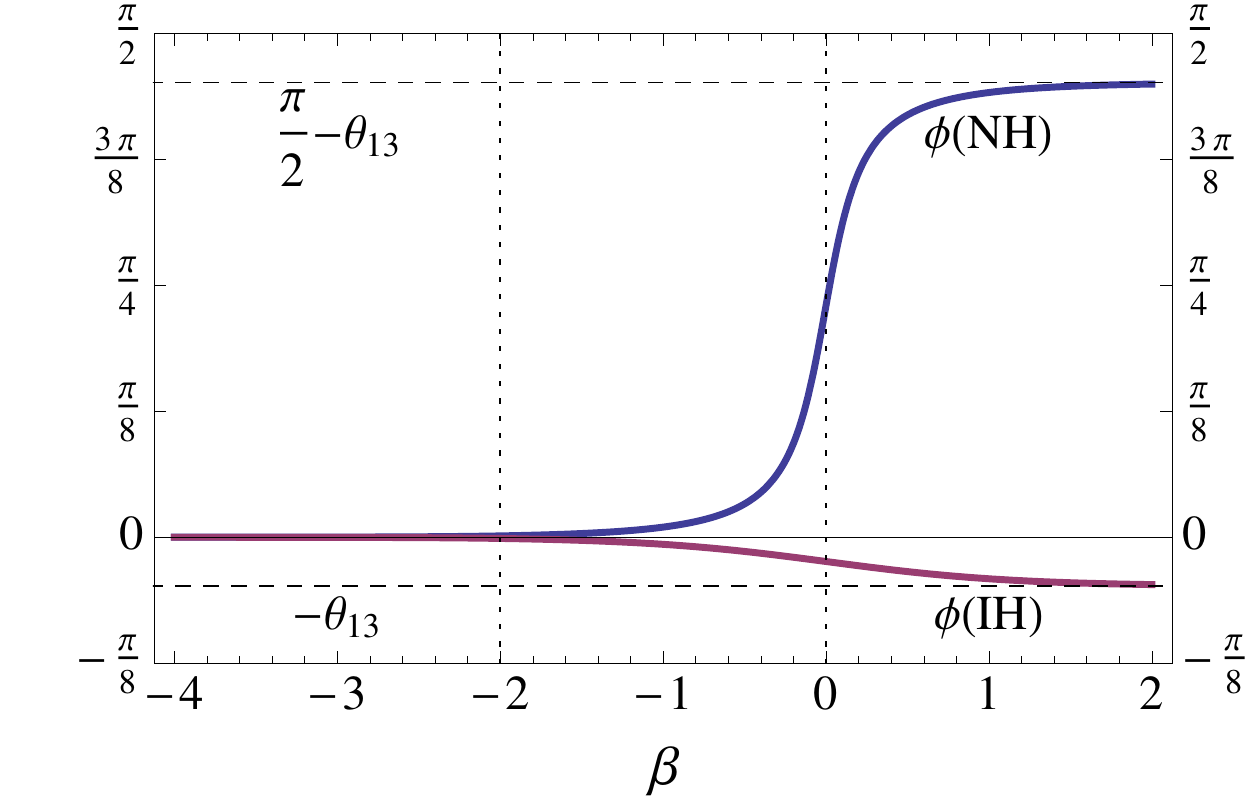}}
\subfigure[]{\includegraphics[height=5cm]{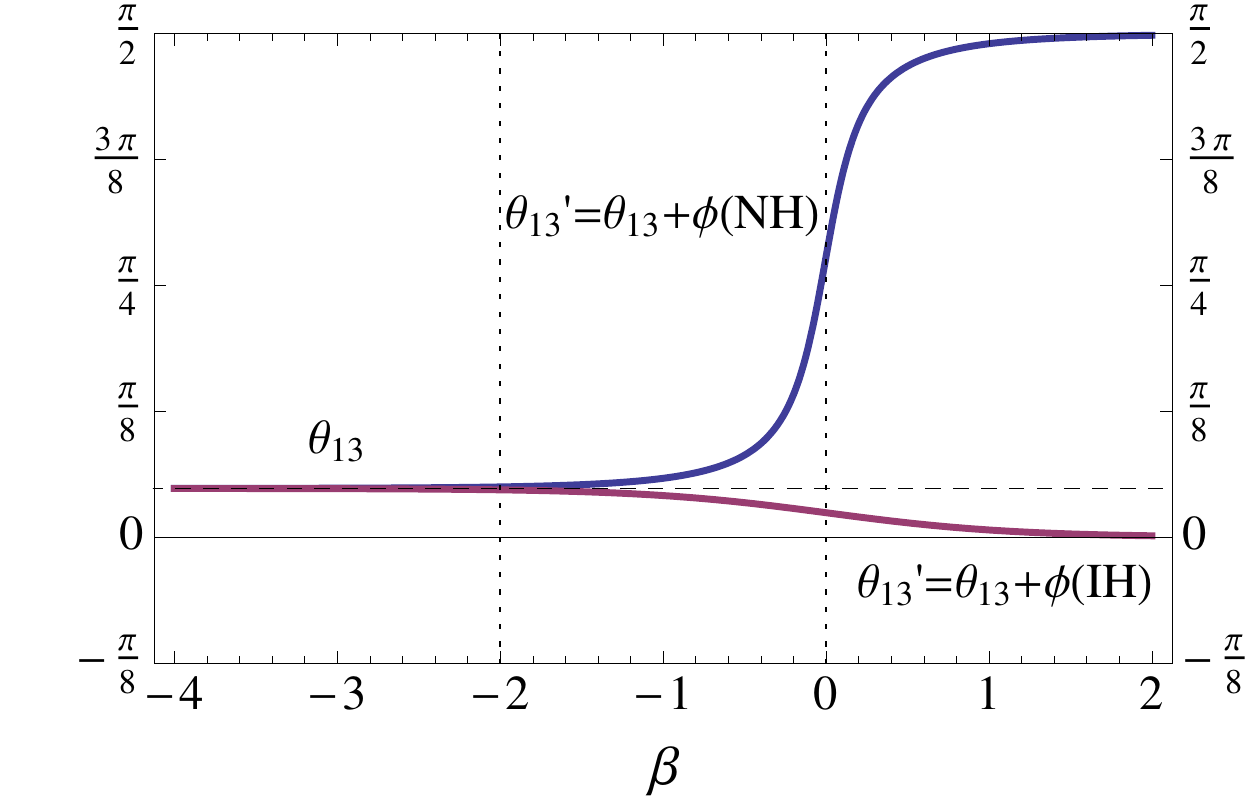}}
\caption{The dependence of (a) $\phi$ and 
(b) $\theta'_{13}=\theta_{13}+\phi$ on 
$\beta=-\log_{\,\epsilon}\left(\hat{a}/|\delta m^2_{31}|\right)$
for the normal (NH) and inverted (IH) mass hierarchies.
}
\label{phiplot}
\end{figure}

\begin{figure}[t]
\subfigure[Normal Hierarchy]{\includegraphics[height=5cm]{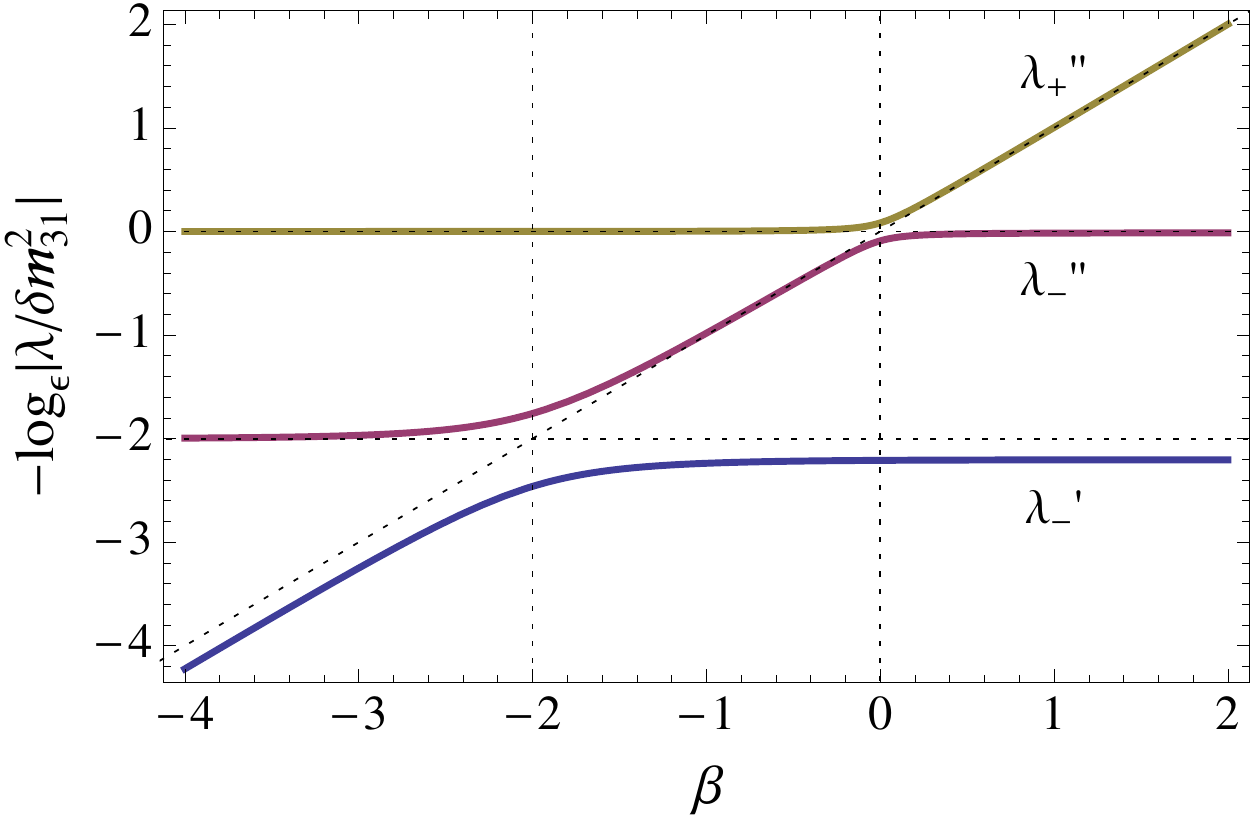}}
\subfigure[Inverted Hierarchy]{\includegraphics[height=5cm]{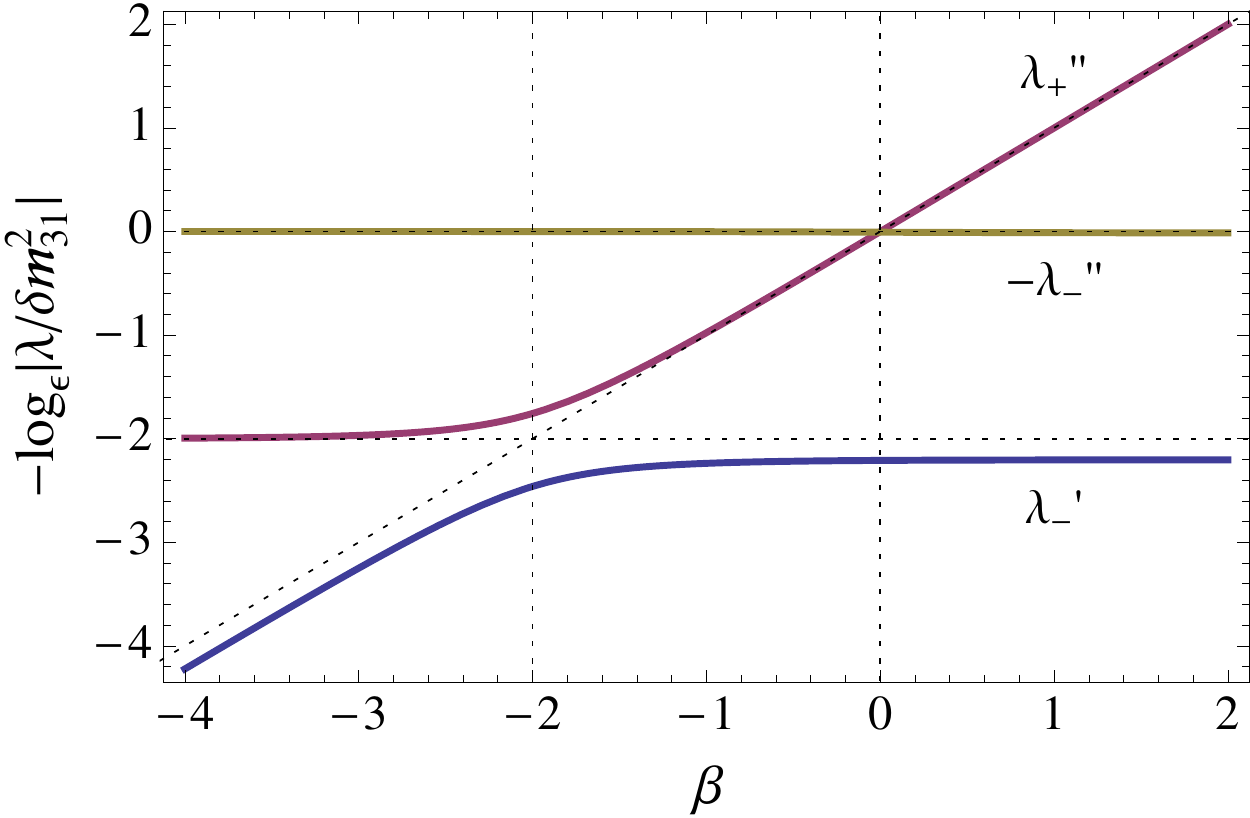}}
\caption{The $\beta$-dependence of $\lambda''_{\pm}$
for the (a) normal and (b) inverted mass hierarchies.
}
\label{lambdadoubleprimepmplot}
\end{figure}

Using $W$, we obtain
\begin{eqnarray}
H_0'''
& = & W^\dagger H_0'' W \cr
& = &
\begin{bmatrix}
\lambda'_- 
& -\hat{a}{c}_{12}' c_{13}s_{13}s_{\phi}  
&  \hat{a}{c}_{12}' c_{13}s_{13}c_{\phi} \\
-\hat{a}{c}_{12}' c_{13}s_{13}s_{\phi} & \lambda''_\mp & 0  \\
 \hat{a}{c}_{12}' c_{13}s_{13}c_{\phi} & 0 & \lambda''_\pm
\end{bmatrix} \;,
\label{Htripleprimedef}
\end{eqnarray}
where the upper signs are for the $\delta m^2_{31}>0$ (normal hierarchy) case and the
lower signs are for the $\delta m^2_{31}<0$ (inverted hierarchy) case, with
\begin{equation}
\lambda''_{\pm} \equiv
   \dfrac{ [ \lambda'_{+} + (\delta m^2_{31}+\hat{a} s_{13}^2) ]
\pm \sqrt{ [ \lambda'_{+} - (\delta m^2_{31}+\hat{a} s_{13}^2) ]^2 
         + 4 ( \hat{a}s'_{12}c_{13}s_{13})^2 }
         }
         { 2 } \;.
\label{lambdadoubleprimeplusminusdef}
\end{equation}
As $\beta$ is increased beyond 0, the $\lambda''_\pm$ asymptote to
\begin{eqnarray}
\lambda''_+ & \rightarrow & \hat{a} + \delta m^2_{31} s_{13}^2 + \delta m^2_{21} s_{12}^2 c_{13}^2 \;,\cr
\lambda''_- & \rightarrow & \delta m^2_{31} c_{13}^2 + \delta m^2_{21} s_{12}^2 s_{13}^2 \;,
\end{eqnarray}
for both mass hierarchies.
Note that $\lambda''_- <0$ for the $\delta m^2_{31}<0$ case.
The $\beta$-dependences of $\lambda''_{\pm}$ are shown in Fig.~\ref{lambdadoubleprimepmplot}.
Order-of-magnitude-wise, we have
\begin{equation}
\begin{array}{lll}
  \lambda''_-  \;=\; |\delta m^2_{31}|\, O(\epsilon^{-\max(\min(\beta,0),-2)})\;,\quad
& \phantom{|}\lambda''_+\phantom{|} \;=\; |\delta m^2_{31}|\, O(\epsilon^{-\max(\beta,0)})\;,\quad 
& \mbox{if $\delta m^2_{31} > 0$}\;, \\
  \lambda''_+  \;=\; |\delta m^2_{31}|\, O(\epsilon^{-\max(\beta,-2)})\;,\quad
&|\lambda''_-| \;=\; |\delta m^2_{31}|\, O(1)\;,\quad 
& \mbox{if $\delta m^2_{31} < 0$}\;.
\end{array}
\end{equation}
In particular, in the range $\beta\agt 0$ we have
\begin{equation}
\begin{array}{lll}
  \lambda''_-  \;=\; |\delta m^2_{31}|\, O(1)\;,\quad
& \phantom{|}\lambda''_+\phantom{|} \;=\; |\delta m^2_{31}|\, O(\epsilon^{-\beta})\;,\qquad 
& \mbox{if $\delta m^2_{31} > 0$}\;, \\
  \lambda''_+  \;=\; |\delta m^2_{31}|\, O(\epsilon^{-\beta})\;,\quad
&|\lambda''_-| \;=\; |\delta m^2_{31}|\, O(1)\;,\qquad 
& \mbox{if $\delta m^2_{31} < 0$}\;.
\end{array}
\end{equation}
For the off-diagonal terms, since 
$\hat{a} c'_{12} = |\delta m^2_{31}| O(\epsilon^2)$, $c_{13} = O(1)$, $s_{13}=O(\epsilon)$, $s_{\phi}=O(1/\epsilon)$, $c_{\phi}=O(\epsilon/1)$,
we have
\begin{eqnarray}
-\hat{a} c'_{12}c_{13}s_{13}s_{\phi} &=& O(\epsilon^3/ \epsilon^4), \nonumber \\
\hat{a} c'_{12}c_{13}s_{13}c_{\phi} &=& O(\epsilon^4/ \epsilon^3).
\end{eqnarray}
Thus, looking at the sizes of the elements of $H'''_0$ in that range we find:
\begin{equation}
H'''_0 \;=\; |\delta m^2_{31}|
\begin{bmatrix}
O(\epsilon^2) & O(\epsilon^3 / \epsilon^4) & O(\epsilon^4 / \epsilon^3) \\
O(\epsilon^3 / \epsilon^4) & O(1/\epsilon^{-\beta})          & 0 \\
O(\epsilon^4 / \epsilon^3) & 0             & O(\epsilon^{-\beta}/1)
\end{bmatrix}
\;,
\end{equation}
where the elements with two entries denote the two
different mass hierarchies,
$O(\mathrm{NH}/\mathrm{IH})$, 
and we can see that further diagonalization only require angles
of order $O(\epsilon^3)$.
Therefore, $H'''_0$ can be considered approximately diagonal.
\end{enumerate}

\subsubsection{$\eta\neq 0$ Case, Third Rotation}

\begin{figure}[t]
\subfigure[Normal Hierarchy]{\includegraphics[height=5cm]{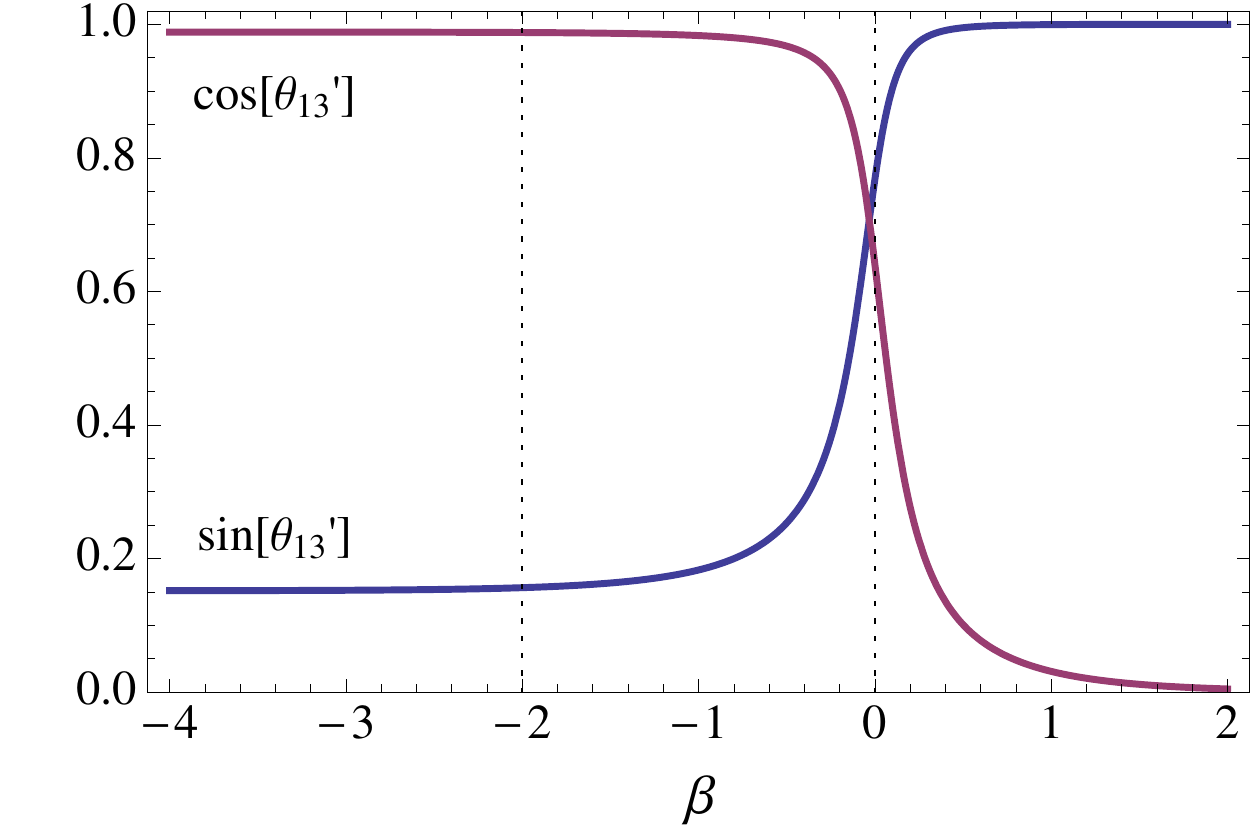}}\quad
\subfigure[Inverted Hierarchy]{\includegraphics[height=5cm]{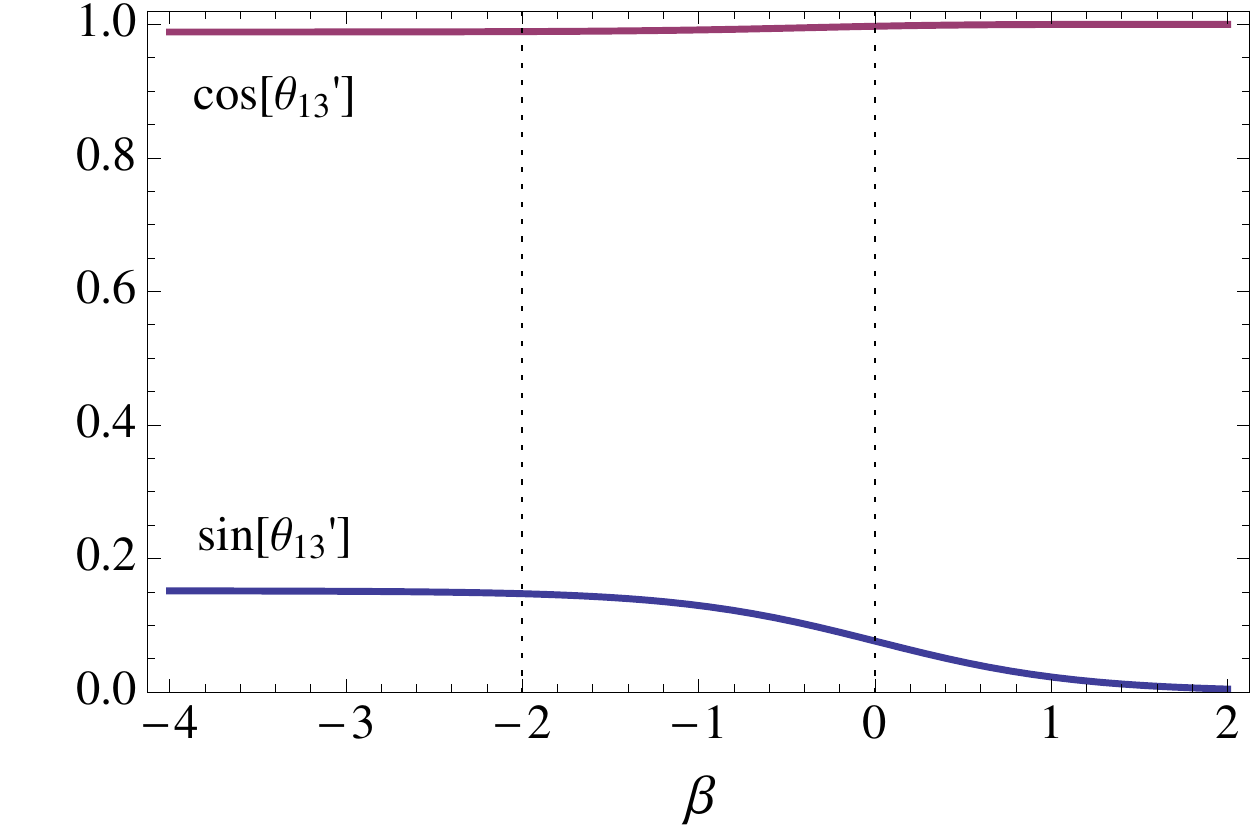}}
\subfigure[Normal Hierarchy]{\includegraphics[height=5cm]{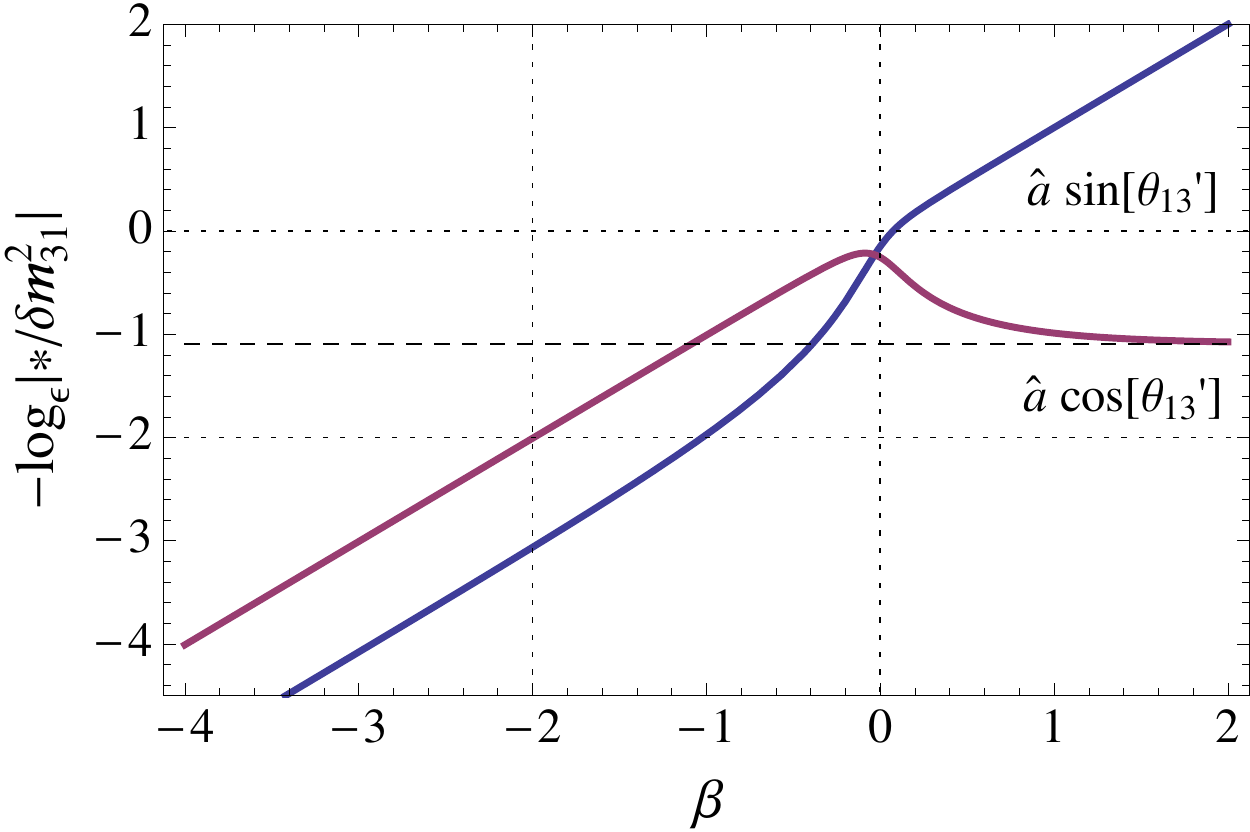}}\quad
\subfigure[Inverted Hierarchy]{\includegraphics[height=5cm]{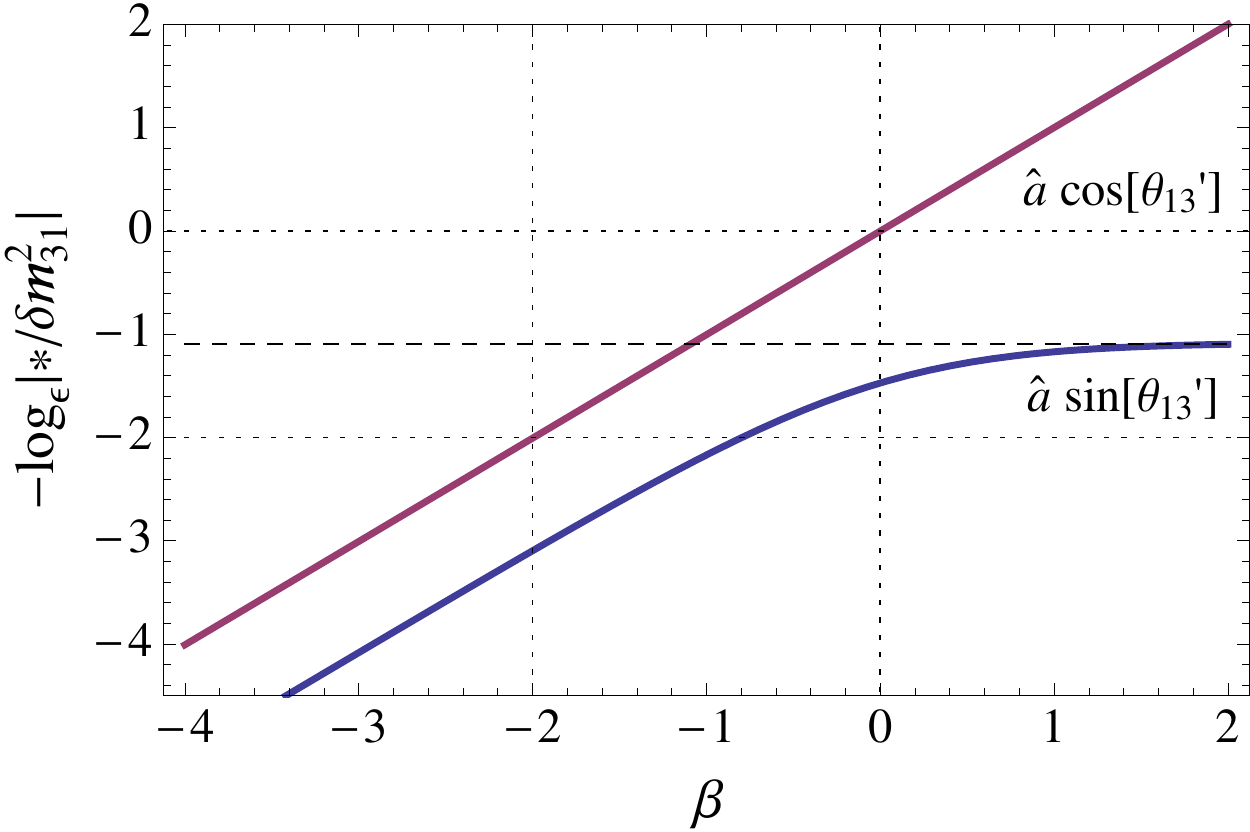}}
\caption{The dependence of $\sin\theta'_{13}$ and
$\cos\theta'_{13}$ on 
$\beta=-\log_{\,\epsilon}\left(\hat{a}/|\delta m^2_{31}|\right)$
for the (a) normal and (b) inverted mass hierarchies.
The dependence of $\hat{a}\sin\theta'_{13}$ and
$\hat{a}\cos\theta'_{13}$ on 
$\beta=-\log_{\,\epsilon}\left(\hat{a}/|\delta m^2_{31}|\right)$
for the (c) normal and (d) inverted mass hierarchies.}
\label{asintheta13primeacostheta13primeplot}
\end{figure}

Let us now consider the $\eta\neq 0$ case.
If we perform the same $(1,2)$ rotation $V$ on 
$H_\eta'=H_0'+\hat{a}\eta M'_\eta$ as we did 
on $H_0'$, the $M'_\eta$ part is transformed to
\begin{eqnarray}
V^\dagger M'_\eta(\theta_{12},\theta_{13},\theta_{23},\delta)V
& = & M'_\eta(\underbrace{\theta_{12}+\varphi}_{\displaystyle =\theta'_{12}},\theta_{13},\theta_{23},\delta)
\cr
& = & M'_\eta(\theta'_{12},\theta_{13},\theta_{23},\delta)
\;.
\end{eqnarray}
Using 
$\theta'_{12}\rightarrow\frac{\pi}{2}$,
$\hat{a}s'_{12}\rightarrow\hat{a}$, and
$\hat{a}c'_{12}\rightarrow O(\epsilon^2)|\delta m^2_{31}|$ as $\beta$ is increased beyond $-2$,
we can approximate
\begin{eqnarray}
\lefteqn{\hat{a}\eta M'_\eta\left(\theta'_{12},\theta_{13},\theta_{23},\delta\right)} \cr
& \approx & \hat{a}\eta M'_\eta\left(\frac{\pi}{2},\theta_{13},\theta_{23},\delta\right)
\cr
& = & \hat{a}\eta
\begin{bmatrix}
\cos(2\theta_{23}) & e^{i\delta} s_{13}\sin(2\theta_{23}) & -e^{i\delta} c_{13}\sin(2\theta_{23}) \\
e^{-i\delta} s_{13}\sin(2\theta_{23}) & -s_{13}^2\cos(2\theta_{23}) & s_{13}c_{13}\cos(2\theta_{23}) \\
-e^{-i\delta} c_{13}\sin(2\theta_{23}) & s_{13}c_{13}\cos(2\theta_{23}) & -c_{13}^2\cos(2\theta_{23})
\end{bmatrix}
\;.
\end{eqnarray}
Performing the $(2,3)$ rotation $W$ next, we find:
\begin{eqnarray}
\lefteqn{W^\dagger M'_\eta\left(\frac{\pi}{2},\theta_{13},\theta_{23},\delta\right) W}
\cr
& = & M'_\eta\Bigl(\frac{\pi}{2},\underbrace{\theta_{13}+\phi}_{\displaystyle =\theta'_{13}},\theta_{23},\delta\Bigr) \cr
& = & M'_\eta\left(\frac{\pi}{2},\theta'_{13},\theta_{23},\delta\right)
\cr
& = &
\begin{bmatrix}
\cos(2\theta_{23}) & e^{i\delta} s_{13}'\sin(2\theta_{23}) & -e^{i\delta} c_{13}'\sin(2\theta_{23}) \\
e^{-i\delta} s_{13}'\sin(2\theta_{23}) & -s_{13}^{\prime 2}\cos(2\theta_{23}) & s_{13}'c_{13}'\cos(2\theta_{23}) \\
-e^{-i\delta} c_{13}'\sin(2\theta_{23}) & s_{13}'c_{13}'\cos(2\theta_{23}) & -c_{13}^{\prime 2}\cos(2\theta_{23})
\end{bmatrix}
\;,
\end{eqnarray}
where $s'_{13}=\sin\theta_{13}'$ and $c'_{13}=\cos\theta_{13}'$.
The angle $\theta'_{13}=\theta_{13}+\phi$ can be calculated directly
without the need to calculate $\phi$ using
\begin{equation}
\tan 2\theta'_{13} \;=\;
\dfrac{(\delta m^2_{31}-\delta m^2_{21}s_{12}^2)\sin 2\theta_{13}}
      {(\delta m^2_{31}-\delta m^2_{21}s_{12}^2)\cos 2\theta_{13} - \hat{a}}
\;,
\end{equation}
and its $\beta$-dependence is shown in Fig.~\ref{phiplot}(b).
As can be seen, $\theta_{13}'$ increases rapidly to $\pi/2$ when $\delta m^2_{31}>0$, 
while damping quickly to zero when $\delta m^2_{31}<0$, once $\beta$ is increased above zero.
Consequently, $\hat{a}\cos\theta'_{13}$ for the $\delta m^2_{31}>0$ case,
and $\hat{a}\sin\theta'_{13}$ for the $\delta m^2_{31}<0$ case plateau to
$c_{13}s_{13}(1-\epsilon^2 s_{12}^2)|\delta m^2_{31}| = O(\epsilon) |\delta m^2_{31}|$
as $\beta$ is increased as shown in Fig.~\ref{asintheta13primeacostheta13primeplot}.
Note that in the $\delta m^2_{31}>0$ case, $\hat{a}\cos\theta'_{13}$ increases to
$O(1)|\delta m^2_{31}|$ in the vicinity of $\beta=0$ before
plateauing to $O(\epsilon) |\delta m^2_{31}|$.  This will cause a slight problem in our
approximation later. We now look at the normal and inverted mass hierarchy 
cases separately.

\begin{enumerate}
\item \textbf{$\bm{\delta m^2_{31}>0}$ Case}

For the $\delta m^2_{31}>0$ case 
$\hat{a}c'_{13}\rightarrow O(\epsilon)|\delta m^2_{31}|$ as $\beta$ is increased
beyond 0.
Therefore, we can approximate
\begin{eqnarray}
H_\eta''' 
& = & W^\dagger V^\dagger H_\eta' V W \cr
& \approx & H_0''' + \hat{a}\eta M'_\eta\left(\frac{\pi}{2},\theta'_{13},\theta_{23},\delta\right) \cr
& \approx &
\begin{bmatrix}
\lambda'_- + \hat{a}\eta\cos(2\theta_{23})
& \hat{a} \eta e^{i\delta} s_{13}' \sin(2\theta_{23})  
&  0 \\
\hat{a}\eta e^{-i\delta} s_{13}' \sin(2\theta_{23}) & \lambda''_- - \hat{a}\eta s_{13}^{\prime 2}\cos(2\theta_{23}) & 0  \\
0 & 0 & \lambda''_+
\end{bmatrix} \;,
\end{eqnarray}
where we have dropped off-diagonal terms of 
order $O(\epsilon^3)|\delta m^2_{31}|$ or smaller.
(This approximation breaks down in the vicinity of $\beta=0$ 
where both $\hat{a}s'_{13}$ and $\hat{a}c'_{13}$ 
are of order $O(1)|\delta m^2_{31}|$.) 
Define the matrix $X$ as
\begin{equation}
X = 
\left[ \begin{array}{ccc} c_\chi              & s_\chi e^{i\delta} & 0 \\
                         -s_\chi e^{-i\delta} & c_\chi             & 0 \\
                          0                   & 0                  & 1
       \end{array}
\right] \;,
\label{Xdef}
\end{equation}
where
$c_\chi = \cos\chi$,
$s_\chi = \sin\chi$,
and
\begin{eqnarray}
\quad
\tan 2\chi 
& \equiv &
\dfrac{ 2\hat{a}\eta s_{13}'\sin(2\theta_{23}) }
       { (\lambda''_{-} - \lambda'_{-}) - \hat{a}\eta(1+s_{13}^{\prime 2})\cos(2\theta_{23})}
\cr
& \approx &
\dfrac{ 2\hat{a}\eta\sin(2\theta_{23}) }
      { [\delta m^2_{31}c_{13}^2-\delta m^2_{21}(c_{12}^2-s_{12}^2 s_{13}^2)] - 2\hat{a}\eta\cos(2\theta_{23})}
\;.
\label{chidef}
\end{eqnarray}
Note that
\begin{equation}
0\;\le\;\chi\;<\;\dfrac{\pi}{2}-\theta_{23}
\quad\mbox{for $\eta>0$,}\quad
-\theta_{23}\;<\;\chi\;\le\;0 
\quad\mbox{for $\eta<0$}\;.
\label{chirange}
\end{equation}
The $\beta$-dependence of $\chi$ is shown in 
Fig.~\ref{chiplot} for several values of $\eta$,
both positive (Fig.~\ref{chiplot}(a)) and 
negative (Fig.~\ref{chiplot}(b)).

\begin{figure}[t]
\subfigure[$\eta>0$]{\includegraphics[height=5cm]{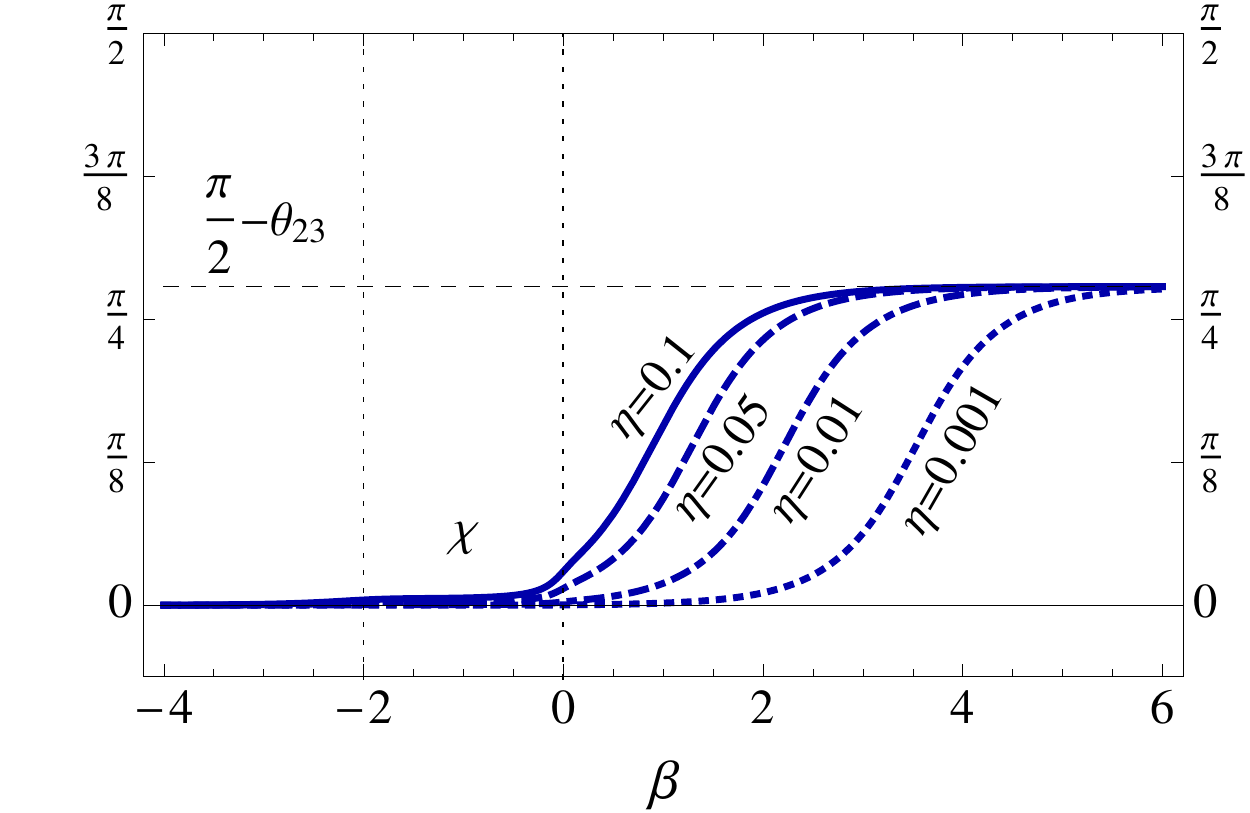}}
\subfigure[$\eta<0$]{\includegraphics[height=5cm]{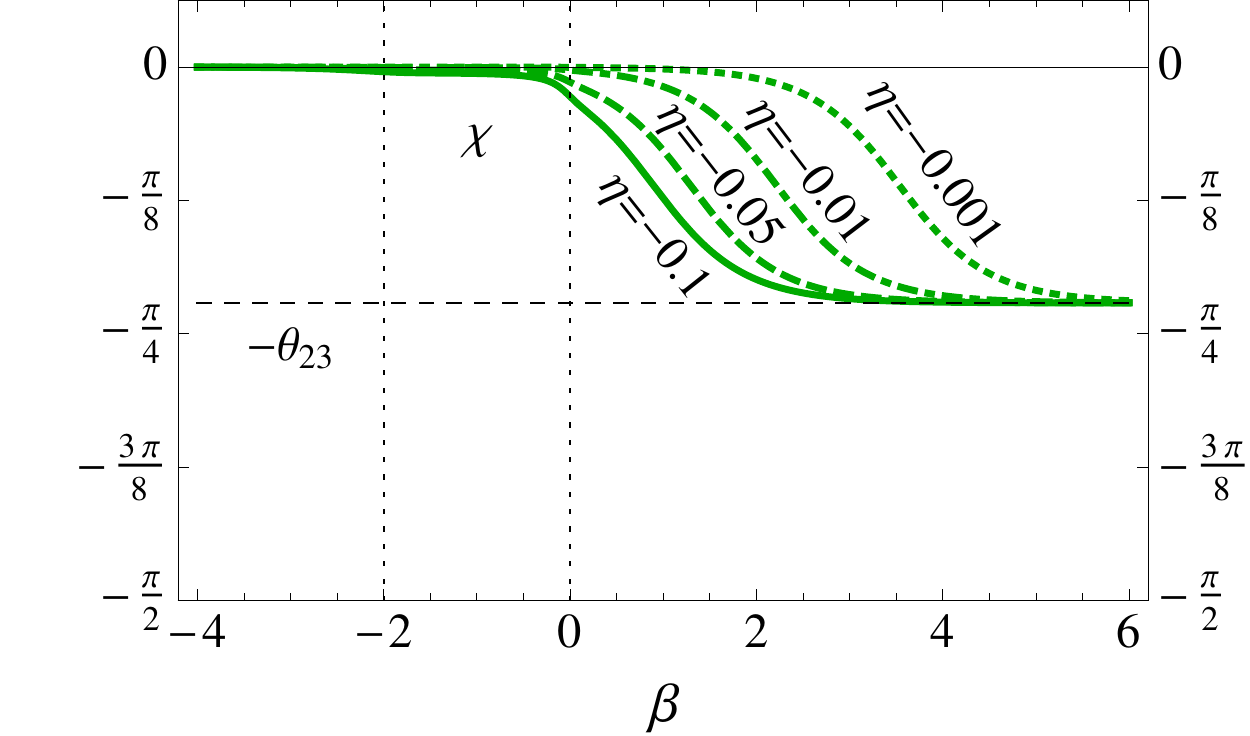}}
\subfigure[$\eta>0$]{\includegraphics[height=5cm]{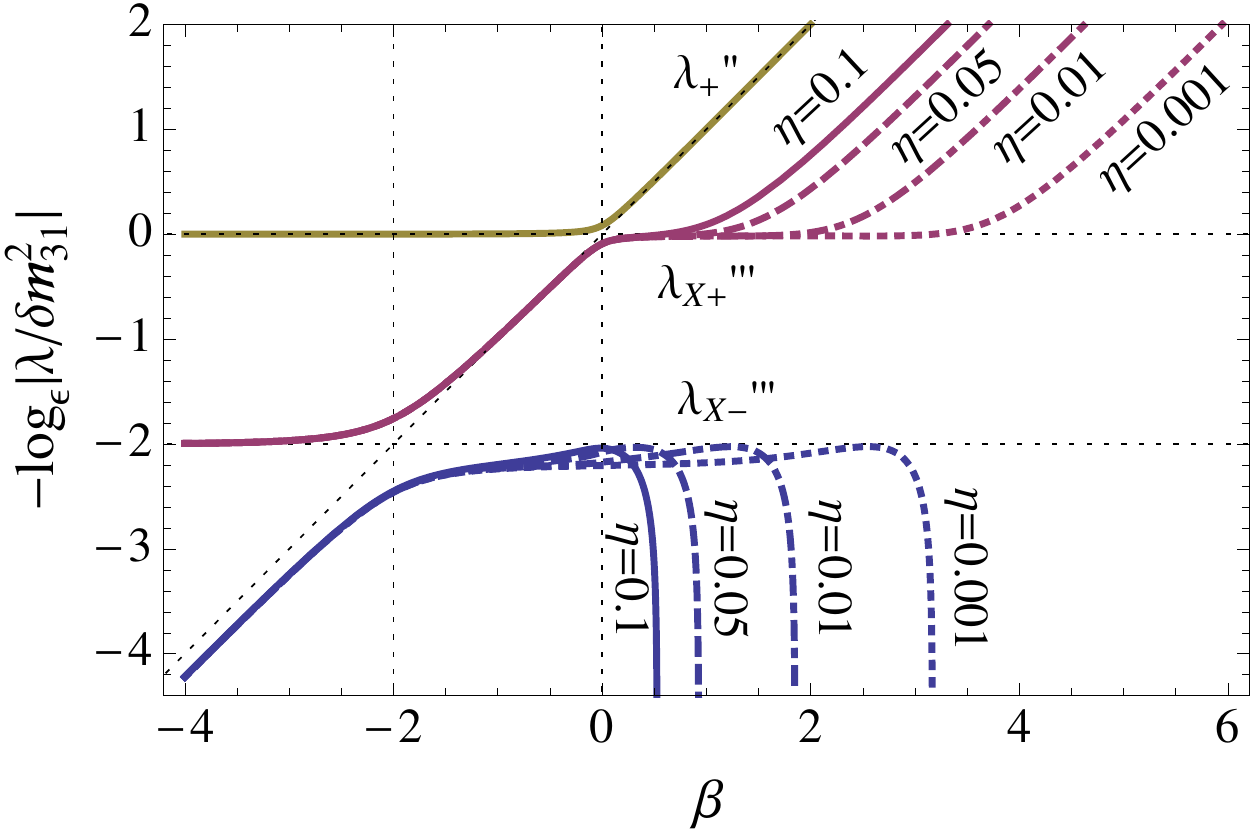}}\qquad
\subfigure[$\eta<0$]{\includegraphics[height=5cm]{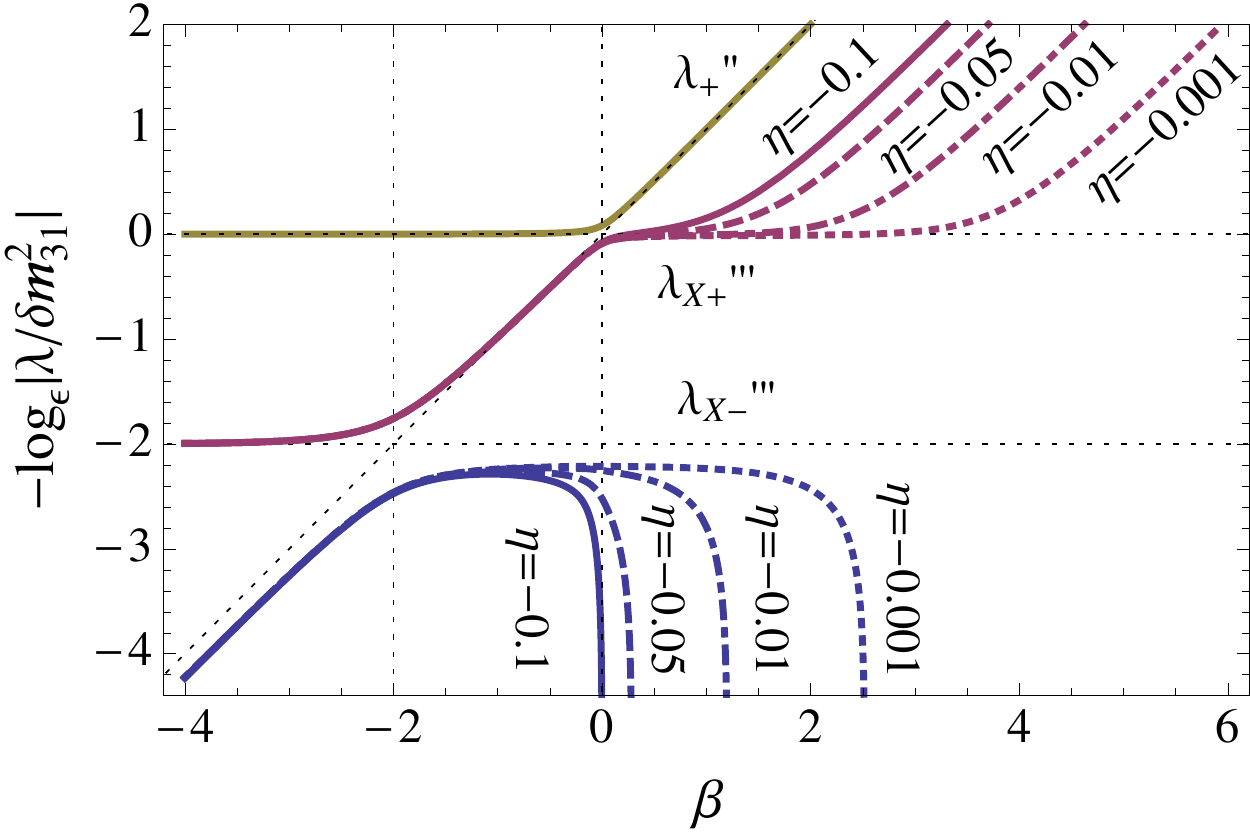}}
\subfigure[$\eta>0$]{\includegraphics[height=5cm]{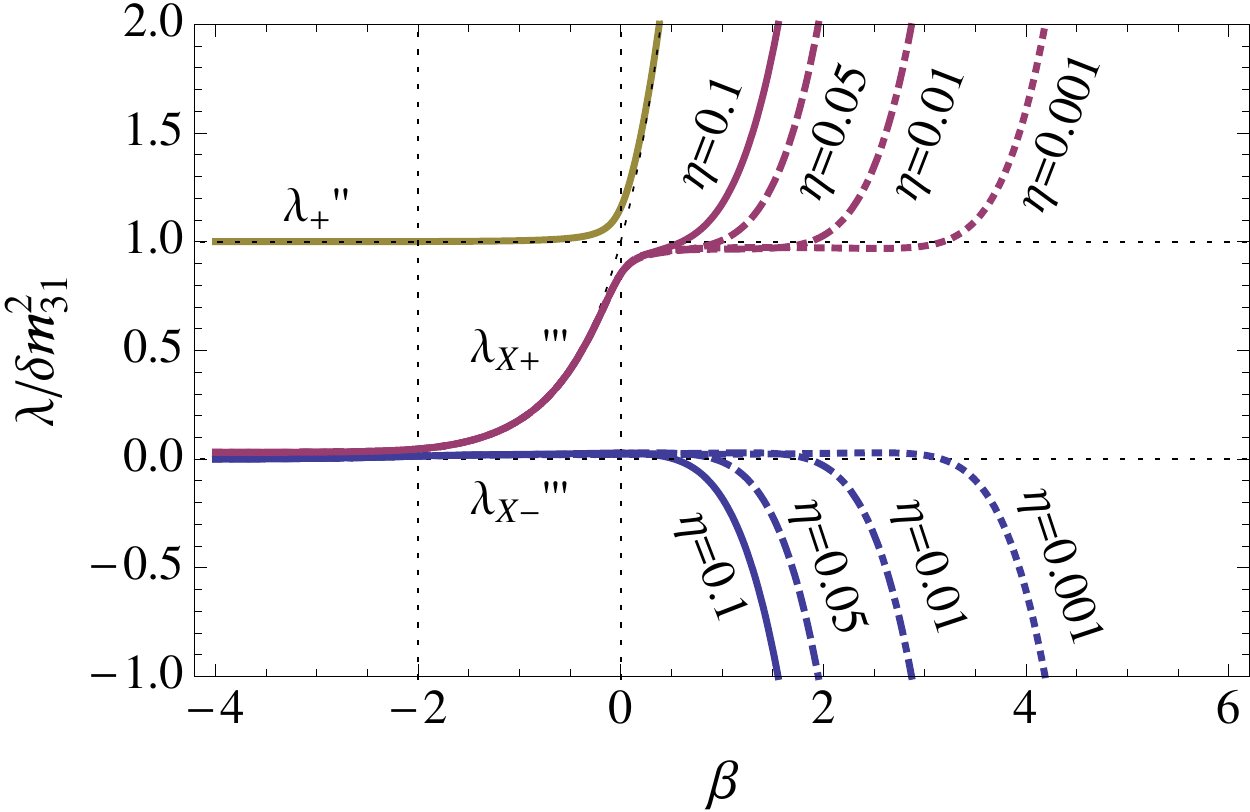}}\quad
\subfigure[$\eta<0$]{\includegraphics[height=5cm]{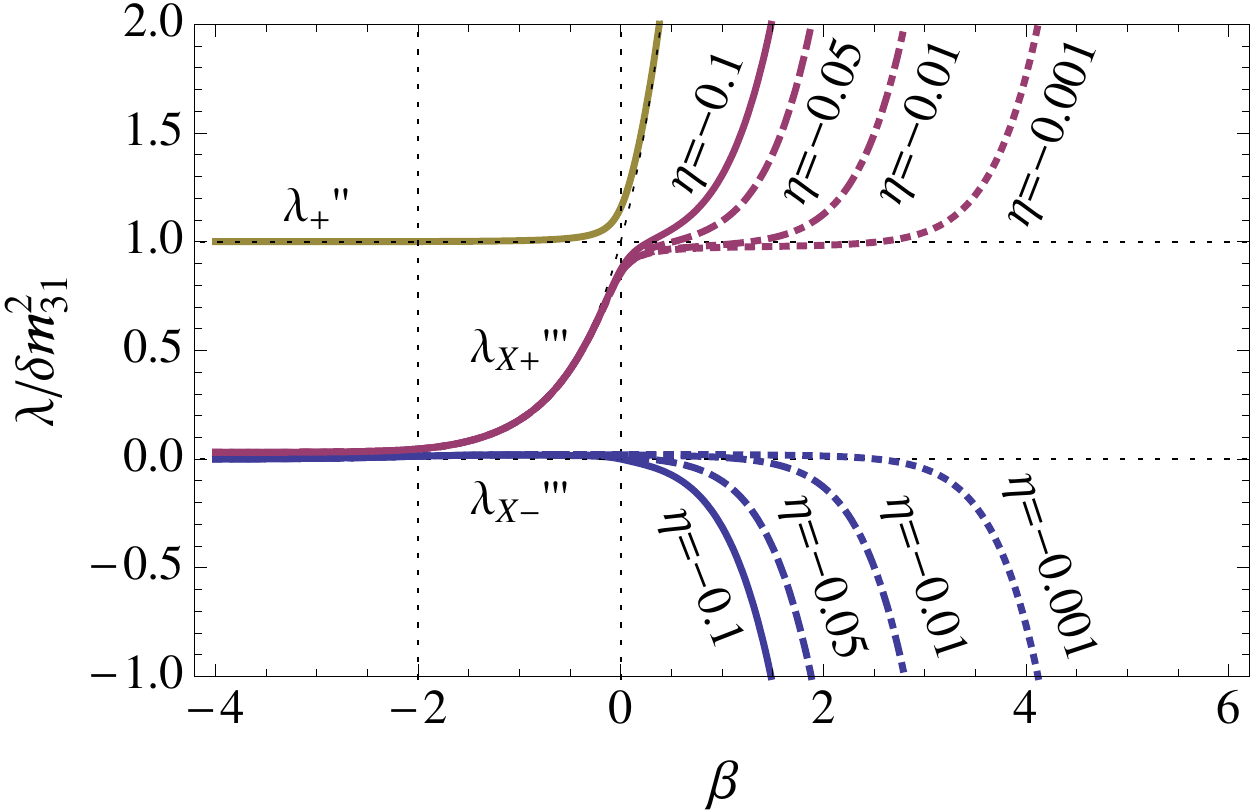}}
\caption{The $\beta$-dependence of $\chi$ and $\lambda'''_{X\pm}$
for several values of $\eta$ with $s_{23}^2=0.4$.}
\label{chiplot}
\end{figure}

\noindent
Using $X$, we find
\begin{equation}
H''''_{\eta+} 
\;=\; X^\dagger H'''_\eta X 
\;\approx\;
\begin{bmatrix}
\lambda'''_{X-} & 0 & 0 \\
0 & \lambda'''_{X+} & 0 \\
0 & 0 & \lambda''_+
\end{bmatrix}
\;,
\end{equation}
where
\begin{eqnarray}
\lefteqn{\lambda'''_{X\pm}} \cr
& \equiv &
   \dfrac{ ( \lambda''_{-} + \lambda'_{-} + \hat{a}\eta c_{13}^{\prime 2}\cos 2\theta_{23} )
\pm \sqrt{ [ \lambda''_{-} - \lambda'_{-} - \hat{a}\eta(1+s_{13}^{\prime 2})\cos 2\theta_{23} ]^2
         + 4( \hat{a}\eta s_{13}' \sin 2\theta_{23} )^2 }
         }
         { 2 } \;.
\cr & &
\label{lambdatripleprimeXplusminusdef}
\end{eqnarray}
Thus, $H_{\eta +}''''$ is approximately diagonal.
The asymptotic forms of $\lambda_{X\pm}'''$ at $\beta\gg 0$ are
\begin{eqnarray}
\lambda'''_{X+} & \rightarrow & \phantom{-}\hat{a}|\eta| +
\left\{
\begin{array}{ll}
 \delta m^2_{31}c_{13}^2 s_{23}^2
+\delta m^2_{21}(c_{12}^2 c_{23}^2 + s_{12}^2 s_{13}^2 s_{23}^2)
\quad & \mbox{for $\eta>0$} \\
 \delta m^2_{31}c_{13}^2 c_{23}^2
+\delta m^2_{21}(c_{12}^2 s_{23}^2 + s_{12}^2 s_{13}^2 c_{23}^2)
& \mbox{for $\eta<0$}
\end{array}
\right.
\cr
\lambda'''_{X-} & \rightarrow & -\hat{a}|\eta| +
\left\{
\begin{array}{ll}
 \delta m^2_{31}c_{13}^2 c_{23}^2
+\delta m^2_{21}(c_{12}^2 s_{23}^2 + s_{12}^2 s_{13}^2 c_{23}^2)
\quad & \mbox{for $\eta>0$} \\
 \delta m^2_{31}c_{13}^2 s_{23}^2
+\delta m^2_{21}(c_{12}^2 c_{23}^2 + s_{12}^2 s_{13}^2 s_{23}^2)
& \mbox{for $\eta<0$}
\end{array}
\right.
\end{eqnarray}
The $\beta$-dependence of $\lambda'''_{X\pm}$ are shown in Figs.~\ref{chiplot}(c) to \ref{chiplot}(f).

\item $\bm{\delta m^2_{31}<0}$ Case

For the $\delta m^2_{31}<0$ case we have
$\hat{a}s'_{13}\rightarrow O(\epsilon)|\delta m^2_{31}|$ as $\beta$ is increased beyond 0.
Therefore, we can approximate
\begin{eqnarray}
H_\eta''' 
& = & W^\dagger V^\dagger H_\eta' V W \cr
& \approx & H_0''' + \hat{a}\eta M'_\eta\left(\frac{\pi}{2},\theta'_{13},\theta_{23},\delta\right) \cr
& \approx &
\begin{bmatrix}
\lambda'_- + \hat{a}\eta\cos(2\theta_{23}) & 0
& -\hat{a}\eta e^{i\delta}c'_{13}\sin(2\theta_{23}) \\
0 & \lambda''_+ & 0 \\
-\hat{a}\eta e^{-i\delta}c_{13}'\sin(2\theta_{23}) & 0 & \lambda''_- - \hat{a}\eta c_{13}^{\prime 2}\cos(2\theta_{23})
\end{bmatrix} \;,
\end{eqnarray}
where we have dropped off-diagonal terms of order $O(\epsilon^3)|\delta m^2_{31}|$ or smaller.
Unlike the $\delta m^2_{31}>0$ case, this approximation is valid in the
vicinity of $\beta=0$ since $\hat{a}s'_{13}$ never exceeds $O(\epsilon)|\delta m^2_{31}|$
for all $\hat{a}$.

Define the matrix $Y$ as
\begin{equation}
Y \;=\; 
\left[ \begin{array}{ccc} c_\psi              & 0 & -s_\psi e^{i\delta} \\
                          0                   & 1 & 0      \\
                          s_\psi e^{-i\delta} & 0 &  c_\psi
       \end{array}
\right] \;,
\label{Ydef}
\end{equation}
where $c_\psi = \cos\psi$, $s_\psi = \sin\psi$, and
\begin{eqnarray}
\tan 2\psi 
& \equiv &
 \dfrac{ 2\hat{a}\eta c'_{13}\sin(2\theta_{23}) }
       { (\lambda''_{-} - \lambda'_{-}) - \hat{a}\eta(1+c_{13}^{\prime 2})\cos(2\theta_{23}) }
\cr
& \approx & 
-\dfrac{ 2\hat{a}\eta\sin(2\theta_{23}) }
{ [|\delta m^2_{31}| c_{13}^2 + \delta m^2_{21}(c_{12}^2-s_{12}^2 s_{13}^2)] + 2\hat{a}\eta\cos(2\theta_{23})}
\;.
\label{psidef}
\end{eqnarray}
Note that
\begin{equation}
-\theta_{23}\;<\;\psi\;\le\;0 
\quad\mbox{for $\eta<0$}\;,
\quad
0\;\le\;\psi\;<\;\dfrac{\pi}{2}-\theta_{23}
\quad\mbox{for $\eta>0$}
\;.
\label{psirange}
\end{equation}
Comparing Eq.~(\ref{chidef}) and Eq.~(\ref{psidef}), we can infer that
$\psi(\eta)\approx\chi(-\eta)$, the small difference due to the $\delta m^2_{21}$ term
in the denominator of the expressions for $\tan 2\chi$ and $\tan 2\psi$.
This can be seen in  Fig.~\ref{psiplot} where the $\beta$-dependence of $\psi$ is shown 
for several values of $\eta$, both positive (Fig.~\ref{psiplot}(a)) 
and negative (Fig.~\ref{psiplot}(b)).

\begin{figure}[t]
\subfigure[$\eta>0$]{\includegraphics[height=5cm]{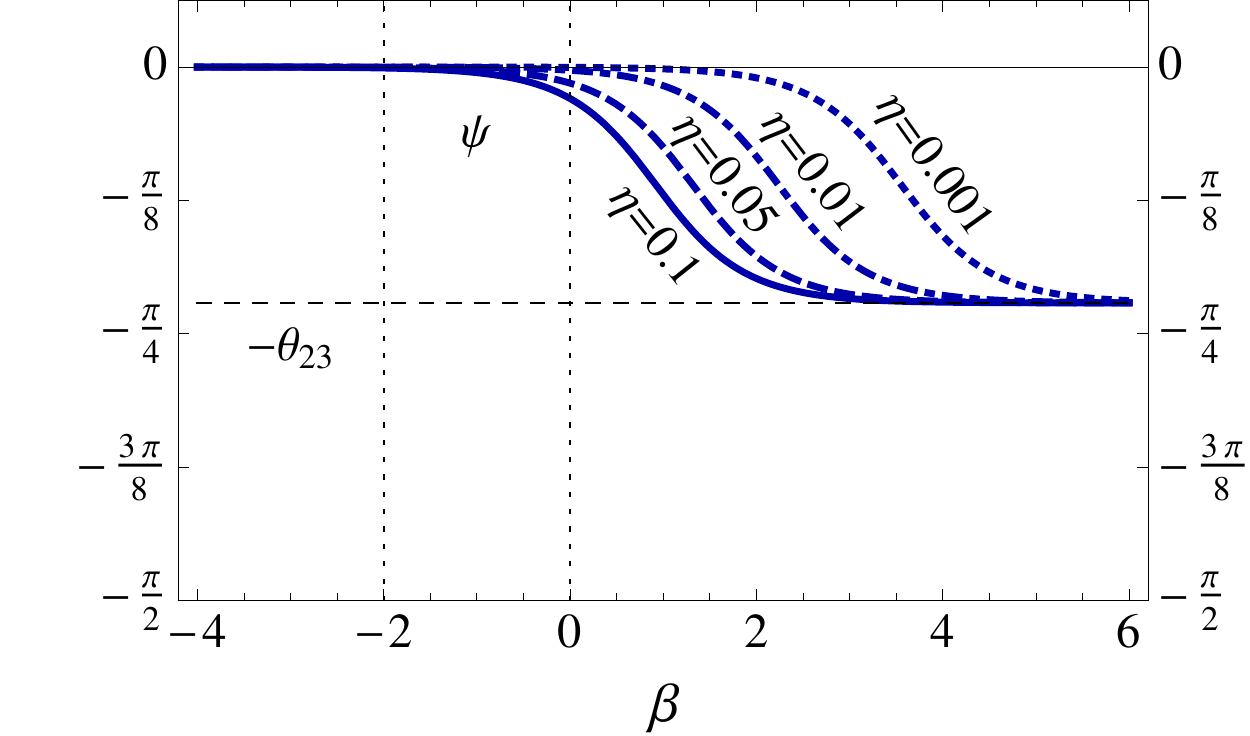}}
\subfigure[$\eta<0$]{\includegraphics[height=5cm]{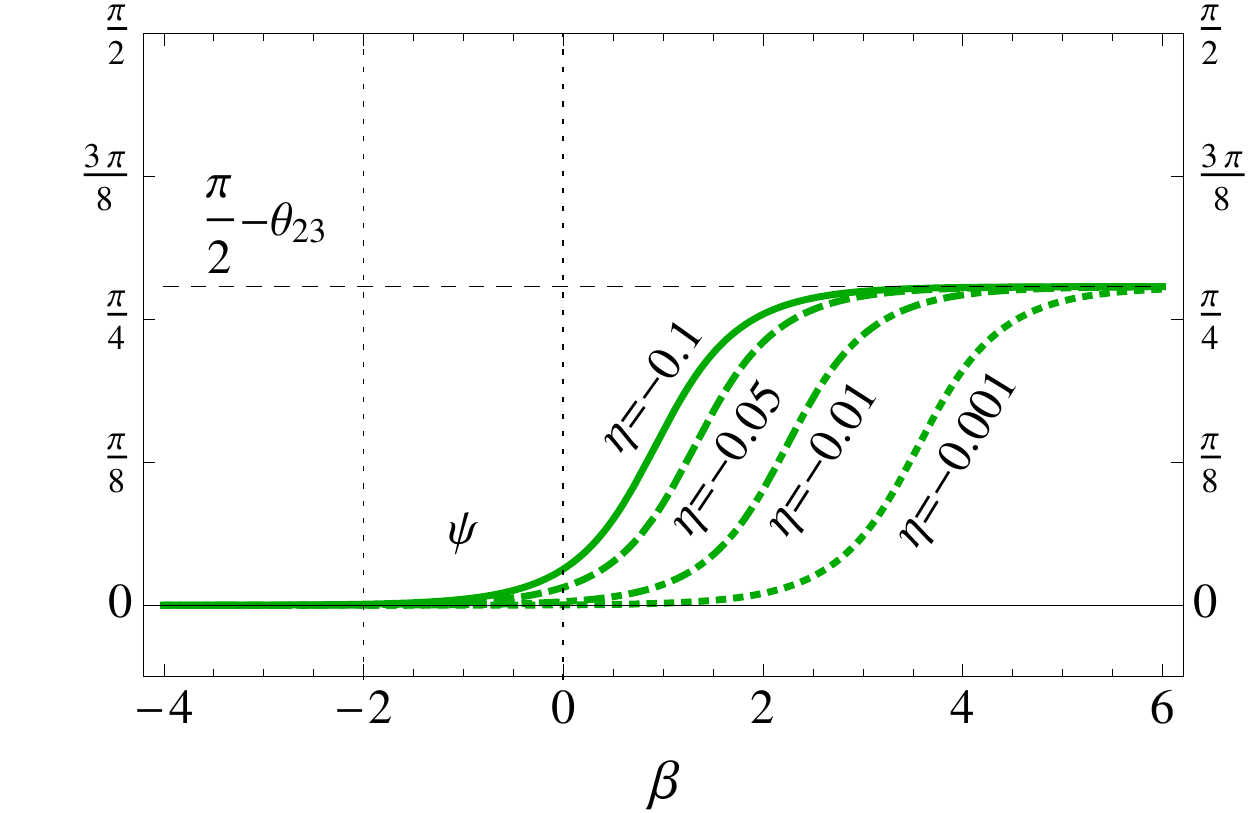}}
\subfigure[$\eta>0$]{\includegraphics[height=5cm]{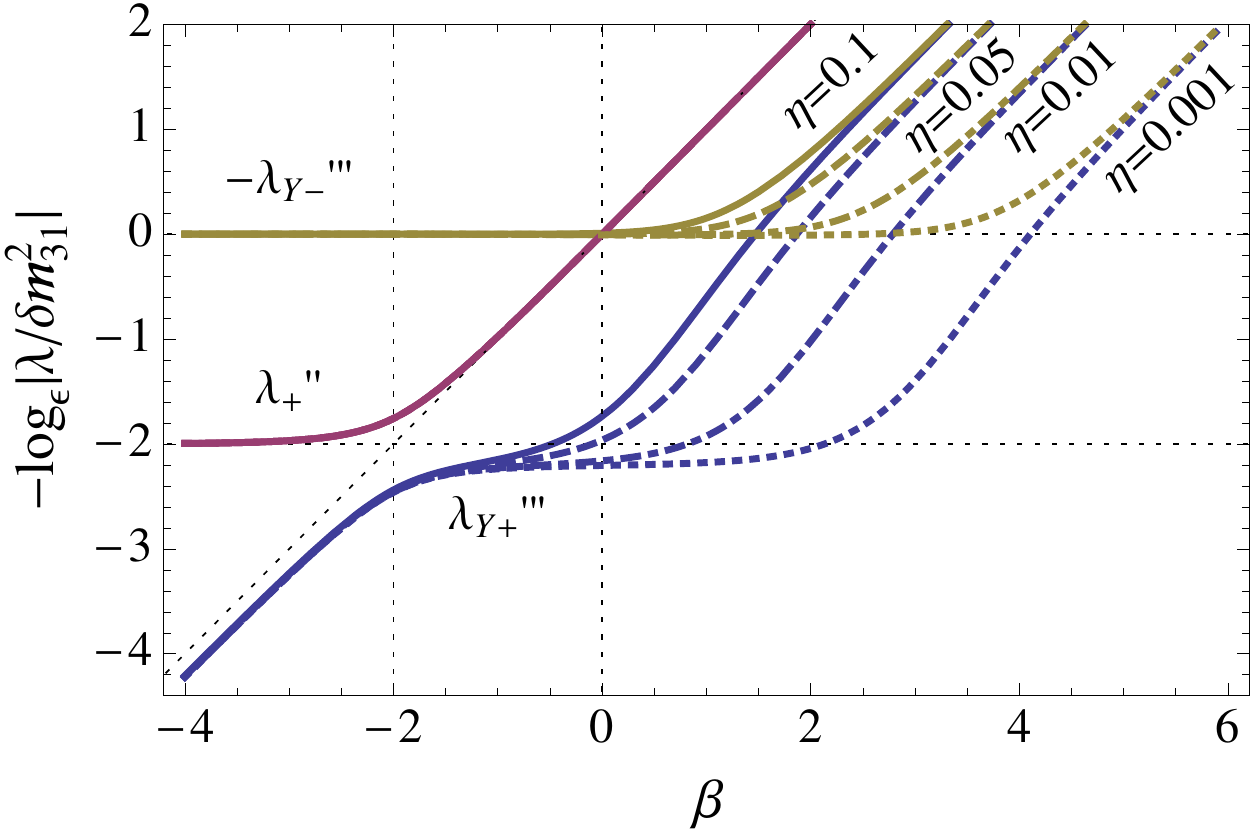}}\qquad
\subfigure[$\eta<0$]{\includegraphics[height=5cm]{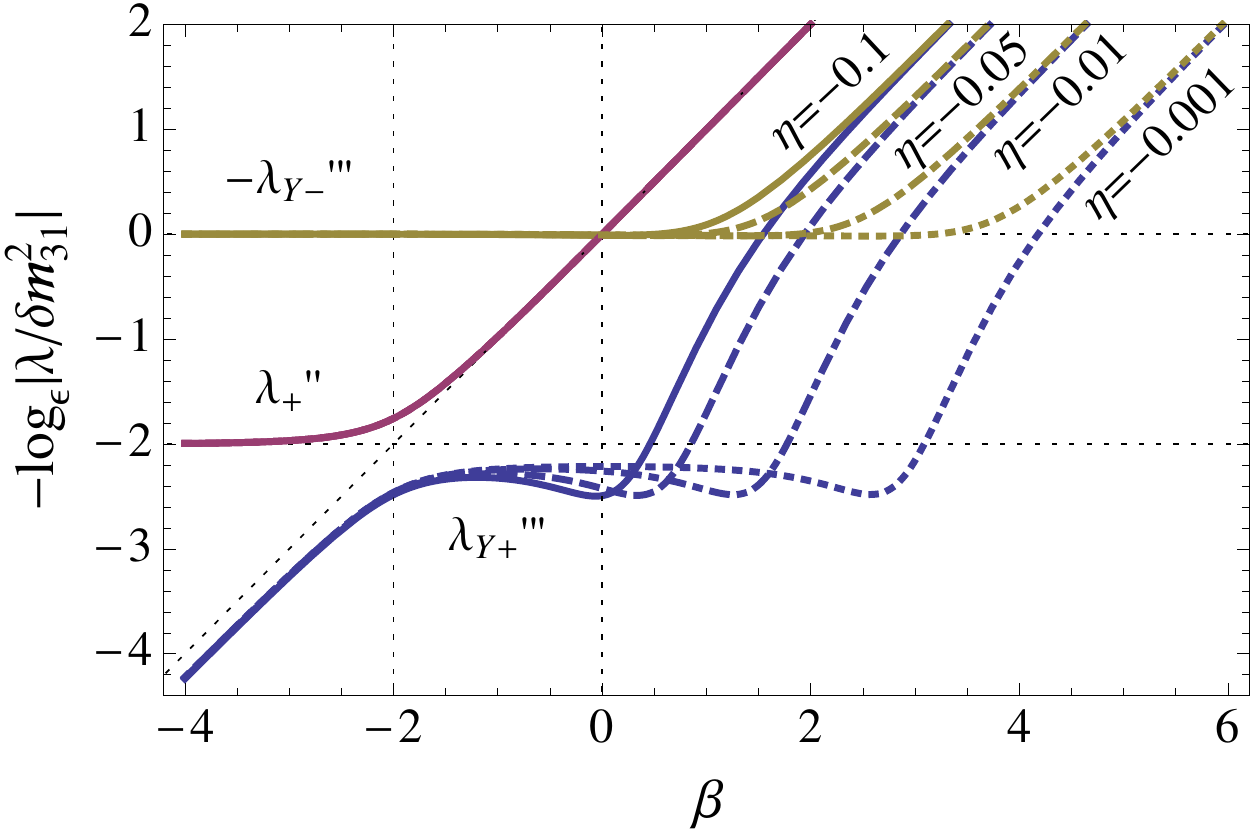}}
\subfigure[$\eta>0$]{\includegraphics[height=5cm]{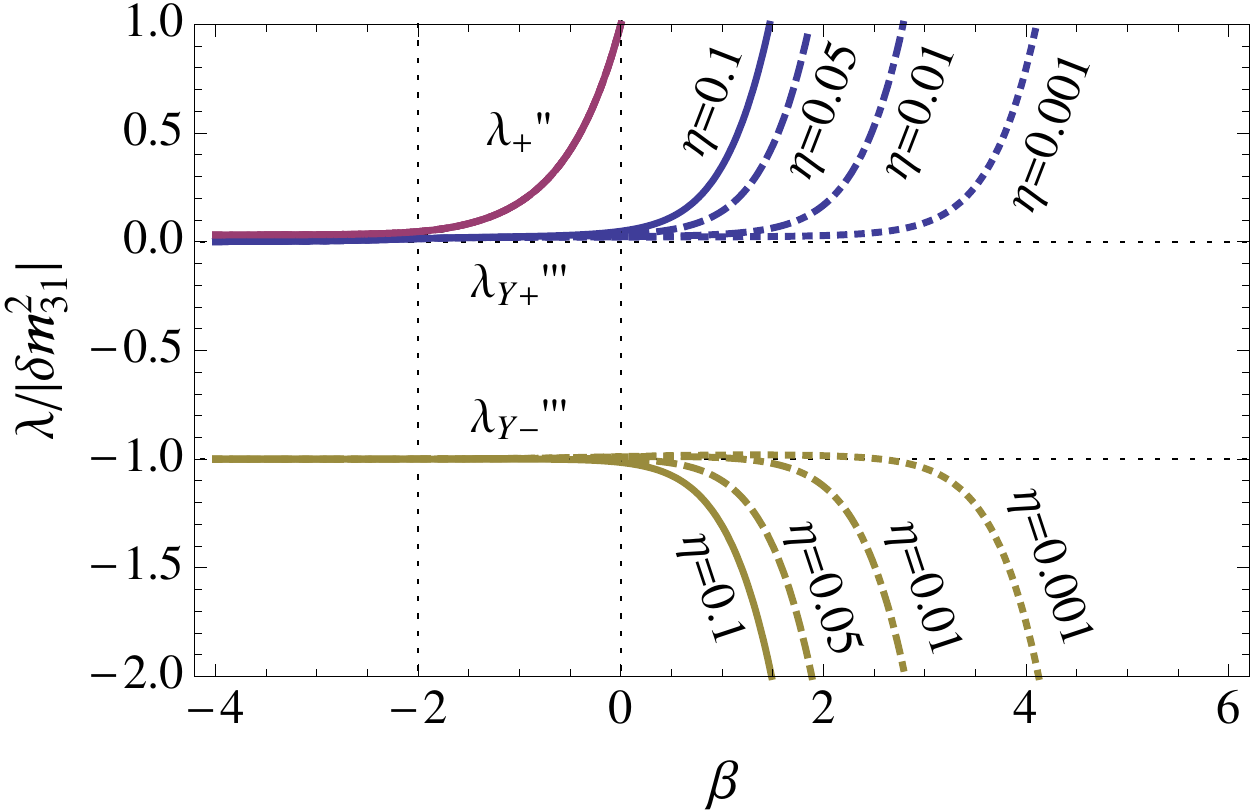}}\quad
\subfigure[$\eta<0$]{\includegraphics[height=5cm]{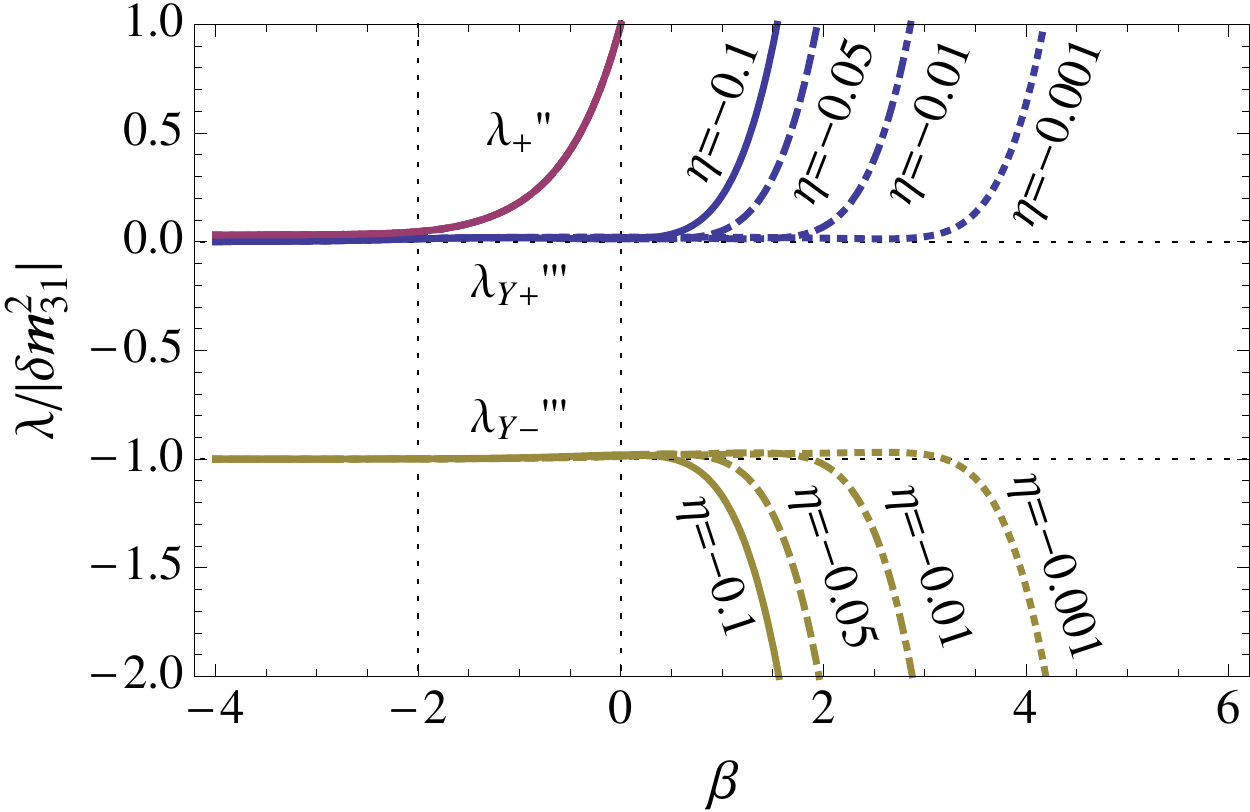}}
\caption{The $\beta$-dependence of $\psi$ and $\lambda'''_{Y\pm}$
for several values of $\eta$ with $s_{23}^2=0.4$.
}
\label{psiplot}
\end{figure}

\noindent
Using $Y$, we find
\begin{equation}
H''''_{\eta -} 
\;=\; Y^\dagger H''' Y 
\;\approx\; 
\left[ \begin{array}{ccc}
\lambda'''_{Y+} & 0 & 0 \\
0 & \lambda''_+ & 0 \\
0 & 0 & \lambda'''_{Y-}
\end{array} \right]\;, 
\end{equation}
where,
\begin{eqnarray}
\lefteqn{\lambda'''_{Y\pm}} \cr
& \equiv &
   \dfrac{ ( \lambda''_{-} + \lambda'_{-} + \hat{a}\eta s_{13}^{\prime 2}\cos 2\theta_{23} )
\pm \sqrt{ [ \lambda''_{-} - \lambda'_{-} - \hat{a}\eta(1+c_{13}^{\prime 2})\cos 2\theta_{23} ]^2
         + 4( \hat{a}\eta c_{13}' \sin 2\theta_{23} )^2 }
         }
         { 2 } \;.
\cr & &
\label{lambdatripleprimeYplusminusdef}
\end{eqnarray}
Thus, $H_{\eta -}''''$ is approximately diagonal.
The asymptotic forms of $\lambda'''_{Y\pm}$ at $\beta\gg 0$ are
\begin{eqnarray}
\lambda'''_{Y+} & \rightarrow & \phantom{-}\hat{a}|\eta| +
\left\{
\begin{array}{ll}
-|\delta m^2_{31}|c_{13}^2 s_{23}^2
+\delta m^2_{21}(c_{12}^2 c_{23}^2 + s_{12}^2 s_{13}^2 s_{23}^2)
\quad & \mbox{for $\eta>0$} \\
-|\delta m^2_{31}|c_{13}^2 c_{23}^2
+\delta m^2_{21}(c_{12}^2 s_{23}^2 + s_{12}^2 s_{13}^2 c_{23}^2)
& \mbox{for $\eta<0$}
\end{array}
\right.
\cr
\lambda'''_{Y-} & \rightarrow & -\hat{a}|\eta| +
\left\{
\begin{array}{ll}
-|\delta m^2_{31}|c_{13}^2 c_{23}^2
+\delta m^2_{21}(c_{12}^2 s_{23}^2 + s_{12}^2 s_{13}^2 c_{23}^2)
\quad & \mbox{for $\eta>0$} \\
-|\delta m^2_{31}|c_{13}^2 s_{23}^2
+\delta m^2_{21}(c_{12}^2 c_{23}^2 + s_{12}^2 s_{13}^2 s_{23}^2)
& \mbox{for $\eta<0$}
\end{array}
\right.
\end{eqnarray}
The $\beta$-dependence of $\lambda'''_{Y\pm}$ are shown in 
Fig.~\ref{psiplot}(c) to Fig.~\ref{psiplot}(f).

\end{enumerate}

\subsection{Effective Mixing Angles for Neutrinos}
\label{effective-mixing-angle-neutrino}

We have discovered that the unitary matrix 
which approximately diagonalizes $H_\eta$ is
$\tilde{U}=UQ_3VWX$ when $\delta m^2_{31}>0$, and
$\tilde{U}=UQ_3VWY$ when $\delta m^2_{31}<0$.
Introducing the notation
\begin{eqnarray}
R_{12}(\theta,\delta) & = &
\begin{bmatrix}
\cos\theta & \sin\theta \,e^{-i\delta} & 0 \\
-\sin\theta \,e^{i\delta} & \cos\theta & 0 \\
0 & 0 & 1
\end{bmatrix}
\;,\cr
R_{13}(\theta,\delta) & = &
\begin{bmatrix}
\cos\theta & 0 & \sin\theta \,e^{-i\delta} \\
0 & 1 & 0 \\
-\sin\theta \,e^{i\delta} & 0 & \cos\theta \\
\end{bmatrix}
\;,\cr
R_{23}(\theta,\delta) & = &
\begin{bmatrix}
1 & 0 & 0 \\
0 & \cos\theta & \sin\theta \,e^{-i\delta} \\
0 & -\sin\theta \,e^{i\delta} & \cos\theta \\
\end{bmatrix}
\;,
\end{eqnarray}
the PMNS matrix $U$ in vacuum can be parametrized as
\begin{eqnarray}
U 
& = & R_{23}(\theta_{23},0)\;R_{13}(\theta_{13},\delta)\;R_{12}(\theta_{12},0)
\phantom{\bigg|} \cr
& = &
\left[ \begin{array}{ccc}
c_{12}c_{13} & s_{12}c_{13} & s_{13} e^{-i\delta} \\
-s_{12}c_{23} - c_{12}s_{13}s_{23}e^{i\delta} &
\phantom{-}c_{12}c_{23} - s_{12}s_{13}s_{23}e^{i\delta} & c_{13}s_{23} \\
\phantom{-}s_{12}s_{23} - c_{12}s_{13}c_{23}e^{i\delta} &
-c_{12}s_{23} - s_{12}s_{13}c_{23}e^{i\delta} & c_{13}c_{23}
\end{array} \right] \;.
\label{UPparam}
\end{eqnarray}
In the following,
we rewrite the mixing matrix in matter $\tilde{U}$ into the analogous form
\begin{equation}
\tilde{U} \;=\; R_{23}(\tilde{\theta}_{23},0)R_{13}(\tilde{\theta}_{13},\tilde{\delta})R_{12}(\tilde{\theta}_{12},0)
\;,
\label{tildeUdef}
\end{equation}
absorbing the extra mixing angles and CP phase into appropriate definitions 
of the `running' parameters $\tilde{\theta}_{12}$, $\tilde{\theta}_{13}$, $\tilde{\theta}_{23}$, 
and $\tilde{\delta}$. Frequent use is made of the relations
\begin{eqnarray}
R_{12}(\theta,\delta)Q_3 & = & Q_3R_{12}(\theta,\delta) \;,\cr
R_{13}(\theta,\delta)Q_3 & = & Q_3R_{13}(\theta,0) \;,\cr
R_{23}(\theta,0)Q_3 & = & Q_3R_{23}(\theta,-\delta) \;,
\label{Q3commutationrelations}
\end{eqnarray}
where $Q_3$ was defined in Eq.~(\ref{Qdef}).

\begin{itemize}
\item \textbf{$\delta m^2_{31}>0$ Case:}

Using Eq.~(\ref{Q3commutationrelations}), it is straightforward to show that
\begin{eqnarray}
\tilde{U} 
& = & UQ_3VWX \cr
& = & 
\underbrace{R_{23}(\theta_{23},0)R_{13}(\theta_{13},\delta)R_{12}(\theta_{12},0)}_{\displaystyle U}Q_3
\underbrace{R_{12}(\varphi,0)}_{\displaystyle V}
\underbrace{R_{23}(\phi,0)}_{\displaystyle W}
\underbrace{R_{12}(\chi,-\delta)}_{\displaystyle X} \cr
& = & R_{23}(\theta_{23},0)Q_3 R_{13}(\theta_{13},0)R_{12}(\theta_{12},0)
R_{12}(\varphi,0)R_{23}(\phi,0)R_{12}(\chi,-\delta) \cr
& = & R_{23}(\theta_{23},0)Q_3 R_{13}(\theta_{13},0)R_{12}(\underbrace{\theta_{12}+\varphi}_{\displaystyle =\theta_{12}'},0)
R_{23}(\phi,0)R_{12}(\chi,-\delta) \cr
& = & R_{23}(\theta_{23},0)Q_3 R_{13}(\theta_{13},0)R_{12}(\theta_{12}',0)
R_{23}(\phi,0)R_{12}(\chi,-\delta) \;,
\end{eqnarray}
where in the last and penultimate lines we have combined the two $12$-rotations into one.
We now commute $R_{23}(\phi,0)R_{12}(\chi,\delta)$ through the other mixing matrices
to the left as follows:

\begin{itemize}
\item \textbf{Step 1:} Commutation of $R_{23}(\phi,0)$ through $R_{12}(\theta_{12}',0)$.

In the range $\beta\agt 0$, the angle $\theta'_{12}$ is approximately equal to $\pi/2$
so we can approximate
\begin{equation}
R_{12}(\theta'_{12},0)
\;\approx\; R_{12}\left(\frac{\pi}{2},0\right)
\;=\; \begin{bmatrix} 0 & 1 & 0 \\ -1 & 0 & 0 \\ 0 & 0 & 1 \end{bmatrix}
\;.
\label{R12theta12primeapprox}
\end{equation}
Note that
\begin{equation}
R_{12}\left(\frac{\pi}{2},0\right)R_{23}(\phi,0)
\;=\; R_{13}(\phi,0)R_{12}\left(\frac{\pi}{2},0\right)
\end{equation}
for any $\phi$.
On the other hand,
in the range $\beta\alt -1$
the angle $\phi$ is negligibly small so we can approximate
\begin{equation}
R_{23}(\phi,0) \;\approx\; R_{23}(0,0)
\;=\; \begin{bmatrix} 1 & 0 & 0 \\ 0 & 1 & 0 \\ 0 & 0 & 1 \end{bmatrix}
\;=\; R_{13}(0,0)
\;.
\end{equation}
Note that
\begin{equation}
R_{12}(\theta'_{12},0)R_{23}(0,0)\;=\;R_{13}(0,0)R_{12}(\theta'_{12},0)
\end{equation}
for any $\theta'_{12}$.
Therefore, for all $\beta$ we have
\begin{equation}
R_{12}(\theta'_{12},0)R_{23}(\phi,0)
\;\approx\; R_{13}(\phi,0)R_{12}(\theta'_{12},0)\;,
\end{equation}
and
\begin{eqnarray}
\tilde{U}
& = & R_{23}(\theta_{23},0)Q_3 R_{13}(\theta_{13},0)R_{12}(\theta_{12}',0)
R_{23}(\phi,0)R_{12}(\chi,-\delta)
\cr
& \approx & R_{23}(\theta_{23},0)Q_3 R_{13}(\theta_{13},0)R_{13}(\phi,0)R_{12}(\theta_{12}',0)R_{12}(\chi,-\delta)
\cr
& = & R_{23}(\theta_{23},0)Q_3 R_{13}(\underbrace{\theta_{13}+\phi}_{\displaystyle =\theta'_{13}},0)R_{12}(\theta_{12}',0)R_{12}(\chi,-\delta)
\cr
& = & R_{23}(\theta_{23},0)Q_3 R_{13}(\theta'_{13},0)R_{12}(\theta'_{12},0)R_{12}(\chi,-\delta)
\;.
\end{eqnarray}
%

\item \textbf{Step 2:} Commutation of $R_{12}(\chi,-\delta)$ through $R_{12}(\theta_{12}',0)$.

In the range $\beta\agt 0$, the angle $\theta'_{12}$ 
is approximately equal to $\pi/2$ as we have noted 
above and we have the approximation given in Eq.~(\ref{R12theta12primeapprox}). 
Note that
\begin{equation}
R_{12}\left(\frac{\pi}{2},0\right)R_{12}(\chi,-\delta)
\;=\; R_{12}(\chi,\delta)R_{12}\left(\frac{\pi}{2},0\right)
\end{equation}
for any $\chi$.
On the other hand, in the range $\beta\alt 0$, 
the angle $\chi$ is negligibly small so we can approximate
\begin{equation}
R_{12}(\chi,-\delta) \;\approx\; R_{12}(0,-\delta) 
\;=\; \begin{bmatrix} 1 & 0 & 0 \\ 0 & 1 & 0 \\ 0 & 0 & 1 \end{bmatrix}
\;=\; R_{12}(0,\delta)
\;=\; R_{23}(0,-\delta)
\;.
\label{R12chideltaapprox}
\end{equation}
Note that
\begin{equation}
R_{12}(\theta'_{12},0)R_{12}(0,-\delta)\;=\; R_{12}(0,\delta)R_{12}(\theta'_{12},0)
\end{equation}
for any $\theta'_{12}$.
Therefore, for all $\beta$ we see that
\begin{equation}
R_{12}(\theta'_{12},0)R_{12}(\chi,-\delta)
\;\approx\; R_{12}(\chi,\delta)R_{12}(\theta'_{12},0)
\;,
\end{equation}
and
\begin{eqnarray}
\tilde{U}
& \approx & R_{23}(\theta_{23},0)Q_3 R_{13}(\theta'_{13},0)R_{12}(\chi,\delta)R_{12}(\theta'_{12},0)
\;.
\end{eqnarray}

\item \textbf{Step 3:} Commutation of $R_{12}(\chi,\delta)$ through $R_{13}(\theta'_{13},0)$.

When $\delta m^2_{31}>0$ we have 
$\theta'_{13}\approx\frac{\pi}{2}$ 
in the range $\beta\agt 1$ so we can approximate
\begin{equation}
R_{13}(\theta'_{13},0) \;\approx\;
R_{13}\left(\frac{\pi}{2},0\right) \;=\;
\begin{bmatrix}
0 & 0 & 1 \\ 0 & 1 & 0 \\ -1 & 0 & 0
\end{bmatrix}
\;.
\end{equation}
Note that
\begin{equation}
R_{13}\left(\frac{\pi}{2},0\right)R_{12}(\chi,\delta)\;=\; R_{23}(\chi,-\delta)R_{13}\left(\frac{\pi}{2},0\right)
\end{equation}
for any $\chi$.
On the other hand, in the range $\beta\alt 0$ the angle $\chi$ was negligibly small so that
we had Eq.~(\ref{R12chideltaapprox}).
Note that
\begin{equation}
R_{13}(\theta'_{13},0)R_{12}(0,\delta)\;=\; R_{23}(0,-\delta)R_{13}(\theta'_{13},0)
\;,
\end{equation}
for any $\theta'_{13}$.
Therefore, for all $\beta$ we see that
\begin{equation}
R_{13}(\theta'_{13},0)R_{12}(\chi,\delta)
\;\approx\; R_{23}(\chi,-\delta)R_{13}(\theta'_{13},0)
\;,
\end{equation}
and using Eq.~(\ref{Q3commutationrelations}) we obtain
\begin{eqnarray}
\tilde{U}
& \approx & R_{23}(\theta_{23},0)Q_3 R_{23}(\chi,-\delta)R_{13}(\theta'_{13},0)R_{12}(\theta'_{12},0)
\cr
& = & R_{23}(\theta_{23},0)R_{23}(\chi,0)R_{13}(\theta'_{13},\delta)R_{12}(\theta'_{12},0)Q_3
\cr
& = & R_{23}(\underbrace{\theta_{23}+\chi}_{\displaystyle =\theta'_{23}},0)R_{13}(\theta'_{13},\delta)R_{12}(\theta'_{12},0)Q_3
\cr
& = & R_{23}(\theta'_{23},0)R_{13}(\theta'_{13},\delta)R_{12}(\theta'_{12},0)Q_3
\;,
\end{eqnarray}
where in the last and penultimate lines 
we have combined the two $23$-rotations into one.
The matrix $Q_3$ on the far right can be absorbed 
into the redefinitions of Majorana phases 
and can be dropped.

\end{itemize}

Thus, we find that the effective mixing matrix $\tilde{U}$ 
in the case $\delta m^2_{31}>0$ can be expressed as 
Eq.~(\ref{tildeUdef}) with the effective mixing angles 
and effective CP-violating phase given approximately by
\begin{eqnarray}
\tilde{\theta}_{12} & \approx & \theta'_{12} \;=\; \theta_{12}+\varphi\;,\cr
\tilde{\theta}_{13} & \approx & \theta'_{13} \;=\; \theta_{13}+\phi\;,\cr
\tilde{\theta}_{23} & \approx & \theta'_{23} \;=\; \theta_{23}+\chi\;,\cr
\tilde{\delta}      & \approx & \delta\;.
\end{eqnarray}
%

\item \textbf{$\delta m^2_{31}<0$ Case:}\\

Using Eq.~(\ref{Q3commutationrelations}), we obtain
\begin{eqnarray}
\tilde{U} 
& = & UQ_3VWY \cr
& = & 
\underbrace{R_{23}(\theta_{23},0)R_{13}(\theta_{13},\delta)R_{12}(\theta_{12},0)}_{\displaystyle U}Q_3
\underbrace{R_{12}(\varphi,0)}_{\displaystyle V}
\underbrace{R_{23}(\phi,0)}_{\displaystyle W}
\underbrace{R_{13}(-\psi,-\delta)}_{\displaystyle Y} \cr
& = & R_{23}(\theta_{23},0)Q_3 R_{13}(\theta_{13},0)R_{12}(\theta_{12}',0)
R_{23}(\phi,0)R_{13}(-\psi,-\delta) \;.
\end{eqnarray}
We now commute $R_{23}(\phi,0)R_{13}(-\psi,-\delta)$ 
through the other mixing matrices to the left and 
re-express $\tilde{U}$ as in Eq.~(\ref{tildeUdef}),
absorbing the extra mixing angles and CP phase 
into $\tilde{\theta}_{12}$, $\tilde{\theta}_{13}$, 
$\tilde{\theta}_{23}$, and $\tilde{\delta}$. 
The first step is the same as the $\delta m^2_{31}>0$ case,
the only difference being the $\beta$-dependence of $\theta'_{13}$, 
which is also shown in Fig.~\ref{phiplot}(b).

\begin{itemize}
\item \textbf{Step 2:} Commutation of $R_{13}(-\psi,-\delta)$ through $R_{12}(\theta_{12}',0)$.

In the range $\beta\agt 0$ the angle $\theta'_{12}$ is 
approximately equal to $\pi/2$ as we have noted previously, 
and we have the approximation given in Eq.~(\ref{R12theta12primeapprox}).
Note that
\begin{equation}
R_{12}\left(\frac{\pi}{2},0\right)R_{13}(-\psi,-\delta)
\;=\; R_{23}(\psi,-\delta)R_{12}\left(\frac{\pi}{2},0\right)
\end{equation}
for any $\psi$.
On the other hand, in the range $\beta\alt 0$ 
the angle $\psi$ is negligibly small so that
\begin{equation}
R_{13}(-\psi,-\delta) \;\approx\; R_{13}(0,-\delta) 
\;=\; \begin{bmatrix} 1 & 0 & 0 \\ 0 & 1 & 0 \\ 0 & 0 & 1 \end{bmatrix}
\;=\; R_{23}(0,-\delta)
\;=\; R_{13}(0,\delta)
\;.
\label{R13psideltaapprox}
\end{equation}
Note that
\begin{equation}
R_{12}(\theta'_{12},0)R_{13}(0,-\delta)
\;=\; R_{23}(0,-\delta)R_{12}(\theta'_{12},0)
\;.
\end{equation}
Therefore, for all $\beta$ we see that
\begin{equation}
R_{12}(\theta'_{12},0)R_{13}(-\psi,-\delta)
\;\approx\; R_{23}(\psi,-\delta)R_{12}(\theta'_{12},0)
\;,
\end{equation}
and 
\begin{eqnarray}
\tilde{U}
& \approx & R_{23}(\theta_{23},0)Q_3 R_{13}(\theta'_{13},0)R_{23}(\psi,-\delta)R_{12}(\theta'_{12},0)
\;.
\end{eqnarray}

\item \textbf{Step 3:} Commutation of $R_{13}(\psi,-\delta)$ through $R_{13}(\theta'_{13},0)$.

When $\delta m^2_{31}<0$ we have $\theta'_{13}\approx 0$ 
in the range $\beta\agt 1$ so that
\begin{equation}
R_{13}(\theta'_{13},0) \;\approx\;
R_{13}(0,0) \;=\;
\begin{bmatrix}
1 & 0 & 0 \\ 0 & 1 & 0 \\ 0 & 0 & 1
\end{bmatrix}
\;.
\end{equation}
Note that
\begin{equation}
R_{13}(0,0)R_{23}(\psi,-\delta)
\;=\; R_{23}(\psi,-\delta)R_{13}(0,0)
\end{equation}
for all $\psi$.
On the other hand, in the range $\beta\alt 0$ 
the angle $\psi$ was negligibly small so that
we had the approximation 
Eq.~(\ref{R13psideltaapprox}).
Note that
\begin{equation}
R_{13}(\theta'_{13},0)R_{23}(0,-\delta)
\;=\; R_{23}(0,-\delta)R_{13}(\theta'_{13},0)
\end{equation}
for all $\theta'_{13}$.
Therefore, for all $\beta$ we see that
\begin{equation}
R_{13}(\theta'_{13},0)R_{23}(\psi,-\delta)
\;\approx\; R_{23}(\psi,-\delta)R_{13}(\theta'_{13},0)
\;,
\end{equation}
and using Eq.~(\ref{Q3commutationrelations}) 
we obtain
\begin{eqnarray}
\tilde{U}
& \approx & R_{23}(\theta_{23},0)Q_3 R_{23}(\psi,-\delta)R_{13}(\theta'_{13},0)R_{12}(\theta'_{12},0)
\cr
& = & R_{23}(\theta_{23},0)R_{23}(\psi,0)R_{13}(\theta'_{13},\delta)R_{12}(\theta'_{12},0)Q_3
\cr
& = & R_{23}(\underbrace{\theta_{23}+\psi}_{\displaystyle =\theta'_{23}},0)R_{13}(\theta'_{13},\delta)R_{12}(\theta'_{12},0)Q_3
\cr
& = & R_{23}(\theta'_{23},0)R_{13}(\theta'_{13},\delta)R_{12}(\theta'_{12},0)Q_3
\;,
\end{eqnarray}
where in the last and penultimate lines 
we have combined the two $23$-rotations into one.
The matrix $Q_3$ on the far right can be absorbed 
into redefinitions of the Majorana phases and can be dropped.

\end{itemize}

Thus, we find that the effective mixing matrix $\tilde{U}$ 
in the case $\delta m^2_{31}<0$ can be expressed as 
Eq.~(\ref{tildeUdef}) with the effective mixing angles 
and effective CP-violating phase given approximately by
\begin{eqnarray}
\tilde{\theta}_{12} & \approx & \theta'_{12} \;=\; \theta_{12}+\varphi\;,\cr
\tilde{\theta}_{13} & \approx & \theta'_{13} \;=\; \theta_{13}+\phi\;,\cr
\tilde{\theta}_{23} & \approx & \theta'_{23} \;=\; \theta_{23}+\psi\;,\cr
\tilde{\delta}      & \approx & \delta\;.
\end{eqnarray}
\end{itemize}

\subsection{Summary of Neutrino Case}

\begin{figure}[t]
\subfigure[Normal Hierarchy]{\includegraphics[height=5cm]{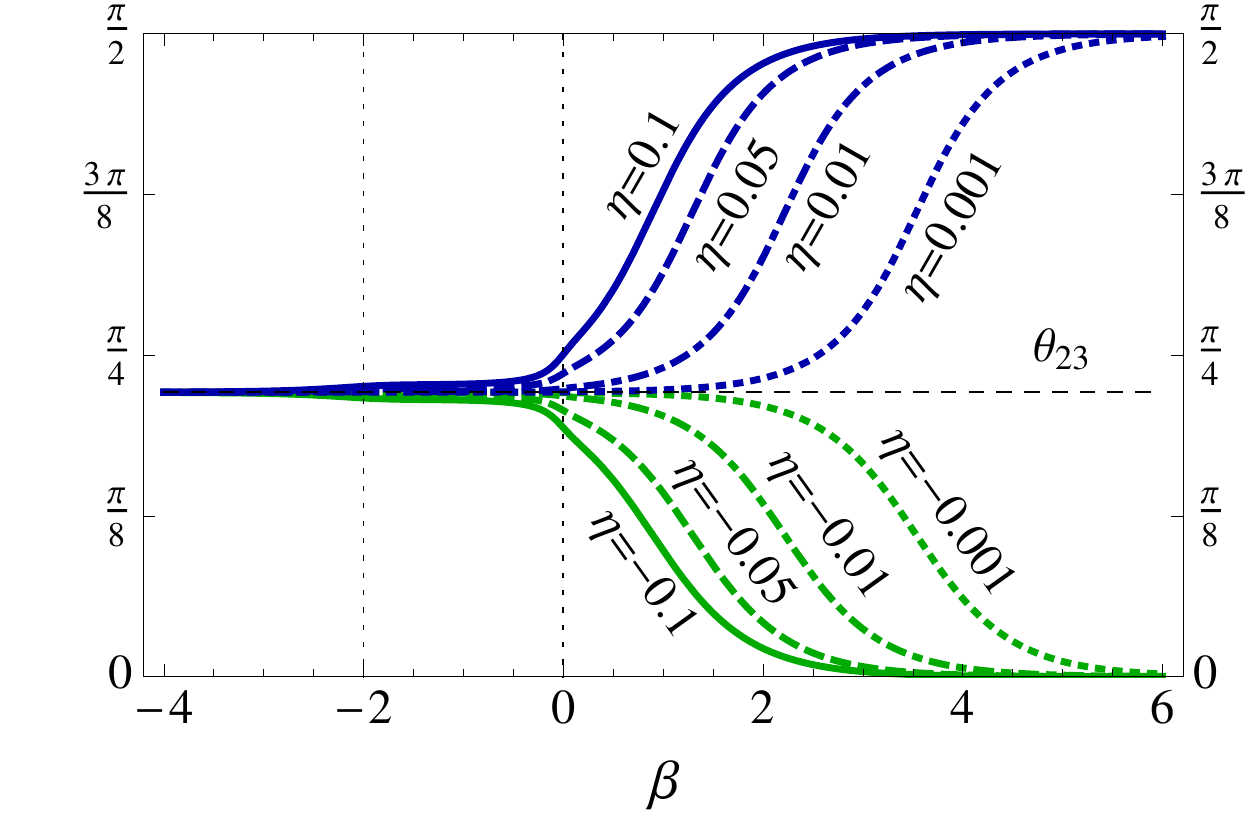}}
\subfigure[Inverted Hierarchy]{\includegraphics[height=5cm]{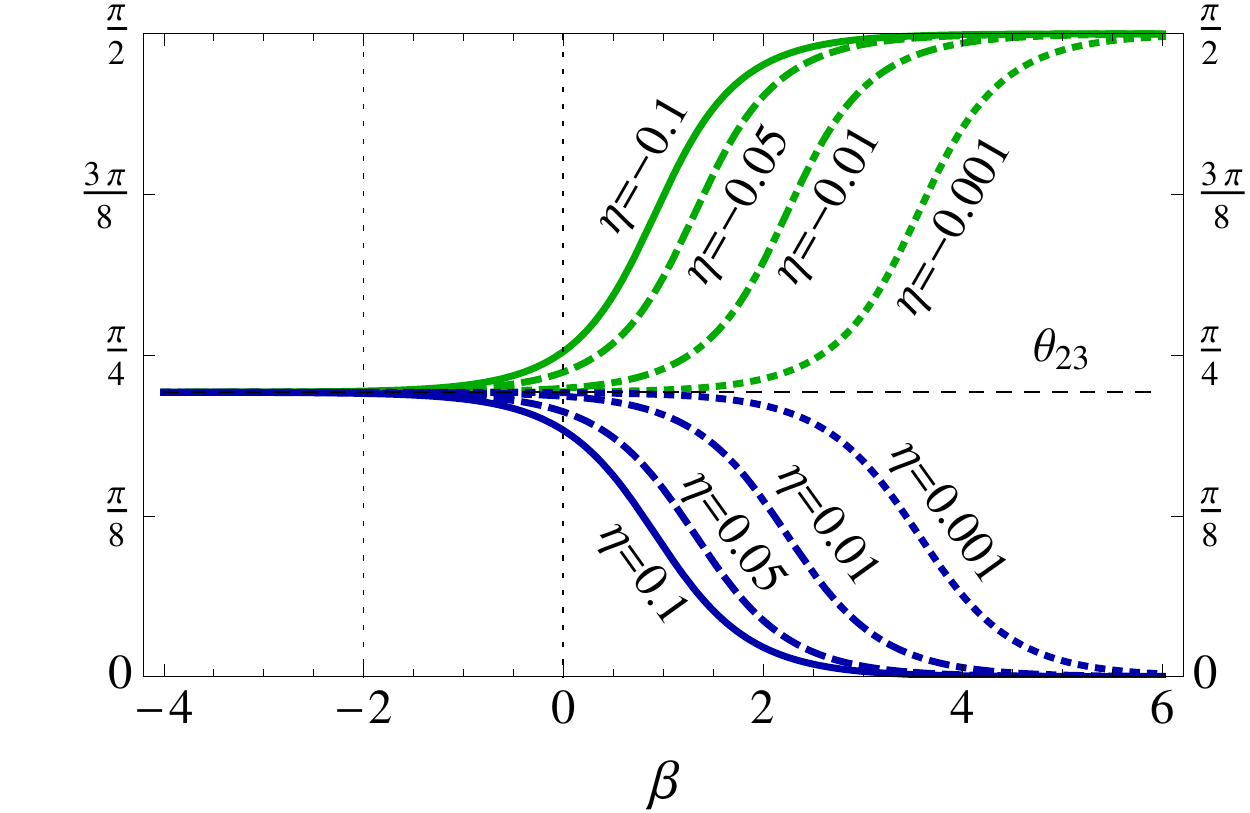}}
\caption{The $\beta$-dependence of $\theta_{23}'$
for the (a) normal and (b) inverted hierarchies
for several values of $\eta$ with $s_{23}^2=0.4$.
}
\label{theta23primeplot}
\end{figure}

To summarize what we have learned, 
inclusion of the $\hat{a}\eta M_\eta$ term in the effective Hamiltonian 
shifts $\theta_{23}$ to $\theta_{23}'=\theta_{23}+\chi$ for the $\delta m^2_{31}>0$ case,
and to $\theta_{23}'=\theta_{23}+\psi$ for the $\delta m^2_{31}<0$ case.
For both cases, $\theta_{23}'$ can be calculated directly without calculating
$\chi$ or $\psi$ first via the expression
\begin{equation}
\tan 2\theta_{23}' \;\approx\; 
\dfrac{[\delta m^2_{31}c_{13}^2-\delta m^2_{21}(c_{12}^2-s_{12}^2 s_{13}^2)]\sin 2\theta_{23}}
      {[\delta m^2_{31}c_{13}^2-\delta m^2_{21}(c_{12}^2-s_{12}^2 s_{13}^2)]\cos 2\theta_{23} -2\hat{a}\eta}
\;.
\label{theta23primeNeutrino}
\end{equation}
Note that as $\beta$ is increased,
$\theta_{23}'$ runs toward $\dfrac{\pi}{2}$ if $\delta m^2_{31}\eta >0$,
while it runs toward $0$ if $\delta m^2_{31}\eta <0$.
The $\beta$-dependence of $\theta_{23}'$ is shown in 
Fig.~\ref{theta23primeplot}.
The CP-violating phase $\delta$ is unaltered 
and maintains its vacuum value.

The running of the effective mass-squared differences are also modified in the range
$\beta \agt 0$.
For the $\delta m^2_{31}>0$ case, $\lambda_1$ and $\lambda_2$ show extra running,
while for the $\delta m^2_{31}<0$ case, it is $\lambda_1$ and $\lambda_3$ that
show extra running, cf. Figs.~\ref{chiplot} and \ref{psiplot}.

\subsection{Discussion at the Probability Level}
\label{comparison}

So far we have focused our attention on how the flavor-diagonal 
NSI parameter $\eta$ modifies the running of the effective mass-squared differences, 
mixing angles, and CP-violating phase as
functions of $\hat{a}=a(1+\zeta)$ for the neutrinos. 
We derived simple analytical expressions 
for these effective parameters using the Jacobi method. 
We have discussed the running of these effective neutrino oscillation parameters for both the 
normal ($\delta m^2_{31}>0$) and inverted ($\delta m^2_{31}<0$) neutrino mass hierarchies. 
The modifications induced by $\eta$ and $\zeta$ in the running of effective oscillation parameters 
in the case of anti-neutrino are discussed in detail in appendix~\ref{anti-neutrino}. 
At this point, we look at how these lepton-flavor-conserving NSI parameters 
alter the neutrino oscillation probabilities for various appearance and disappearance channels. 

In the three-flavor scenario, the neutrino oscillation probabilities in 
vacuum\footnote{We follow the conventions, notations, and the expressions for
various neutrino oscillation probabilities in vacuum as given in the appendices 
A.1 and A.2 of Ref.~\cite{Agarwalla:2013tza}.} for the disappearance channel 
(initial and final flavors are same) take the form
\begin{eqnarray}
P(\nu_\alpha\rightarrow\nu_\alpha)
& = & 1 - 4\, |U_{\alpha 2}|^2 \left( 1 - |U_{\alpha 2}|^2 \right)
            \sin^2\frac{\Delta_{21}}{2}
        - 4\, |U_{\alpha 3}|^2 \left( 1 - |U_{\alpha 3}|^2 \right)
            \sin^2\frac{\Delta_{31}}{2} \cr
& &  \phantom{1}
        + 2\, |U_{\alpha 2}|^2 |U_{\alpha 3}|^2
          \left( 4\sin^2\frac{\Delta_{21}}{2}\sin^2\frac{\Delta_{31}}{2}
                + \sin\Delta_{21}\sin\Delta_{31}
          \right) \;,
\label{Palphatoalpha}
\end{eqnarray}
and for the appearance channel (initial and final flavors are different) we have
\begin{eqnarray}
P(\nu_\alpha \rightarrow \nu_\beta)
& = & 4\, |U_{\alpha 2}|^2 |U_{\beta 2}|^2 \sin^2\frac{\Delta_{21}}{2}
     +4\, |U_{\alpha 3}|^2 |U_{\beta 3}|^2 \sin^2\frac{\Delta_{31}}{2} \cr
&   & +2\;\Re\left( U^*_{\alpha 3}U_{\beta 3}U_{\alpha 2}U^*_{\beta 2}\right)
      \left(4\sin^2\frac{\Delta_{21}}{2}\sin^2\frac{\Delta_{31}}{2}
           +\sin\Delta_{21}\sin\Delta_{31}
      \right) \cr 
&   & -4\,\Im\left(U^*_{\alpha 3}U_{\beta 3}U_{\alpha 2}U^*_{\beta 2}\right)
      \left( \sin^2\frac{\Delta_{21}}{2}\sin\Delta_{31}
            -\sin^2\frac{\Delta_{31}}{2}\sin\Delta_{21}
      \right) \;.    
\label{Palphatonotalpha}       
\end{eqnarray}
In the above equations, we define
\begin{equation}
\Delta_{ij} 
\;\equiv\; \dfrac{\delta m_{ij}^2}{2E} L 
\;=\; 2.534\;
\biggl(\dfrac{\delta m_{ij}^2}{\mathrm{eV}^2}\biggr)
\biggl(\dfrac{\mathrm{GeV}}{E}\biggr)
\biggl(\dfrac{L}{\mathrm{km}}\biggr)
\;,\quad
\delta m_{ij}^2
\,\equiv\, m_i^2-m_j^2\;.
\end{equation}
The transition probabilities in Eq.~(\ref{Palphatoalpha}) and Eq.~(\ref{Palphatonotalpha}) 
contain several elements of the PMNS matrix $U$ which are expressed in terms of three 
mixing angles $\theta_{12}$, $\theta_{23}$, $\theta_{13}$, and a CP-violating phase $\delta$ 
as shown in Eq.~(\ref{UPparam}). The oscillation probabilities for the anti-neutrinos are obtained
by replacing $U_{\alpha i}$ with its complex conjugate.

The neutrino oscillation probabilities in the presence of matter are obtained by replacing the 
vacuum expressions of the elements of the mixing matrix $U$ and the mass-square differences 
$\Delta_{ij}$ with their effective `running' values in 
matter \cite{Wolfenstein:1977ue,Mikheev:1986gs,Mikheev:1986wj} :
\begin{equation}
U_{\alpha i}\;\rightarrow\;\tilde{U}_{\alpha i}\;(\theta_{12}\rightarrow\theta_{12}', 
\theta_{13}\rightarrow\theta_{13}', \theta_{23}\rightarrow\theta_{23}')\;,\quad
\Delta_{ij}\;\rightarrow\;\tilde{\Delta}_{ij} = \dfrac{\lambda_i - \lambda_j}{2E} L \;,
\end{equation}
and for the anti-neutrinos
\begin{equation}
U_{\alpha i}\;\rightarrow\;\edlit{U}_{\alpha i}\;(\theta_{12}\rightarrow\overline{\theta}_{12}', 
\theta_{13}\rightarrow\overline{\theta}_{13}', \theta_{23}\rightarrow\overline{\theta}_{23}')\;,\quad
\Delta_{ij}\;\rightarrow\;\edlit{\Delta}_{ij} = \dfrac{\overline{\lambda}_i - \overline{\lambda}_j}{2E} L \;.
\end{equation}
To demonstrate the accuracy (or lack thereof in special cases) of our approximate 
analytical results, we compare the oscillation probabilities calculated with our 
approximate effective running mixing angles and mass-squared differences with 
those calculated numerically for the same baseline and line-averaged constant 
matter density along it. For the mixing angles and mass-squared differences in 
vacuum, we use the benchmark values from Ref.~\cite{GonzalezGarcia:2012sz} 
as listed in Table~\ref{tab:bench}. In some plots, we take different values of
$\sin^2\theta_{23}$ and $\delta$ which we mention explicitly in the figure 
legends and captions. In this paper, all the plots 
(except in appendix~\ref{Constant-vs-varying-density})
are generated considering the line-averaged constant matter density for a given 
baseline which has been estimated using the 
Preliminary Reference Earth Model (PREM) \cite{PREM:1981}.
In appendix~\ref{Constant-vs-varying-density}, we compare
the exact numerical probabilities with line-averaged constant Earth density 
and varying Earth density profile for 8770 km and 10000 km baselines.

\begin{figure}[t]
\includegraphics[width=7.5cm,height=5.5cm]{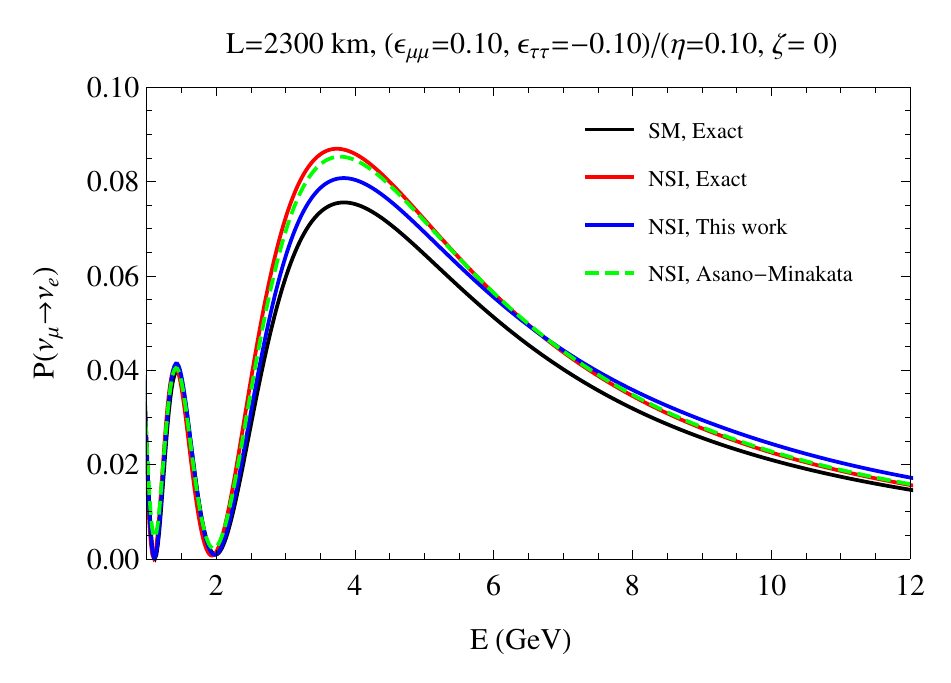}
\includegraphics[width=7.5cm,height=5.5cm]{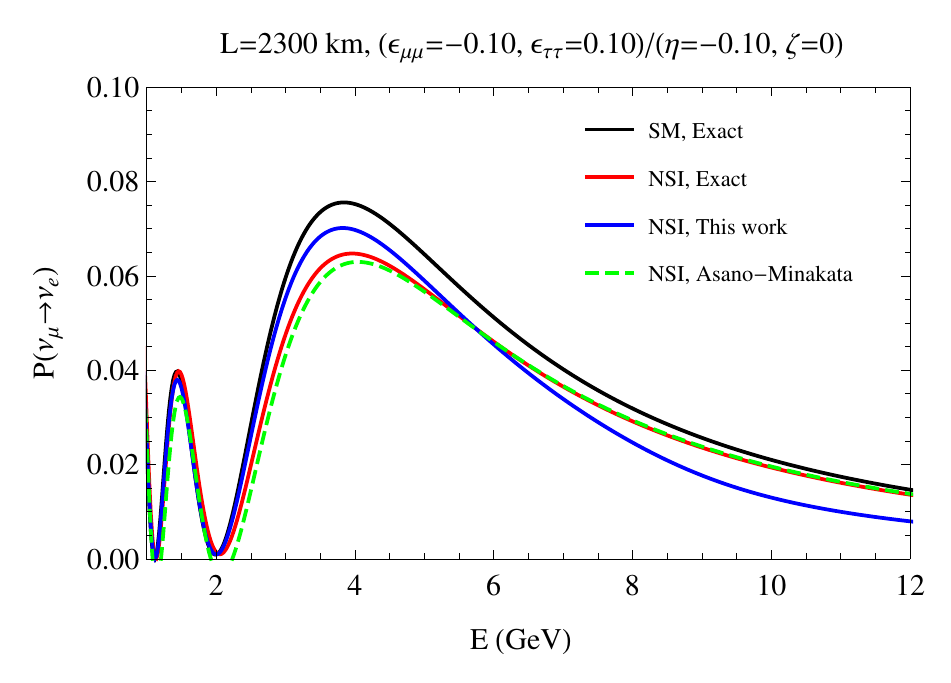}
\includegraphics[width=7.5cm,height=5.5cm]{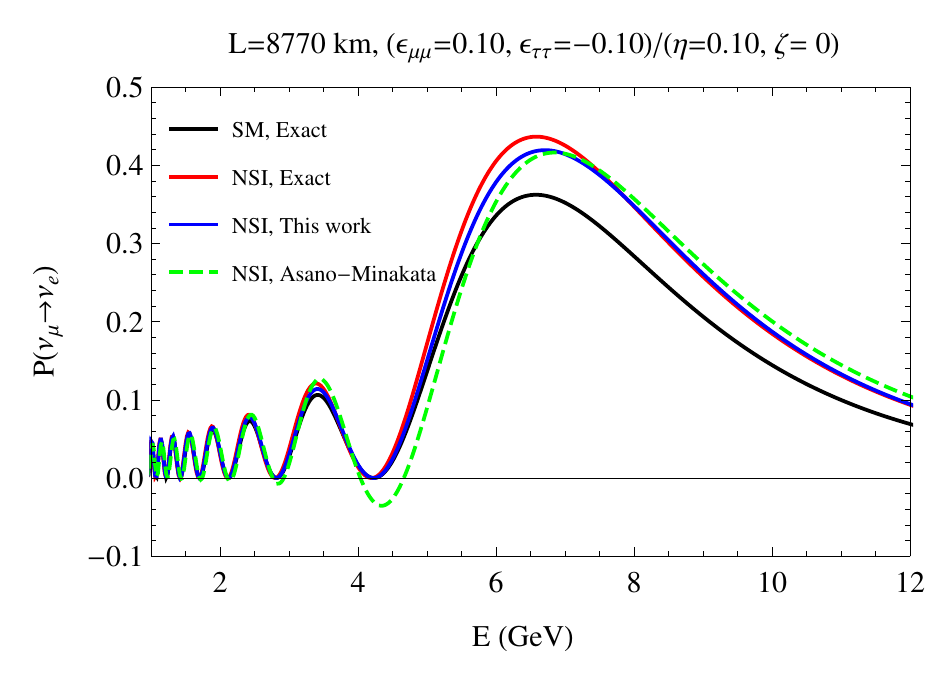}
\includegraphics[width=7.5cm,height=5.5cm]{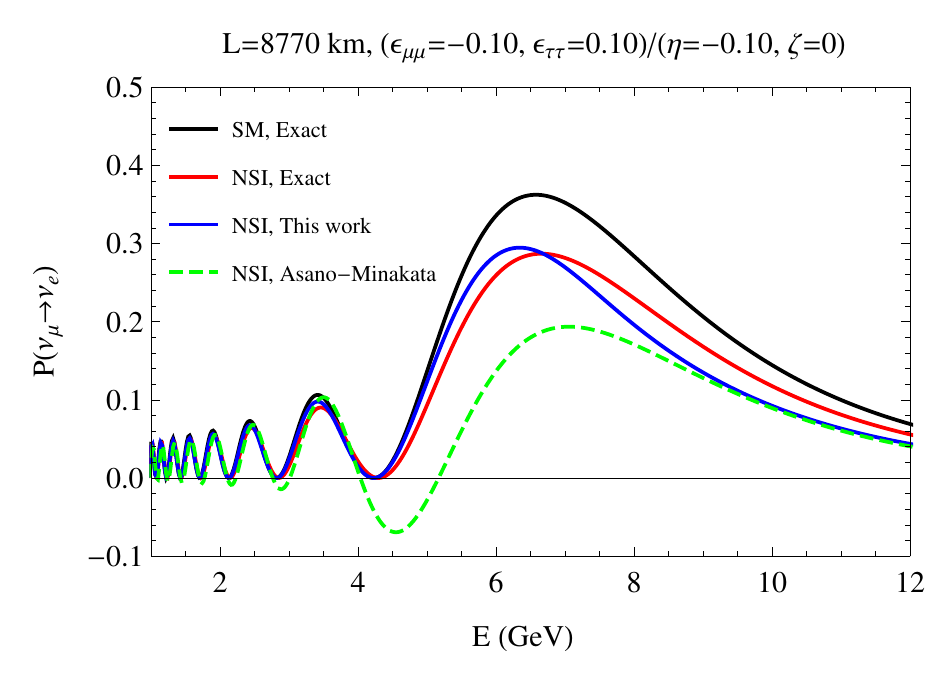}
\caption{$\nu_\mu\rightarrow\nu_e$ transition probability as a function of neutrino energy $E$ in GeV for 2300 km 
(8770 km) baseline in upper (lower) panels. We compare the analytical expressions of this work and 
Asano-Minakata \cite{Asano:2011nj} against the exact numerical result assuming $\eta=0.1, \zeta=0$ 
(left panels) and $\eta=-0.1, \zeta=0$ (right panels). The solid black curves portray the standard three-flavor
oscillation probabilities in matter without NSI's. In all the panels, we consider $\theta_{23}=40^{\circ}$, 
$\delta=0^{\circ}$, and normal mass hierarchy.}
\label{fig:comparison}
\end{figure}
 
In Fig.~\ref{fig:comparison}, we present our approximate $\nu_\mu\rightarrow\nu_e$ oscillation
probabilities (blue curves) as a function of the neutrino energy against the exact numerical 
results (red curves) considering\footnote{We take $\zeta=0$ in our plots since we expect it to be hidden 
in the uncertainties in the matter density and neutrino energy.} $\eta=0.1$, $\zeta=0$ (left panels) 
and $\eta=-0.1$, $\zeta=0$ (right panels). 
The upper panels are drawn for the baseline of $L=2300\,\mathrm{km}$, which 
corresponds to the distance between CERN and Pyh\"asalmi \cite{Agarwalla:2011hh,Stahl:2012exa,Agarwalla:2014tca} 
with the line-averaged constant matter density of $\rho=3.54\,\mathrm{g/cm^3}$.
In the lower panels, we give the probabilities for the baseline of $L=8770\,\mathrm{km}$, which is the distance
from CERN to Kamioka \cite{Agarwalla:2012zu} assuming $\rho=4.33\,\mathrm{g/cm^3}$.
Here, in all the panels, we assume $\sin^2\theta_{23} = 0.41~(\theta_{23}=40^{\circ})$, $\delta=0^{\circ}$, 
and normal mass hierarchy ($\delta m^2_{31}>0$). To see the differences in the oscillation probability 
caused by the NSI parameters, we also give the exact numerical SM three-flavor 
oscillation probabilities in matter in the absence of NSI's which are depicted by the solid 
black curves with figure legend `SM, Exact.' 
It has been already shown in 
Ref.~\cite{Agarwalla:2013tza} that
our approximate expressions for the $\eta=0$ case
match extremely well with the exact numerical results for all these baselines and energies.
We also compare our results with the approximate expressions of 
Asano and Minakata\footnote{For comparison, we take Eq.~(36) of Ref.~\cite{Asano:2011nj} 
where the authors adopted the perturbation method \cite{Kikuchi:2008vq,Minakata:2009sr}
to obtain the analytical expressions for oscillation probability in presence of NSI's 
for large $\theta_{13}$. The same analytical expressions are given in a more detailed fashion
in Eqs.~(8) to (13) in Ref.~\cite{Rahman:2015vqa}.} \cite{Asano:2011nj} (dashed green curves). 
The correspondence between our $\eta$ and $\zeta$ and
the NSI parameters $\varepsilon_{ee}$, $\varepsilon_{\mu\mu}$, and $\varepsilon_{\tau\tau}$ 
used in Refs.~\cite{Asano:2011nj,Rahman:2015vqa} can be obtained via
Eqs.~(\ref{eq:epsilons}), (\ref{eq:a-and-ahat}), and (\ref{eq:eta-zeta}) which suggest the
changes : $a \rightarrow \hat{a} \equiv a(1+\zeta)$,
$\varepsilon_{ee} \rightarrow 0$, $\varepsilon_{\mu\mu} \rightarrow \eta$, 
and $\varepsilon_{\tau\tau} \rightarrow -\eta$. We can see from Fig.~\ref{fig:comparison} 
that for the 2300~km baseline, the Asano-Minakata expressions give better matches compared to our 
results, while for the 8770~km baseline, 
our expressions agree better with the exact numerical results.

\begin{figure}[t]
\includegraphics[width=7.5cm,height=5.5cm]{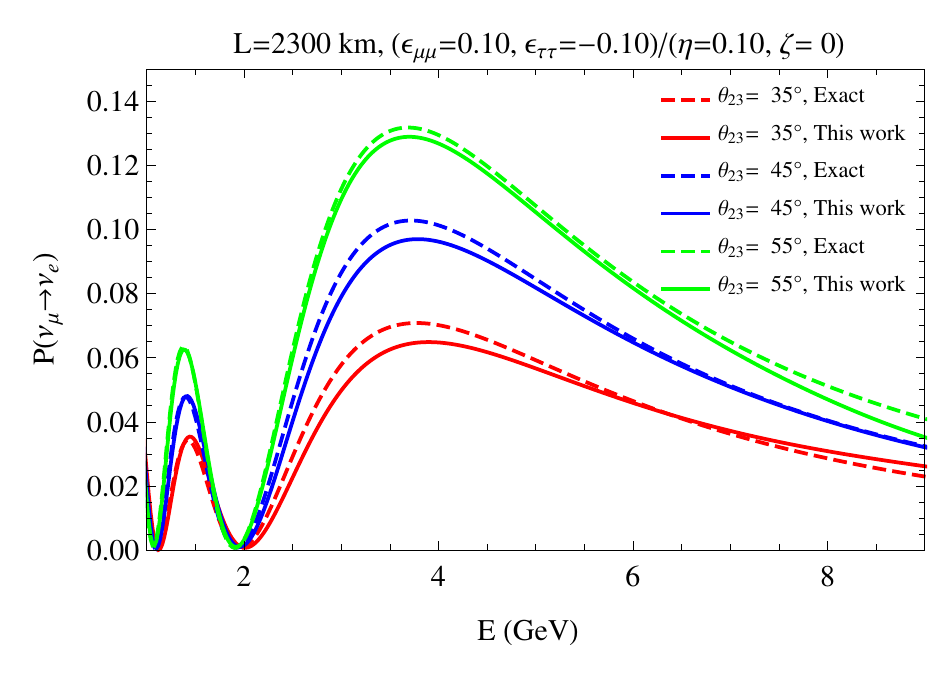}
\includegraphics[width=7.5cm,height=5.5cm]{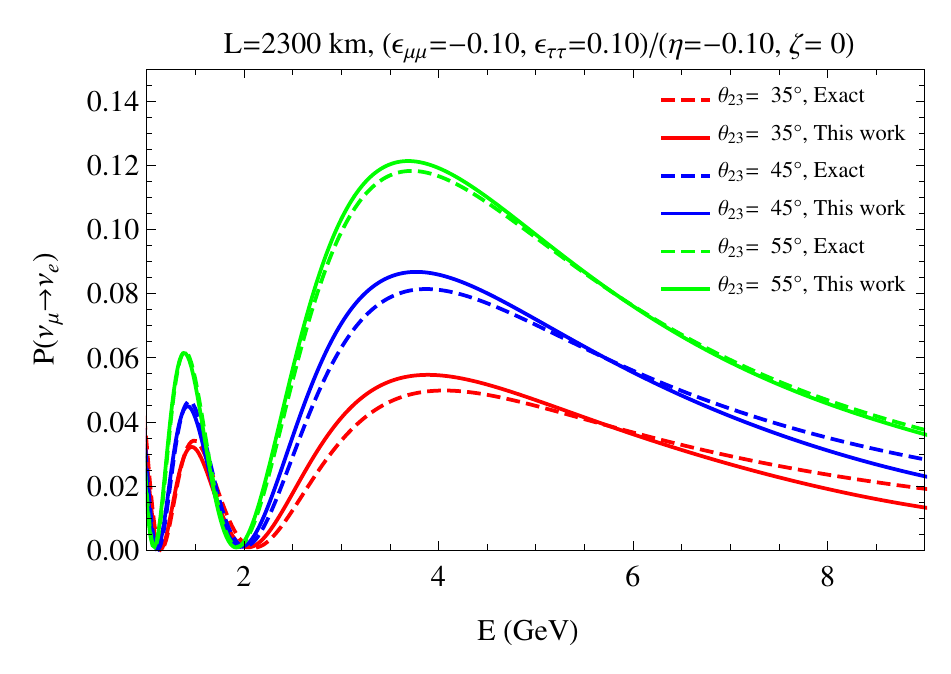}
\includegraphics[width=7.5cm,height=5.5cm]{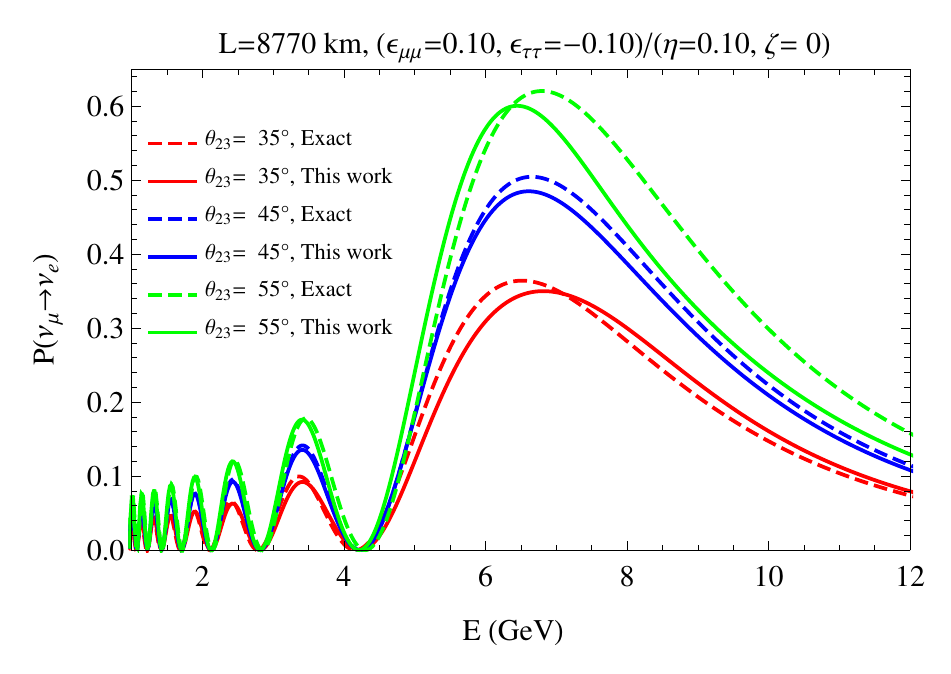}
\includegraphics[width=7.5cm,height=5.5cm]{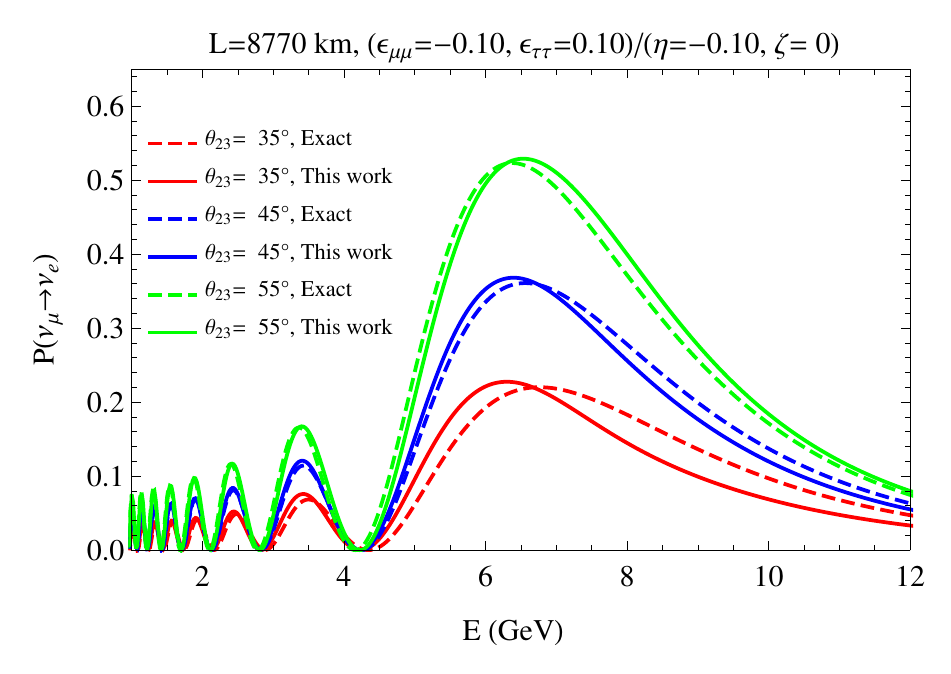}
\caption{Comparison of our analytical expressions (solid curves) 
to the exact numerical results (dashed curves) for various values 
of $\theta_{23}$ assuming $\delta=0^{\circ}$ and 
$\delta m^2_{31}>0$. Upper (lower) panels are 
for 2300 km (8770 km) baseline.}
\label{fig:comparison-theta23}
\end{figure}

\begin{figure}[t]
\includegraphics[width=7.5cm,height=5.5cm]{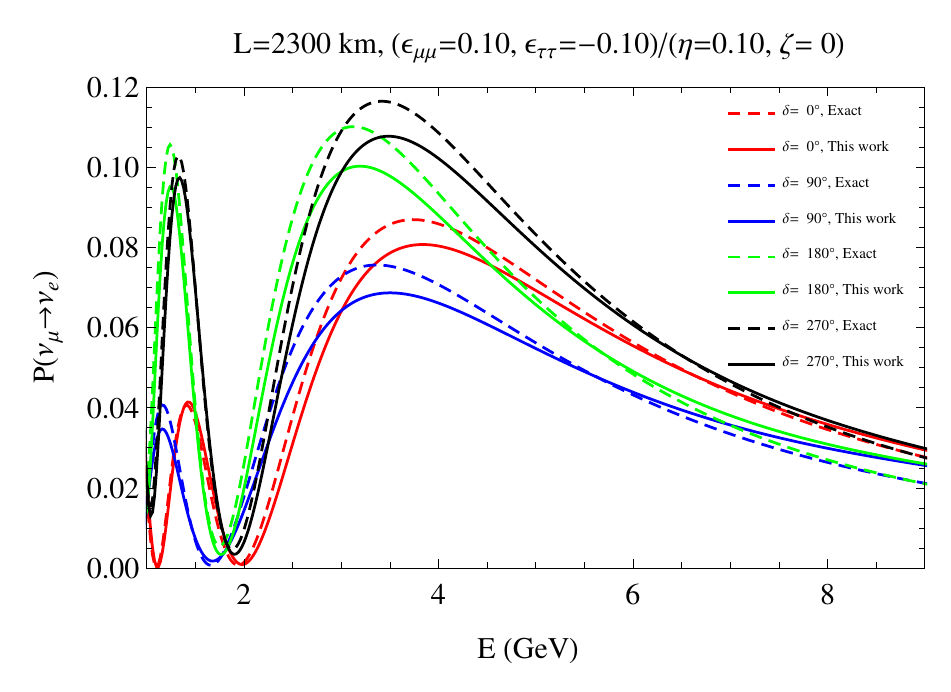}
\includegraphics[width=7.5cm,height=5.5cm]{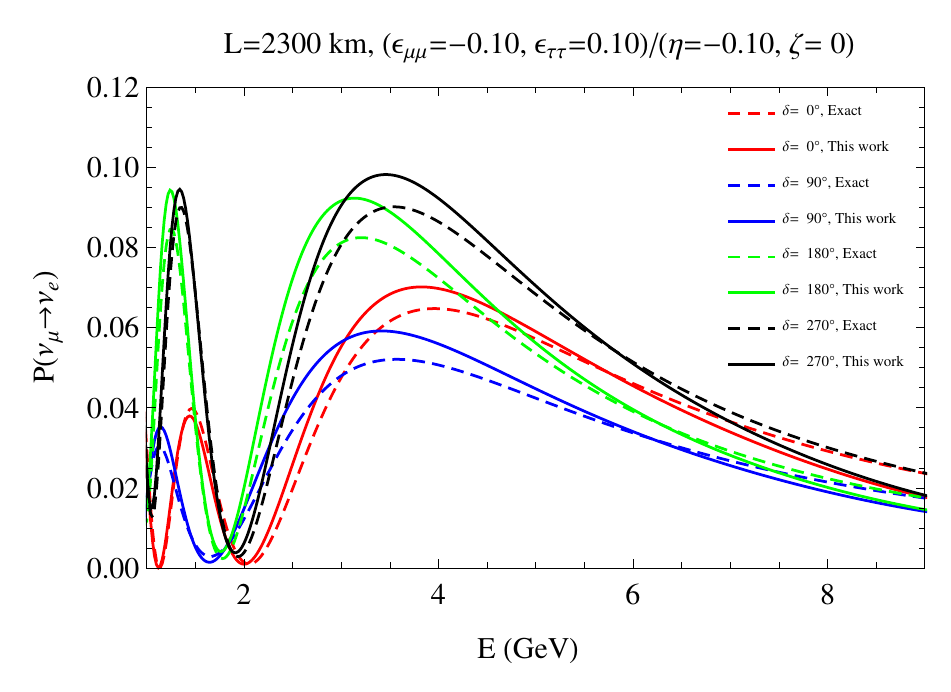}
\caption{Comparison of our analytical expressions (solid curves) 
to the exact numerical results (dashed curves) for four different values 
of the CP-violating phase $\delta$ at 2300 km assuming 
$\theta_{23}=40^{\circ}$ and $\delta m^2_{31}>0$.}
\label{fig:comparison-CP}
\end{figure}

The accuracy of our analytical approximations as compared to the exact numerical results 
for different vacuum values of $\theta_{23}$ is demonstrated in Fig.~\ref{fig:comparison-theta23}. 
Here we consider the minimum (35$^{\circ}$) and maximum (55$^{\circ}$) values of 
$\theta_{23}$ which are allowed in the 3$\sigma$ range \cite{GonzalezGarcia:2012sz}. 
We also present the results for the maximal mixing choice. All the plots in Fig.~\ref{fig:comparison-theta23} 
have been generated assuming $\delta=0^{\circ}$ and normal mass hierarchy ($\delta m^2_{31}>0$).
We consider the same choices of $\eta$ and $\zeta$ as in Fig.~\ref{fig:comparison} and results
are given for 2300~km (upper panels) and 8770~km (lower panels) baselines. As is evident, our
approximation provides satisfactory match with exact numerical results for different values of 
$\theta_{23}$. 

Fig.~\ref{fig:comparison-CP} compares our approximate probability expressions 
(solid curves) against the exact numerical results (dashed curves) assuming four different values 
of the CP-violating phase $\delta$ (0, 90, 180, and 270 degrees) at 2300 km. Here, we consider 
$\theta_{23}=40^{\circ}$ and $\delta m^2_{31}>0$. These plots clearly show that our approximate 
expressions work quite well even for non-zero $\delta$ and can predict almost accurate $L/E$ 
patterns of the oscillation probability for finite $\delta$ and $\eta$. It also suggests that one can 
explain qualitatively the possible correlations and degeneracies between $\delta$ and $\eta$ 
using these analytical expressions which cannot be tackled with numerical studies.

\begin{figure}[t]
\subfigure[$\eta=0.1,\zeta=0$]{\includegraphics[width=7.5cm,height=5.5cm]{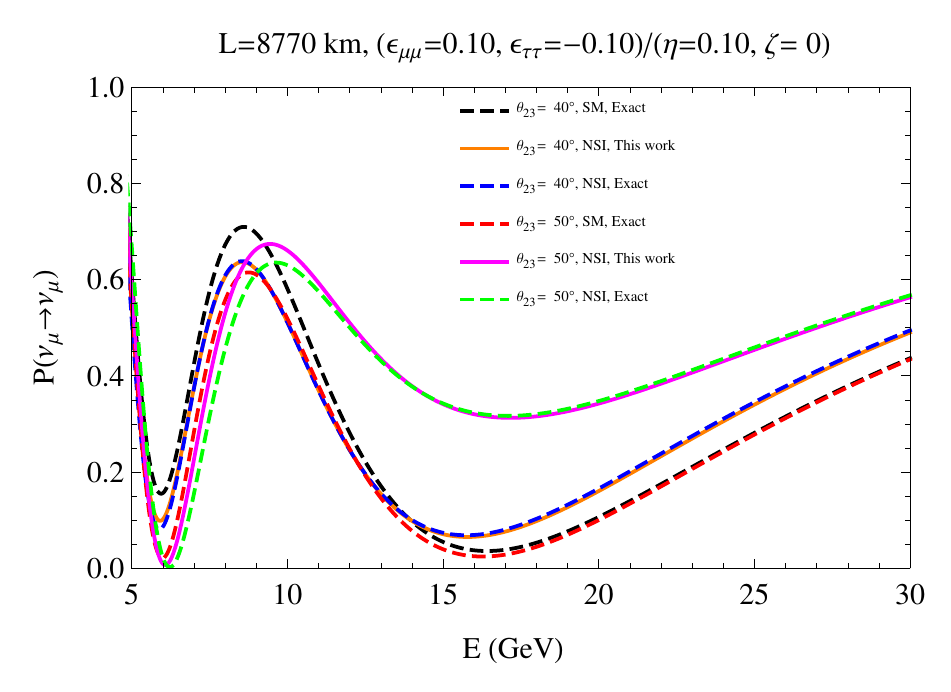}}
\subfigure[$\eta=-0.1,\zeta=0$]{\includegraphics[width=7.5cm,height=5.5cm]{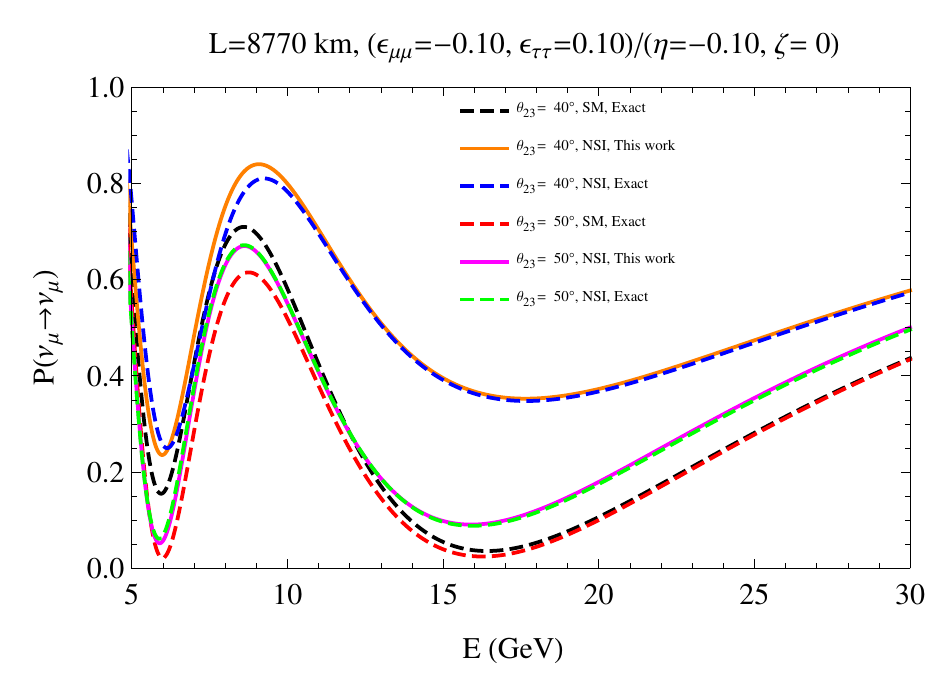}}
\caption{$\nu_\mu\rightarrow\nu_\mu$ survival probability as a function of neutrino energy $E$ 
in GeV for two different values of $\theta_{23}$ at 8770 km baseline. Comparison between the 
analytical and numerical results assuming $\eta=0.1, \zeta=0$ (left panel) and $\eta=-0.1, \zeta=0$ 
(right panel). The standard three-flavor oscillation probabilities in matter without NSI's are also shown.
In both the panels, we assume $\delta=0^{\circ}$ and $\delta m^2_{31}>0$.}
\label{fig:comparison-survival}
\end{figure}

In Fig.~\ref{fig:comparison-survival}, we plot the $\nu_\mu\rightarrow\nu_\mu$ survival probability 
in the presence of NSI for two different vacuum values of $\theta_{23}$ (40$^{\circ}$ and 50$^{\circ}$) at 
8770~km. 
We show the matching between the analytical and numerical results assuming 
$\eta=0.1, \zeta=0$ (left panel) and $\eta=-0.1, \zeta=0$ (right panel). We also give the exact 
numerical standard three-flavor oscillation probabilities in matter without NSI's so that one can 
compare them with the finite $\eta$ case. In both the panels, we assume $\delta=0^{\circ}$ and 
$\delta m^2_{31}>0$. Fig.~\ref{fig:comparison-survival} shows that our approximate expressions 
match quite nicely with the numerical results. Note that at higher energies, the impact of NSI's are 
quite large in the $\nu_\mu\rightarrow\nu_\mu$ survival channel and there is a substantial difference 
in the standard and NSI probabilities for both the choices of $\theta_{23}$ which can be 
probed in future long-baseline \cite{Kopp:2008ds,Coloma:2011rq} and atmospheric 
\cite{Chatterjee:2014gxa,Mocioiu:2014gua,Choubey:2014iia} neutrino oscillation experiments.

Another important point to be noted that in the absence of NSI, the standard probability curves
for both the values of $\theta_{23}$ almost overlap with each other at higher energies, whereas
with NSI, there is a large separation between them. 
It immediately suggests that the corrections in the probability expressions due to the NSI terms depend significantly on whether the 
vacuum value of $\theta_{23}$ lies below or above 45$^{\circ}$ \cite{Fogli:1996pv}.
Fig.~\ref{fig:comparison-survival} also 
indicates that there are degeneracies between {\it the octant of $\theta_{23}$} and 
{\it the sign of NSI parameter $\eta$} 
for a given choice of hierarchy. For an example, 
$P_{\mu\mu} (\theta_{23} = 50^{\circ}, \eta = 0.1)$ in the left panel is almost same with 
$P_{\mu\mu} (\theta_{23} = 40^{\circ}, \eta = -0.1)$ in the right panel for the energies
above 12 GeV or so. Again, $P_{\mu\mu} (\theta_{23} = 40^{\circ}, \eta = 0.1)$ in the left panel 
matches quite well with $P_{\mu\mu} (\theta_{23} = 50^{\circ}, \eta = -0.1)$ in the right panel. 
These kinds of degeneracies can be well explained qualitatively with the help of our analytical 
expressions. We discuss this issue in detail in the next section which is one of the highlights 
of this work.

\section{Possible Applications of Analytical Expressions}
\label{sec:applications}

In this section, we discuss the utility of our analytical probability expressions to determine the 
conditions for which the impact of the NSI parameter $\eta$ becomes significant. We also
give simple and compact analytical expressions to show the possible correlations and 
degeneracies between $\theta_{23}$ and $\eta$ under such situations. We begin our
discussion with electron neutrinos.

\subsection{$\nu_e\rightarrow\nu_\alpha$ Oscillation Channels}
\label{nue-to-nu-alpha}

\begin{figure}[t]
\subfigure[Normal and Inverted Hierarchy]{\includegraphics[height=5cm]{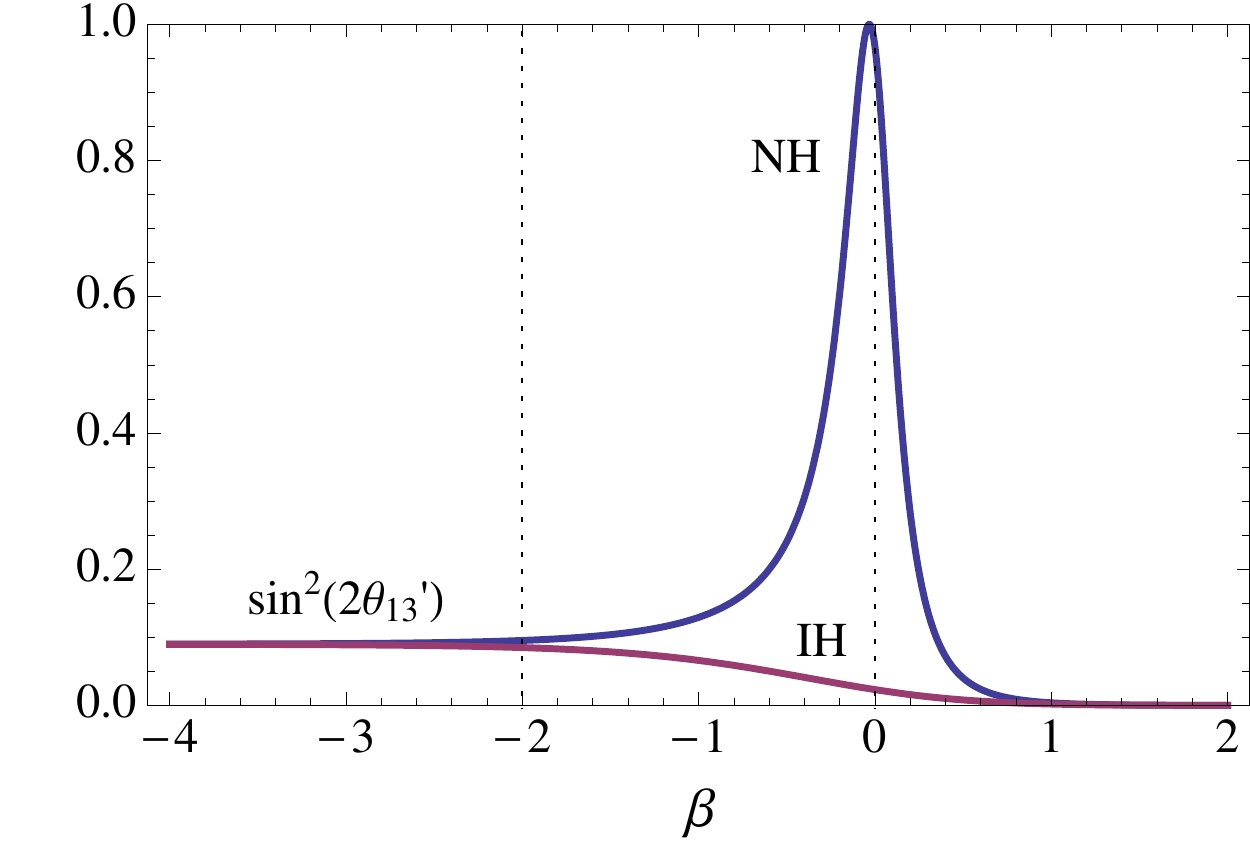}}
\subfigure[Normal Hierarchy]{\includegraphics[height=5cm]{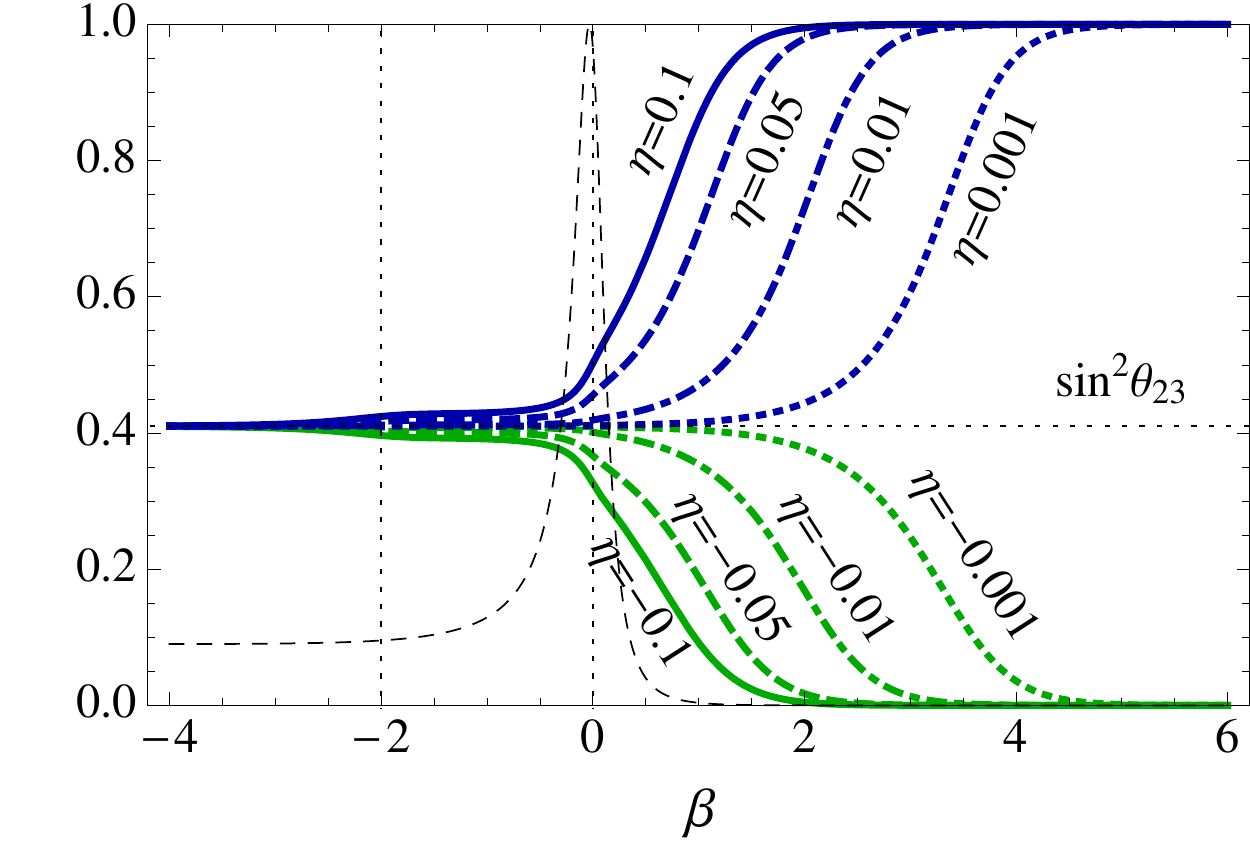}}
\caption{$\beta$-dependence of (a) $\sin^2(2\theta'_{13})$ for the normal 
($\delta m^2_{31}>0$) and inverted ($\delta m^2_{31}<0$) hierarchies, and (b) that of 
$s^{\prime 2}_{23}$ for various values of $\eta$ for the normal hierarchy case.
The dashed peak in (b) indicates the behavior of $\sin^2(2\theta'_{13})$ when
$\delta m^2_{31}>0$.}
\label{sinsquared2theta13plot}
\end{figure}

Let us first consider the $\nu_e\rightarrow\nu_\alpha$ ($\alpha=e,\mu,\tau$) oscillation 
channels in matter in the presence of the NSI parameter $\eta$. Since we expect the effect 
of $\eta$ to become important in the range $\beta\agt 0$, we set $s'_{12}\approx 1$, 
$c'_{12}\approx 0$ (which is valid in the range $\beta\gg -2$, see 
Fig.~\ref{s12primec12primeplot}(a)), which leads to the following simple expressions:
\begin{eqnarray}
\tilde{P}(\nu_e\rightarrow\nu_e)
& \approx &
1 - \sin^2(2\theta'_{13})\;\sin^2\dfrac{\tilde{\Delta}_{32}}{2}
\;,
\label{Pee}
\\
\tilde{P}(\nu_e \rightarrow \nu_\mu)
& \approx &
s^{\prime 2}_{23}\,\sin^2(2\theta'_{13})\;\sin^2\dfrac{\tilde{\Delta}_{32}}{2}
\;,
\label{Pemu}   
\\    
\tilde{P}(\nu_e \rightarrow \nu_\tau)
& \approx &
c^{\prime 2}_{23}\,\sin^2(2\theta'_{13})\;\sin^2\dfrac{\tilde{\Delta}_{32}}{2}
\;.
\label{Petau}
\end{eqnarray}
Recall that when $\delta m^2_{31}>0$, the effective mixing angle $\theta_{13}'$ 
increases monotonically toward $\pi/2$, going through $\pi/4$ around $\beta\sim 0$, 
while in the $\delta m^2_{31}<0$ case, it decreases monotonically toward $0$, 
cf. Fig.~\ref{phiplot}(b). This will cause $\sin^2(2\theta'_{13})$ to peak prominently 
around $\beta\sim 0$ for the normal hierarchy case, but not for the inverted hierarchy 
case as shown in Fig.~\ref{sinsquared2theta13plot}(a). As discussed in 
Ref.~\cite{Agarwalla:2013tza}, demanding that $\sin^2(\tilde{\Delta}_{32}/2)$ also 
peaks at the same energy leads to the requirements of $L\sim 10000\,\mathrm{km}$ 
and $E\sim 7\,\mathrm{GeV}$. Thus, measuring $\tilde{P}(\nu_e\rightarrow\nu_e)$ 
survival probability at this baseline and energy will allow us to discriminate between 
the normal and inverted mass hierarchies irrespective of the value of 
$\theta_{23}$ \cite{Agarwalla:2006gz}. Also, around $\beta\sim 0$, $\tilde{\Delta}_{32}$ 
is not affected by $\eta$ provided $\eta\alt 0.1$ (see middle and bottom panels of 
Fig.~\ref{chiplot}), allowing this channel to determine the mass hierarchy free from any
NSI effect.

\begin{figure}[t]
\subfigure[]{\includegraphics[height=5cm]{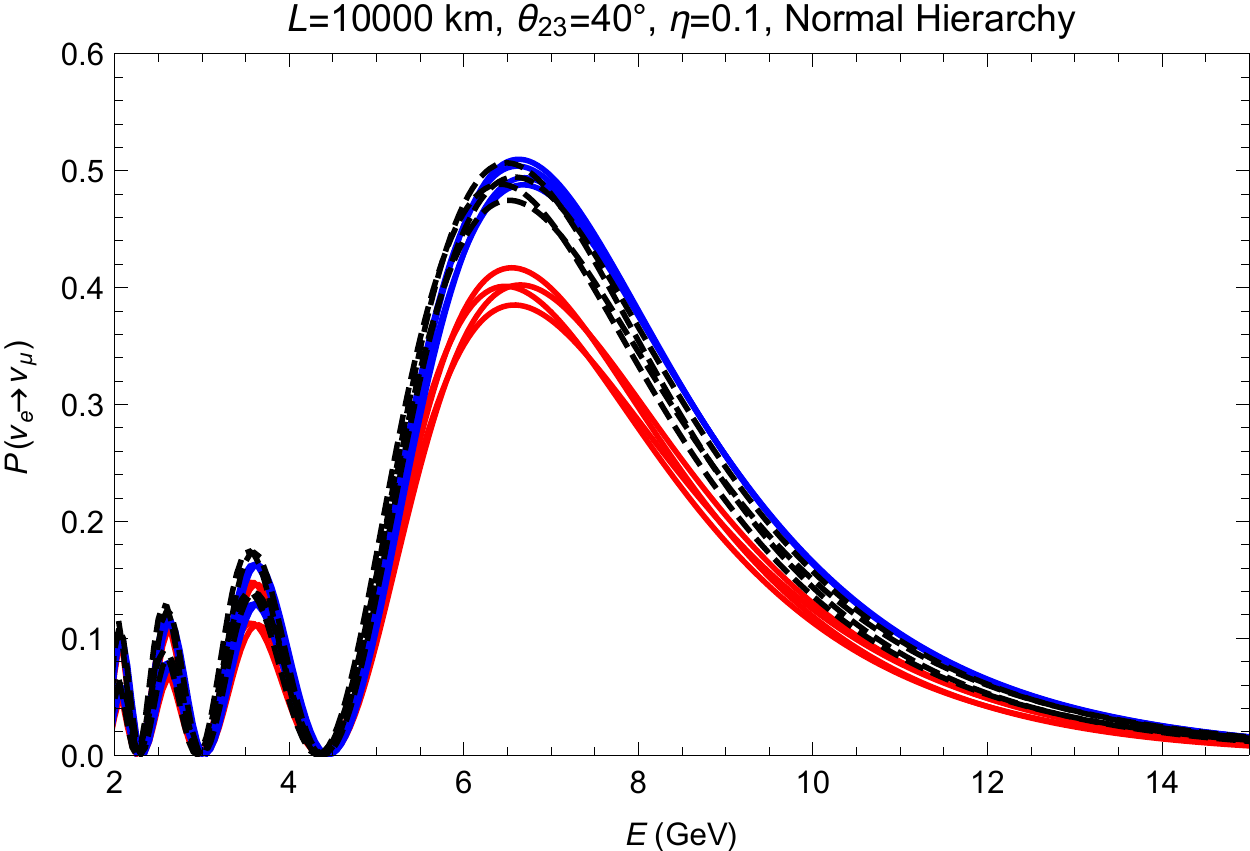}}
\subfigure[]{\includegraphics[height=5cm]{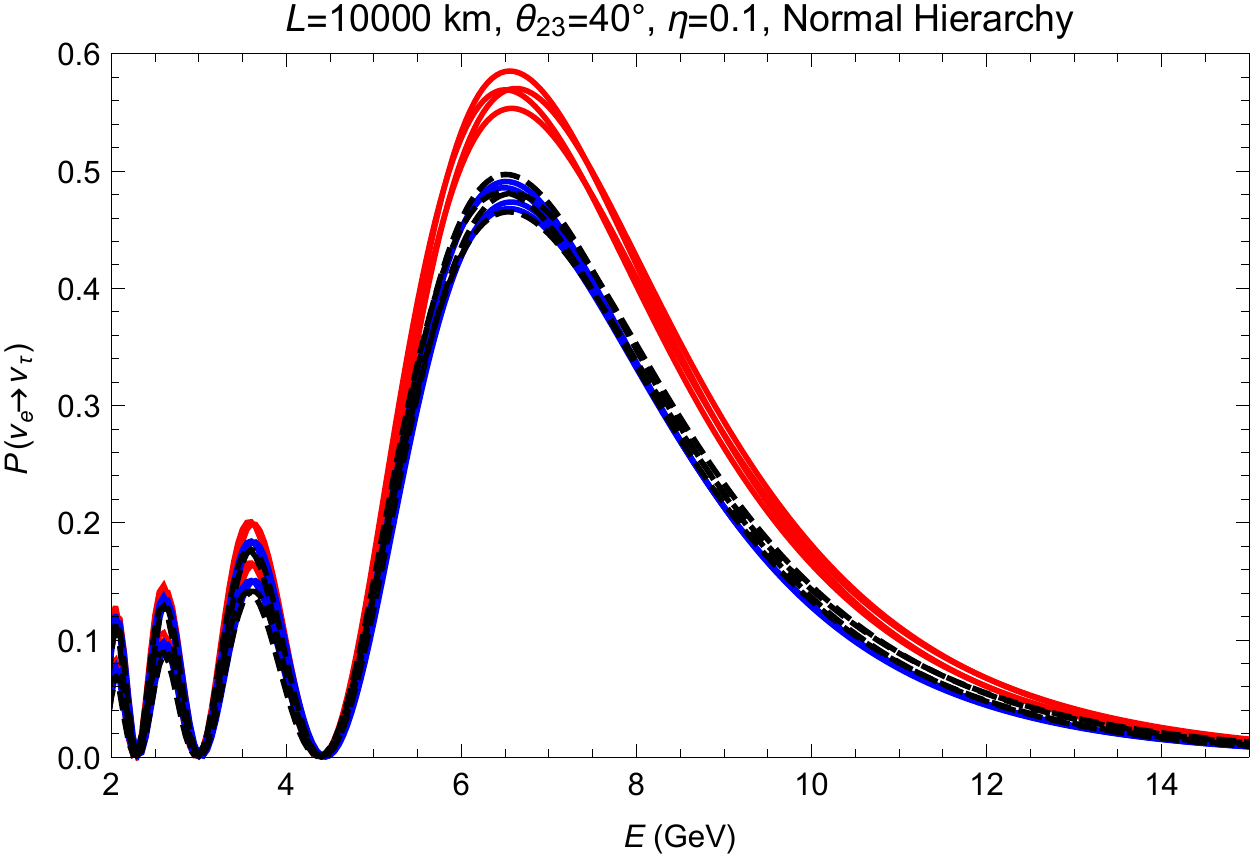}}
\caption{The oscillation probabilities (a) $P(\nu_e\rightarrow\nu_\mu)$
and (b) $P(\nu_e\rightarrow\nu_\tau)$ for the normal hierarchy case with 
$s_{23}^2=0.4$. The red lines are the standard oscillation probabilities with 
$\eta=0$ and $\delta=0, 90, 180,$ and $270$ degrees. The blue lines are our 
approximate analytical results with $\eta=0.1$ and the black dashed lines 
are the numerically calculated probabilities with $\eta=0.1$. In both the cases,
we plot the probabilities for four values of $\delta=0, 90, 180,$ and $270$ 
degrees. Note that $P(\nu_e\rightarrow\nu_\mu)$ is enhanced by about $\eta$, 
while $P(\nu_e\rightarrow\nu_\tau)$ is suppressed by an equal amount.}
\label{e2muAnde2tau}
\end{figure}

To see the effect of $\eta$, we need to observe the running of $\theta_{23}'$.
This could be visible in the $\nu_e\rightarrow\nu_\mu$ and 
$\nu_e\rightarrow\nu_\tau$ appearance channels for the normal hierarchy case 
as a change in the heights of the oscillation peaks around $E\sim 7\,\mathrm{GeV}$ 
provided $\theta'_{23}$ deviates sufficiently from the vacuum value of $\theta_{23}$ 
at that energy. The running of $s^{\prime 2}_{23}$ for various values of $\eta$ is 
depicted in Fig.~\ref{sinsquared2theta13plot}(b). At $\hat{a}\sim\delta m^2_{31}$, 
Eq.~(\ref{theta23primeNeutrino}) tells us that
\begin{equation}
s^{\prime 2}_{23} \;\approx\; s^2_{23} + \eta\;,
\label{shift-in-theta23}
\end{equation}
showing the possible shift in $\theta_{23}$ due to $\eta$ in a simple and compact 
fashion which clearly establishes the merit of our analytical approximation.  
Eq.~(\ref{shift-in-theta23}) also shows compactly possible correlations 
and degeneracies between $\theta_{23}$ and $\eta$. 
Such correlations and degeneracies could also be found numerically, but 
the reason for those features will not be so transparent. 
Note that Eq.~(\ref{shift-in-theta23}) suggests that a positive value of $\eta$ 
would enhance $\tilde{P}(\nu_e\rightarrow\nu_\mu)$ while suppressing 
$\tilde{P}(\nu_e\rightarrow\nu_\tau)$, and a negative $\eta$ would do the opposite.
Fig.~\ref{e2muAnde2tau} confirms this feature where we plot the 
$\nu_e\rightarrow\nu_\mu$ ($\nu_e\rightarrow\nu_\tau$) transition probability in the 
left (right) panel. 
In both panels, the standard oscillation probabilities in matter 
without NSI's for four different values of CP phase\footnote{Fig.~\ref{e2muAnde2tau}
shows that the impact of the CP phase $\delta$ is quite weak around $\beta\sim 0$. 
In this region, $\theta'_{12}$ approaches to $\pi/2$ so that $s'_{12}\approx 1$ and
$c'_{12}\approx 0$. Therefore, the Jarlskog Invariant \cite{Jarlskog:1985ht} in matter
$J = s'_{12}c'_{12}s'_{13}c'^{2}_{13}s'_{23}c'_{23}\sin\delta$ almost approaches to 
zero diminishing the effect of $\delta$. This argument works even in the presence 
of the NSI parameter $\eta$. Our simple and compact approximate probability 
expressions given by Eq.~(\ref{Pemu}) and Eq.~(\ref{Petau}) also validate this point 
as there are no $\delta$-dependent terms in these expressions.}
$\delta$ are given by the solid red lines. Blue lines (black dashed lines) depict the 
approximate analytical (exact numerical) probabilities with $\eta = 0.1$ and four 
different values of $\delta$. Fig.~\ref{e2muAnde2tau} infers that the effect of $\eta$ 
is visible in these channels provided the vacuum value of $\theta_{23}$ is sufficiently 
well known and $\eta$ is also large enough.

\subsection{$\nu_\mu\rightarrow\nu_\alpha$ Oscillation Channels}
\label{numu-to-nu-alpha}

\begin{figure}[t]
\subfigure[Normal Hierarchy]{\includegraphics[height=5cm]{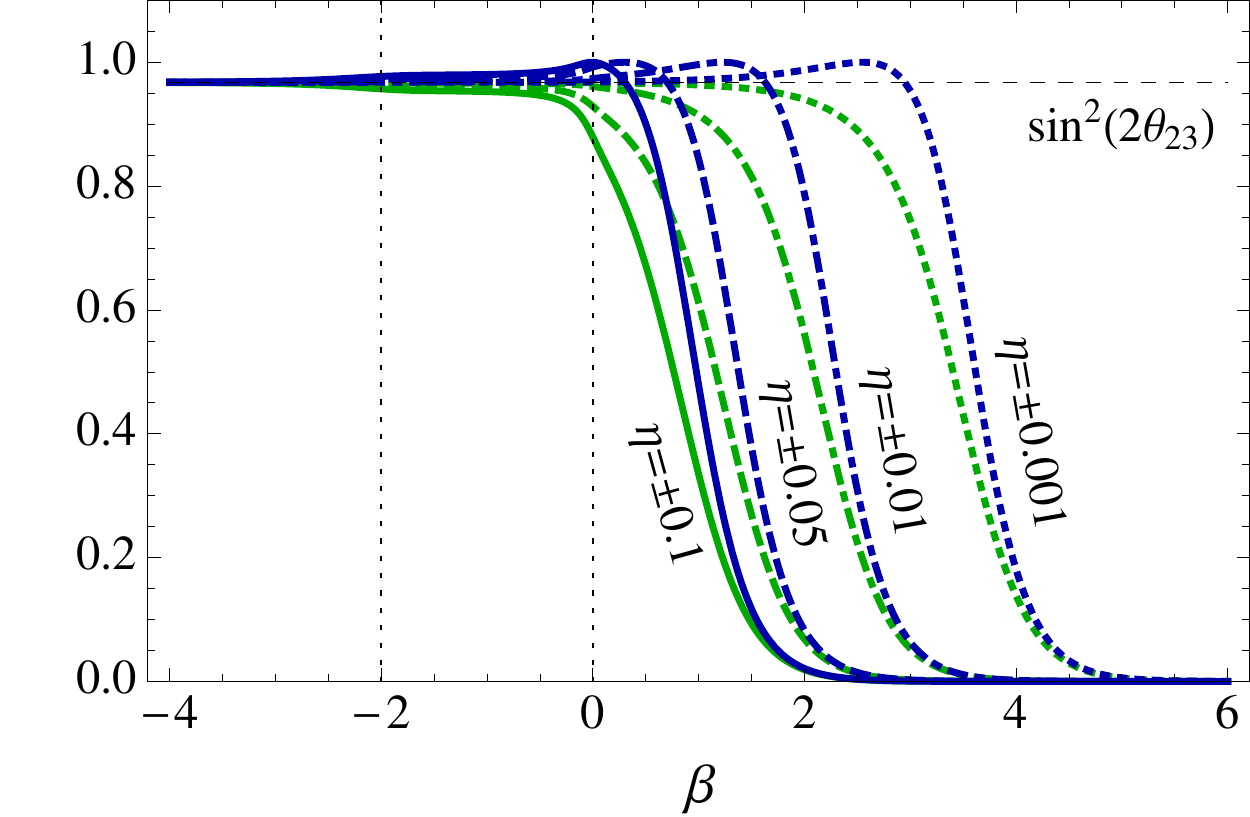}}
\subfigure[Inverted Hierarchy]{\includegraphics[height=5cm]{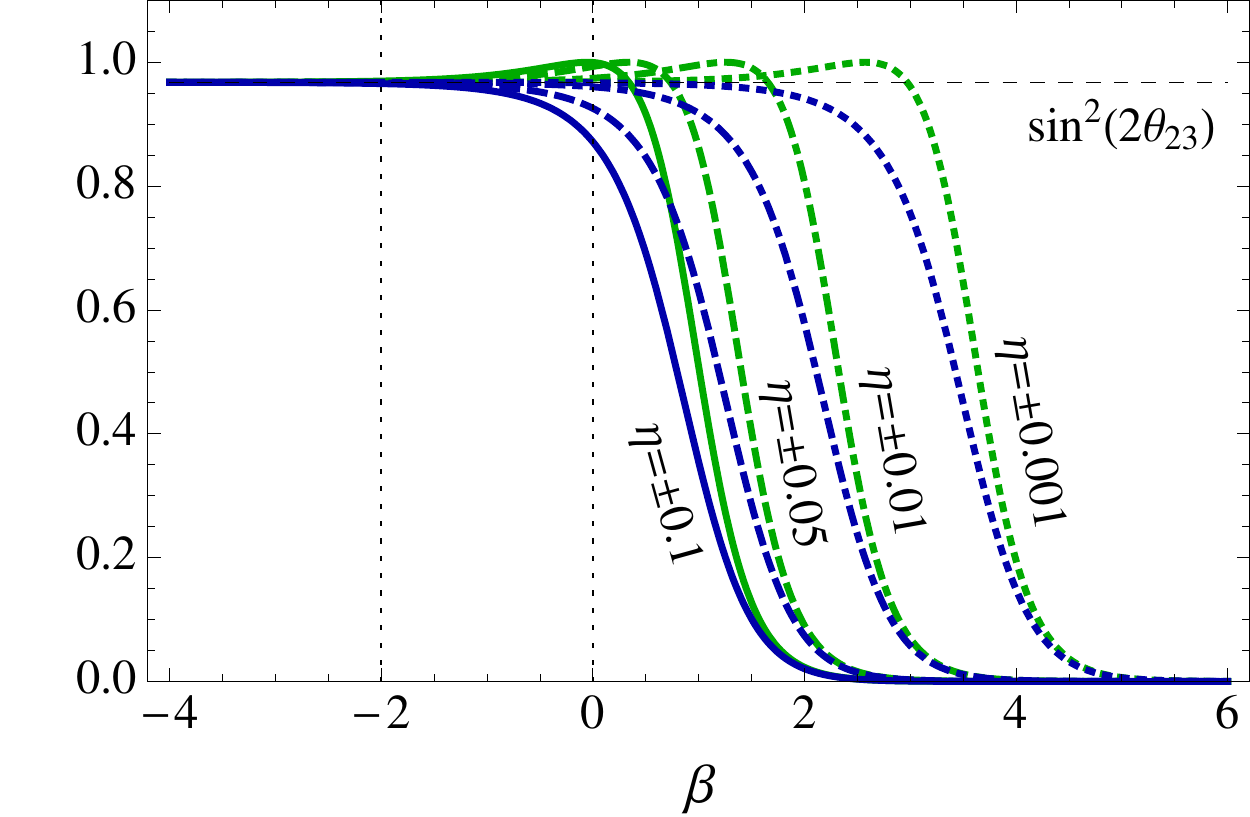}}
\caption{The running of $\sin^2(2\theta'_{23})$ for various values of 
$\eta$ for the (a) normal and (b) inverted mass hierarchies with
$s^2_{23}=0.4$. The blue/green lines indicate positive/negative values 
of $\eta$.}
\label{SinSquared2theta23primePlot}
\end{figure}

Next, let us consider $\nu_\mu\rightarrow\nu_\alpha$ $(\alpha=\mu,\tau)$
oscillation channels. In addition to assuming $\beta\gg -2$, which allows us 
to set $s'_{12}\approx 1$, $c'_{12}\approx 0$, we further restrict our 
consideration to the range $\beta\agt 1$, which allows us to set 
$s'_{13}\approx 1$, $c'_{13}\approx 0$ or $s'_{13}\approx 0$, $c'_{13}\approx 1$
depending on whether $\delta m^2_{31}>0$ or $\delta m^2_{31}<0$ 
(see Fig.~\ref{phiplot}(b)). With these conditions, we obtain the following 
simple expressions: 
\begin{eqnarray}
\tilde{P}({\nu}_\mu\rightarrow{\nu}_\mu)
& \approx &
1
- 4 s_{13}^{\prime 2} s_{23}^{\prime 2}(1  - s_{13}^{\prime 2} s_{23}^{\prime 2} )
 \sin^2\frac{\tilde{\Delta}_{21}}{2}
- 4 c_{13}^{\prime 2} s_{23}^{\prime 2} (1 - c_{13}^{\prime 2} s_{23}^{\prime 2})
\sin^2\frac{\tilde{\Delta}_{31}}{2} \cr
& & \qquad\qquad\quad
+ 2 s_{13}^{\prime 2} c_{13}^{\prime 2} s_{23}^{\prime 4} 
\left( 4\sin^2\dfrac{\tilde{\Delta}_{21}}{2}\sin^2\dfrac{\tilde{\Delta}_{31}}{2}
      + \sin\tilde{\Delta}_{21} \sin\tilde{\Delta}_{31} \right) 
\label{Pmumu1}
\\
& & \xrightarrow{\beta\agt 1} \;
\left\{
\begin{array}{ll}
1 - \sin^2(2\theta_{23}') \sin^2\dfrac{\tilde{\Delta}_{21}}{2}
\qquad & \qquad (\delta m^2_{31}>0) \\
1 - \sin^2(2\theta_{23}') \sin^2\dfrac{\tilde{\Delta}_{31}}{2}
\qquad & \qquad (\delta m^2_{31}<0) 
\end{array}
\right.
\;, 
\label{Pmumu2} 
\\
& & \phantom{[]}\cr
\tilde{P}(\nu_\mu \rightarrow \nu_\tau)
& \approx &
\sin(2\theta'_{23})
\left[
 s^{\prime 4}_{13} \sin^2\frac{\tilde{\Delta}_{21}}{2}
+c^{\prime 4}_{13} \sin^2\frac{\tilde{\Delta}_{31}}{2} 
\right.
\cr
& & \qquad\qquad\quad
\left.
+\;s_{13}^{\prime 2}c^{\prime 2}_{13}
      \left(2\sin^2\frac{\tilde{\Delta}_{21}}{2}\sin^2\frac{\tilde{\Delta}_{31}}{2}
           +\dfrac{1}{2}\sin\tilde{\Delta}_{21}\sin\tilde{\Delta}_{31}
      \right) 
\right]
\qquad
\label{Pmutau1}
\\
& & \xrightarrow{\beta\agt 1} \;
\left\{
\begin{array}{ll}
\sin^2(2\theta_{23}') \sin^2\dfrac{\tilde{\Delta}_{21}}{2}
\qquad & \qquad (\delta m^2_{31}>0) \\
\sin^2(2\theta_{23}') \sin^2\dfrac{\tilde{\Delta}_{31}}{2}
\qquad & \qquad (\delta m^2_{31}<0) 
\end{array}
\right.
\;.
\label{Pmutau2} 
\end{eqnarray}
In the absence of $\eta$, $\theta'_{23}$ does not run and 
$\sin^2(2\theta'_{23})$ will maintain its vacuum value close to
one. In the presence of a non-zero $\eta$, however, $\theta'_{23}$ 
will run towards either $\dfrac{\pi}{2}$ or $0$ depending on the sign 
of $\delta m^2_{31}\eta$, as was shown in Fig.~\ref{theta23primeplot},
and $\sin^2(2\theta'_{23})$ will run toward zero in both cases.
This is depicted in Fig.~\ref{SinSquared2theta23primePlot}.

\begin{figure}[t]
\subfigure[]{\includegraphics[height=5cm]{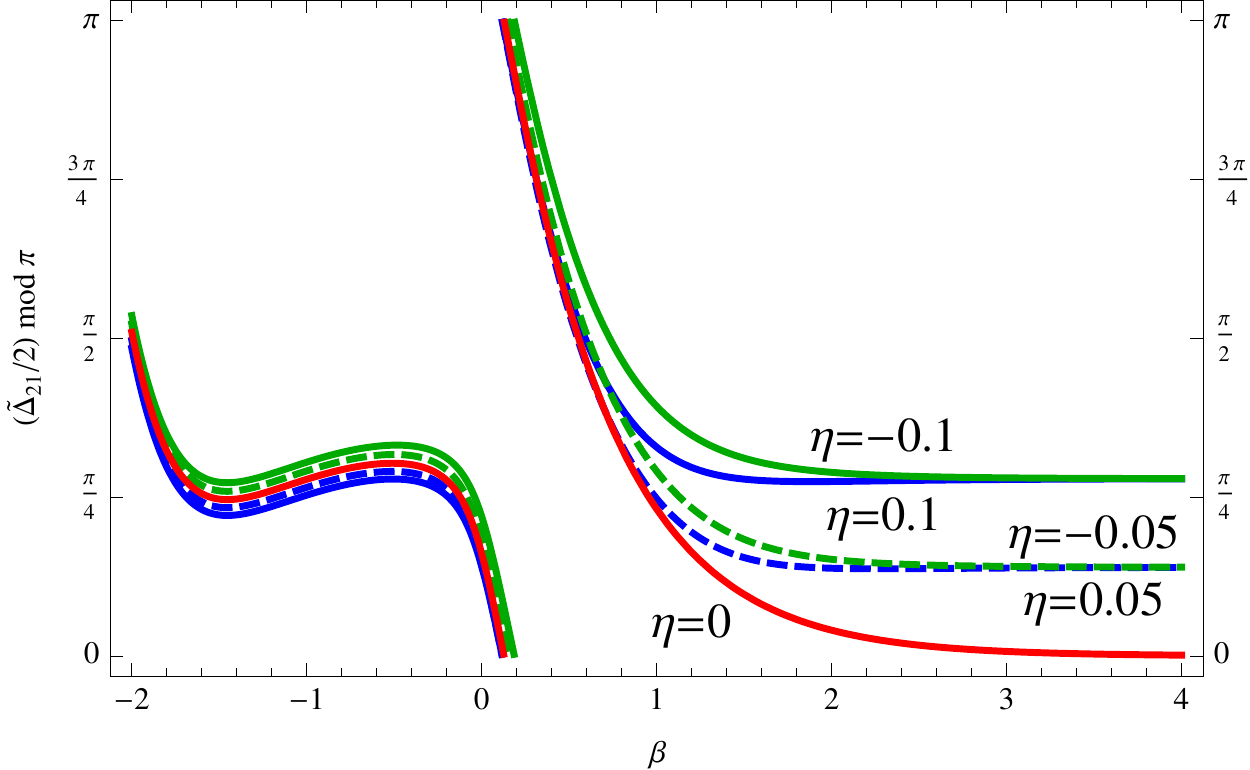}}
\subfigure[]{\includegraphics[height=5cm]{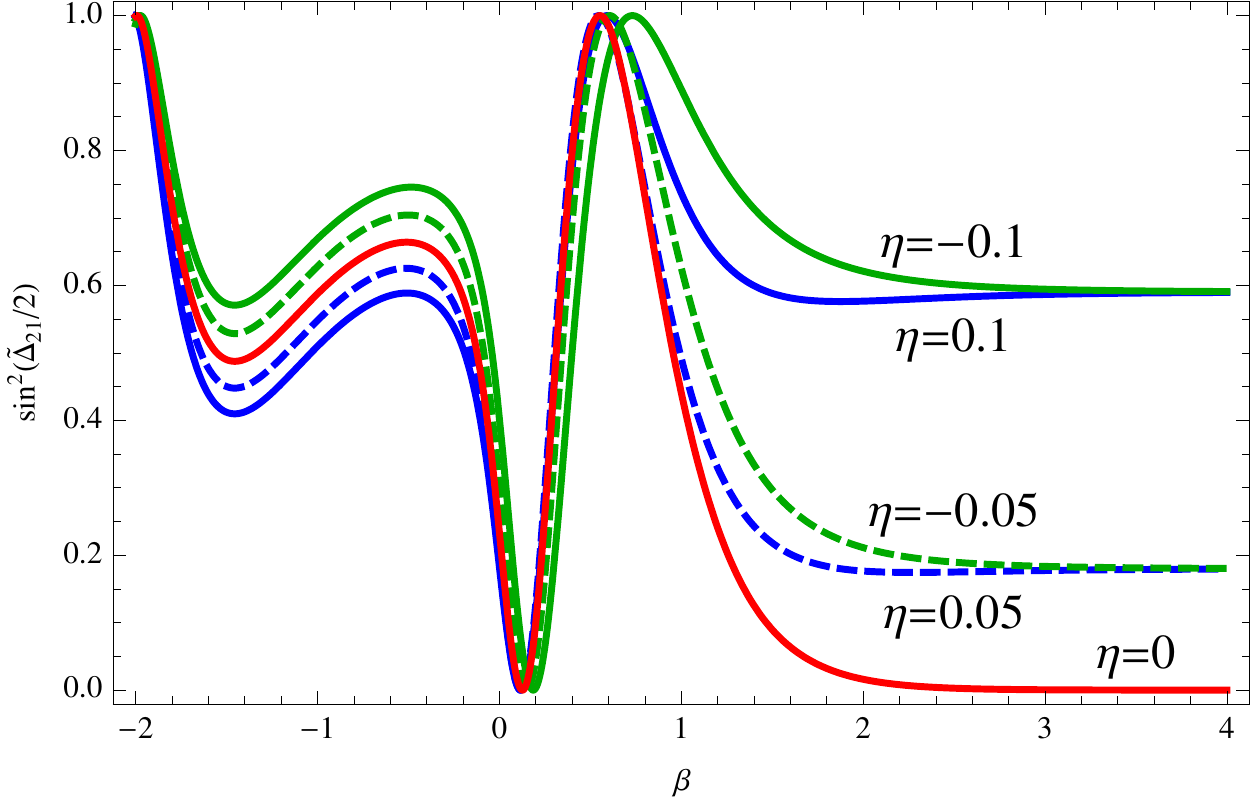}}
\caption{The running of (a) $\tilde{\Delta}_{21}/2$ and
(b) $\sin^2(\tilde{\Delta}_{21}/2)$ for various values of $\eta$ for the normal 
hierarchy case with $L=10,000\,\mathrm{km}$ and $\rho=4.53\,\mathrm{g/cm^3}$.
The blue/green lines indicate positive/negative values of $\eta$,
while the red lines are drawn for $\eta=0$ case.}
\label{tildeDelta21Plot}
\end{figure}

Let us see how the $\Delta$ factors in Eq.~(\ref{Pmumu2}) and 
Eq.~(\ref{Pmutau2}) behave in the range $\beta\agt 1$. For the 
normal hierarchy case, we have
\begin{equation}
\delta\lambda_{21}
\;\approx\;
\left\{
\begin{array}{llll}
\lambda_{X+}'''&-\;\lambda_{X-}'''
& \approx\; 2\hat{a}|\eta| & \qquad (\eta\neq 0) \\
\lambda_{-}''&-\;\lambda_{-}'
& \approx\; \delta m^2_{31}c_{13}^2 \qquad & \qquad (\eta=0)
\end{array}
\right. 
\;,
\end{equation}
while for the inverted hierarchy case, they take the form
\begin{equation}
\delta\lambda_{31}
\;\approx\;
\left\{
\begin{array}{llll}
\lambda_{Y-}''' &-\;\lambda_{Y+}'''
& \approx\; -2\hat{a}|\eta| & \qquad (\eta\neq 0) \\
\lambda_{-}''&-\;\lambda_{-}'
& \approx\; -|\delta m^2_{31}|c_{13}^2 \qquad & \qquad (\eta=0)
\end{array}
\right.
\;. 
\end{equation}
Since the sign of $\delta\lambda_{ij}$ does not affect the value of
$\sin^2(\tilde{\Delta}_{ij}/2)$, both mass hierarchies lead to the same 
asymptotic oscillation probabilities. We will therefore only consider 
the normal hierarchy case in the following. Recalling 
(see Eq.~(\ref{matter-term})) that 
\begin{equation}
\hat{a} \;=\; 2\sqrt{2}G_F N_e E
\;=\; 7.6324\times 10^{-5}(\mathrm{eV}^2)
\left(\dfrac{\rho}{\mathrm{g/cm^3}}\right)
\left(\dfrac{E}{\mathrm{GeV}}\right)
\;,
\end{equation}
(where we have set $\zeta=0$)
and
\begin{eqnarray}
\tilde{\Delta}_{ij}
& = & 2.534
\left(\dfrac{\delta\lambda_{ij}}{\mathrm{eV}^2}\right)
\left(\dfrac{\mathrm{GeV}}{E}\right)
\left(\dfrac{L}{\mathrm{km}}\right)
\cr
& = & 1.934
\left(\dfrac{\delta\lambda_{ij}}{\hat{a}}\right)
\left(\dfrac{\rho}{\mathrm{g/cm^3}}\right)
\left(\dfrac{L}{10^{4}\,\mathrm{km}}\right)
\;,
\label{DeltaIJ}
\end{eqnarray}
we find
\begin{eqnarray}
\lefteqn{
\dfrac{\tilde{\Delta}_{21}}{2}
\;=\; \dfrac{\delta\lambda_{21}}{4E}L
}
\cr
& \approx &
\left\{
\begin{array}{lll}
2\,|\eta|
\left(\dfrac{\rho}{\mathrm{g/cm^3}}\right)
\left(\dfrac{L}{10^{4}\,\mathrm{km}}\right)
&
& \qquad (\eta\neq 0) \\
\left(\dfrac{\delta m^2_{31}}{\hat{a}}\right)
\left(\dfrac{\rho}{\mathrm{g/cm^3}}\right)
\left(\dfrac{L}{10^{4}\,\mathrm{km}}\right)
& \approx \; 30
\left(\dfrac{\mathrm{GeV}}{E}\right)
\left(\dfrac{L}{10^4\,\mathrm{km}}\right)
& \qquad (\eta = 0)
\end{array}
\right. \;. 
\cr
& &
\label{Delta21Over2}
\end{eqnarray}
Therefore, for fixed baseline $L$ and matter density $\rho$,
as the neutrino energy $E$ is increased, $\tilde{\Delta}_{21}/2$ 
damps to zero when $\eta=0$, but asymptotes to a constant value 
proportional to $|\eta|$ when $\eta\neq 0$. This is demonstrated in 
Fig.~\ref{tildeDelta21Plot} for the baseline $L=10000\,\mathrm{km}$
with average matter density $\rho=4.53\,\mathrm{g/cm^3}$.
Consequently, when $\eta=0$, the factor $\sin^2(2\theta_{23}')$ stays 
constant while $\sin^2(\tilde{\Delta}_{21}/2)$ damps to zero
as we increase $E$, while in the $\eta\neq 0$ case, the factor 
$\sin^2(2\theta_{23}')$ damps to zero while $\sin^2(\tilde{\Delta}_{21}/2)$ 
asymptotes to a constant value as $E$ is increased.
In either case, the $\nu_\mu\rightarrow\nu_\tau$ oscillation probability is 
suppressed at high energy. 

\begin{figure}[t]
\subfigure[$s^2_{23}=0.4$]{\includegraphics[width=7.5cm,height=5.5cm]{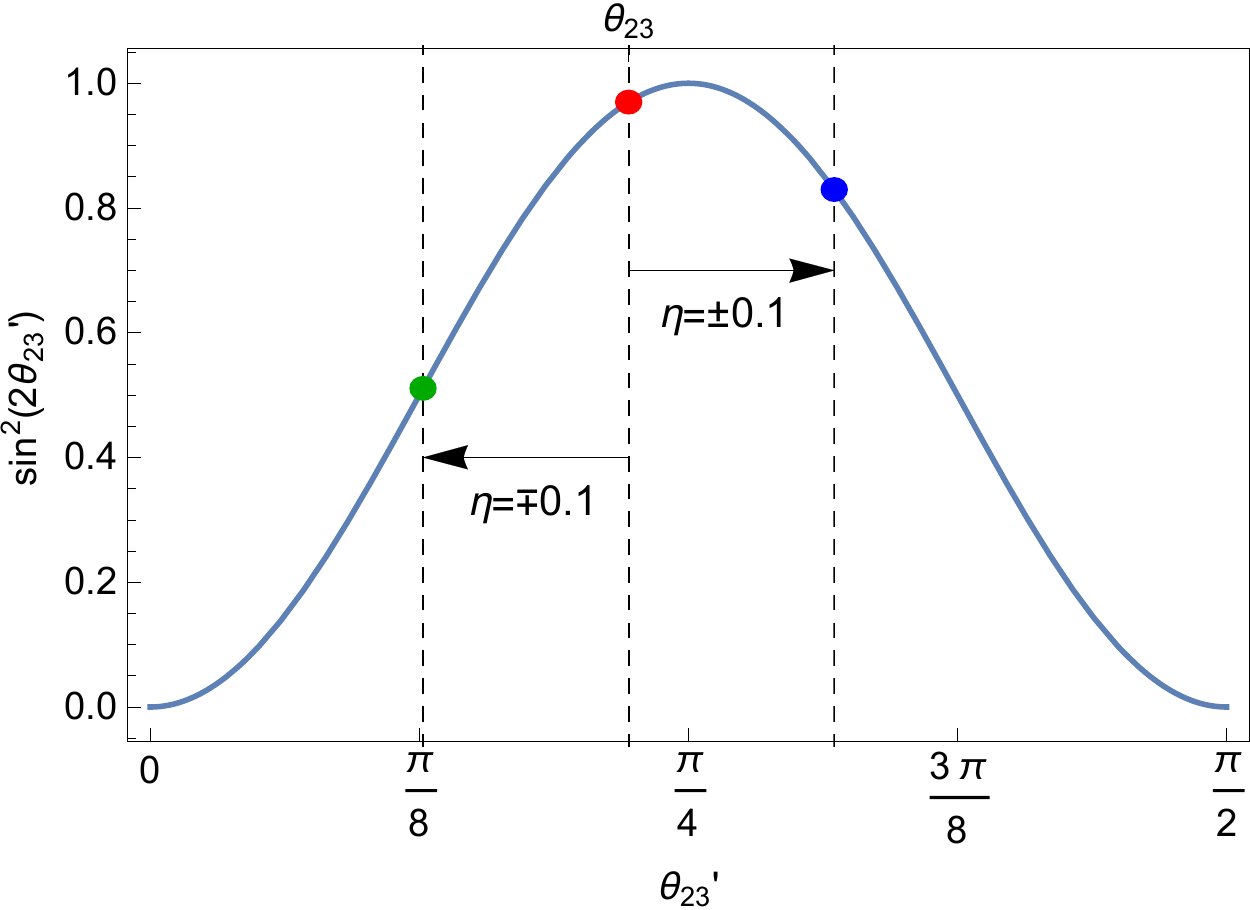}}
\subfigure[$s^2_{23}=0.6$]{\includegraphics[width=7.5cm,height=5.5cm]{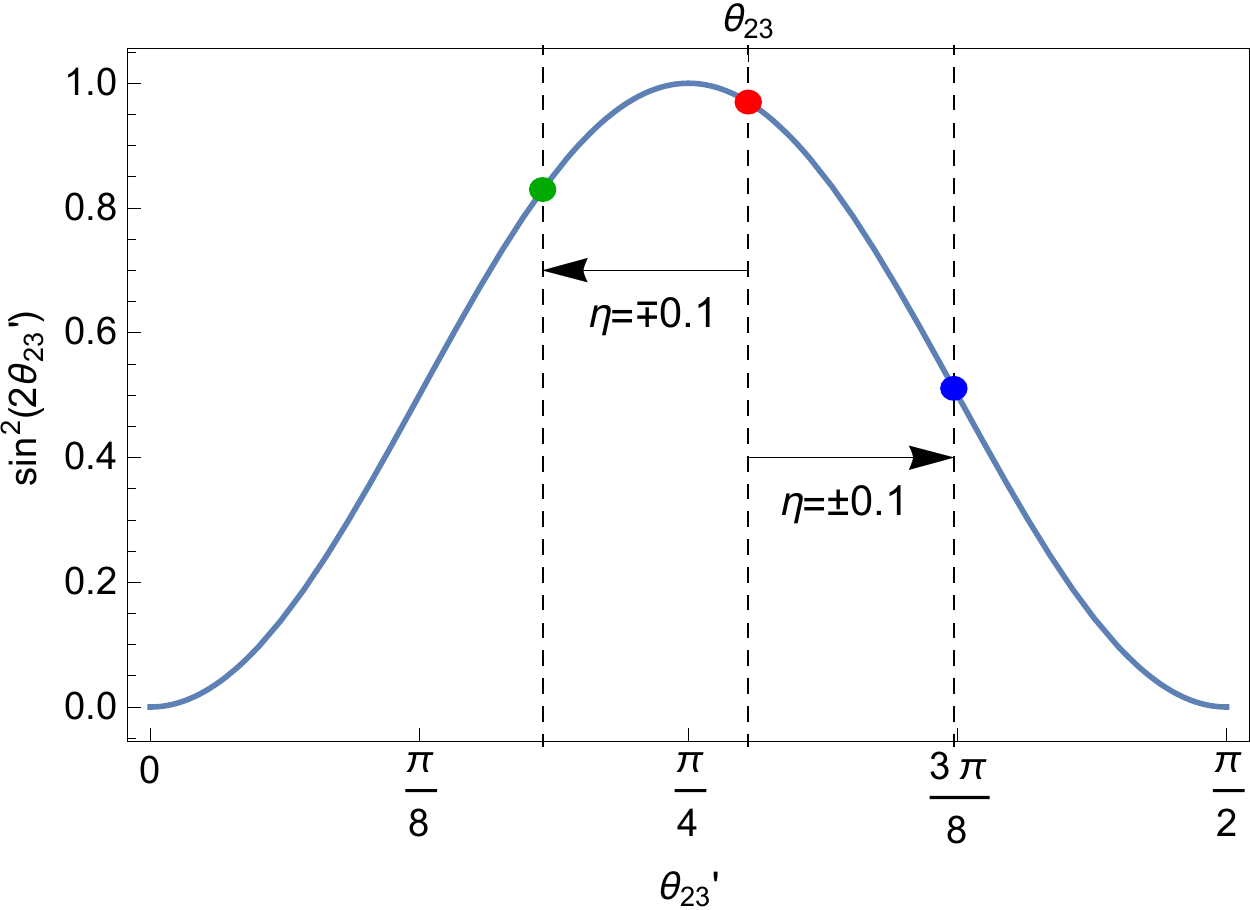}}
\caption{$\theta_{23}'$ and $\sin^2(2\theta'_{23})$ as a function of positive and negative values of $\eta$
for a fixed $s^2_{23}=0.4$ ($s^2_{23}=0.6$) in the left (right) panel. The upper (lower) signs are applicable
for the normal (inverted) hierarchy case.}
\label{theta23-prime}
\end{figure}

\begin{figure}[t]
\subfigure[$P(\nu_\mu\rightarrow\nu_\tau)$]{\includegraphics[height=5cm]{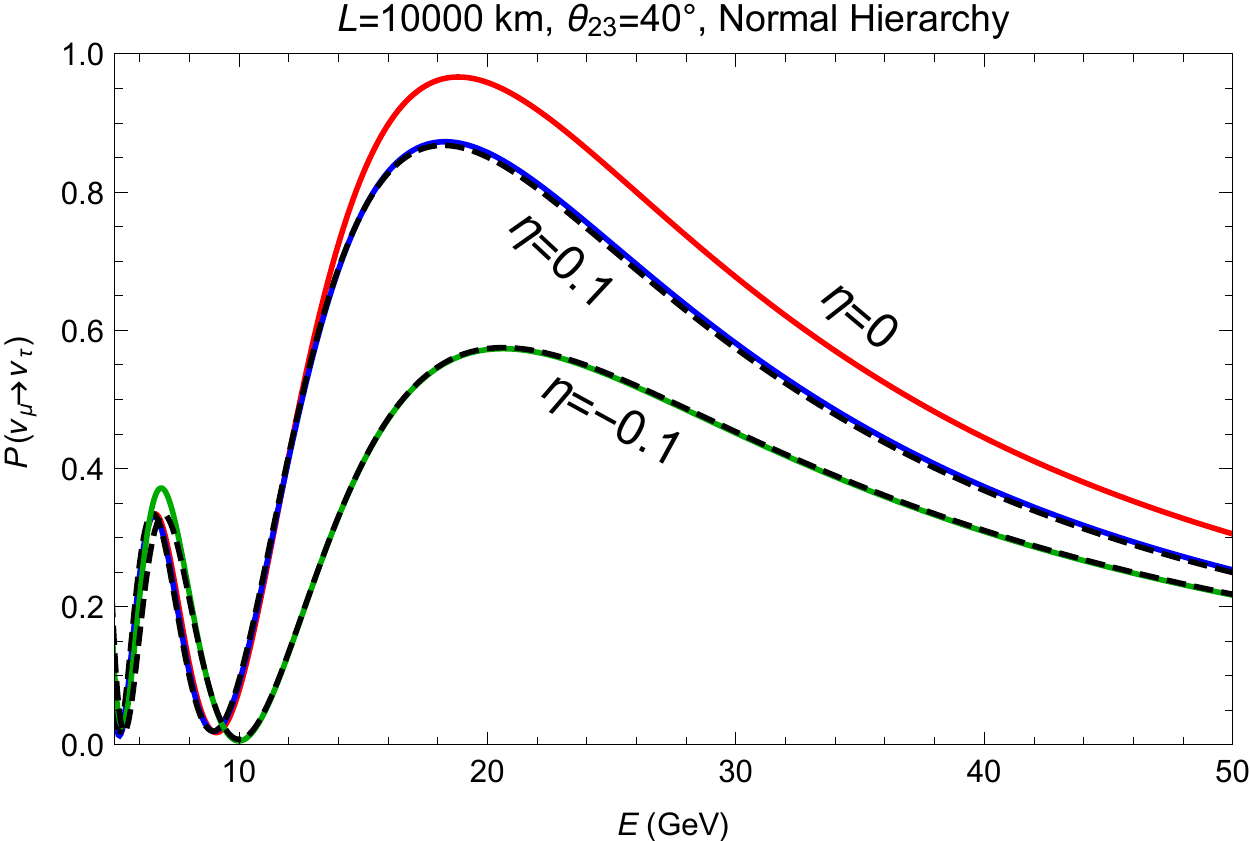}}
\subfigure[$P(\nu_\mu\rightarrow\nu_\mu)$]{\includegraphics[height=5cm]{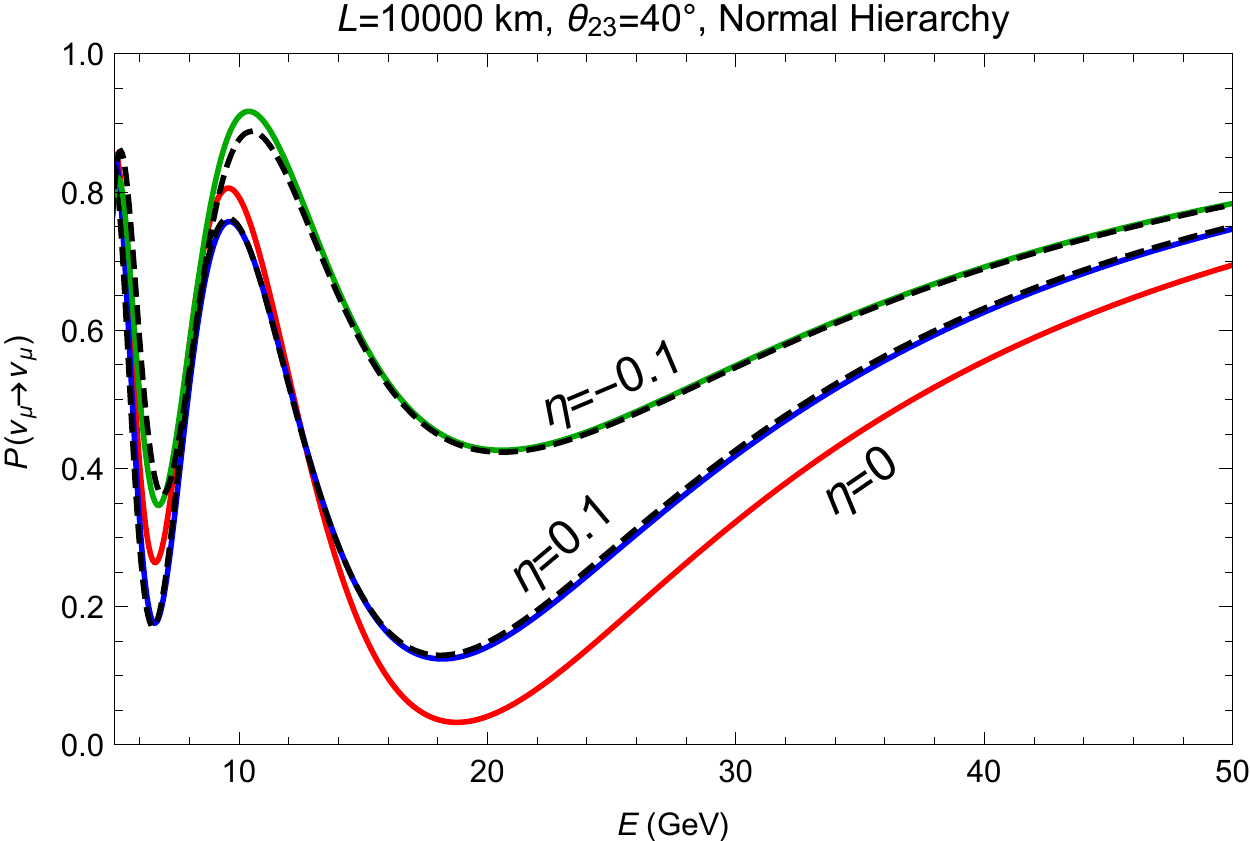}}
\caption{(a) $\nu_\mu\rightarrow\nu_\tau$ and (b) $\nu_\mu\rightarrow\nu_\mu$
oscillation probabilities for $s^2_{23}=0.4$, $L=10,000\,\mathrm{km}$, and
$\rho=4.53\,\mathrm{g/cm^3}$. The red lines are the standard oscillation probabilities 
with $\eta=0$. The solid blue (green) lines are our approximate analytical results with 
$\eta=0.1$ ($\eta=-0.1$). The black dashed lines are the numerically calculated 
probabilities. In both the panels, we take $\delta=0^{\circ}$ and $\delta m^2_{31}>0$.}
\label{mu2muAndmu2tau}
\end{figure}

The difference between the $\eta=0$ and $\eta\neq 0$ cases could manifest 
itself at the $\nu_\mu\rightarrow\nu_\tau$ oscillation peak which happens at 
\begin{equation}
\dfrac{\pi}{2}\;=\;\dfrac{\tilde{\Delta}_{21}}{2}
\quad\rightarrow\quad
E \;\approx\; 20\,\mathrm{GeV}\;,\;
\beta\;\approx\; 0.6\;,
\end{equation}
for $L=10000\,\mathrm{km}$, $\rho=4.53\,\mathrm{g/cm^3}$ case.
$\beta\approx 0.6$ is on the borderline of the applicability of our
$\beta\agt 1$ approximation. Nevertheless, let us examine what our 
approximation suggests. Expanding Eq.~(\ref{theta23primeNeutrino}) 
for small $\hat{a}\eta$, we find 
\begin{equation}
\theta'_{23} \;=\; \theta_{23} 
+ \dfrac{\hat{a}\sin(2\theta_{23})}{\delta m^2_{31}c_{13}^2} \eta 
+ \cdots
\end{equation}
which at $\beta\approx 0.6$ yields
\begin{equation}
\theta'_{23} \;\approx\; \theta_{23} \pm 3\eta
\;.
\end{equation}
In the above equation, upper (lower) sign is applicable for the normal 
(inverted) hierarchy case. It suggests that there is a degeneracy 
between the choices of sign of $\delta m^2_{31}$ and the sign of $\eta$
which give rise to same amount of corrections in $\theta_{23}$.
To observe the shift in the oscillation probability, we keep terms up 
to order $\eta^2$ since the linear term is suppressed due to the fact 
that $\theta_{23}$ is close to $\pi/4$. Therefore, we have
\begin{eqnarray}
\sin(2\theta'_{23}) 
& = & \sin(2\theta_{23}) 
+ 2\cos(2\theta_{23}) \delta\theta_{23} 
- 2\sin(2\theta_{23})(\delta\theta_{23})^2 + \cdots  
\cr
& \approx & \sin(2\theta_{23}) 
\pm  6 \cos(2\theta_{23})\eta
- 18 \sin(2\theta_{23})\eta^2\;.
\end{eqnarray}
For the benchmark value of $s^2_{23}$ = 0.4 (0.6) in the lower (higher) octant, 
above equation can be written as 
\begin{equation}
\sin(2\theta'_{23})
\;\approx\;
\left\{
\begin{array}{llll}
\sin(2\theta_{23})  \pm  \eta  - 18\eta^2 \qquad  (s^2_{23}=0.4) \\
\sin(2\theta_{23})  \mp  \eta  - 18\eta^2 \qquad  (s^2_{23}=0.6)
\end{array}
\right.
\;,
\label{sin2theta23-prime}
\end{equation}
where upper (lower) signs are for the normal (inverted) hierarchy. 
The above equations clearly reveal that for a given choice of hierarchy, 
there are degenerate solutions\footnote{We have already seen this 
degeneracy in the $P(\nu_\mu\rightarrow\nu_\mu)$ oscillation
channel in Fig.~\ref{fig:comparison-survival}.}
of {\it the octant of $\theta_{23}$} and {\it the sign of NSI parameter $\eta$} 
($s^2_{23}=0.4, \eta=\pm 0.1$ and $s^2_{23}=0.6, \eta=\mp 0.1$),
giving rise to same value of $\sin(2\theta'_{23})$.
Fig.~\ref{theta23-prime} shows the variation in $\theta_{23}'$ and 
$\sin^2(2\theta'_{23})$ as a function of positive and negative values of $\eta$ 
for a fixed $s^2_{23}=0.4$ ($s^2_{23}=0.6$) in the left (right) panel. 
The upper (lower) signs correspond to the normal (inverted) hierarchy 
scenario. We can see from the left panel of Fig.~\ref{theta23-prime} 
that assuming normal hierarchy and $s^2_{23}=0.4$, the value of 
$\sin^2(2\theta'_{23})$ gets reduced by substantial amount for 
$\eta = -0.1$ case as compared to $\eta = 0.1$ as given by 
Eq.~(\ref{sin2theta23-prime}). It means that negative values of $\eta$, 
which shifts $\theta_{23}$ further away from $\pi/4$, would lead to a 
larger suppression of the $\nu_\mu\rightarrow\nu_\tau$ oscillation 
probability, and enhancement of the $\nu_\mu\rightarrow\nu_\mu$
survival probability. The left and right panels of 
Fig.~\ref{mu2muAndmu2tau} exactly show this behaviour where
we plot the approximate analytical and exact numerical 
$\nu_\mu\rightarrow\nu_\tau$ (left panel) and 
$\nu_\mu\rightarrow\nu_\mu$ (right panel) probabilities 
for $\eta=\pm 0.1$ assuming $s^2_{23}=0.4$, $\delta m^2_{31}>0$ 
and $L=10000\,\mathrm{km}$, $\rho=4.53\,\mathrm{g/cm^3}$.
The situation gets reversed completely for cases in which 
$s^2_{23} > 0.5$ which is quite evident from the right panel of
Fig.~\ref{theta23-prime} where we consider $s^2_{23}=0.6$.
All these observations in Fig.~\ref{theta23-prime} and 
Fig.~\ref{mu2muAndmu2tau} suggest that our approximate 
calculations are valid. 

\section{Summary and Conclusions}
\label{summary-conclusions}

Analytical studies of the neutrino oscillation probabilities are inevitable 
to understand how neutrino interactions with matter modify the mixing
angles and mass-squared differences in a complicated manner in a 
three-flavor framework. In previous papers \cite{Agarwalla:2013tza,Honda:2006hp}, 
we showed that the neutrino oscillation
probabilities in matter can be well understood if we allow the mixing
angles and mass-squared differences in the standard parametrization 
to `run' with the matter effect parameter $a=2\sqrt{2}G_F N_e E$, 
where $N_e$ is the electron density in matter and $E$ is the neutrino 
energy. We managed to derive simple and compact analytical 
approximations to these running parameters using the 
Jacobi method. We found that for large $\theta_{13}$,
the entire matter effect could be absorbed into the running of the 
effective mass-squared differences and the
effective mixing angles $\theta_{12}$ and $\theta_{13}$, while neglecting 
the running of the mixing angle $\theta_{23}$ and the CP-violating 
phase $\delta$.

In this paper, we extended our analysis to study how the running 
of the neutrino oscillation parameters in matter would be altered in the
presence of NSI's of neutrinos with the matter fermions.  
Such NSI's are predicted in most of the new physics models that attempt to explain 
the non-zero neutrino masses, as well as in a wealth of various other BSM models.
There, the NSI's are simply the effective four-fermion interactions 
at the energy scales relevant for neutrino oscillation experiments
that remain when the heavy mediator fields of the full theory are integrated out. 
These NSI's 
give rise to new neutral-current type interactions of neutrinos 
(both flavor-conserving and flavor-violating) during their 
propagation through matter on top of the SM interactions, 
causing the change in the effective mass matrix for the neutrinos 
which ultimately affect the running of the oscillation 
parameters and hence change the oscillation probabilities
between different neutrino flavors. These sub-leading new physics
effects in the probability due to NSI's can be probed in upcoming 
long-baseline and atmospheric neutrino oscillation experiments.   

In this work, we restricted our attention to the matter effect of
flavor-conserving, non-universal NSI's 
of the neutrino, relegating the discussion of the
flavor-violating NSI case to a separate paper \cite{AgarwallaKaoSunTakeuchi:2015}.
The relevant linear combinations of the flavor-diagonal NSI's were
$\eta = (\varepsilon_{\mu\mu}-\varepsilon_{\tau\tau})/2$ and 
$\zeta = \varepsilon_{ee}-(\varepsilon_{\mu\mu}+\varepsilon_{\tau\tau})/2$, 
where a non-zero $\zeta$ led to a rescaling of the matter-effect parameter 
$a\rightarrow\hat{a}=a(1+\zeta)$, while a non-zero $\eta$ led to 
non-trivial modifications on how the running oscillation parameters 
depend on $\hat{a}$.

Utilizing the Jacobi method, as in Refs.~\cite{Agarwalla:2013tza,Honda:2006hp},
we obtained approximate analytical expressions for the effective neutrino oscillation parameters 
to study how they `run' with the rescaled matter-effect parameter $\hat{a}$, and to explore the role of non-zero $\eta$ in neutrino oscillation. 
We found that in addition to the two rotations, which were required for the 
SM matter interaction and were absorbed into 
effective values of $\theta_{12}$ and $\theta_{13}$, a third 
rotation was needed to capture the effects of $\eta$, which could be absorbed into the effective value of $\theta_{23}$.
Thus, within the neutrino mixing matrix, 
the effect of $\eta$ appears as a shift in the 
effective mixing angle $\theta_{23}$, while the SM 
matter effects show up as shifts in $\theta_{12}$ and $\theta_{13}$.
The CP-violating phase $\delta$ remains unaffected and maintains its vacuum value. 
The running of all the effective neutrino oscillation parameters 
were presented 
for both the normal and inverted neutrino mass hierarchies. 
The changes caused by $\eta$ in the running 
of the effective oscillation parameters for the anti-neutrino case
are discussed in detail in appendix~\ref{anti-neutrino}.

We have also studied the impact of the lepton-flavor-conserving 
NSI parameters on the neutrino oscillation probabilities for 
various appearance and the disappearance channels.
To demonstrate the accuracy (or lack thereof in special cases) 
of our approximate analytical expressions, we compared
the oscillation probabilities estimated with our approximate 
effective `running' mixing angles and mass-squared differences 
with those calculated numerically for the same choices of benchmark 
oscillation parameters, energy, baseline, and line-averaged 
constant matter density along it. 
We found that our approximation provided satisfactory matches with exact numerical results in light of
large $\theta_{13}$ for different values of $\theta_{23}$, 
CP-violating phase $\delta$, and for positive and negative
values of the NSI parameter $\eta$.  
A comparison of our results with the approximate expressions 
of Asano and Minakata \cite{Asano:2011nj} for the
$\nu_\mu\rightarrow\nu_e$ appearance channel has also been presented.

Finally, we examined the merit of our analytical probability 
expressions to identify the situations at which the impact 
of the NSI's become compelling. 
It was found that at higher baselines and energies, the impact of $\eta$
can be quite significant in the $\nu_\mu\rightarrow\nu_\mu$ survival channel 
if $|\eta|$ is of the order of its current experimental upper bound. 
A considerable difference between the SM and NSI 
probabilities can be seen irrespective of the vacuum value of $\theta_{23}$, and
the sign of the NSI parameter $\eta$. 
We note that this feature may be  
explorable with the upcoming 50~kiloton magnetized iron 
calorimeter detector at the India-based Neutrino Observatory (INO), 
which aims to detect atmospheric neutrinos and anti-neutrinos 
separately over a wide range of energies and
path lengths \cite{KhatunChatterjeeThakoreAgarwalla:2015}.
Using our analytical approach, we showed in a very 
simple and compact fashion that the corrections in $\theta_{23}$ 
due to the $\eta$ depend significantly on whether 
the vacuum value of $\theta_{23}$ lies below or above 45$^{\circ}$, 
suggesting a possible degeneracy between the octant of 
$\theta_{23}$ and the sign of $\eta$ for a 
given choice of mass hierarchy.

\subsubsection*{Acknowledgments}

We would like to thank Minako Honda and Naotoshi Okamura 
for their contributions to Ref.~\cite{Honda:2006gv}, the predecessor to 
this work. Helpful conversations with Sabya Sachi Chatterjee, Arnab Dasgupta, 
Shunsaku Horiuchi, Gail McLaughlin, and Matthew Rave are gratefully acknowledged. 
We thank the referee for useful suggestions.
SKA was supported by the DST/INSPIRE Research Grant [IFA-PH-12],
Department of Science \& Technology, India.
TT is grateful for the hospitality of the Kavli-IPMU during his
sabbatical year from fall 2012 to summer 2013 where portions of this work 
was performed, and where he was supported by the World Premier International
Research Center Initiative (WPI Initiative), MEXT, Japan.

\newpage

\begin{appendix}

\section{Effective Mixing Angles and Effective Mass-Squared Differences \\ -- Anti-Neutrino Case}
\label{anti-neutrino}

In this appendix, we study the matter effect due to
the anti-neutrino NSI's. We again utilize the 
Jacobi method to estimate how the NSI parameter 
$\eta$ alters the `running' of the effective mixing angles, 
effective mass-squared differences, and the effective 
CP-violating phase $\delta$ in matter for the anti-neutrinos. 
Like the neutrino case, we also present here a  
comparison between our approximate 
analytical probability expressions and exact numerical 
calculations towards the end of this appendix.

\subsection{Differences from the Neutrino Case}

For the anti-neutrinos, the effective Hamiltonian is given by
\begin{equation}
\overline{H}_\eta 
\;=\;
\edlit{U}{}^*
\left[ \begin{array}{ccc} \overline{\lambda}_1 & 0 & 0 \\
                          0 & \overline{\lambda}_2 & 0 \\
                          0 & 0 & \overline{\lambda}_3
       \end{array}
\right]
\edlit{U}{}^T
= 
\underbrace{
U^*
\left[ \begin{array}{ccc} 0 & 0 & 0 \\
                          0 & \delta m^2_{21} & 0 \\
                          0 & 0 & \delta m^2_{31}
       \end{array}
\right]
U^T 
-\hat{a}
\underbrace{
\left[ \begin{array}{ccc}  1 & 0 & 0 \\
                           0 & 0 & 0 \\
                           0 & 0 & 0 
       \end{array}
\right]
}_{\displaystyle = M_a}
}_{\displaystyle = \overline{H}_0}
-\hat{a}\eta
\underbrace{
\left[ \begin{array}{ccc}  0 & 0 & 0 \\
                           0 & 1 & 0 \\
                           0 & 0 & -1 
       \end{array}
\right]
}_{\displaystyle = M_\eta}
\;.
\end{equation}
The differences from the neutrino case are the reversal of signs of
the CP-violating phase $\delta$ (and thus the complex conjugation of the
PMNS matrix $U$), and the matter interaction parameter $\hat{a}=a(1+\zeta)$. 
We denote the matter effect corrected diagonalization matrix as 
$\edlit{U}$ (note the mirror image tilde on top) to distinguish it from that
for the neutrinos.
%

\subsection{Diagonalization of the Effective Hamiltonian}

\subsubsection{Change to the Mass Eigenbasis in Vacuum}

Using the matrix $Q_3$ from Eq.~(\ref{Qdef}),
we begin by partially diagonalize the effective Hamiltonian $\overline{H}_\eta$ as
\begin{eqnarray}
\overline{H}_\eta' 
& = & Q_3 U^T \overline{H}_\eta U^* Q_3^* \cr
& = &
\underbrace{
\left[ \begin{array}{ccc} 0 & 0 & 0 \\
                          0 & \delta m^2_{21} & 0 \\
                          0 & 0 & \delta m^2_{31}
       \end{array}
\right] 
-\hat{a} 
\underbrace{
Q_3 U^T 
\underbrace{
\left[ \begin{array}{ccc} 1 & 0 & 0 \\
                          0 & 0 & 0 \\
                          0 & 0 & 0 
       \end{array}
\right]
}_{\displaystyle M_a}
U^* Q_3^*
}_{\displaystyle \equiv \overline{M}_a'(\theta_{12},\theta_{13},\theta_{23})}
}_{\displaystyle \equiv \overline{H}_0'}
-\hat{a}\eta
\underbrace{
Q_3 U^T 
\underbrace{\left[ \begin{array}{ccc} 0 & 0 & 0 \\
                          0 & 1 & 0 \\
                          0 & 0 & -1 
       \end{array}
\right]}_{\displaystyle M_\eta}
U^* Q_3^*
}_{\displaystyle \equiv \overline{M}_\eta'(\theta_{12},\theta_{13},\theta_{23},\delta)}
\;,
\end{eqnarray}
where
\begin{eqnarray}
\overline{M}'_a(\theta_{12},\theta_{13},\theta_{23})
\;=\; 
Q_3
\left[ \begin{array}{ccc} U_{e1}U_{e1}^* & U_{e1}U_{e2}^* & U_{e1}U_{e3}^* \\
                          U_{e2}U_{e1}^* & U_{e2}U_{e2}^* & U_{e2}U_{e3}^* \\
                          U_{e3}U_{e1}^* & U_{e3}U_{e2}^* & U_{e3}U_{e3}^* 
       \end{array}
\right] Q_3^*
& = &
\left[ \begin{array}{ccc}
       c_{12}^2 c_{13}^2    & c_{12}s_{12}c_{13}^2 & c_{12}c_{13}s_{13} \\
       c_{12}s_{12}c_{13}^2 & s_{12}^2 c_{13}^2    & s_{12}c_{13}s_{13} \\
       c_{12}c_{13}s_{13}   & s_{12}c_{13}s_{13}   & s_{13}^2
       \end{array}
\right]
\cr
& = & M'_a(\theta_{12},\theta_{13},\theta_{23}) 
\;, \phantom{\Bigg|}
\end{eqnarray}
and
\begin{eqnarray}
\overline{M}'_{\eta}(\theta_{12},\theta_{13},\theta_{23},\delta) 
& = & 
Q_3\left\{
\left[ \begin{array}{ccc} U_{\mu 1}U_{\mu 1}^* & U_{\mu 1}U_{\mu 2}^* & U_{\mu 1}U_{\mu 3}^* \\
                          U_{\mu 2}U_{\mu 1}^* & U_{\mu 2}U_{\mu 2}^* & U_{\mu 2}U_{\mu 3}^* \\
                          U_{\mu 3}U_{\mu 1}^* & U_{\mu 3}U_{\mu 2}^* & U_{\mu 3}U_{\mu 3}^* 
       \end{array}
\right] - 
\left[ \begin{array}{ccc} U_{\tau 1}U_{\tau 1}^* & U_{\tau 1}U_{\tau 2}^* & U_{\tau 1}U_{\tau 3}^* \\
                          U_{\tau 2}U_{\tau 1}^* & U_{\tau 2}U_{\tau 2}^* & U_{\tau 2}U_{\tau 3}^* \\
                          U_{\tau 3}U_{\tau 1}^* & U_{\tau 3}U_{\tau 2}^* & U_{\tau 3}U_{\tau 3}^* 
       \end{array}
\right] \right\} Q_3^* 
\cr
& = &
\left[ \begin{array}{l}
\sin(2\theta_{12})\sin(2\theta_{23})s_{13}\cos\delta 
+(s_{12}^2-c_{12}^2 s_{13}^2)\cos(2\theta_{23}) 
\\
(s_{12}^2 e^{+i\delta} -c_{12}^2 e^{-i\delta}) s_{13}\sin(2\theta_{23})
-(1+s_{13}^2)s_{12}c_{12}\cos(2\theta_{23}) 
\\
-s_{12}c_{13}\sin(2\theta_{23})e^{+i\delta} + c_{12}s_{13}c_{13}\cos(2\theta_{23}) 
\end{array} \right.  \cr
&   & \qquad\qquad
\begin{array}{l}
(s_{12}^2 e^{-i\delta} - c_{12}^2 e^{+i\delta}) s_{13}\sin(2\theta_{23}) 
-(1+s_{13}^2)s_{12}c_{12}\cos(2\theta_{23}) 
\\
- \sin(2\theta_{12})\sin(2\theta_{23})s_{13}\cos\delta 
+(c_{12}^2-s_{12}^2 s_{13}^2)\cos(2\theta_{23}) 
\\
\phantom{-}c_{12}c_{13}\sin(2\theta_{23})e^{+i\delta} + s_{12}s_{13}c_{13}\cos(2\theta_{23}) 
\end{array} \cr
&   & \qquad\qquad\qquad\qquad
\left. \begin{array}{l}
-s_{12}c_{13}\sin(2\theta_{23})e^{-i\delta} + c_{12}s_{13}c_{13}\cos(2\theta_{23}) \\
\phantom{-}c_{12}c_{13}\sin(2\theta_{23})e^{-i\delta} + s_{12}s_{13}c_{13}\cos(2\theta_{23}) \\
 -c_{13}^2\cos(2\theta_{23})
\end{array} \right] 
\cr
& = & M_\eta^{\prime *}(\theta_{12},\theta_{13},\theta_{23},\delta) 
\;.
\end{eqnarray}
%

\subsubsection{$\eta=0$ Case, First and Second Rotations}

As in the neutrino case, we will first approximately 
diagonalize $\overline{H}'_0$ and then add on the 
$\hat{a}\eta\overline{M}'_\eta$ term later.
The Jacobi method applied to $\overline{H}'_0$ is as follows:

\begin{enumerate}
\item{First Rotation}

Define the matrix $\overline{V}$ as
\begin{equation}
\overline{V} \;=\; 
\begin{bmatrix} \overline{c}_{\varphi} &  \overline{s}_{\varphi} & 0 \\
	                     -\overline{s}_{\varphi} &  \overline{c}_{\varphi} & 0 \\
	                      0 & 0 & 1
\end{bmatrix}
\;,
\label{antiVdef}
\end{equation}
where
\begin{equation}
\overline{c}_{\varphi} \,=\,\cos\overline{\varphi}\;,\quad
\overline{s}_{\varphi} \,=\,\sin\overline{\varphi}\;,\quad
\tan 2\overline{\varphi} \,\equiv\, 
-
\dfrac{\hat{a} c_{13}^2\sin2\theta_{12}}{\delta m^2_{21}+\hat{a} c_{13}^2\cos2\theta_{12}}\;,\quad
\left(-\theta_{12}<\overline{\varphi}\le 0\right)\;.
\label{antiphi1def}
\end{equation}
Then,
\begin{eqnarray}
\overline{H}''_0 
& = & \overline{V}^\dagger \overline{H}'_0 \overline{V} 
\cr
& = &
\begin{bmatrix}
\overline{\lambda}'_- & 0 & -\hat{a}{\overline{c}}_{12}' c_{13}s_{13} \\
0 & \overline{\lambda}'_+ & -\hat{a}{\overline{s}}_{12}' c_{13}s_{13} \\
-\hat{a}{\overline{c}}_{12}' c_{13}s_{13} & 
-\hat{a}{\overline{s}}_{12}' c_{13}s_{13} &
-\hat{a} s_{13}^2 + \delta m^2_{31}
\end{bmatrix}
\;,
\cr & &
\label{antiHdoubleprimedef}
\end{eqnarray}
where
\begin{equation}
{\overline{c}}_{12}' = \cos\bar{\theta}_{12}' \;,\quad
{\overline{s}}_{12}' = \sin\bar{\theta}_{12}' \;,\quad
{\bar{\theta}}_{12}' = \theta_{12} + \overline{\varphi} \;,
\label{antitheta12primedef}
\end{equation}
and
\begin{equation}
\overline{\lambda}'_{\pm}
\;=\; \dfrac{ (\delta m^2_{21} - \hat{a} c_{13}^2)
          \pm\sqrt{ (\delta m^2_{21}+\hat{a} c_{13}^2)^2 - 4 \hat{a} c_{13}^2 s_{12}^2 \delta m^2_{21} }
        }
        { 2 }\;.
\label{antilambdaprimeplusminusdef}
\end{equation}
The angle $\bar{\theta}'_{12}=\theta_{12}+\overline{\varphi}$ can be calculated directly without
calculating $\overline{\varphi}$ via
\begin{equation}
\tan 2\bar{\theta}'_{12} \;=\;
\dfrac{\delta m^2_{21}\sin 2\theta_{12}}{\delta m^2_{21}\cos 2\theta_{12}+\hat{a}c_{13}^2}
\;,\qquad
\left(0<\bar{\theta}'_{12}\le \theta_{12}\right)
\;.
\end{equation}
The dependences of $\bar{\theta}'_{12}$ and $\overline{\lambda}'_{\pm}$ on $\beta$
are plotted in Fig.~\ref{theta12primebarplot}.

\begin{figure}[t]
\subfigure[]{\includegraphics[height=5.1cm]{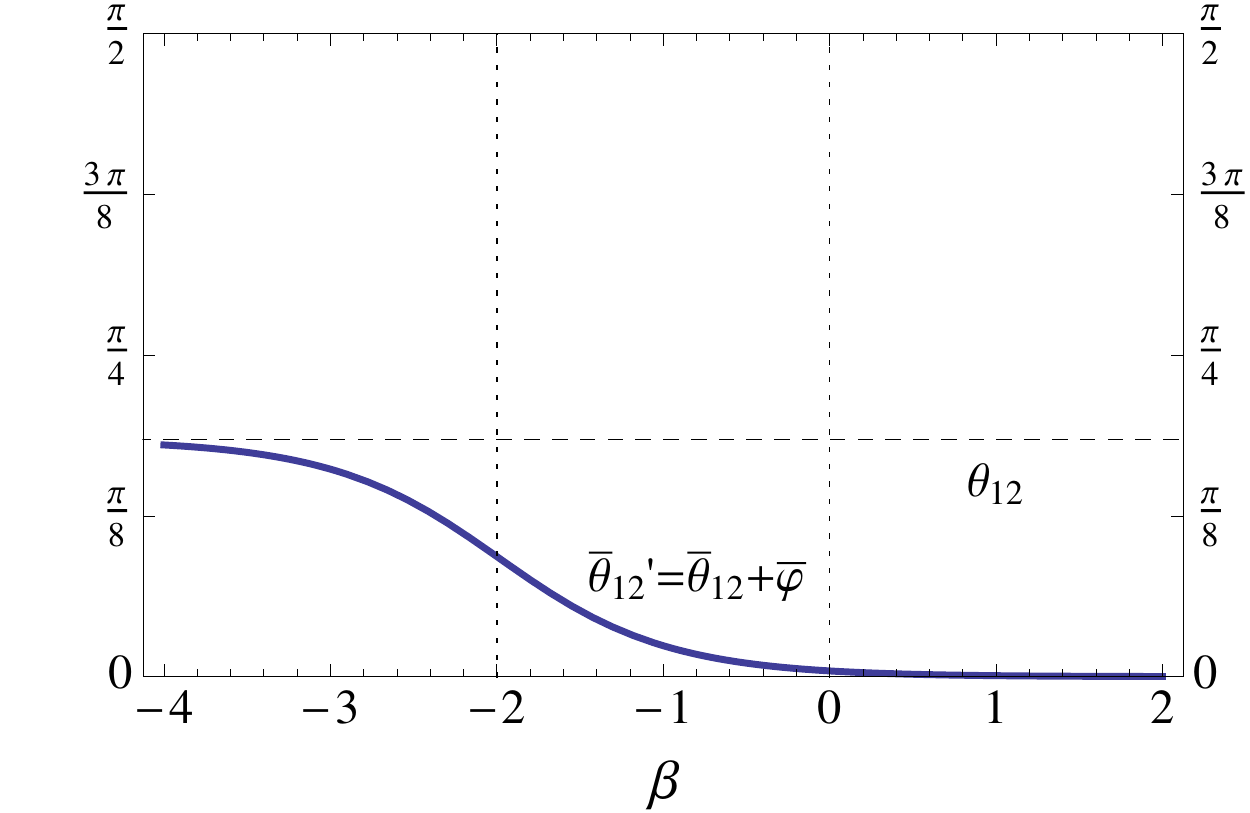}}
\subfigure[]{\includegraphics[height=5cm]{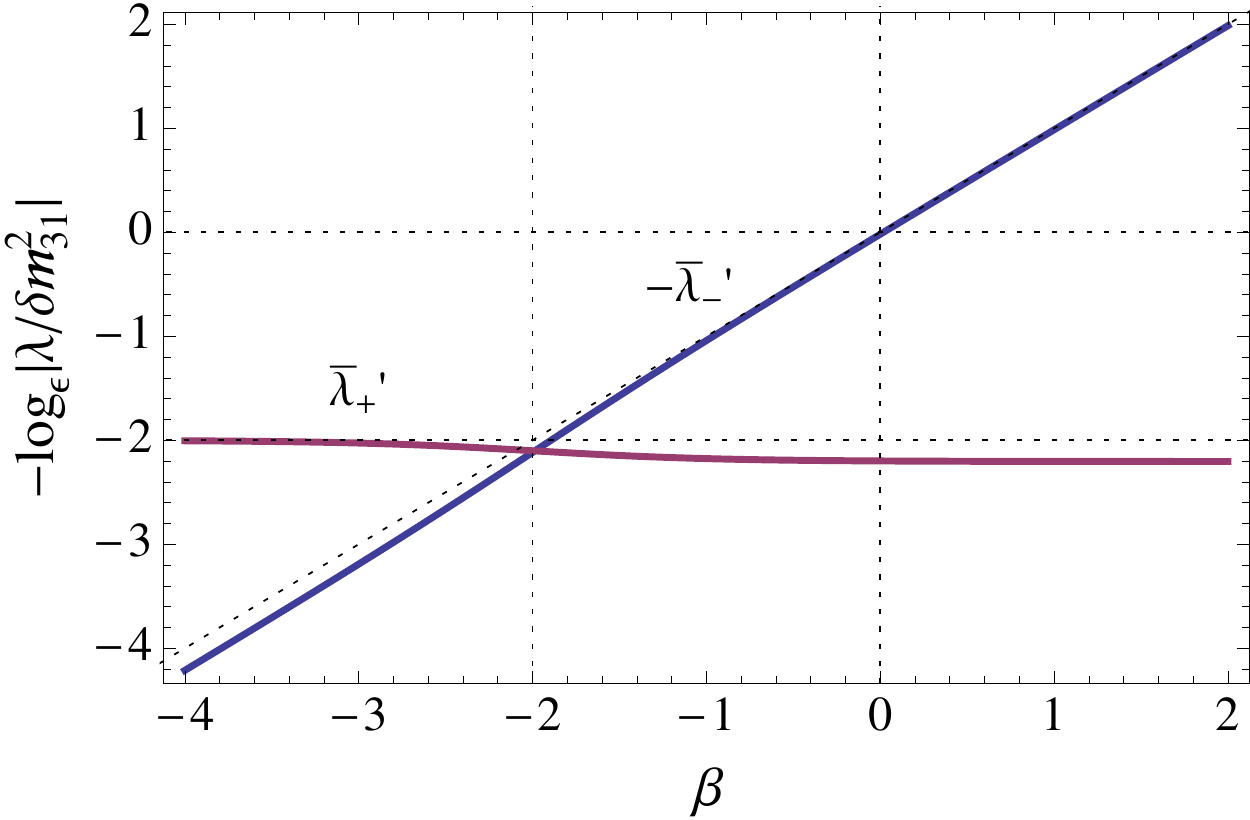}}
\caption{(a) The dependence of $\overline{\theta}'_{12}$ 
on $\beta=-\log_\epsilon(a/|\delta m^2_{31}|)$. 
(b) The $\beta$-dependence of $\overline{\lambda}'_{\pm}$. }
\label{theta12primebarplot}
\end{figure}

Note that in contrast to the neutrino case, $\bar{\theta}'_{12}$ 
decreases monotonically from $\theta_{12}$ to zero as $\beta$ is increased.
The $\beta$-dependences of $\overline{s}'_{12}=\sin\overline{\theta}'_{12}$ and
$\overline{c}'_{12}=\cos\overline{\theta}'_{12}$ are shown 
in Fig.~\ref{s12primebarc12primebarplot}(a). As $\beta$ is increased 
beyond $\beta=-2$, $\overline{c}'_{12}$ grows quickly to one
while $\overline{s}'_{12}$ damps quickly to zero.
The product $\hat{a}\overline{s}'_{12}$ stops increasing 
around $\beta=-2$ and plateau's to the asymptotic value of 
$\delta m^2_{21}s_{12}c_{12}/c_{13}^2 = |\delta m^2_{31}|O(\epsilon^2)$
as shown in Fig.~\ref{s12primebarc12primebarplot}(b).  
That is:
\begin{eqnarray}
\hat{a}\overline{s}'_{12} & = & |\delta m^2_{31}|\,O(\epsilon^{-\min(\beta,-2)}) \;\le\; |\delta m^2_{31}|\,O(\epsilon^2)\;,
\cr
\hat{a}\overline{c}'_{12} & = & |\delta m^2_{31}|\,O(\epsilon^{-\beta}) \;.
\end{eqnarray}
Note also that the scales of $\overline{\lambda}'_{\pm}$ are given simply by
\begin{eqnarray}
\overline{\lambda}'_- & = & -O(\hat{a}) \;=\; -|\delta m^2_{31}|\,O(\epsilon^{-\beta})\;,\cr 
\overline{\lambda}'_+ & = &  O(\delta m^2_{21}) \;=\; |\delta m^2_{31}|\,O(\epsilon^2)\;,
\end{eqnarray}
since no level crossing occurs in this case.

\begin{figure}[t]
\subfigure[]{\includegraphics[height=5cm]{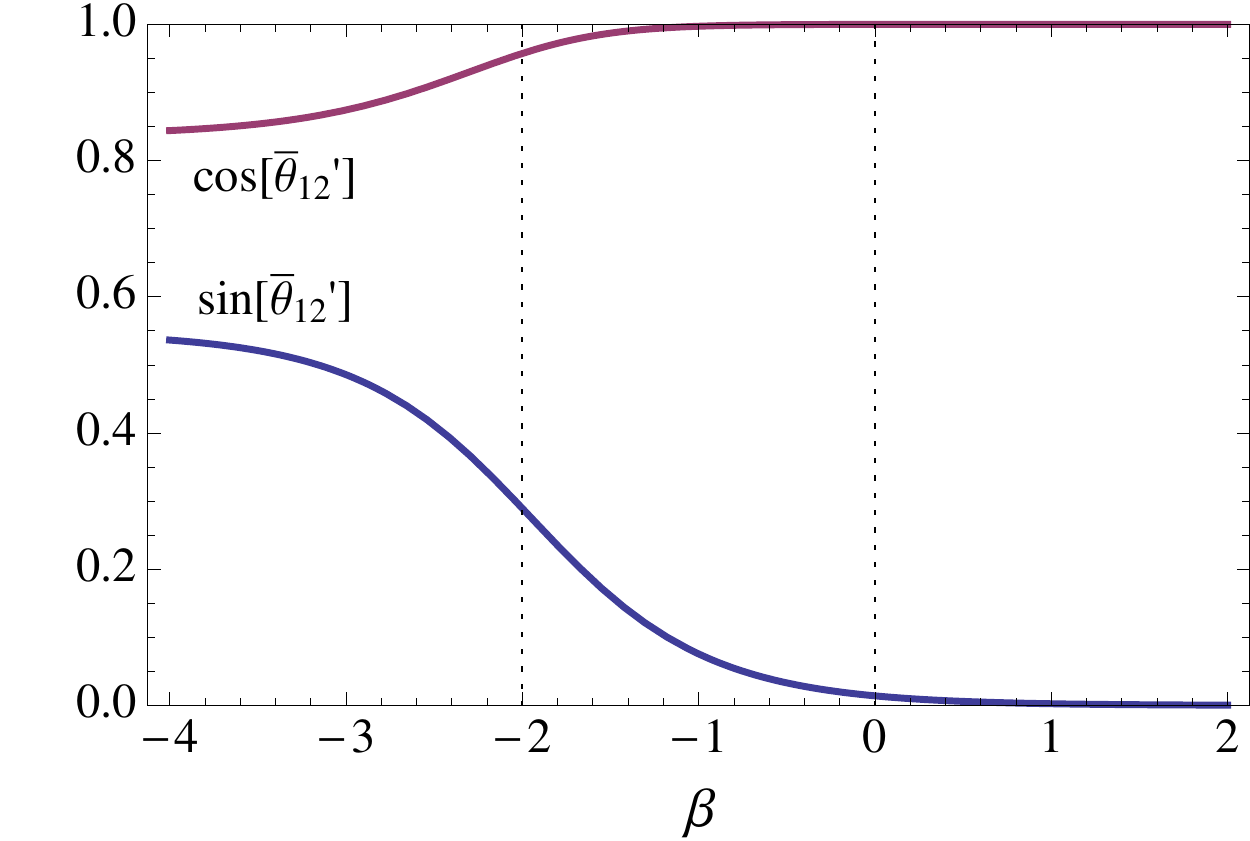}}
\subfigure[]{\includegraphics[height=5cm]{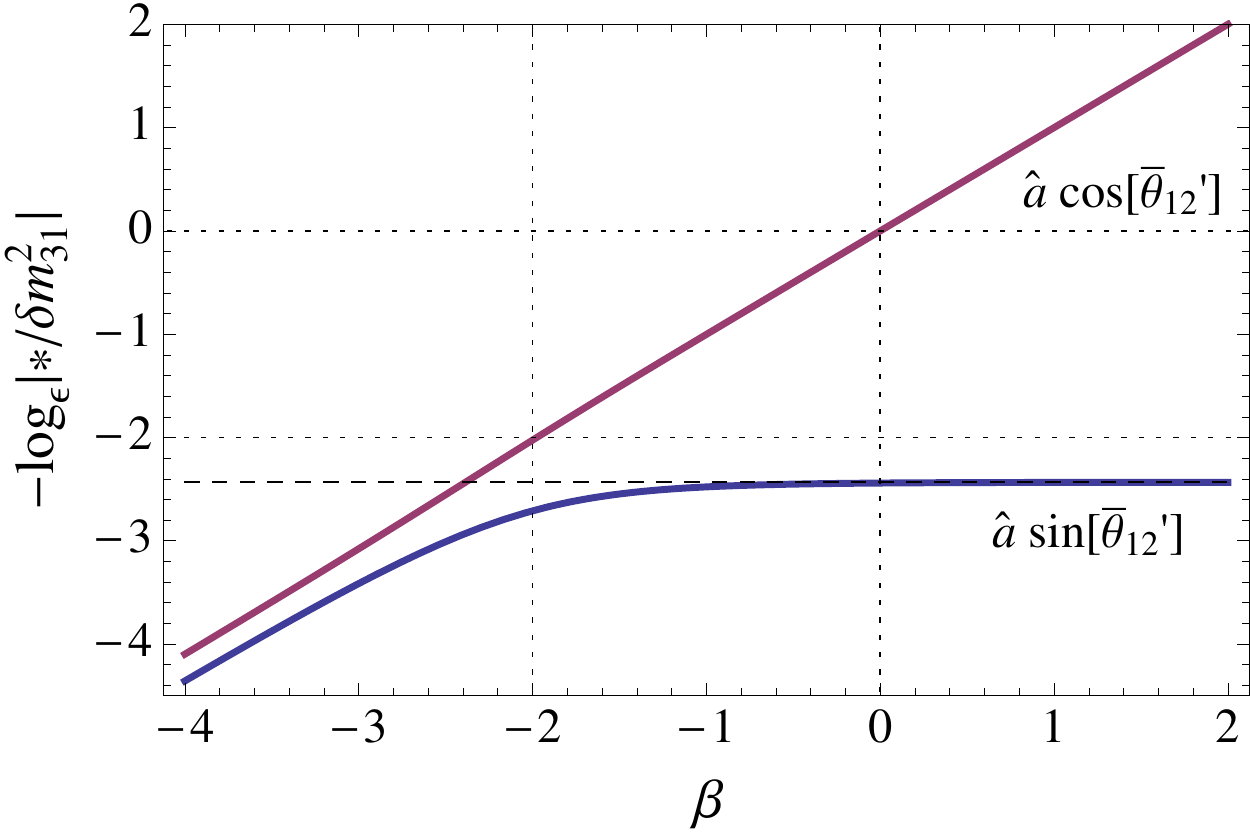}}
\caption{(a) The $\beta$-dependence of 
$\overline{s}'_{12}=\sin\overline{\theta}'_{12}$ and 
$\overline{c}'_{12}=\cos\overline{\theta}'_{12}$. 
(b) The $\beta$-dependence of $a\overline{s}'_{12}$ 
and $a\overline{c}'_{12}$. The asymptotic value of 
$a\overline{s}'_{12}$ is 
$\delta m^2_{21}s_{12}c_{12}/c_{13}^2 \approx 
0.014\,|\delta m^2_{31}| = O(\varepsilon^2|\delta m^2_{31}|)$.}
\label{s12primebarc12primebarplot}
\end{figure}

\item{Second Rotation}

Since $\hat{a}\overline{c}_{12}'$ continues to increase with 
$\beta$ while $\hat{a}\overline{s}_{12}'$ does not, we perform 
a $(1,3)$ rotation on $\overline{H}''_0$ next.

Define the matrix $\overline{W}$ as
\begin{equation}
\overline{W} = 
\begin{bmatrix} 
 \overline{c}_\phi & 0 & \overline{s}_\phi \\
 0 &  1 &  0 \\
-\overline{s}_\phi & 0 &  \overline{c}_\phi 
\end{bmatrix}
\;,
\label{antiWdef}
\end{equation}
where
$\overline{c}_{\phi} = \cos\overline{\phi}$,
$\overline{s}_{\phi} = \sin\overline{\phi}$, and
\begin{equation}
\tan 2\overline{\phi}
\;\equiv\; 
-\dfrac{2\hat{a}\overline{c}_{12}'c_{13}s_{13}}
       {\delta m^2_{31}-\hat{a} s_{13}^2 - \overline{\lambda}'_-}
\;\approx\;
-\dfrac{\hat{a}\sin 2\theta_{13}}{(\delta m^2_{31}-\delta m^2_{21}s_{12}^2)+\hat{a}\cos 2\theta_{13}} 
\;.
\label{antiphi2def}
\end{equation}
The angle $\overline{\phi}$ is in the fourth quadrant when 
$\delta m^2_{31} > 0$ (normal hierarchy),
and in the first quadrant when $\delta m^2_{31} < 0$ (inverted hierarchy).
The $\beta$-dependence of $\overline{\phi}$ is shown in 
Fig.~\ref{phibarplot}(a) for both mass hierarchies.

\begin{figure}[t]
\subfigure[]{\includegraphics[height=5cm]{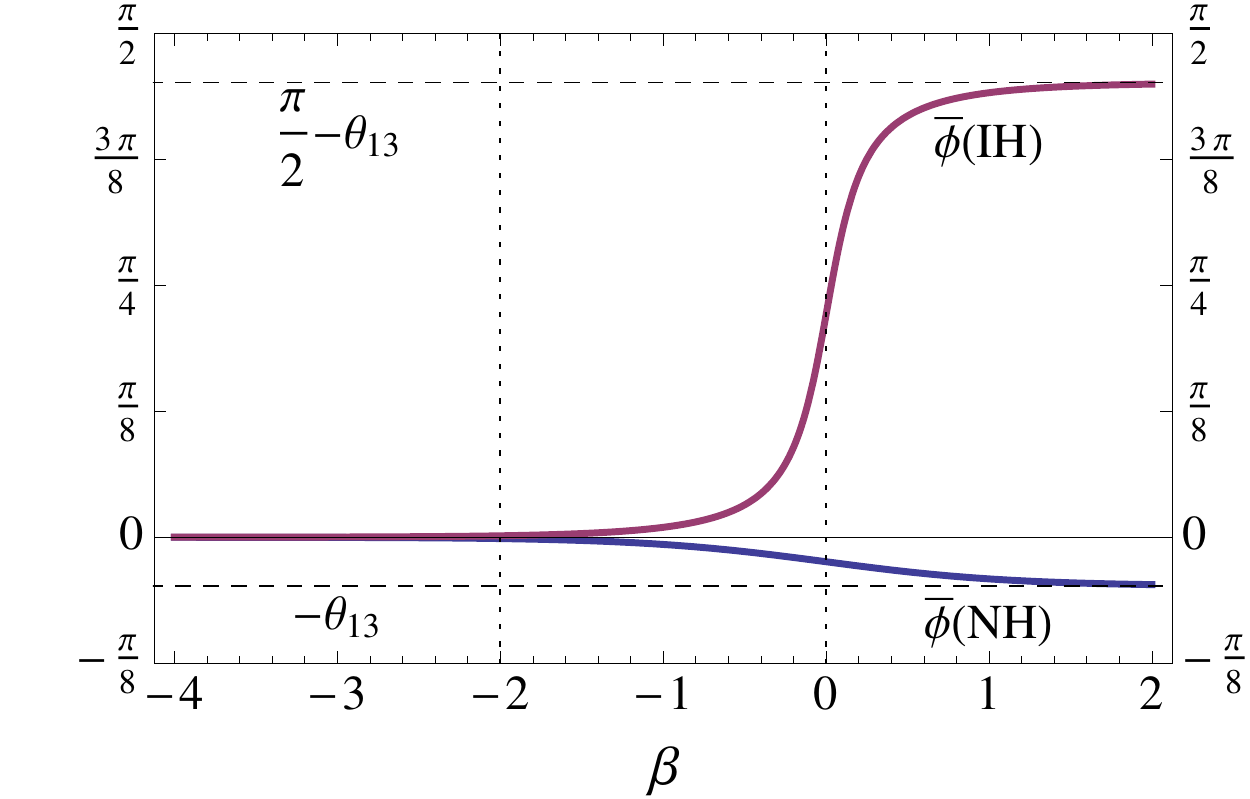}}
\subfigure[]{\includegraphics[height=5cm]{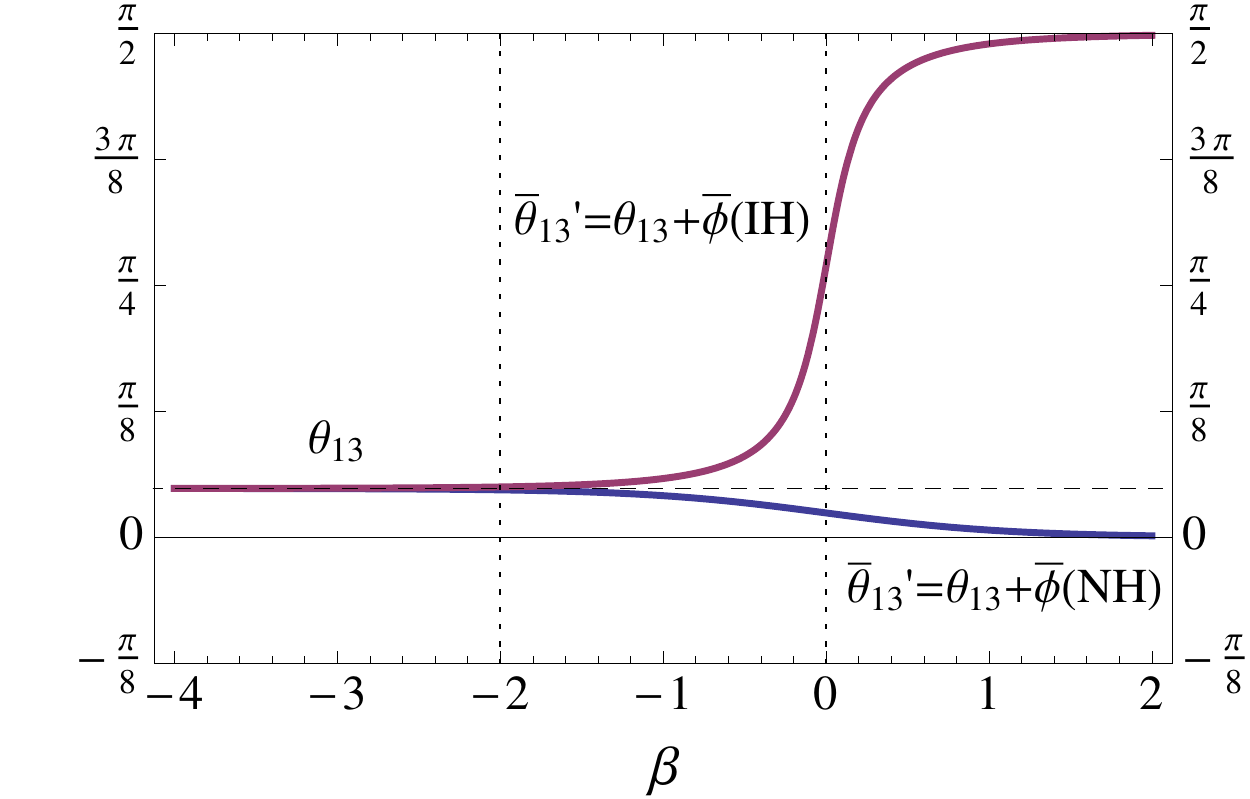}}
\caption{The dependence of (a) $\overline{\phi}$ and
(b) $\overline{\theta}_{13}'=\theta_{13}+\overline{\phi}$ on 
$\beta=-\log_{\,\epsilon}\left(\hat{a}/|\delta m^2_{31}|\right)$
for the normal (NH) and inverted (IH) mass hierarchies.}
\label{phibarplot}
\end{figure}

Using $\overline{W}$, we obtain
\begin{eqnarray}
\overline{H}_0'''
& = & \overline{W}^\dagger \overline{H}_0'' \overline{W} \cr
& = &
\left[ \begin{array}{ccc}
\overline{\lambda}''_\mp
& \hat{a}\overline{s}_{12}'c_{13}s_{13}\overline{s}_{\phi} 
& 0 \\
\hat{a}\overline{s}_{12}'c_{13}s_{13}\overline{s}_{\phi} & 
\overline{\lambda}'_+ & 
-\hat{a}\overline{s}_{12}'c_{13}s_{13}\overline{c}_{\phi} \\
0 & -\hat{a}\overline{s}_{12}'c_{13}s_{13}\overline{c}_{\phi} & 
\overline{\lambda}''_\pm
\end{array} \right]  \;,
\label{antiHtripleprimedef}
\end{eqnarray}
where the upper signs are for the $\delta m^2_{31}>0$ case and
the lower signs are for the $\delta m^2_{31}<0$ case, with
\begin{equation}
\overline{\lambda}''_{\pm} \;\equiv\;
   \dfrac{ [(\delta m^2_{31}-\hat{a} s_{13}^2)+ \overline{\lambda}'_{-}  ]
\pm \sqrt{ [(\delta m^2_{31}-\hat{a} s_{13}^2)- \overline{\lambda}'_{-}  ]^2 
         + 4(\hat{a}\overline{c}'_{12}s_{13}c_{13})^2 }
         }
         { 2 } 
\;.
\label{antilambdadoubleprimeplusminusdef}
\end{equation}
As $\beta$ is increased beyond 0, the $\overline{\lambda}''_\pm$
asymptote to
\begin{eqnarray} 
\overline{\lambda}''_+ & \rightarrow & 
\delta m^2_{31}c_{13}^2 + \delta m^2_{21} s_{12}^2 s_{13}^2
\;,\cr
\overline{\lambda}''_- & \rightarrow & -\hat{a}+\delta m^2_{31}s_{13}^2 + \delta m^2_{21}s_{12}^2 c_{13}^2 
\;,
\end{eqnarray}
for both mass hierarchies.
Note that $\overline{\lambda}''_- <0<\overline{\lambda}''_+$ 
for the $\delta m^2_{31}>0$ case, while
both $\overline{\lambda}''_\pm <0$ 
for the $\delta m^2_{31}<0$ case.
The $\beta$-dependences of $\overline{\lambda}''_{\pm}$ 
are shown in Fig.~\ref{lambdabardoubleprimepmplot}.
Order-of-magnitude-wise, we have
\begin{equation}
\begin{array}{lll}
  \overline{\lambda}''_- \;=\; -|\delta m^2_{31}|\, O(\epsilon^{-\beta})\;,\quad
& \overline{\lambda}''_+ \;=\; \phantom{+}|\delta m^2_{31}|\, O(1)\;,\qquad 
& \mbox{if $\delta m^2_{31} > 0$}\;, \\
  \overline{\lambda}''_+ \;=\; -|\delta m^2_{31}|\, O(\epsilon^{-\min(\beta,0)})\;,\quad
& \overline{\lambda}''_- \;=\; -|\delta m^2_{31}|\, O(\epsilon^{-\max(\beta,0)})\;,\qquad 
& \mbox{if $\delta m^2_{31} < 0$}\;.
\end{array}
\end{equation}
In particular, in the range $\beta\agt 0$, we have
\begin{equation}
\begin{array}{lll}
  \overline{\lambda}''_- \;=\; -|\delta m^2_{31}|\, O(\epsilon^{-\beta})\;,\quad
& \overline{\lambda}''_+ \;=\; \phantom{+}|\delta m^2_{31}|\, O(1)\;,\qquad 
& \mbox{if $\delta m^2_{31} > 0$}\;, \\
  \overline{\lambda}''_+ \;=\; -|\delta m^2_{31}|\, O(1)\;,\quad
& \overline{\lambda}''_- \;=\; -|\delta m^2_{31}|\, O(\epsilon^{-\beta})\;,\qquad 
& \mbox{if $\delta m^2_{31} < 0$}\;.
\end{array}
\end{equation}
For the off-diagonal terms, we have 
$\hat{a}\overline{s}_{12}'= |\delta m^2_{31}| O(\epsilon^2)$,
$c_{13}=O(1)$, $s_{13}=O(\epsilon)$, and
\begin{equation}
\begin{array}{lll}
  \overline{s}_{\phi}=O(\epsilon),\quad
& \overline{c}_{\phi}=O(1),\qquad 
& \mbox{if $\delta m^2_{31} > 0$}\;, \\
  \overline{s}_{\phi}=O(1)\;,\quad
& \overline{c}_{\phi}=O(\epsilon)\;,\qquad 
& \mbox{if $\delta m^2_{31} < 0$}\;.
\end{array}
\end{equation}
%

\begin{figure}[t]
\subfigure[]{\includegraphics[height=5cm]{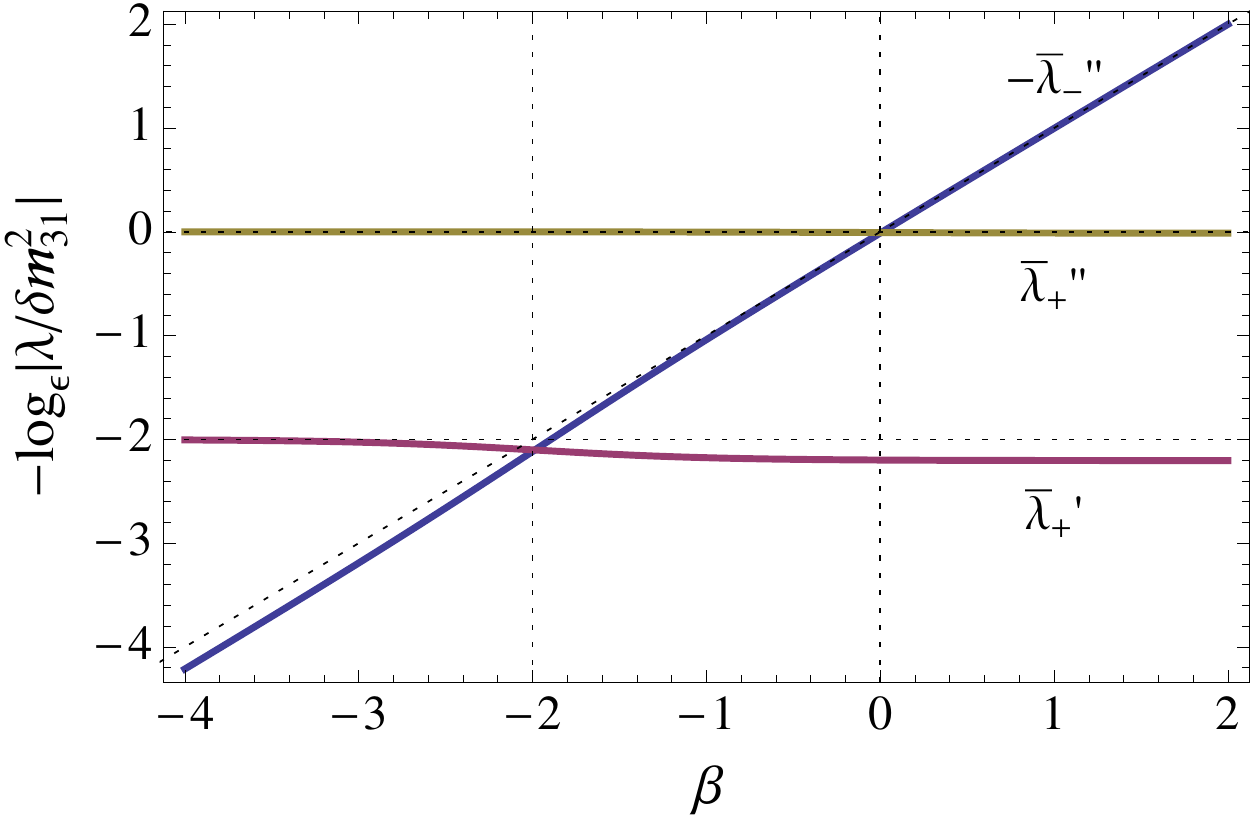}}
\subfigure[]{\includegraphics[height=5cm]{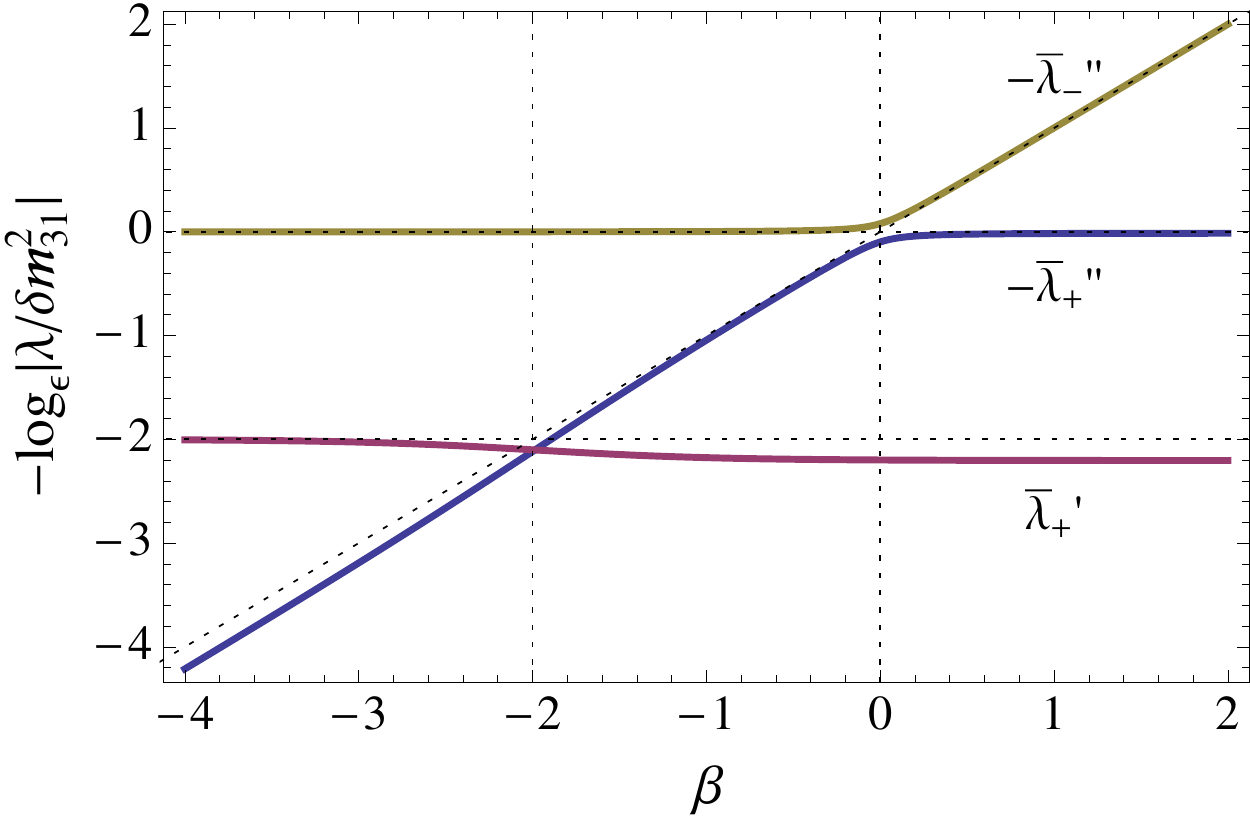}}
\caption{The $\beta$-dependence of $\overline{\lambda}''_{\pm}$
for the (a) normal and (b) inverted mass hierarchies.}
\label{lambdabardoubleprimepmplot}
\end{figure}

Thus, looking at the elements of $\overline{H}_0'''$ in that range, 
we find:
\begin{equation}
\overline{H}_0'''\;=\; |\delta m^2_{31}|
\begin{bmatrix} 
       O(\epsilon^{-\beta}/1)  & O(\epsilon^{4}/\epsilon^{3}) & 0  \\
       O(\epsilon^{4}/\epsilon^{3}) & O(\epsilon^{2}) & O(\epsilon^{3}/\epsilon^{4}) \\
       0                     & O(\epsilon^{3}/\epsilon^{4}) & O(1/\epsilon^{-\beta})
\end{bmatrix}
\;,
\end{equation}
where the elements with two entries denote the two different
mass hierarchies, $O(\mathrm{NH}/\mathrm{IH})$,
and we see that further diagonalization require angle of order 
$O(\epsilon^3)$.  Therefore, $\overline{H}_0'''$ is
approximately diagonal.

\end{enumerate}

\subsubsection{$\eta\neq 0$ Case, Third Rotation}

\begin{figure}[t]
\subfigure[Normal Hierarchy]{\includegraphics[height=5cm]{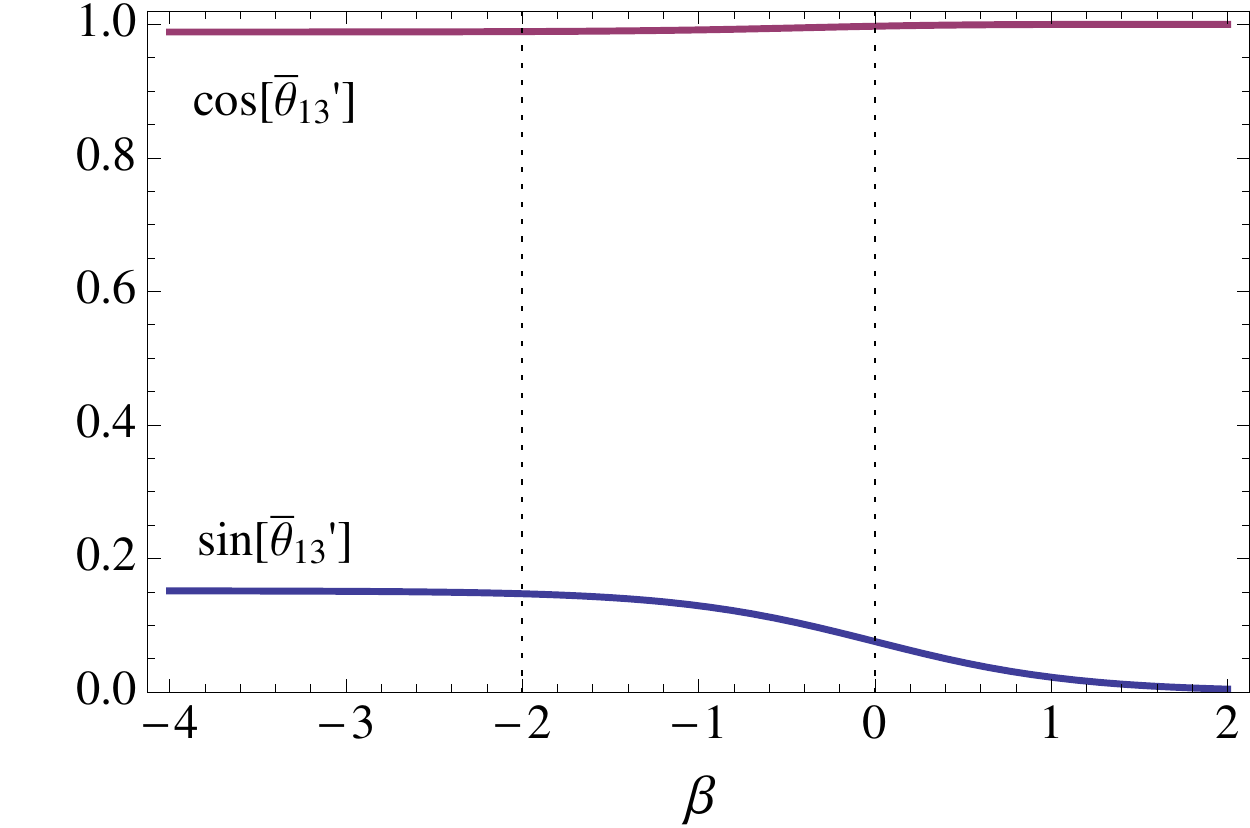}}
\subfigure[Inverted Hierarchy]{\includegraphics[height=5cm]{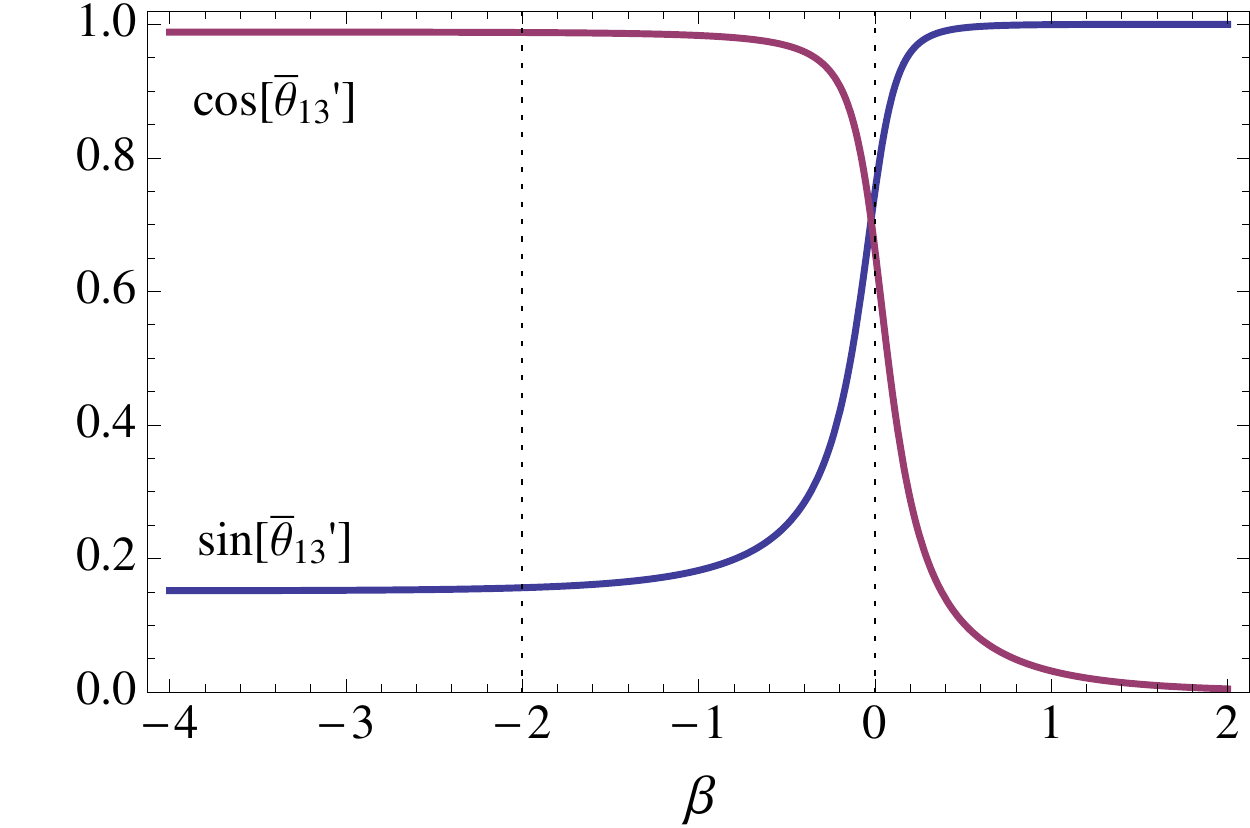}}
\subfigure[Normal Hierarchy]{\includegraphics[height=5cm]{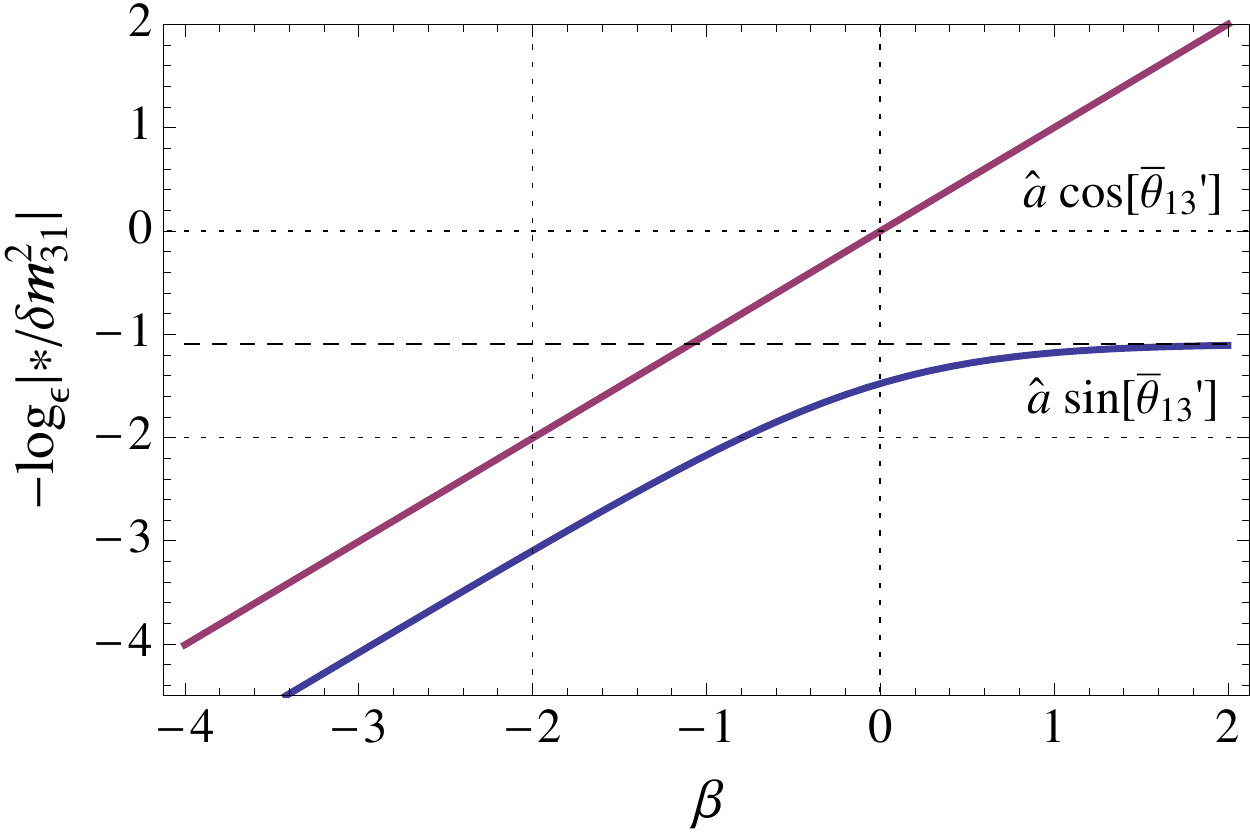}}
\subfigure[Inverted Hierarchy]{\includegraphics[height=5cm]{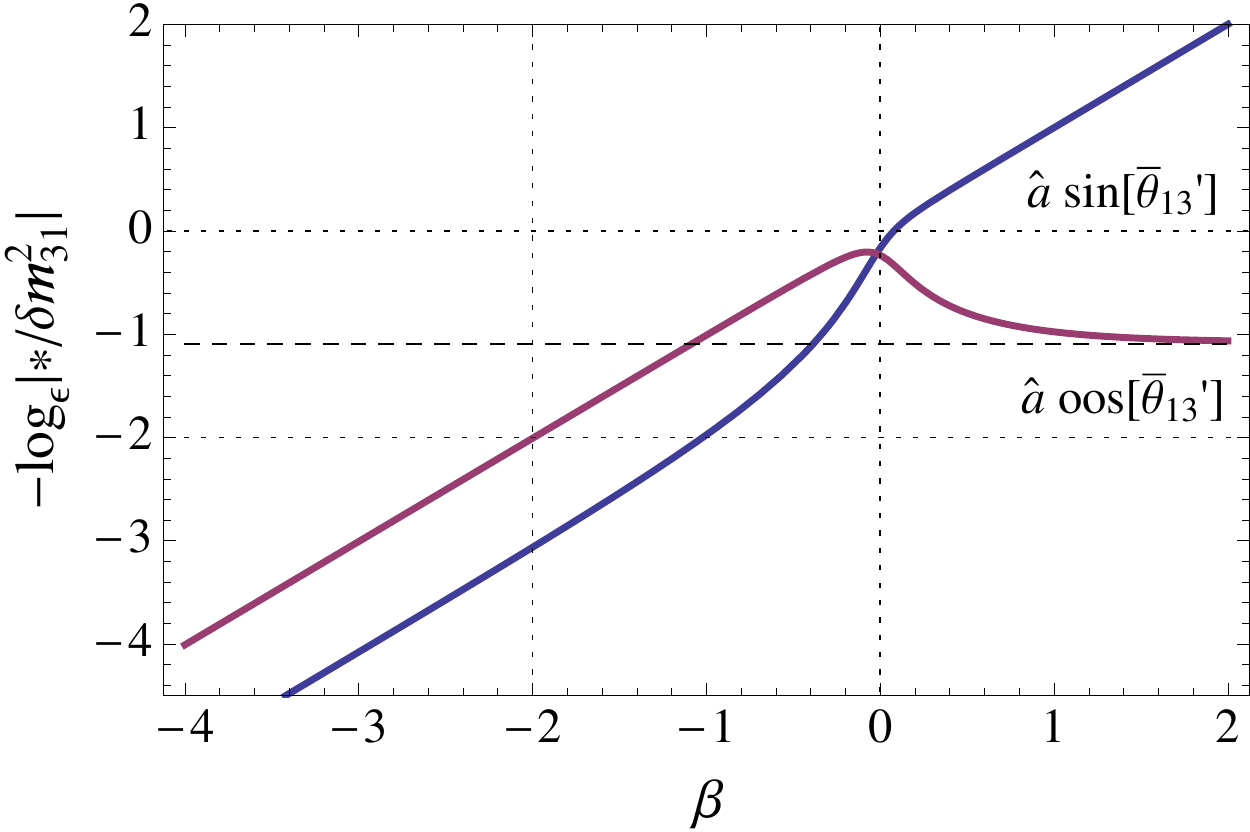}}
\caption{The dependence of $\sin\theta'_{13}$ and
$\cos\theta'_{13}$ on 
$\beta=-\log_{\,\epsilon}\left(\hat{a}/|\delta m^2_{31}|\right)$
for the (a) normal and (b) inverted mass hierarchies.
The dependence of $\hat{a}\sin\theta'_{13}$ and
$\hat{a}\cos\theta'_{13}$ on 
$\beta=-\log_{\,\epsilon}\left(\hat{a}/|\delta m^2_{31}|\right)$
for the (c) normal and (d) inverted mass hierarchies.}
\label{asintheta13primebaracostheta13primebarplot}
\end{figure}

Next, we consider the $\eta\neq 0$ case.
Performing the same $(1,2)$ rotation $\overline{V}$
on $\overline{H}_\eta'=\overline{H}_0'-\hat{a}\eta\overline{M}_\eta'$
as we did on $\overline{H}_0'$, the $\overline{M}_\eta'$ part is transformed
to
\begin{eqnarray}
\overline{V}^\dagger \overline{M}_\eta'(\theta_{12},\theta_{13},\theta_{23},\delta) \overline{V}
& = & \overline{M}_\eta'(\underbrace{\theta_{12}+\overline{\varphi}}_{\displaystyle =\overline{\theta}_{12}'},\theta_{13},\theta_{23},\delta) \cr
& = & \overline{M}_\eta'(\overline{\theta}_{12}',\theta_{13},\theta_{23},\delta)\;.
\end{eqnarray}
Using $\overline{\theta}_{12}'\rightarrow 0$, $\hat{a}\overline{c}_{12}'\rightarrow \hat{a}$,
$\hat{a}\overline{s}_{12}'\rightarrow O(\epsilon^2)|\delta m^2_{31}|$ 
as $\beta$ is increased beyond $-2$, we can approximate
\begin{eqnarray}
\lefteqn{\hat{a}\eta \overline{M}_\eta'(\overline{\theta}_{12}',\theta_{13},\theta_{23},\delta)}
\cr
& \approx & \hat{a}\eta \overline{M}_\eta'(0,\theta_{13},\theta_{23},\delta) \cr
& = & \hat{a}\eta
\begin{bmatrix}
-s_{13}^2\cos(2\theta_{23}) & -e^{i\delta}s_{13}\sin(2\theta_{23}) & s_{13}c_{13}\cos(2\theta_{23}) \\
-e^{-i\delta}s_{13}\sin(2\theta_{23}) & \cos(2\theta_{23}) & e^{-i\delta}c_{13}\sin(2\theta_{23}) \\
s_{13}c_{13}\cos(2\theta_{23}) & e^{i\delta}c_{13}\sin(2\theta_{23}) & -c_{13}^2\cos(2\theta_{23})
\end{bmatrix}
\;.
\end{eqnarray}
Performing the $(1,3)$ rotation $\overline{W}$ next, we find
\begin{eqnarray}
\lefteqn{
\overline{W}^\dagger \overline{M}_\eta'(0,\theta_{13},\theta_{23},\delta)\overline{W}
} \cr
& = &
\overline{M}_\eta'(0,\underbrace{\theta_{13}+\overline{\phi}}_{\displaystyle \overline{\theta}_{13}'},\theta_{23},\delta) \cr
& = &
\overline{M}_\eta'(0,\overline{\theta}'_{13},\theta_{23},\delta)
\cr
& = &
\begin{bmatrix}
-\overline{s}_{13}^{\prime 2}\cos(2\theta_{23}) & -e^{i\delta}\overline{s}_{13}'\sin(2\theta_{23}) & \overline{s}_{13}'\overline{c}_{13}'\cos(2\theta_{23}) \\
-e^{-i\delta}\overline{s}_{13}'\sin(2\theta_{23}) & \cos(2\theta_{23}) & e^{-i\delta}\overline{c}_{13}'\sin(2\theta_{23}) \\
\overline{s}_{13}'\overline{c}_{13}'\cos(2\theta_{23}) & e^{i\delta}\overline{c}_{13}'\sin(2\theta_{23}) & -\overline{c}_{13}^{\prime 2}\cos(2\theta_{23})
\end{bmatrix}
\;,
\end{eqnarray}
where 
$\overline{s}'_{13}=\sin\overline{\theta}_{13}'$,
$\overline{c}'_{13}=\cos\overline{\theta}_{13}'$.
The angle $\overline{\theta}_{13}'=\theta_{13}+\overline{\phi}$ can be 
calculated directly without the need to calculate $\overline{\phi}$ using
\begin{equation}
\tan 2\overline{\theta}_{13}' \;=\;
\dfrac{(\delta m^2_{31}-\delta m^2_{21}s_{12}^2)\sin(2\theta_{13})}
      {(\delta m^2_{31}-\delta m^2_{21}s_{12}^2)\cos(2\theta_{13})+\hat{a}}
\;,
\end{equation}
and its $\beta$-dependence is shown in Fig.~\ref{phibarplot}(b).
In contrast to the neutrino case, $\overline{\theta}_{13}'$ increases rapidly to $\pi/2$
when $\delta m^2_{31}<0$, while damping to zero when $\delta m^2_{31}>0$ once 
$\beta$ is increased above zero.
Consequently, $\hat{a}\cos\overline{\theta}_{13}'$ for the $\delta m^2_{31}<0$ case, 
and $\hat{a}\sin\overline{\theta}_{13}'$ for the $\delta m^2_{31}>0$ case plateau to
$c_{13}s_{13}(1-\epsilon^2 s_{12}^2)|\delta m^2_{31}| = O(\epsilon)|\delta m^2_{31}|$ 
as $\beta$ is increased as shown in Fig.~\ref{asintheta13primebaracostheta13primebarplot}.
Note that in the $\delta m^2_{31}<0$ case, $\hat{a}\cos\overline{\theta}_{13}'$
increased to $O(1)|\delta m^2_{31}|$ in the vicinity of $\beta=0$ before
plateauing to $O(\epsilon)|\delta m^2_{31}|$.
As in the neutrino case with $\delta m^2_{31}>0$, this will cause a slight problem
with our approximation later.
We now look at the normal and inverted mass hierarchy cases separately.

\begin{enumerate}
\item {$\bm{\delta m^2_{31}>}0$ \textbf{Case}}

\begin{figure}[t]
\subfigure[$\eta>0$]{\includegraphics[height=5cm]{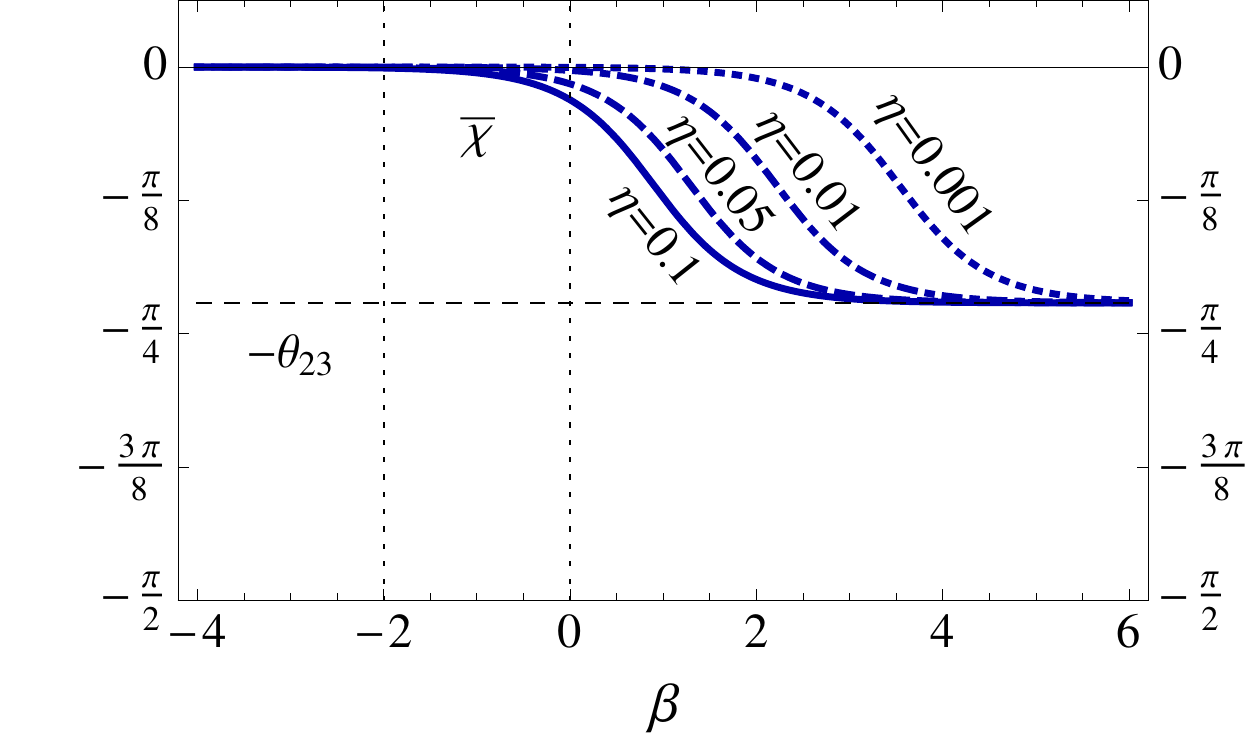}}
\subfigure[$\eta<0$]{\includegraphics[height=5cm]{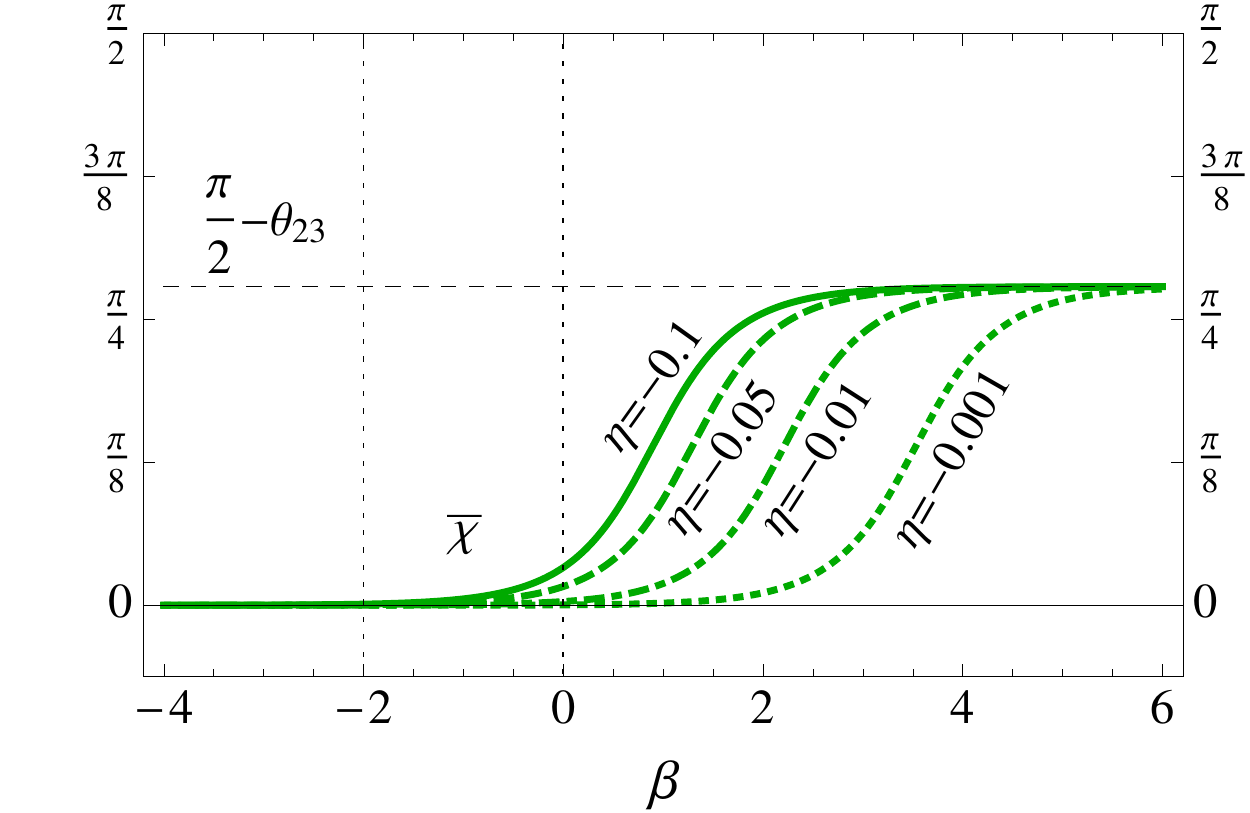}}
\subfigure[$\eta>0$]{\includegraphics[height=5cm]{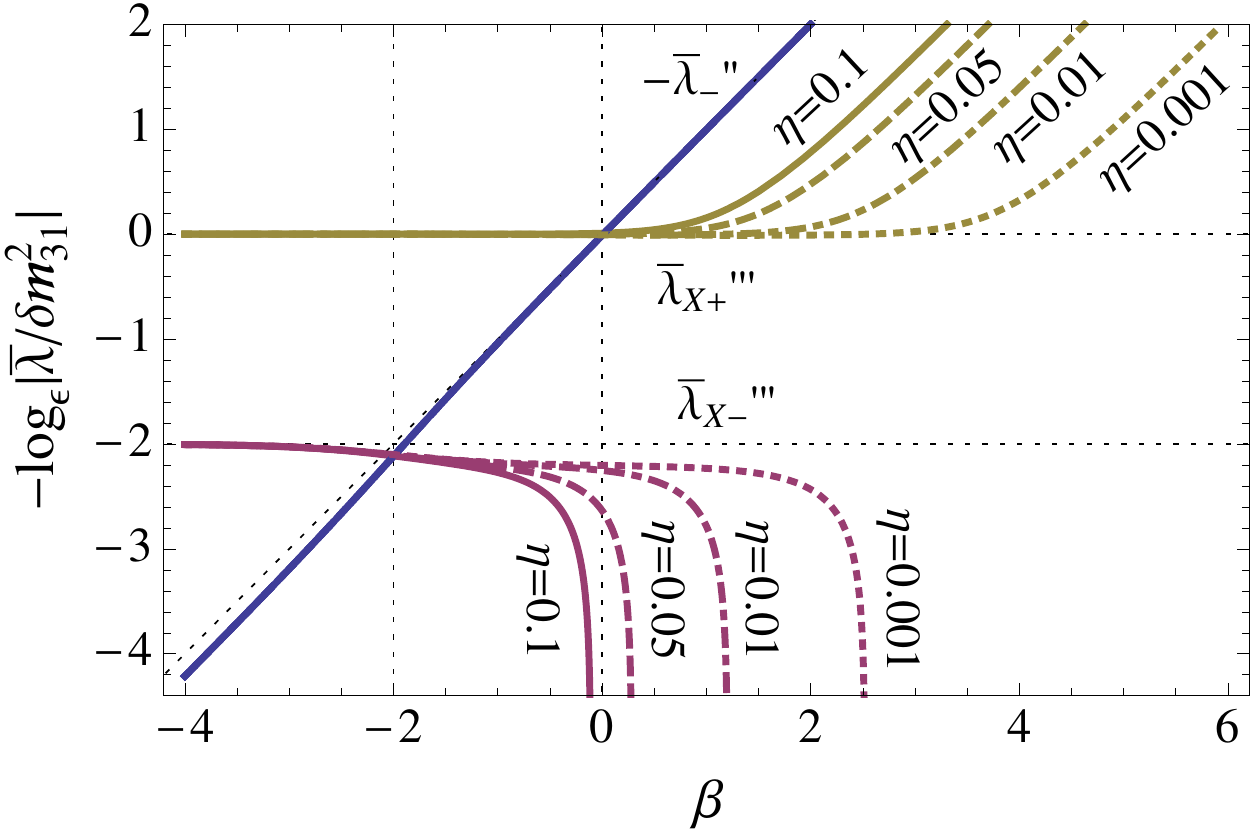}}\qquad
\subfigure[$\eta<0$]{\includegraphics[height=5cm]{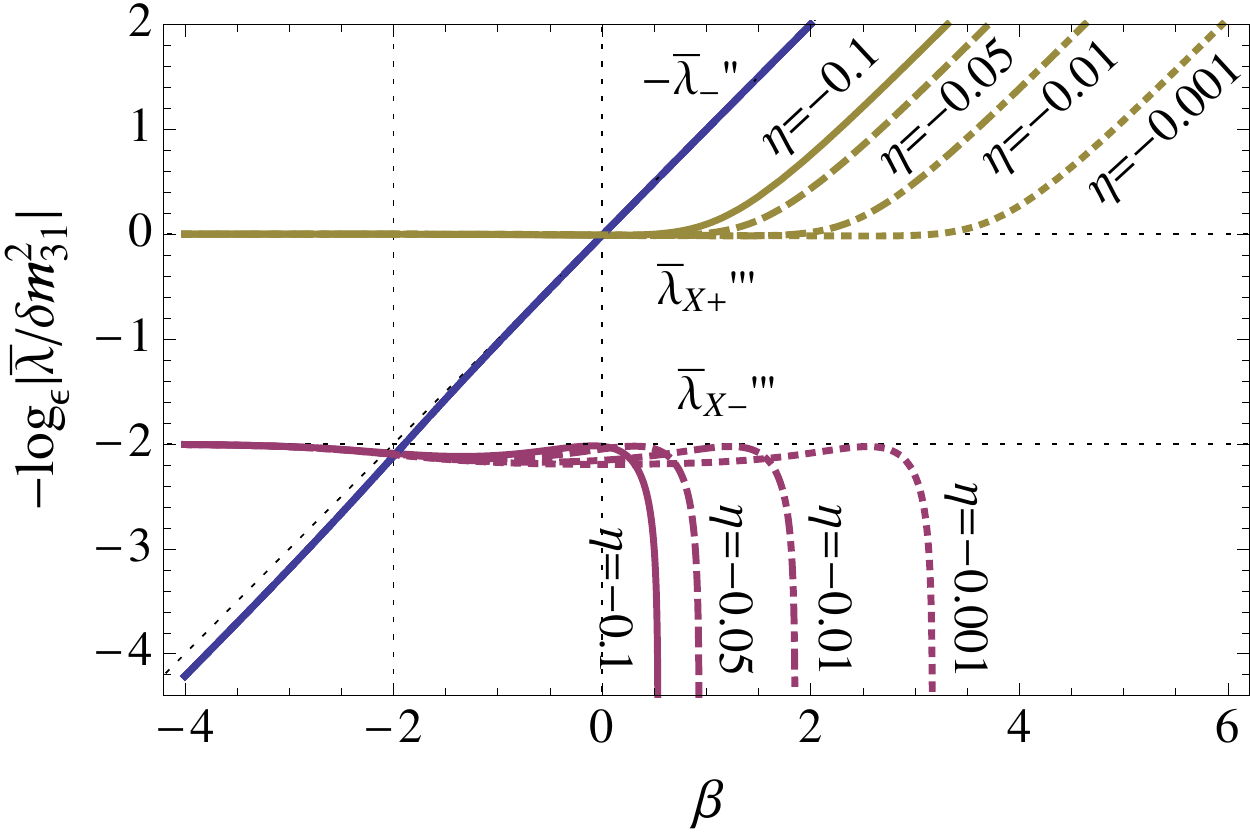}}
\subfigure[$\eta>0$]{\includegraphics[height=5cm]{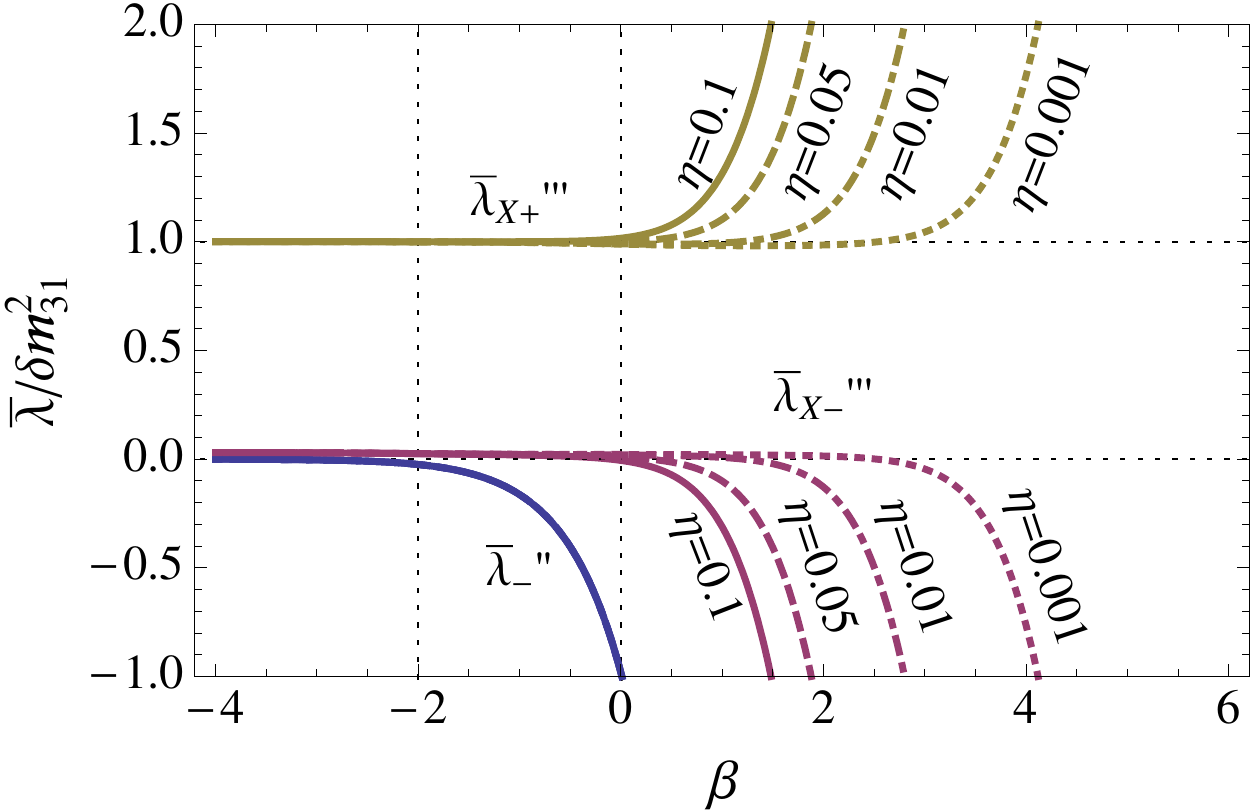}}\quad
\subfigure[$\eta<0$]{\includegraphics[height=5cm]{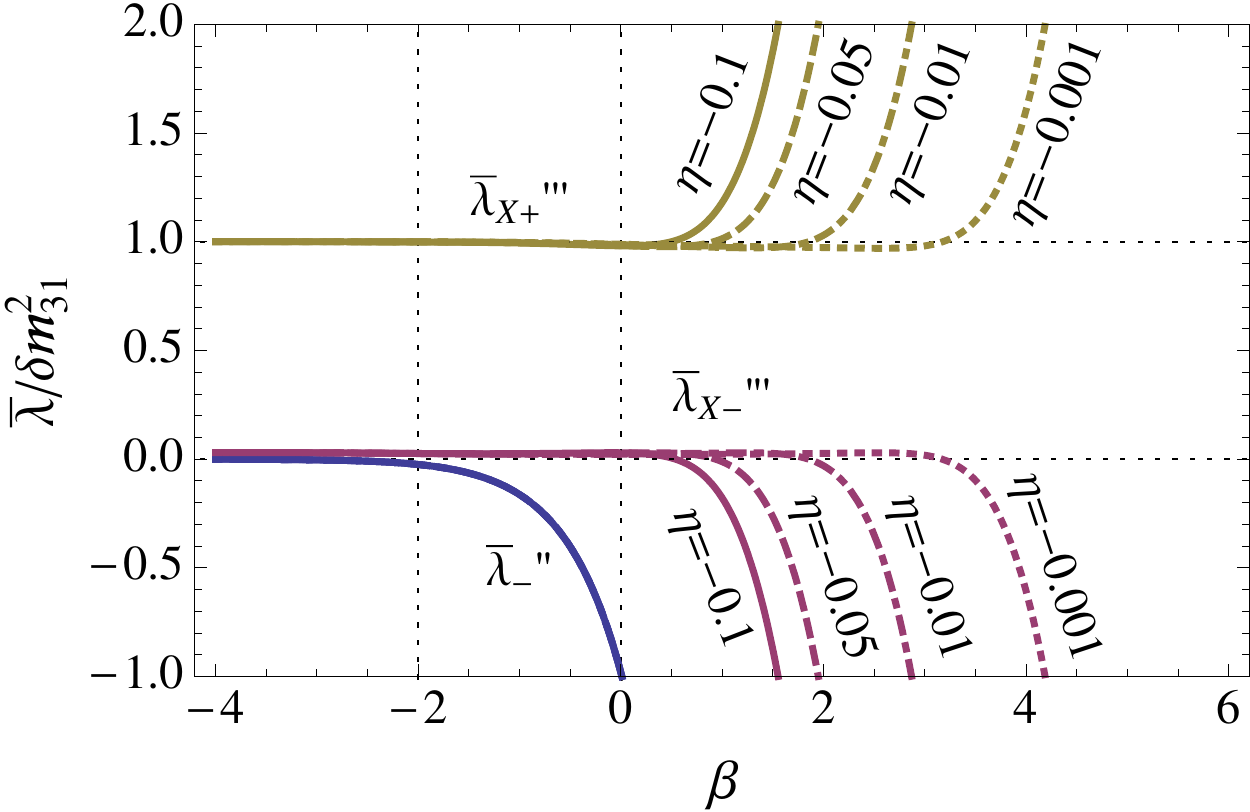}}
\caption{Normal hierarchy case. The $\beta$-dependence 
of $\overline{\chi}$ and $\overline{\lambda}'''_{X\pm}$
for several values of $\eta$ with $s_{23}^2=0.4$.}
\label{chibarplot}
\end{figure}

In the $\delta m^2_{31}>0$ case 
$\hat{a}\overline{s}_{13}'\rightarrow O(\epsilon)|\delta m^2_{31}|$
as $\beta$ is increased beyond 0.  Therefore, we can approximate
\begin{eqnarray}
H_\eta''' 
& = & \overline{W}^\dagger\overline{V}^\dagger \overline{H}_\eta' \overline{V}\overline{W} \cr
& \approx & \overline{H}_0''' - \hat{a}\eta \overline{M}_\eta'(0,\overline{\theta}_{13}',\theta_{23},\delta) \cr
& \approx &
\begin{bmatrix}
\overline{\lambda}''_- & 0 & 0 \\
0 & \overline{\lambda}_+' -\hat{a}\eta\cos(2\theta_{23}) & -\hat{a}\eta e^{-i\delta}\overline{c}_{13}'\sin(2\theta_{23}) \\
0 & -\hat{a}\eta e^{i\delta}\overline{c}_{13}'\sin(2\theta_{23}) & \overline{\lambda}''_+ +\hat{a}\eta\overline{c}_{13}^{\prime 2}\cos(2\theta_{23}) 
\end{bmatrix}
\;,
\end{eqnarray}
where we have dropped off-diagonal 
terms of order $O(\epsilon^3)|\delta m^2_{31}|$ 
or smaller.
Define the matrix $\overline{X}$ as
\begin{equation}
\overline{X} \;=\; 
\begin{bmatrix} 1 & 0 & 0 \\
                          0 & \overline{c}_\chi & \overline{s}_\chi e^{-i\delta}\\
                          0 & -\overline{s}_\chi e^{+i\delta} & \overline{c}_\chi
\end{bmatrix}
\;,
\end{equation}
where 
$\overline{c}_\chi = \cos\overline{\chi}$,
$\overline{s}_\chi = \sin\overline{\chi}$,
and
\begin{eqnarray}
\tan 2\overline{\chi} 
& \equiv &
\dfrac{ -2\hat{a}\eta\overline{c}_{13}'\sin(2\theta_{23}) }
       { (\overline{\lambda}''_{+} - \overline{\lambda}'_{+})+\hat{a}\eta(1+\overline{c}_{13}^{\prime 2})\cos(2\theta_{23}) }
\cr  
& \approx &
-\dfrac{2\hat{a}\eta\sin(2\theta_{23})}
{[\delta m^2_{31}c_{13}^2-\delta m^2_{21}(c_{12}^2-s_{12}^2 s_{13}^2)]+2\hat{a}\eta\cos(2\theta_{23})}
\;.
\end{eqnarray}
Note that
\begin{equation}
-\theta_{23}\;<\;\overline{\chi}\;\le\;0\quad\mbox{for $\eta>0$}\;,\qquad
0\;\le\;\overline{\chi}\;<\;\dfrac{\pi}{2}-\theta_{23}\quad\mbox{for $\eta<0$}\;.
\end{equation}
The $\beta$-dependence of $\overline{\chi}$ is shown in Fig.~\ref{chibarplot} for
several values of $\eta$, both positive (Fig.~\ref{chibarplot}(a)) and negative
(Fig.~\ref{chibarplot}(b)).

Using $\overline{X}$, we find
\begin{equation}
\overline{H}''''_{\eta +} 
\;=\; \overline{X}^\dagger \overline{H}_\eta''' \overline{X} 
\;\approx\;
\begin{bmatrix}
\overline{\lambda}''_- & 0 & 0 \\
0 & \overline{\lambda}'''_{X-} & 0 \\
0 & 0 & \overline{\lambda}'''_{X+}
\end{bmatrix}\;, 
\end{equation}
where
\begin{eqnarray}
\lefteqn{
\overline{\lambda}'''_{X\pm}
} \cr
& \equiv &
   \dfrac{ ( \overline{\lambda}''_{+} + \overline{\lambda}'_{+} - \hat{a}\eta\overline{s}_{13}^{\prime 2}\cos 2\theta_{23})
\pm \sqrt{ [ \overline{\lambda}''_{+} - \overline{\lambda}'_{+} + \hat{a}\eta(1+\overline{c}_{13}^{\prime 2})\cos 2\theta_{23}]^2 
         + 4 (\hat{a}\eta\overline{c}_{13}'\sin 2\theta_{23})^2 }
         }
         { 2 } \;.
\cr
& &
\label{antilambdatripleprimeplusminusdef}
\end{eqnarray}
Thus, $\overline{H}_{\eta +}''''$ is approximately diagonal.
The asymptotic forms of $\overline{\lambda}_{X\pm}'''$ at $\beta\gg 0$ are
\begin{eqnarray}
\overline{\lambda}'''_{X+} & \rightarrow & \phantom{-}\hat{a}|\eta| +
\left\{
\begin{array}{ll}
 \delta m^2_{31}c_{13}^2 c_{23}^2
+\delta m^2_{21}(c_{12}^2 s_{23}^2 + s_{12}^2 s_{13}^2 c_{23}^2)
\quad & \mbox{for $\eta>0$} \\
 \delta m^2_{31}c_{13}^2 s_{23}^2
+\delta m^2_{21}(c_{12}^2 c_{23}^2 + s_{12}^2 s_{13}^2 s_{23}^2)
& \mbox{for $\eta<0$}
\end{array}
\right.
\cr
\overline{\lambda}'''_{X-} & \rightarrow & -\hat{a}|\eta| +
\left\{
\begin{array}{ll}
 \delta m^2_{31}c_{13}^2 s_{23}^2
+\delta m^2_{21}(c_{12}^2 c_{23}^2 + s_{12}^2 s_{13}^2 s_{23}^2)
\quad & \mbox{for $\eta>0$} \\
 \delta m^2_{31}c_{13}^2 c_{23}^2
+\delta m^2_{21}(c_{12}^2 s_{23}^2 + s_{12}^2 s_{13}^2 c_{23}^2)
& \mbox{for $\eta<0$}
\end{array}
\right.
\end{eqnarray}
Note that $\overline{\lambda}'''_{X\pm}$ have
the same asymptotics as $\lambda'''_{X\pm}$ 
for the neutrino case except with the 
$\eta>0$ and $\eta<0$ cases reversed. 
The $\beta$-dependence of 
$\overline{\lambda}_{X\pm}'''$ are shown in 
Figs.~\ref{chibarplot}(c) to (f).

\item {$\bm{\delta m^2_{31}<}0$ \textbf{Case}}

\begin{figure}[t]
\subfigure[$\eta>0$]{\includegraphics[height=5cm]{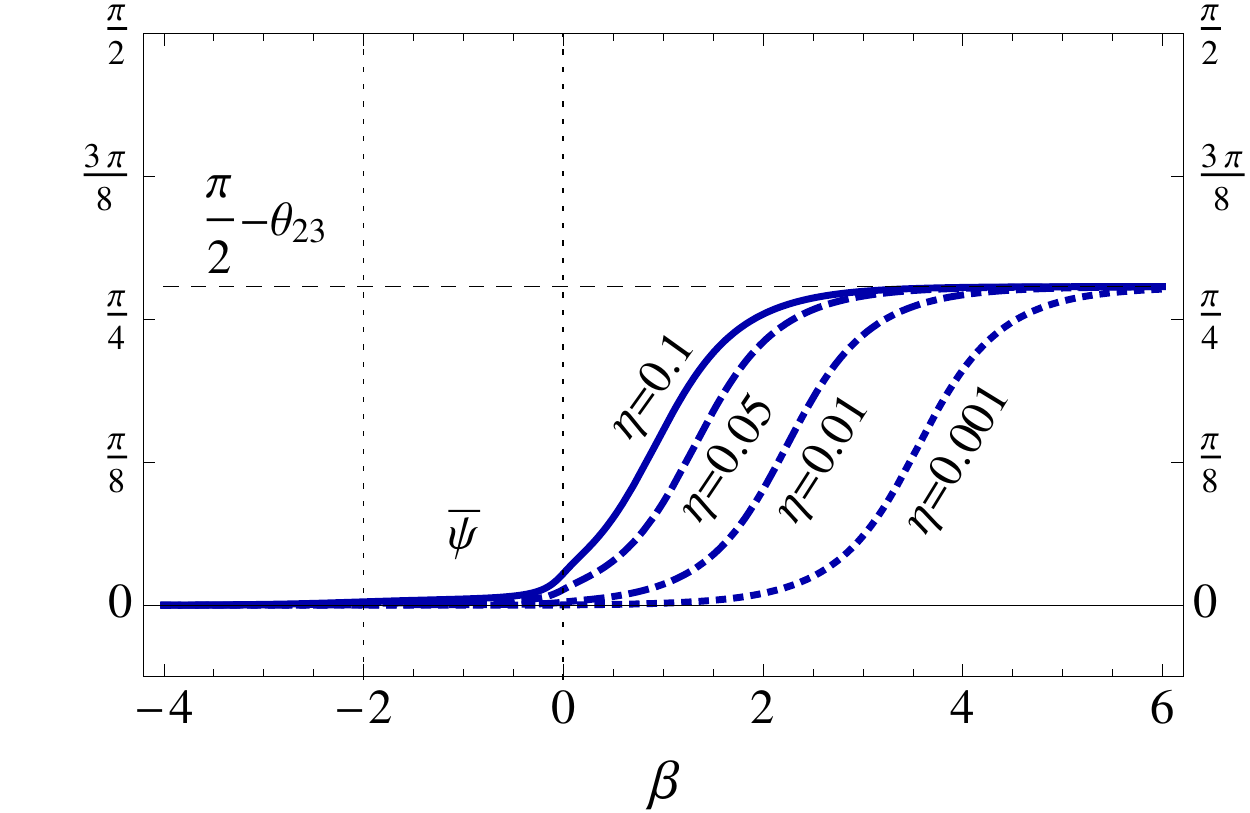}}
\subfigure[$\eta<0$]{\includegraphics[height=5cm]{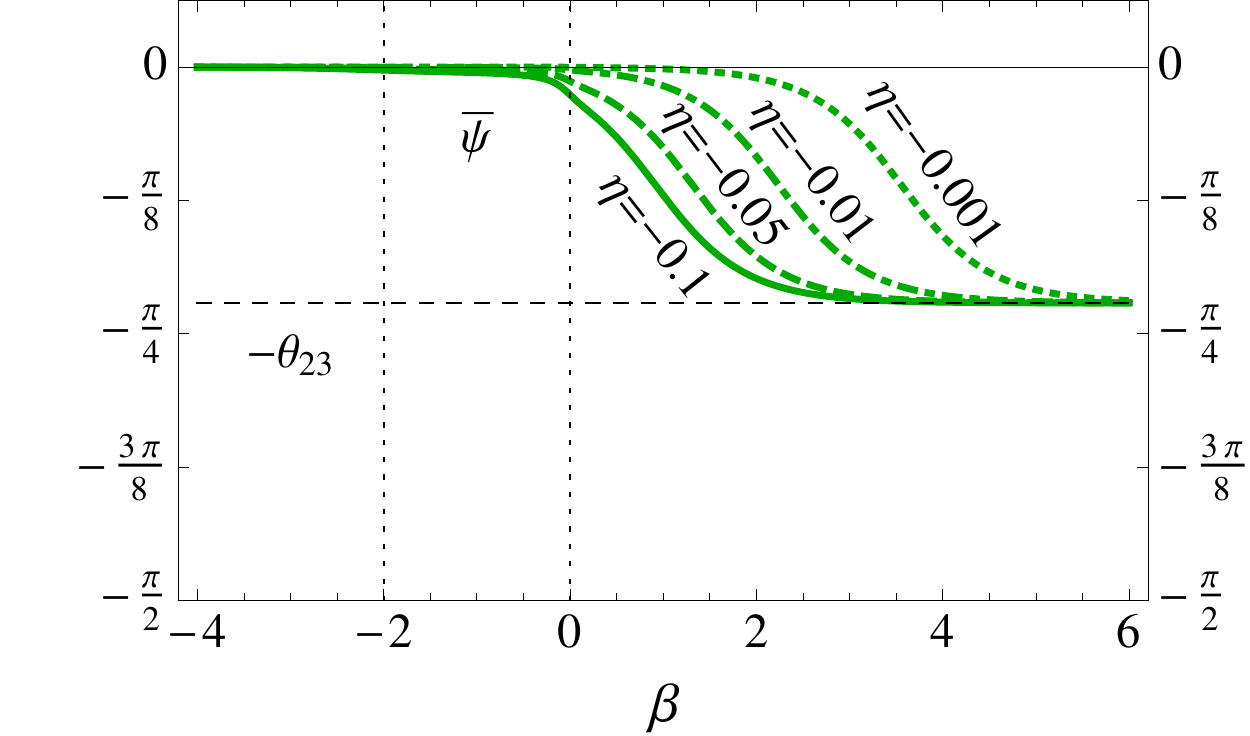}}
\subfigure[$\eta>0$]{\includegraphics[height=5cm]{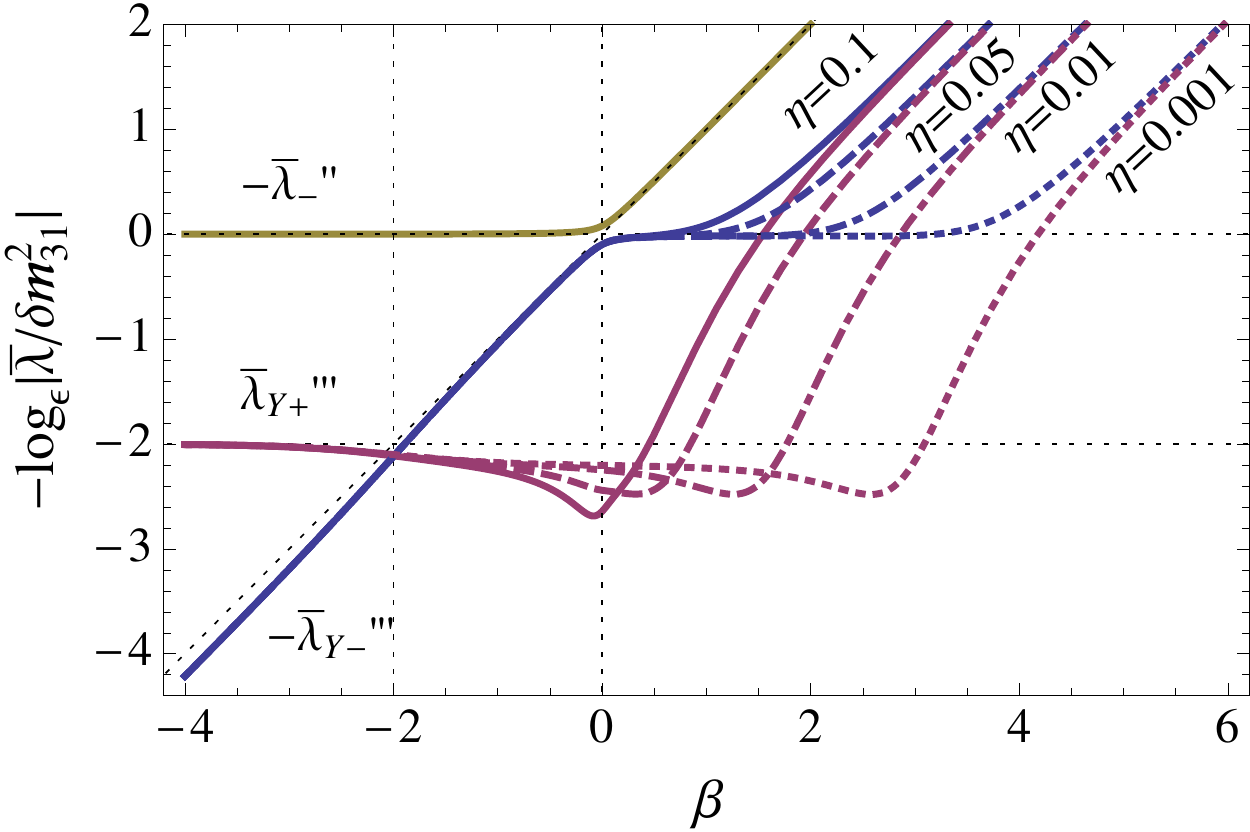}}\qquad
\subfigure[$\eta<0$]{\includegraphics[height=5cm]{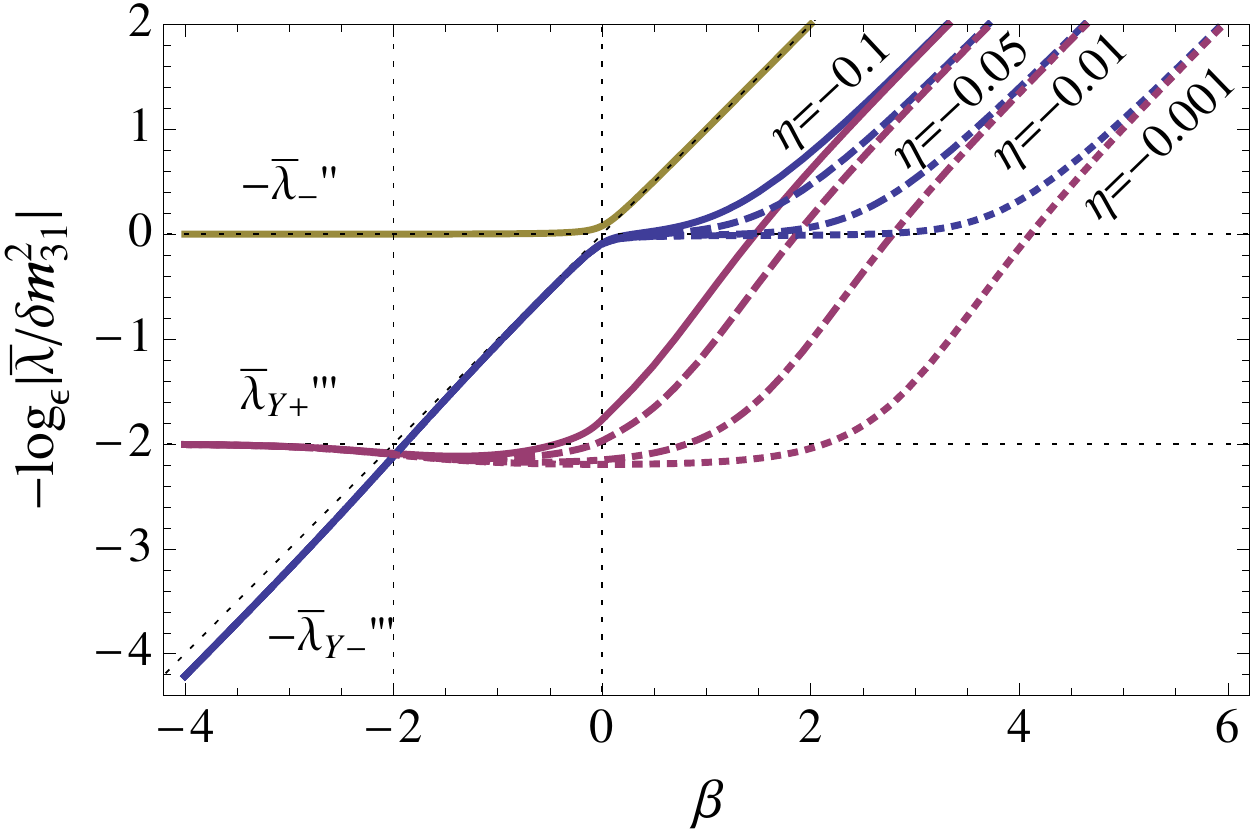}}
\subfigure[$\eta>0$]{\includegraphics[height=5cm]{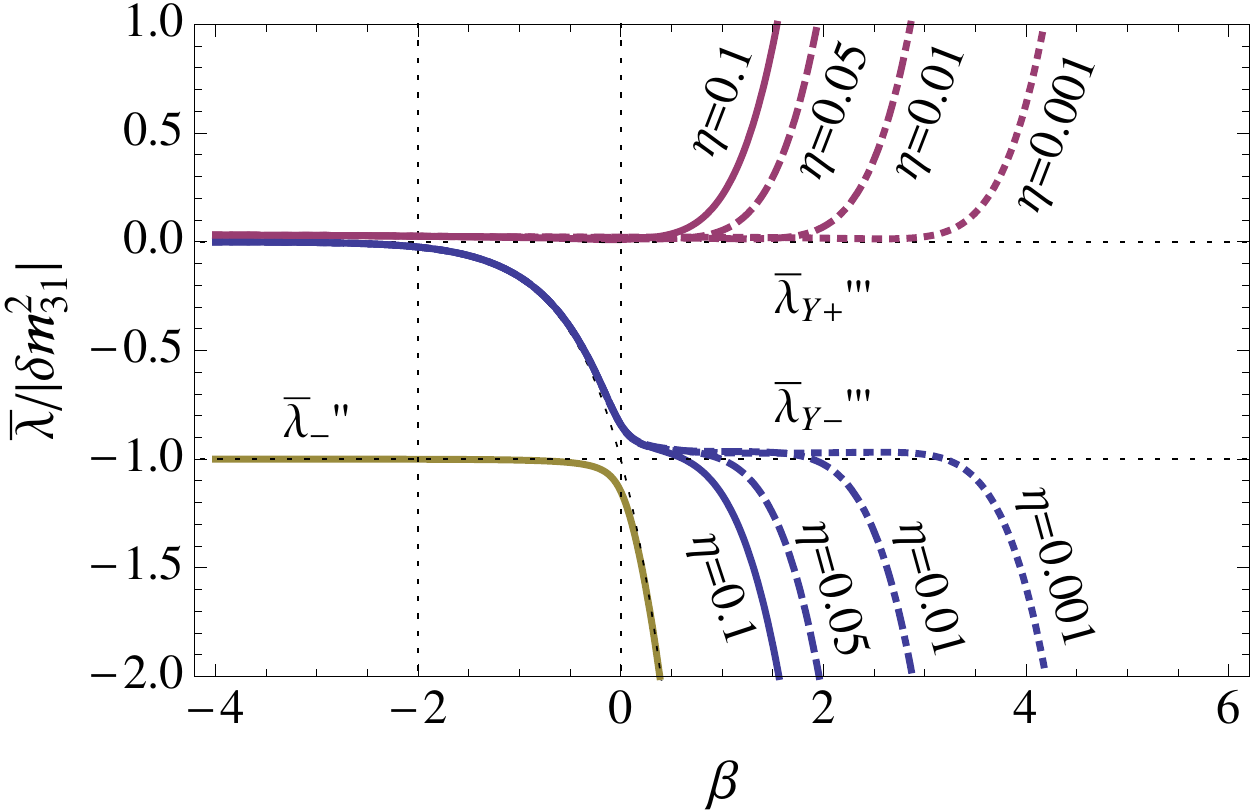}}\quad
\subfigure[$\eta<0$]{\includegraphics[height=5cm]{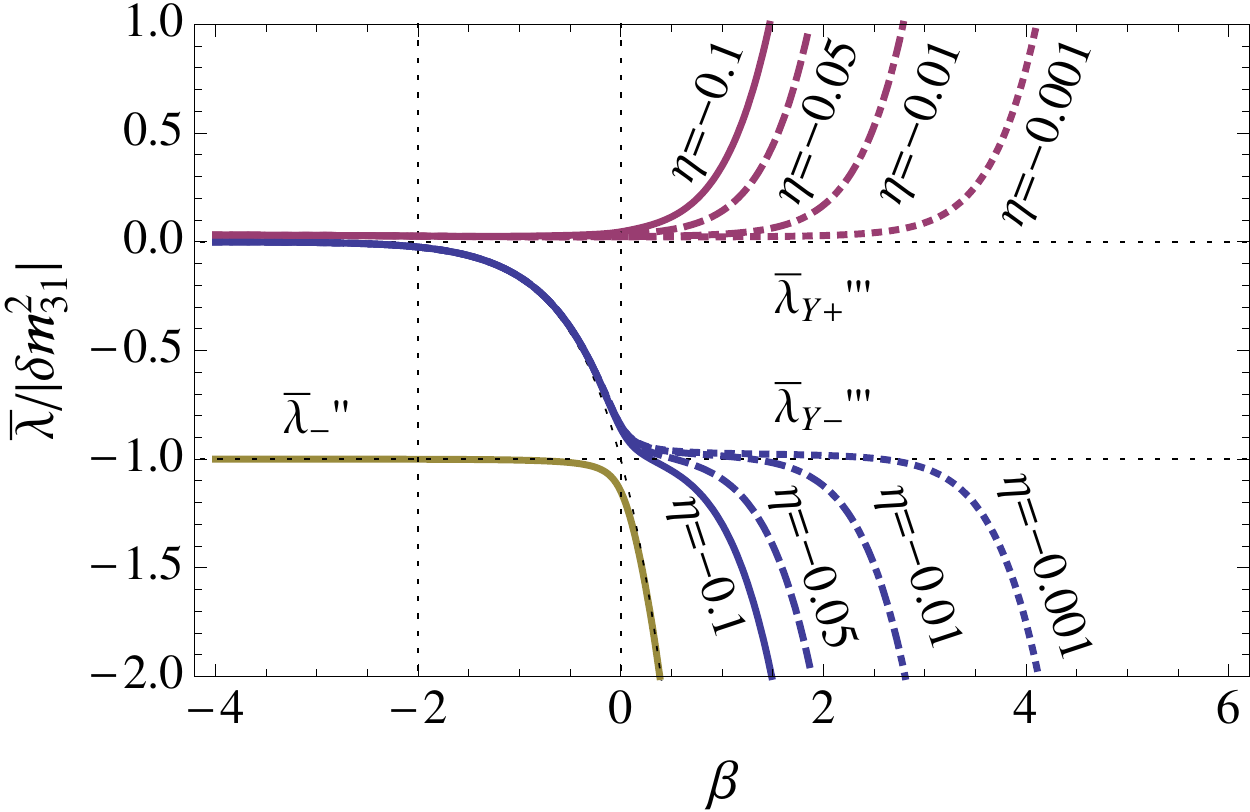}}
\caption{Inverted hierarchy case. The $\beta$-dependence of 
$\overline{\psi}$ and $\overline{\lambda}'''_{Y\pm}$
for several values of $\eta$ with $s_{23}^2=0.4$.}
\label{psibarplot}
\end{figure}

In the $\delta m^2_{31}<0$ case 
$\hat{a}\overline{c}_{13}'\rightarrow O(\epsilon)|\delta m^2_{31}|$
as $\beta$ is increased beyond 0.  Therefore, we can approximate
\begin{eqnarray}
H_\eta''' 
& = & \overline{W}^\dagger\overline{V}^\dagger \overline{H}_\eta' \overline{V}\overline{W} \cr
& \approx & \overline{H}_0''' - \hat{a}\eta \overline{M}_\eta'(0,\overline{\theta}_{13}',\theta_{23},\delta) \cr
& \approx &
\begin{bmatrix}
\overline{\lambda}''_+ + \hat{a}\eta\overline{s}_{13}^{\prime 2}\cos(2\theta_{23}) & \hat{a}\eta e^{i\delta}\overline{s}_{13}'\sin(2\theta_{23}) & 0 \\
\hat{a}\eta e^{-i\delta}\overline{s}_{13}'\sin(2\theta_{23}) & \overline{\lambda}_+' -\hat{a}\eta\cos(2\theta_{23}) & 0 \\
0 & 0 & \overline{\lambda}''_-  
\end{bmatrix}
\;,
\end{eqnarray}
where we have dropped off-diagonal terms 
of order $O(\epsilon^3)|\delta m^2_{31}|$ or smaller.
Define the matrix $\overline{Y}$ as
\begin{equation}
\overline{Y} = 
\left[ \begin{array}{ccc} \overline{c}_\psi & \overline{s}_\psi e^{+i\delta} & 0 \\
                         -\overline{s}_\psi e^{-i\delta} & \overline{c}_\psi & 0      \\
                          0 & 0 & 1
       \end{array}
\right] \;,
\end{equation}
where 
$\overline{c}_\psi = \cos\overline{\psi}$,
$\overline{s}_\psi = \sin\overline{\psi}$,
and
\begin{eqnarray}
\tan 2\overline{\psi}
& \equiv &
-\dfrac{ 2\hat{a}\eta\overline{s}_{13}'\sin(2\theta_{23}) }
       { (\overline{\lambda}''_{+} - \overline{\lambda}'_{+})+\hat{a}\eta(1+\overline{s}_{13}^{\prime 2})\cos(2\theta_{23}) }
\cr
& \approx &
-\dfrac{2\hat{a}\eta\sin(2\theta_{23})}
{[\delta m^2_{31}c_{13}^2-\delta m^2_{21}(c_{12}^2-s_{12}^2 s_{13}^2)]+2\hat{a}\eta\cos(2\theta_{23})}
\;.
\end{eqnarray}
Note that
\begin{equation}
0\;\le\;\overline{\psi}\;<\;\dfrac{\pi}{2}-\theta_{23}\quad\mbox{for $\eta>0$}
\;,
\qquad
-\theta_{23}\;<\;\overline{\psi}\;\le\;0\quad\mbox{for $\eta<0$}
\;.
\end{equation}
The $\beta$-dependence of $\overline{\psi}$ 
is shown in Fig.~\ref{psibarplot} for several values 
of $\eta$, both positive (Fig.~\ref{psibarplot}(a)) 
and negative (Fig.~\ref{psibarplot}(b)).

Using $\overline{Y}$, we find
\begin{equation}
\overline{H}''''_{\eta-} 
\;=\; \overline{Y}^\dagger \overline{H}_\eta''' \overline{Y} 
\;\approx\; 
\left[ \begin{array}{ccc}
\overline{\lambda}'''_{Y-} & 0 & 0 \\
0 & \overline{\lambda}'''_{Y+} & 0 \\
0 & 0 & \overline{\lambda}''_-
\end{array} \right]\;, 
\end{equation}
where
\begin{eqnarray}
\lefteqn{
\overline{\lambda}'''_{Y\pm}
} \cr
& \equiv &
   \dfrac{ ( \overline{\lambda}''_{+} + \overline{\lambda}'_{+} - \hat{a}\eta\overline{c}_{13}^{\prime 2}\cos 2\theta_{23})
\pm \sqrt{ [ \overline{\lambda}''_{+} - \overline{\lambda}'_{+} + \hat{a}\eta(1+\overline{s}_{13}^{\prime 2})\cos 2\theta_{23}]^2 
         + 4 (\hat{a}\eta\overline{s}_{13}'\sin 2\theta_{23})^2 }
         }
         { 2 } \;.
\cr
& &
\label{antilambdatripleprimeplusminusIHdef}
\end{eqnarray}
Therefore, $\overline{H}''''_{\eta-}$ is approximately diagonal.
The asymptotic forms of $\overline{\lambda}_{Y\pm}'''$ at $\beta\gg 0$ are
\begin{eqnarray}
\overline{\lambda}'''_{Y+} & \rightarrow & \phantom{-}\hat{a}|\eta| +
\left\{
\begin{array}{ll}
-|\delta m^2_{31}|c_{13}^2 c_{23}^2
+\delta m^2_{21}(c_{12}^2 s_{23}^2 + s_{12}^2 s_{13}^2 c_{23}^2)
\quad & \mbox{for $\eta>0$} \\
-|\delta m^2_{31}|c_{13}^2 s_{23}^2
+\delta m^2_{21}(c_{12}^2 c_{23}^2 + s_{12}^2 s_{13}^2 s_{23}^2)
& \mbox{for $\eta<0$}
\end{array}
\right.
\cr
\overline{\lambda}'''_{Y-} & \rightarrow & -\hat{a}|\eta| +
\left\{
\begin{array}{ll}
-|\delta m^2_{31}|c_{13}^2 s_{23}^2
+\delta m^2_{21}(c_{12}^2 c_{23}^2 + s_{12}^2 s_{13}^2 s_{23}^2)
\quad & \mbox{for $\eta>0$} \\
-|\delta m^2_{31}|c_{13}^2 c_{23}^2
+\delta m^2_{21}(c_{12}^2 s_{23}^2 + s_{12}^2 s_{13}^2 c_{23}^2)
& \mbox{for $\eta<0$}
\end{array}
\right.
\end{eqnarray}
Note that $\overline{\lambda}'''_{Y\pm}$ have the same asymptotics 
as $\lambda'''_{Y\pm}$ for the neutrino case except 
with the $\eta>0$ and $\eta<0$ cases reversed. 
The $\beta$-dependence of $\overline{\psi}$ and
$\overline{\lambda}'''_{Y\pm}$ are shown 
in Fig.~\ref{psibarplot}.

\end{enumerate}

\subsection{Effective Mixing Angles for Anti-Neutrinos}

We have discovered that the unitary matrix which approximately diagonalizes $\overline{H}$ is
$\edlit{U}{}^*=U^*Q_3^*\overline{V}\,\overline{W}\,\overline{X}$ when $\delta m^2_{31}>0$, and
$\edlit{U}{}^*=U^*Q_3^*\overline{V}\,\overline{W}\,\overline{Y}$ when $\delta m^2_{31}<0$.
Taking the complex conjugate, these are respectively
$\edlit{U}{}=UQ_3\overline{V}\,\overline{W}\,\overline{X}^*$ when $\delta m^2_{31}>0$, and
$\edlit{U}{}=UQ_3\overline{V}\,\overline{W}\,\overline{Y}^*$ when $\delta m^2_{31}<0$.

In the following,
we rewrite the mixing matrix in matter $\edlit{U}{}$ into the form
\begin{equation}
\edlit{U} \;=\; R_{23}(\edlit{\theta}_{23},0)R_{13}(\edlit{\theta}_{13},\edlit{\delta})R_{12}(\edlit{\theta}_{12},0)
\;,
\label{edlitUdef}
\end{equation}
absorbing the extra mixing angles and CP phase into appropriate 
definitions of the `running' parameters 
$\edlit{\theta}_{12}$, $\edlit{\theta}_{13}$, $\edlit{\theta}_{23}$, and $\edlit{\delta}$. 
As in the neutrino case, frequent use is made 
of Eq.~(\ref{Q3commutationrelations}).

\begin{itemize}
\item \textbf{$\delta m^2_{31}>0$ Case:}

Using Eq.~(\ref{Q3commutationrelations}), 
it is straightforward to show that
\begin{eqnarray}
\edlit{U} 
& = & UQ_3\overline{V}\,\overline{W}\,\overline{X}^* \cr
& = & \underbrace{R_{23}(\theta_{23},0)R_{13}(\theta_{13},\delta)R_{12}(\theta_{12},0)}_{\displaystyle U}Q_3
\underbrace{R_{12}(\overline{\varphi},0)}_{\displaystyle \overline{V}}
\underbrace{R_{13}(\overline{\phi},0)}_{\displaystyle \overline{W}}
\underbrace{R_{23}(\overline{\chi},-\delta)}_{\displaystyle \overline{X}^*} \cr
& = & R_{23}(\theta_{23},0)Q_3 R_{13}(\theta_{13},0)R_{12}(\theta_{12},0)
R_{12}(\overline{\varphi},0)R_{13}(\overline{\phi},0)R_{23}(\overline{\chi},-\delta) \cr
& = & R_{23}(\theta_{23},0)Q_3 R_{13}(\theta_{13},0)R_{12}(\underbrace{\theta_{12}+\overline{\varphi}}_{\displaystyle =\overline{\theta}_{12}'},0)
R_{13}(\overline{\phi},0)R_{23}(\overline{\chi},-\delta) \cr
& = & R_{23}(\theta_{23},0)Q_3 R_{13}(\theta_{13},0)R_{12}(\overline{\theta}_{12}',0)
R_{13}(\overline{\phi},0)R_{23}(\overline{\chi},-\delta) \;,
\end{eqnarray}
where in the last and penultimate lines we have combined 
the two $12$-rotations into one.
We now commute $R_{13}(\overline{\phi},0)R_{23}(\overline{\chi},\delta)$ 
through the other mixing matrices to the left as follows:

\begin{itemize}
\item \textbf{Step 1:} Commutation of $R_{13}(\overline{\phi},0)$ 
through $R_{12}(\overline{\theta}_{12}',0)$.

In the range $\beta\agt 0$, the angle $\theta'_{12}$ 
is approximately equal to zero, so we can approximate
\begin{equation}
R_{12}(\theta'_{12},0)
\;\approx\; R_{12}(0,0)
\;=\; \begin{bmatrix} 1 & 0 & 0 \\ 0 & 1 & 0 \\ 0 & 0 & 1 \end{bmatrix}
\;.
\label{R12theta12primebarapprox}
\end{equation}
Note that
\begin{equation}
R_{12}(0,0)R_{13}(\overline{\phi},0) \;=\; R_{13}(\overline{\phi},0)R_{12}(0,0)
\end{equation}
for any $\overline{\phi}$.
On the other hand,
in the range $\beta\alt -1$
the angle $\overline{\phi}$ is negligibly small, so we can approximate
\begin{equation}
R_{13}(\overline{\phi},0) \;\approx\; R_{13}(0,0)
\;=\; \begin{bmatrix} 1 & 0 & 0 \\ 0 & 1 & 0 \\ 0 & 0 & 1 \end{bmatrix}
\;.
\end{equation}
Note that
\begin{equation}
R_{12}(\overline{\theta}'_{12},0)R_{13}(0,0)\;=\;R_{13}(0,0)R_{12}(\overline{\theta}'_{12},0)
\end{equation}
for any $\overline{\theta}'_{12}$.
Therefore, for all $\beta$ we have
\begin{equation}
R_{12}(\overline{\theta}'_{12},0)R_{13}(\overline{\phi},0)
\;\approx\; R_{13}(\overline{\phi},0)R_{12}(\overline{\theta}'_{12},0)\;,
\end{equation}
and
\begin{eqnarray}
\edlit{U}
& = & R_{23}(\theta_{23},0)Q_3 R_{13}(\theta_{13},0)R_{12}(\overline{\theta}_{12}',0)
R_{13}(\overline{\phi},0)R_{23}(\overline{\chi},-\delta)
\cr
& \approx & R_{23}(\theta_{23},0)Q_3 R_{13}(\theta_{13},0)R_{13}(\overline{\phi},0)R_{12}(\overline{\theta}_{12}',0)R_{23}(\overline{\chi},-\delta)
\cr
& = & R_{23}(\theta_{23},0)Q_3 R_{13}(\underbrace{\theta_{13}+\overline{\phi}}_{\displaystyle =\overline{\theta}'_{13}},0)R_{12}(\overline{\theta}_{12}',0)R_{23}(\overline{\chi},-\delta)
\cr
& = & R_{23}(\theta_{23},0)Q_3 R_{13}(\overline{\theta}'_{13},0)R_{12}(\overline{\theta}'_{12},0)R_{23}(\overline{\chi},-\delta)
\;,
\end{eqnarray}
where in the last and penultimate lines we have combined the two $13$-rotations into one.
The $\beta$-dependence of $\overline{\theta}'_{13}$ was shown in Fig.~\ref{phibarplot}(b).

\item \textbf{Step 2:} Commutation of $R_{23}(\overline{\chi},\delta)$ through $R_{12}(\overline{\theta}_{12}',0)$.

In the range $\beta\agt 0$, the angle $\overline{\theta}'_{12}$ 
is approximately equal to zero as we have noted above and 
we have the approximation given in Eq.~(\ref{R12theta12primebarapprox}).
Note that
\begin{equation}
R_{12}(0,0)R_{23}(\overline{\chi},-\delta)
\;=\; R_{23}(\overline{\chi},-\delta)R_{12}(0,0)
\end{equation}
for any $\overline{\chi}$.
On the other hand, in the range $\beta\alt 0$, the angle $\overline{\chi}$ is negligibly small so we can approximate
\begin{equation}
R_{23}(\overline{\chi},-\delta) \;\approx\; R_{23}(0,-\delta) 
\;=\; \begin{bmatrix} 1 & 0 & 0 \\ 0 & 1 & 0 \\ 0 & 0 & 1 \end{bmatrix}
\;.
\label{R12chibardeltaapprox}
\end{equation}
Note that
\begin{equation}
R_{12}(\overline{\theta}'_{12},0)R_{23}(0,\delta)\;=\; R_{23}(0,\delta)R_{12}(\overline{\theta}'_{12},0)
\end{equation}
for any $\overline{\theta}'_{12}$.
Therefore, for all $\beta$ we see that
\begin{equation}
R_{12}(\overline{\theta}'_{12},0)R_{23}(\overline{\chi},-\delta)
\;\approx\; R_{23}(\overline{\chi},-\delta)R_{12}(\overline{\theta}'_{12},0)
\;,
\end{equation}
and
\begin{eqnarray}
\edlit{U}
& \approx & R_{23}(\theta_{23},0)Q_3 R_{13}(\overline{\theta}'_{13},0)R_{23}(\overline{\chi},-\delta)R_{12}(\overline{\theta}'_{12},0)
\;.
\end{eqnarray}

\item \textbf{Step 3:} Commutation of $R_{23}(\overline{\chi},-\delta)$ through $R_{13}(\overline{\theta}'_{13},0)$.

When $\delta m^2_{31}>0$ we have 
$\overline{\theta}'_{13}\approx 0$ 
in the range $\beta\agt 1$
so we can approximate
\begin{equation}
R_{13}(\overline{\theta}'_{13},0) \;\approx\;
R_{13}(0,0) \;=\;
\begin{bmatrix}
1 & 0 & 0 \\ 0 & 1 & 0 \\ 0 & 0 & 1
\end{bmatrix}
\;.
\end{equation}
Note that
\begin{equation}
R_{13}(0,0)R_{23}(\overline{\chi},-\delta)\;=\; R_{23}(\overline{\chi},-\delta)R_{13}(0,0)
\end{equation}
for any $\overline{\chi}$.
On the other hand, in the range $\beta\alt 0$ the angle $\overline{\chi}$ was negligibly small so that
we had Eq.~(\ref{R12chibardeltaapprox}).
Note that
\begin{equation}
R_{13}(\overline{\theta}'_{13},0)R_{12}(0,-\delta)\;=\; R_{23}(0,-\delta)R_{13}(\overline{\theta}'_{13},0)
\;,
\end{equation}
for any $\overline{\theta}'_{13}$.
Therefore, for all $\beta$ we see that
\begin{equation}
R_{13}(\overline{\theta}'_{13},0)R_{23}(\overline{\chi},-\delta)
\;\approx\; R_{23}(\overline{\chi},-\delta)R_{13}(\overline{\theta}'_{13},0)
\;,
\end{equation}
and using Eq.~(\ref{Q3commutationrelations}) we obtain
\begin{eqnarray}
\edlit{U}
& \approx & R_{23}(\theta_{23},0)Q_3 R_{23}(\overline{\chi},-\delta)R_{13}(\overline{\theta}'_{13},0)R_{12}(\overline{\theta}'_{12},0)
\cr
& = & R_{23}(\theta_{23},0)R_{23}(\overline{\chi},0)R_{13}(\overline{\theta}'_{13},\delta)R_{12}(\overline{\theta}'_{12},0)Q_3
\cr
& = & R_{23}(\underbrace{\theta_{23}+\overline{\chi}}_{\displaystyle =\overline{\theta}'_{23}},0)R_{13}(\overline{\theta}'_{13},\delta)R_{12}(\overline{\theta}'_{12},0)Q_3
\cr
& = & R_{23}(\overline{\theta}'_{23},0)R_{13}(\overline{\theta}'_{13},\delta)R_{12}(\overline{\theta}'_{12},0)Q_3
\;,
\end{eqnarray}
where in the last and penultimate lines 
we have combined the two $23$-rotations into one.
The matrix $Q_3$ on the far right can be absorbed 
into the redefinitions of Majorana phases 
and can be dropped.

\end{itemize}

Thus, we find that the effective mixing matrix $\edlit{U}$ 
in the case $\delta m^2_{31}>0$ can be expressed as 
Eq.~(\ref{edlitUdef}) with the effective mixing angles 
and effective CP-violating phase given approximately by
\begin{eqnarray}
\edlit{\theta}_{12} & \approx & \overline{\theta}'_{12} \;=\; \theta_{12}+\overline{\varphi}\;,\cr
\edlit{\theta}_{13} & \approx & \overline{\theta}'_{13} \;=\; \theta_{13}+\overline{\phi}\;,\cr
\edlit{\theta}_{23} & \approx & \overline{\theta}'_{23} \;=\; \theta_{23}+\overline{\chi}\;,\cr
\edlit{\delta}      & \approx & \delta\;.
\end{eqnarray}
%

\item \textbf{$\delta m^2_{31}<0$ Case:}

Using Eq.~(\ref{Q3commutationrelations}), we obtain
\begin{eqnarray}
\edlit{U} 
& = & UQ_3\overline{V}\,\overline{W}\,\overline{Y}^* \cr
& = & 
\underbrace{R_{23}(\theta_{23},0)R_{13}(\theta_{13},\delta)R_{12}(\theta_{12},0)}_{\displaystyle U}Q_3
\underbrace{R_{12}(\overline{\varphi},0)}_{\displaystyle \overline{V}}
\underbrace{R_{13}(\overline{\phi},0)}_{\displaystyle \overline{W}}
\underbrace{R_{12}(\overline{\psi},\delta)}_{\displaystyle \overline{Y}^*} \cr
& = & R_{23}(\theta_{23},0)Q_3 R_{13}(\theta_{13},0)R_{12}(\overline{\theta}_{12}',0)
R_{13}(\overline{\phi},0)R_{12}(\overline{\psi},\delta) \;.
\end{eqnarray}
We now commute $R_{13}(\overline{\phi},0)R_{12}(\overline{\psi},\delta)$ 
through the other mixing matrices to the left and re-express 
$\edlit{U}$ as in Eq.~(\ref{edlitUdef}),
absorbing the extra mixing angles and CP phase 
into $\edlit{\theta}_{12}$, $\edlit{\theta}_{13}$, $\edlit{\theta}_{23}$, 
and $\edlit{\delta}$. 
The first step is the same as the $\delta m^2_{31}>0$ case,
the only difference being the $\beta$-dependence of $\overline{\theta}'_{13}$, 
which is also shown in Fig.~\ref{phibarplot}(b).

\begin{itemize}
\item \textbf{Step 2:} Commutation of $R_{12}(\overline{\psi},\delta)$ through $R_{12}(\overline{\theta}_{12}',0)$.

In the range $\beta\agt 0$ the angle $\overline{\theta}'_{12}$ 
is approximately equal to zero as we have noted previously, 
and we have the approximation given in Eq.~(\ref{R12theta12primebarapprox}).
Note that
\begin{equation}
R_{12}(0,0)R_{12}(\overline{\psi},\delta)
\;=\; R_{12}(\overline{\psi},\delta)R_{12}(0,0)
\end{equation}
for any $\overline{\psi}$.
On the other hand, in the range $\beta\alt 0$ 
the angle $\overline{\psi}$ is negligibly small so that
\begin{equation}
R_{12}(\overline{\psi},\delta) \;\approx\; R_{12}(0,\delta) 
\;=\; \begin{bmatrix} 1 & 0 & 0 \\ 0 & 1 & 0 \\ 0 & 0 & 1 \end{bmatrix}
\;.
\label{R13psibardeltaapprox}
\end{equation}
Note that
\begin{equation}
R_{12}(\overline{\theta}'_{12},0)R_{12}(0,\delta)
\;=\; R_{12}(0,\delta)R_{12}(\overline{\theta}'_{12},0)
\end{equation}
for any $\overline{\theta}'_{12}$.
Therefore, for all $\beta$ we see that
\begin{equation}
R_{12}(\overline{\theta}'_{12},0)R_{12}(\overline{\psi},\delta)
\;\approx\; R_{12}(\overline{\psi},\delta)R_{12}(\overline{\theta}'_{12},0)
\;,
\end{equation}
and 
\begin{equation}
\edlit{U}
\;\approx\; 
R_{23}(\theta_{23},0)Q_3 R_{13}(\overline{\theta}'_{13},0)R_{12}(\overline{\psi},\delta)R_{12}(\overline{\theta}'_{12},0)
\;.
\end{equation}

\item \textbf{Step 3:} Commutation of $R_{12}(\overline{\psi},\delta)$ through $R_{13}(\overline{\theta}'_{13},0)$.

When $\delta m^2_{31}<0$ we have $\overline{\theta}'_{13}\approx \frac{\pi}{2}$ 
in the range $\beta\agt 1$ so that
\begin{equation}
R_{13}(\theta'_{13},0) \;\approx\;
R_{13}\left(\frac{\pi}{2},0\right) \;=\;
\begin{bmatrix}
0 & 0 & 1 \\ 0 & 1 & 0 \\ -1 & 0 & 0
\end{bmatrix}
\;.
\end{equation}
Note that
\begin{equation}
R_{13}\left(\frac{\pi}{2},0\right)R_{12}(\overline{\psi},\delta)
\;=\; R_{23}(\overline{\psi},-\delta)R_{13}\left(\frac{\pi}{2},0\right)
\end{equation}
for all $\overline{\psi}$.
On the other hand, in the range $\beta\alt 0$ the angle $\overline{\psi}$ 
was negligibly small so that we had the approximation Eq.~(\ref{R13psideltaapprox}).
Note that
\begin{equation}
R_{13}(\overline{\theta}'_{13},0)R_{12}(0,\delta)
\;=\; R_{23}(0,-\delta)R_{13}(\overline{\theta}'_{13},0)
\end{equation}
for all $\overline{\theta}'_{13}$.
Therefore, for all $\beta$ we see that
\begin{equation}
R_{13}(\overline{\theta}'_{13},0)R_{12}(\overline{\psi},\delta)
\;\approx\; R_{23}(\overline{\psi},-\delta)R_{13}(\overline{\theta}'_{13},0)
\;,
\end{equation}
and using Eq.~(\ref{Q3commutationrelations}) we obtain
\begin{eqnarray}
\edlit{U}
& \approx & R_{23}(\theta_{23},0)Q_3 R_{23}(\overline{\psi},-\delta)R_{13}(\overline{\theta}'_{13},0)R_{12}(\overline{\theta}'_{12},0)
\cr
& = & R_{23}(\theta_{23},0)R_{23}(\overline{\psi},0)R_{13}(\overline{\theta}'_{13},\delta)R_{12}(\overline{\theta}'_{12},0)Q_3
\cr
& = & R_{23}(\underbrace{\theta_{23}+\overline{\psi}}_{\displaystyle =\overline{\theta}'_{23}},0)R_{13}(\overline{\theta}'_{13},\delta)R_{12}(\overline{\theta}'_{12},0)Q_3
\cr
& = & R_{23}(\overline{\theta}'_{23},0)R_{13}(\overline{\theta}'_{13},\delta)R_{12}(\overline{\theta}'_{12},0)Q_3
\;,
\end{eqnarray}
where in the last and penultimate lines 
we have combined the two $23$-rotations into one.
The matrix $Q_3$ on the far right can be absorbed 
into redefinitions of the Majorana phases 
and can be dropped.

\end{itemize}

Thus, we find that the effective mixing matrix $\edlit{U}$ 
for anti-neutrinos in the case $\delta m^2_{31}<0$ 
can be expressed as Eq.~(\ref{edlitUdef}) with the 
effective mixing angles and effective CP-violating 
phase given approximately by
\begin{eqnarray}
\edlit{\theta}_{12} & \approx & \overline{\theta}'_{12} \;=\; \theta_{12}+\overline{\varphi}\;,\cr
\edlit{\theta}_{13} & \approx & \overline{\theta}'_{13} \;=\; \theta_{13}+\overline{\phi}\;,\cr
\edlit{\theta}_{23} & \approx & \overline{\theta}'_{23} \;=\; \theta_{23}+\overline{\psi}\;,\cr
\edlit{\delta}      & \approx & \delta\;.
\end{eqnarray}

\end{itemize}

\subsection{Summary of Anti-Neutrino Case}

\begin{figure}[t]
\subfigure[Normal Hierarchy]{\includegraphics[height=5cm]{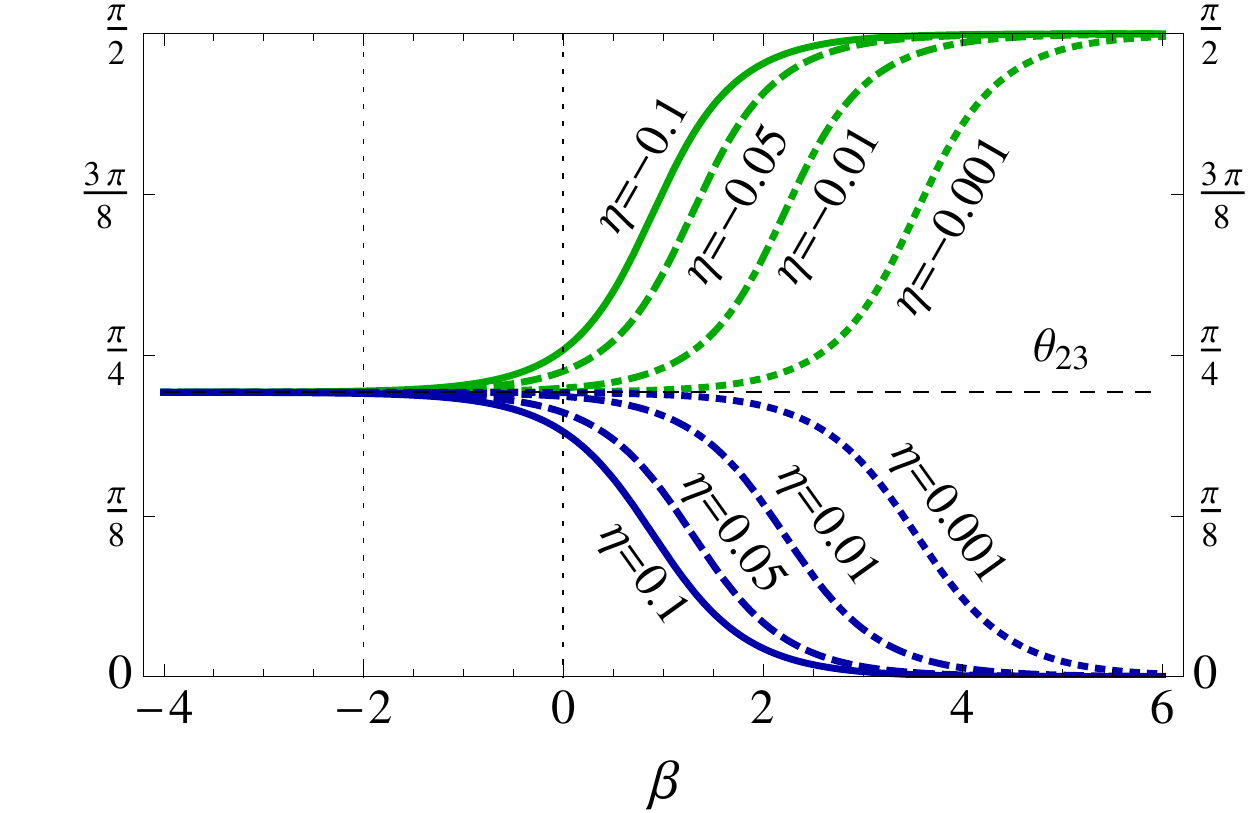}}
\subfigure[Inverted Hierarchy]{\includegraphics[height=5cm]{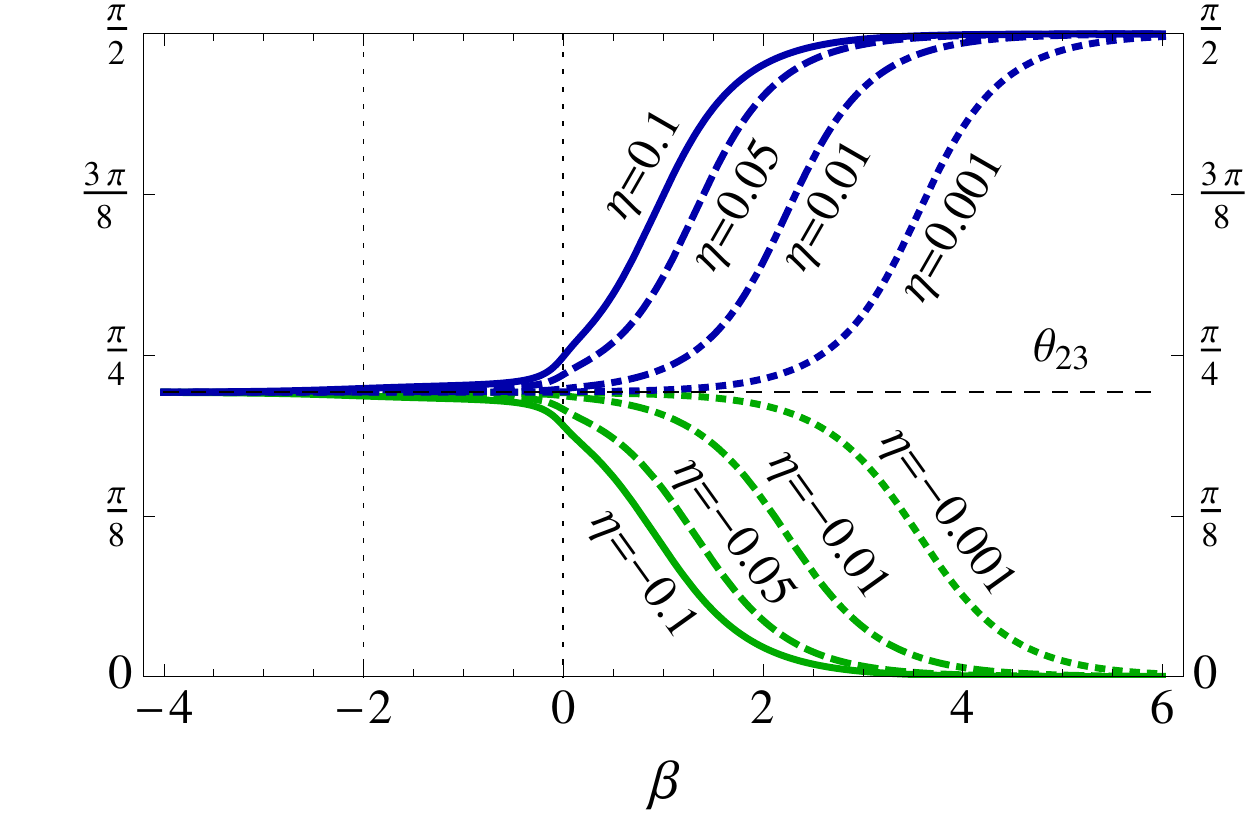}}
\caption{The $\beta$-dependence of $\overline{\theta}_{23}'$
for the (a) normal and (b) inverted hierarchies
for several values of $\eta$ with $s_{23}^2=0.4$.
}
\label{theta23primebarplot}
\end{figure}

To summarize what we have learned, the 
inclusion of the $\hat{a}\eta M_\eta$ term in the effective Hamiltonian shifts
$\theta_{23}$ to $\overline{\theta}_{23}'=\theta_{23}+\overline\chi$ for the $\delta m^2_{31}>0$ case,
and to $\overline{\theta}_{23}'=\theta_{23}+\overline\psi$ for the $\delta m^2_{31}<0$ case.
For both cases, $\theta_{23}'$ can be calculated directly without calculating
$\overline\chi$ or $\overline\psi$ first via the expression
\begin{equation}
\tan 2\overline{\theta}_{23}' \;\approx\; 
\dfrac{[\delta m^2_{31}c_{13}^2-\delta m^2_{21}(c_{12}^2-s_{12}^2 s_{13}^2)]\sin 2\theta_{23}}
      {[\delta m^2_{31}c_{13}^2-\delta m^2_{21}(c_{12}^2-s_{12}^2 s_{13}^2)]\cos 2\theta_{23} +2\hat{a}\eta}
\;.
\label{theta23primeNeutrinobar}
\end{equation}
Note that as $\beta$ is increased,
$\overline{\theta}_{23}'$ runs toward $\dfrac{\pi}{2}$ 
if $\delta m^2_{31}\eta <0$, while it runs toward $0$ 
if $\delta m^2_{31}\eta >0$. The $\beta$-dependence 
of $\overline{\theta}_{23}'$ is shown in 
Fig.~\ref{theta23primebarplot}. The CP-violating phase 
$\delta$ remains unaffected and maintains its 
vacuum value.

The running of the effective mass-squared differences are also modified in the range
$\beta \agt 0$.
For the $\delta m^2_{31}>0$ case, $\overline\lambda_2$ and $\overline\lambda_3$ show extra running,
while for the $\delta m^2_{31}<0$ case, it is $\overline\lambda_1$ and $\overline\lambda_2$ that
show extra running, cf. Figs.~\ref{chibarplot} and \ref{psibarplot}.

\subsection{Discussion at the Probability Level}
\label{comparison-anti-neutrino}

\begin{figure}[t]
\includegraphics[width=7.5cm,height=5.5cm]{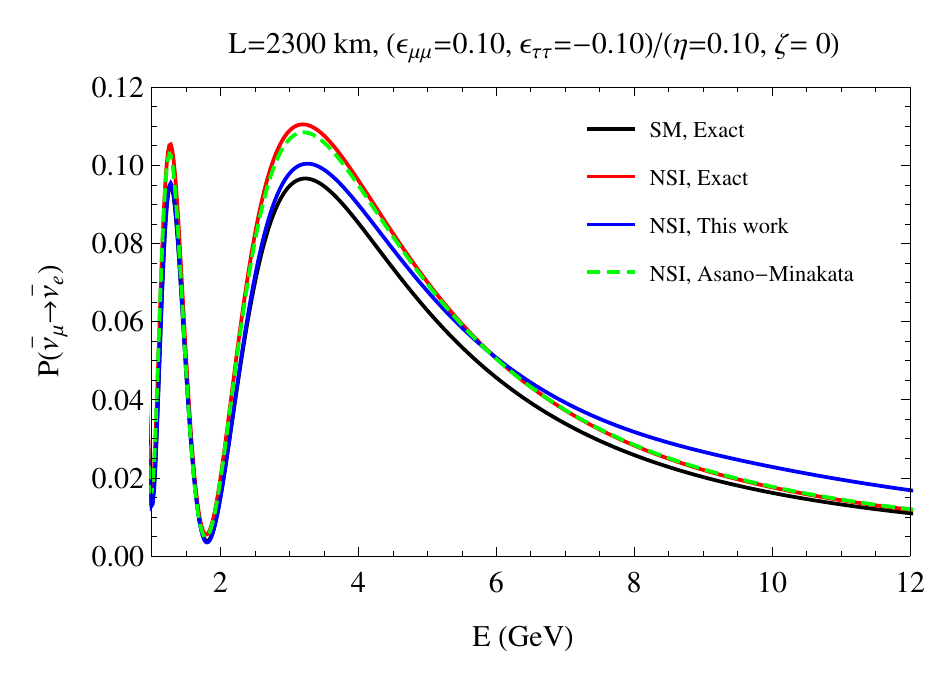}
\includegraphics[width=7.5cm,height=5.5cm]{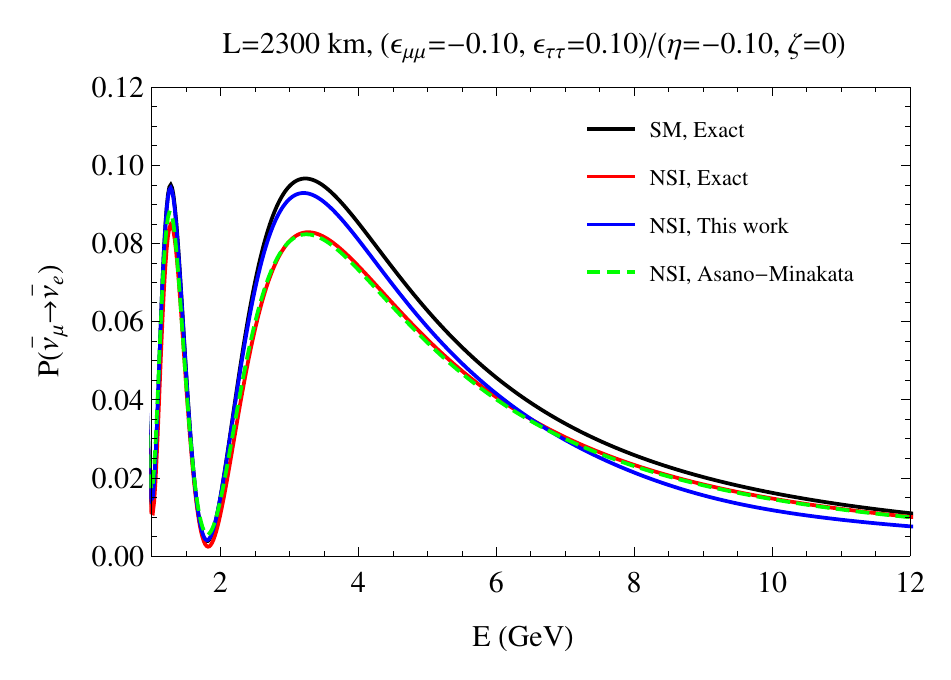}
\includegraphics[width=7.5cm,height=5.5cm]{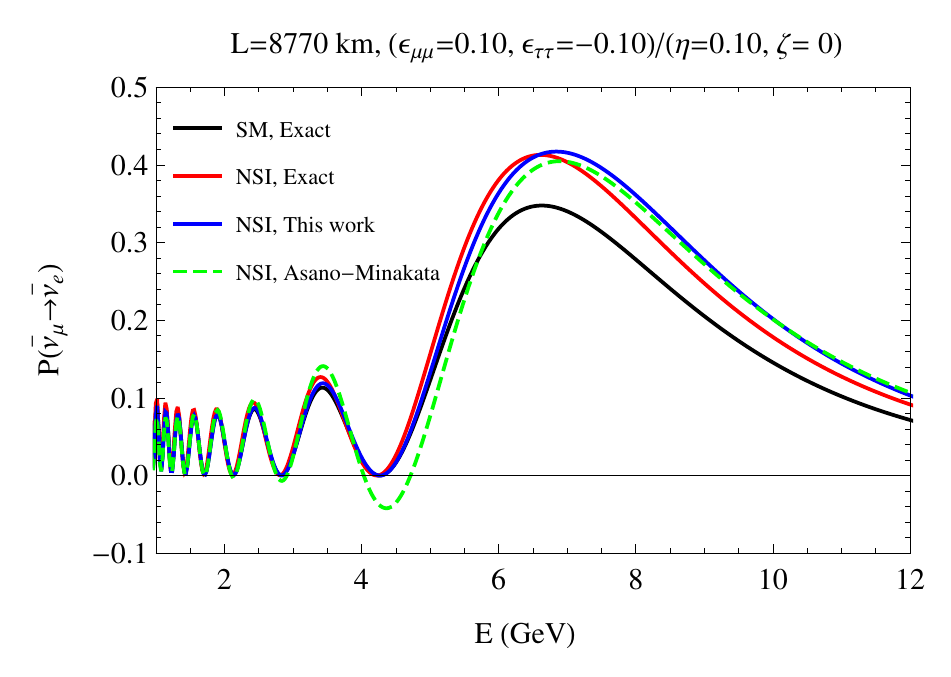}
\includegraphics[width=7.5cm,height=5.5cm]{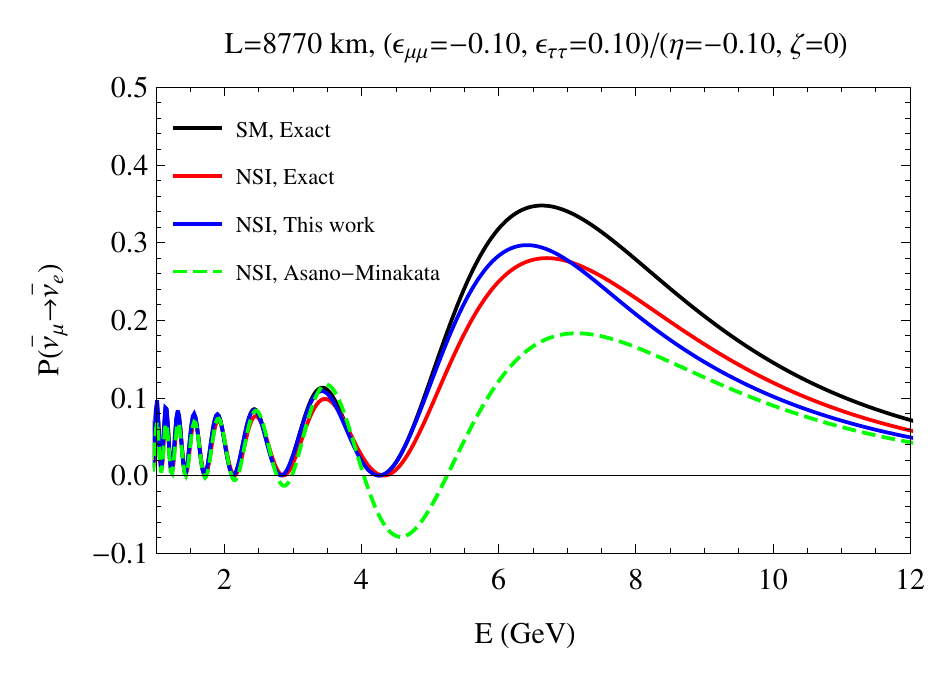}
\caption{$\bar\nu_\mu\rightarrow\bar\nu_e$ transition probability as 
a function of anti-neutrino energy $E$ in GeV for 2300 km 
(8770 km) baseline in upper (lower) panels. We compare the analytical expressions of this work and 
Asano-Minakata \cite{Asano:2011nj} against the exact numerical result assuming $\eta=0.1, \zeta=0$ 
(left panels) and $\eta=-0.1, \zeta=0$ (right panels). The solid black curves portray the standard three-flavor
oscillation probabilities in matter without NSI's. In all the panels, we consider $\theta_{23}=40^{\circ})$, 
$\delta=0^{\circ}$, and inverted mass hierarchy.}
\label{fig:comparison-anti-neutrino}
\end{figure}

Now, we demonstrate how these lepton-flavor-conserving 
NSI parameters affect the anti-neutrino oscillation probability 
for the various appearance and the disappearance channels. 
In Fig.~\ref{fig:comparison-anti-neutrino}, we compare 
our approximate $\bar\nu_\mu\rightarrow\bar\nu_e$ oscillation
probabilities (blue curves) as a function of the neutrino energy 
against the exact numerical results (red curves) assuming
$\eta=0.1, \zeta=0$ (left panels) and $\eta=-0.1, \zeta=0$ (right panels). 
The upper (lower) panels are drawn for 2300 km (8770 km) baseline. 
Here, in all the panels, we assume $\theta_{23}=40^{\circ}$, 
$\delta=0^{\circ}$, and inverted mass hierarchy ($\delta m^2_{31}<0$).
We also compare our results with the approximate expressions of 
Asano and Minakata \cite{Asano:2011nj} (dashed green curves).

\begin{figure}[t]
\includegraphics[width=7.5cm,height=5.5cm]{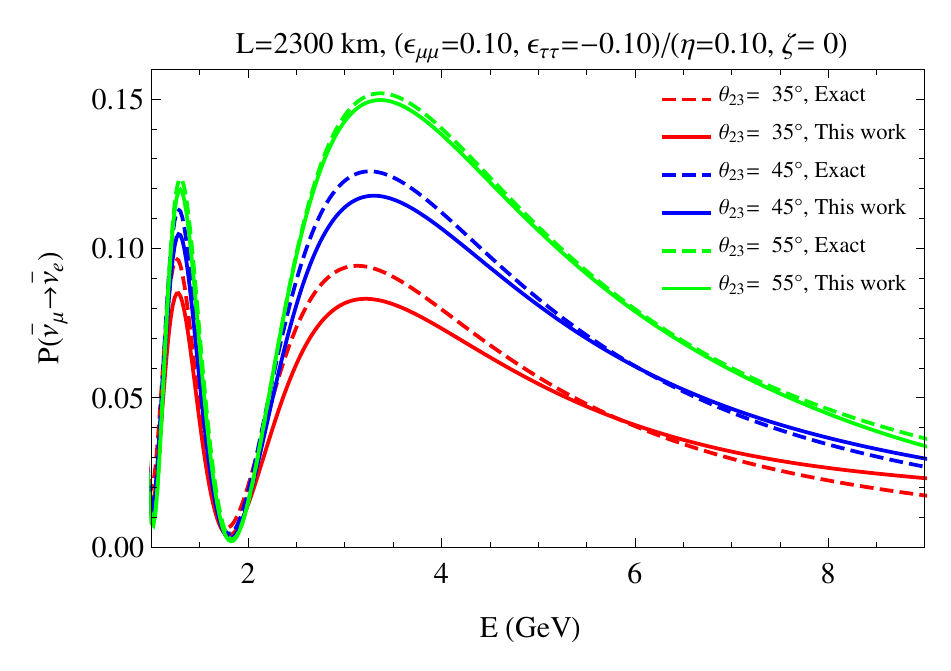}
\includegraphics[width=7.5cm,height=5.5cm]{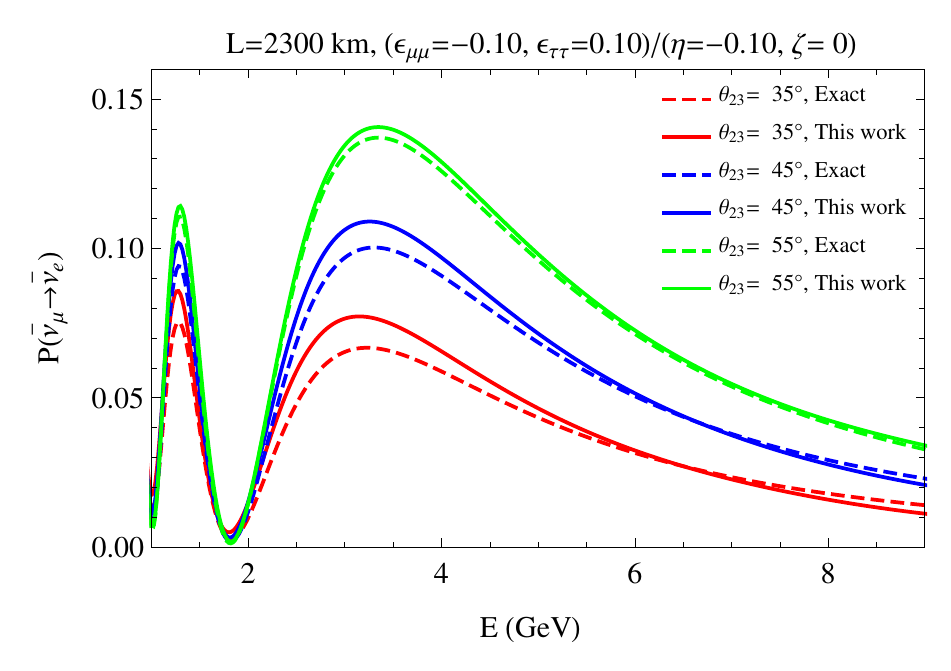}
\includegraphics[width=7.5cm,height=5.5cm]{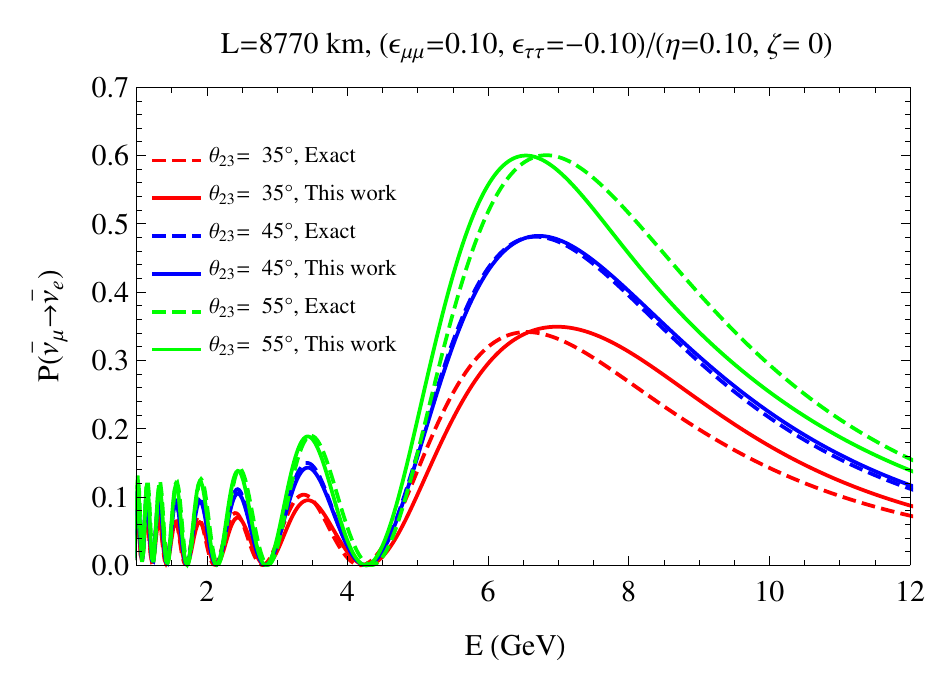}
\includegraphics[width=7.5cm,height=5.5cm]{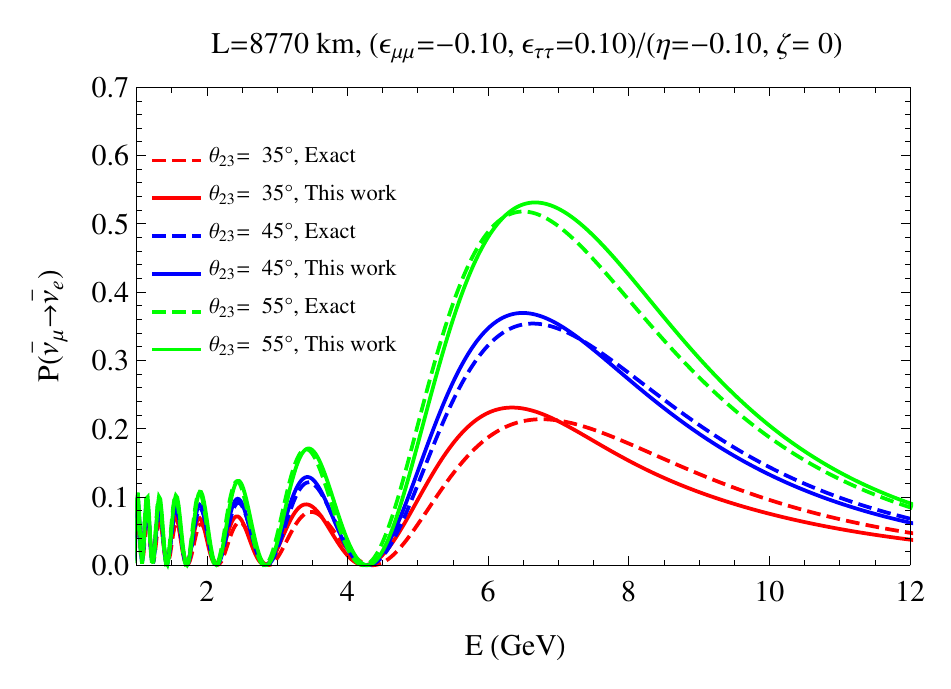}
\caption{Comparison of our analytical expressions (solid curves) to the exact numerical results 
(dashed curves) in case of anti-neutrino for various values of $\theta_{23}$. We assume $\delta=0^{\circ}$ 
and $\delta m^2_{31}<0$. Upper (lower) panels are for 2300 km (8770 km) baseline.}
\label{fig:comparison-theta23-anti-neutrino}
\end{figure}

\begin{figure}[t]
\includegraphics[width=7.5cm,height=5.5cm]{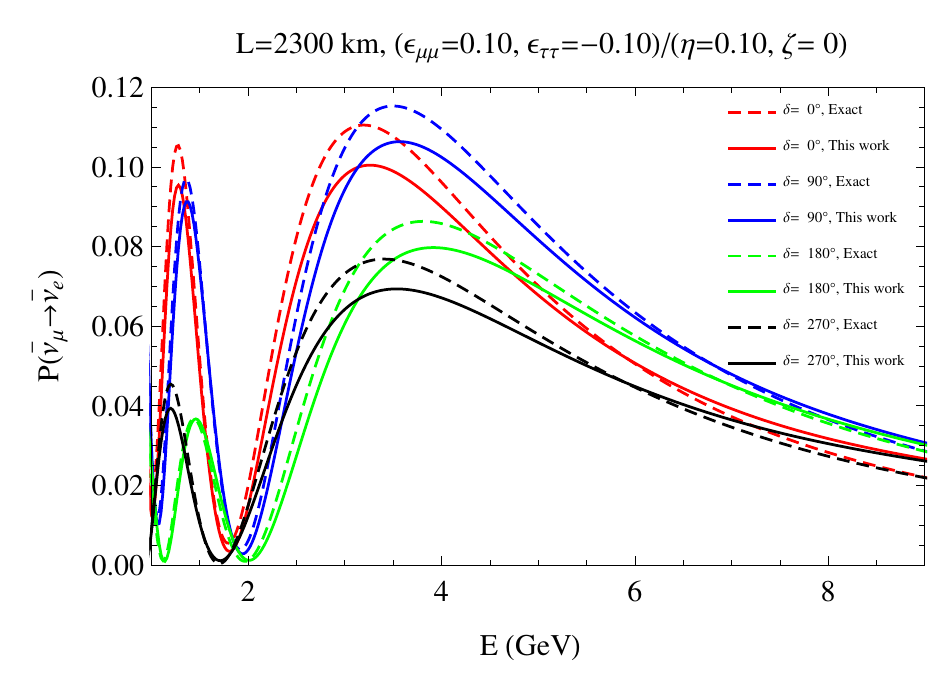}
\includegraphics[width=7.5cm,height=5.5cm]{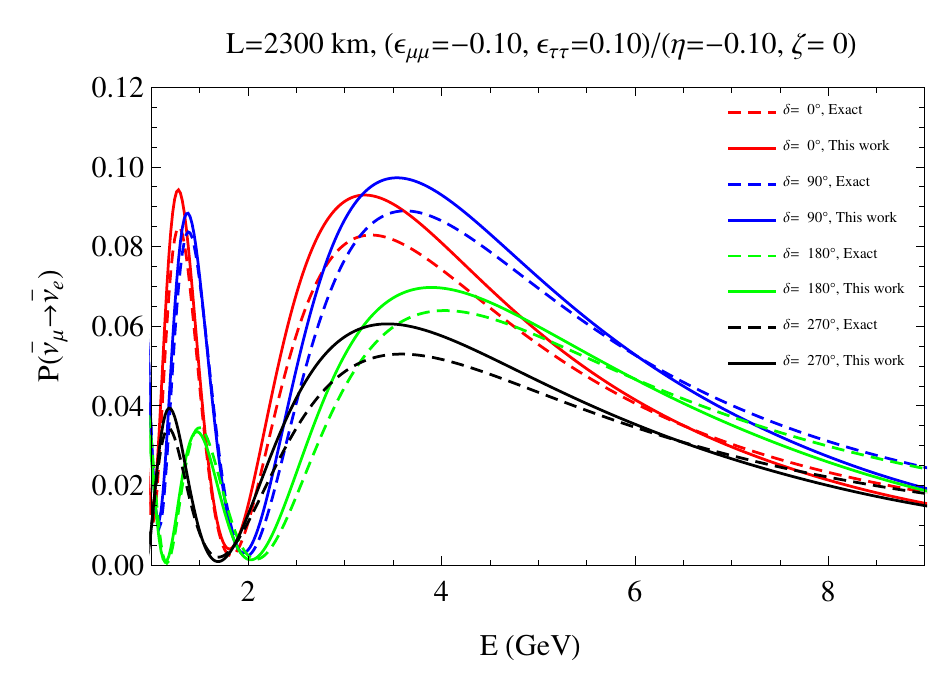}
\caption{Comparison of our analytical expressions (solid curves) 
to the exact numerical results (dashed curves) in case of anti-neutrino 
for different values of the CP-violating phase $\delta$ at 2300 km. 
To generate these plots, we consider $\theta_{23}=40^{\circ}$ 
and $\delta m^2_{31}<0$.}
\label{fig:comparison-CP-anti-neutrino}
\end{figure}

The accuracy of our analytical approximations as compared to 
the exact numerical results for different values of $\theta_{23}$ 
is shown in Fig.~\ref{fig:comparison-theta23-anti-neutrino}. 
All the plots in Fig.~\ref{fig:comparison-theta23-anti-neutrino} 
have been generated assuming $\delta=0^{\circ}$ and 
inverted mass hierarchy ($\delta m^2_{31}<0$).
We take the same choices of $\eta$ and $\zeta$ like in 
Fig.~\ref{fig:comparison-anti-neutrino} and results
are given for 2300 km (upper panels) and 8770 km 
(lower panels) baselines. We find that our approximation 
provides satisfactory match with exact numerical results 
for different values of $\theta_{23}$. 
Fig.~\ref{fig:comparison-CP-anti-neutrino} presents 
a comparison of our approximate probability expressions 
(solid curves) against the exact numerical results (dashed curves) 
assuming four different values of the CP-violating phase 
$\delta$ at 2300 km. Here, we take $\theta_{23}=40^{\circ}$ 
and $\delta m^2_{31}<0$.

\begin{figure}[t]
\subfigure[$\eta=0.1,\zeta=0$]{\includegraphics[width=7.5cm,height=5.5cm]{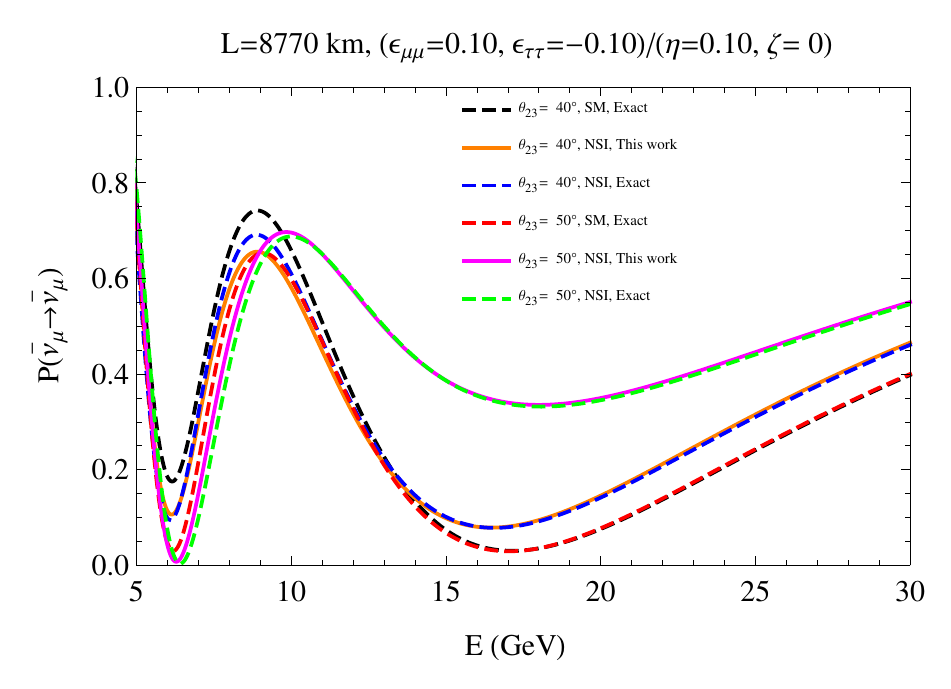}}
\subfigure[$\eta=-0.1,\zeta=0$]{\includegraphics[width=7.5cm,height=5.5cm]{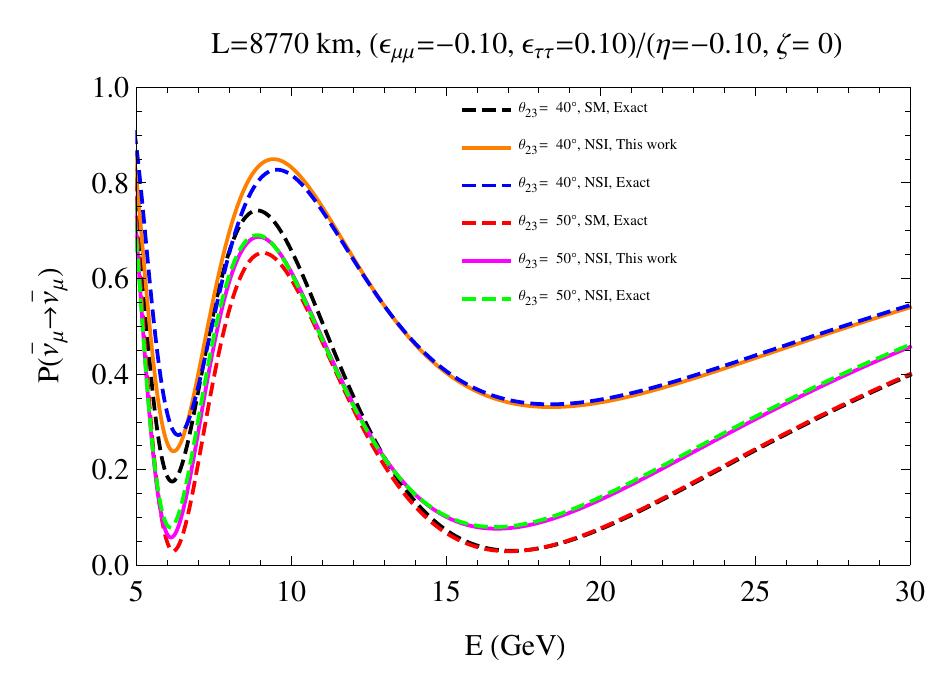}}
\caption{$\bar\nu_\mu\rightarrow\bar\nu_\mu$ survival probability as a function 
of anti-neutrino energy $E$ in GeV for two different values of $\theta_{23}$ 
at 8770 km baseline. Comparison between the analytical and numerical results assuming
$\eta=0.1, \zeta=0$ (left panel) and $\eta=-0.1, \zeta=0$ (right panel). 
The standard three-flavor oscillation probabilities in matter without NSI's are also shown.
In both the panels, we assume $\delta=0^{\circ}$ and $\delta m^2_{31}<0$.}
\label{fig:comparison-survival-anti-neutrino}
\end{figure}

In Fig.~\ref{fig:comparison-survival-anti-neutrino}, we depict 
the $\bar\nu_\mu\rightarrow\bar\nu_\mu$ survival probability 
in the presence of NSI for two different values of $\theta_{23}$ 
(40$^{\circ}$ and 50$^{\circ}$) at 8770 km baseline. 
We present the matching between the analytical and numerical 
results assuming $\eta=0.1, \zeta=0$ (left panel) and 
$\eta=-0.1, \zeta=0$ (right panel). In both the panels, we consider
$\delta=0^{\circ}$ and $\delta m^2_{31}<0$. 
Fig.~\ref{fig:comparison-survival-anti-neutrino} portrays that 
our approximate expressions match quite nicely with the 
numerical results. 

\section{Comparing Probabilities with Constant \& Varying Earth Density Profile}
\label{Constant-vs-varying-density}

\begin{figure}[t]
\includegraphics[width=7.5cm,height=5.5cm]{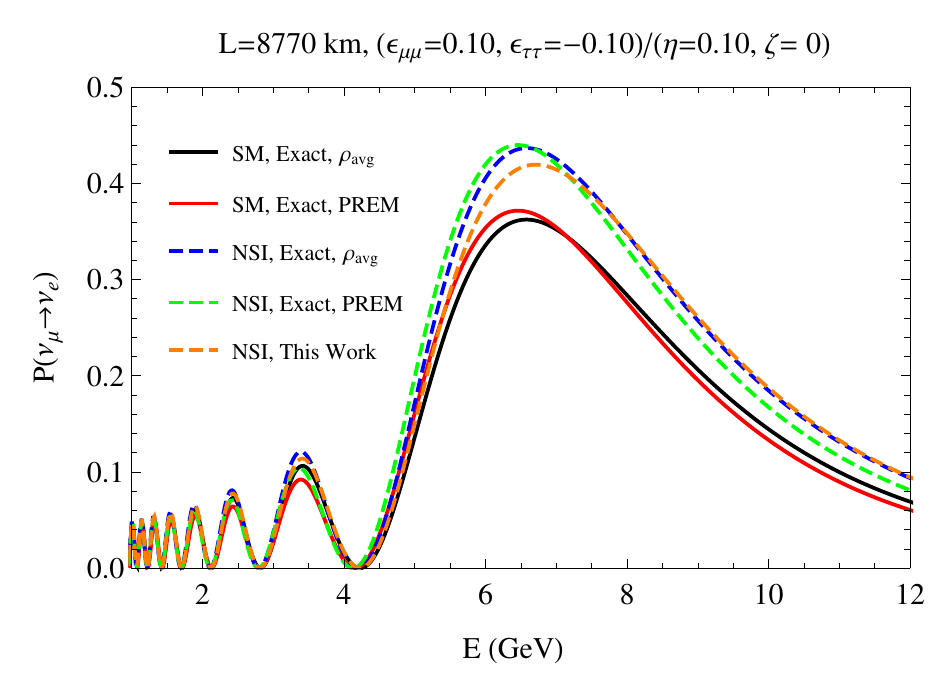}
\includegraphics[width=7.5cm,height=5.5cm]{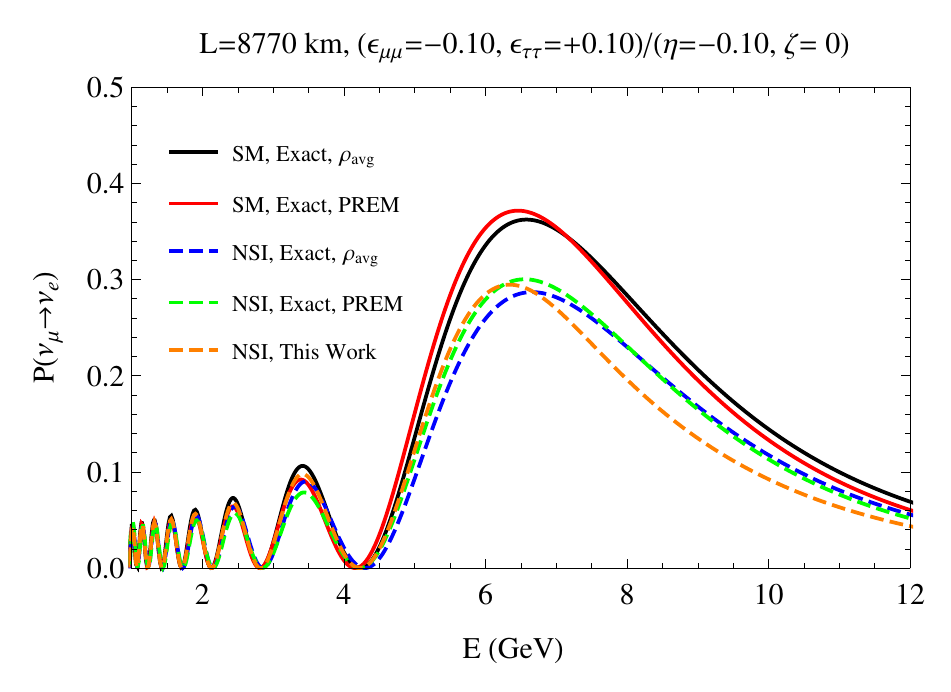}
\includegraphics[width=7.5cm,height=5.5cm]{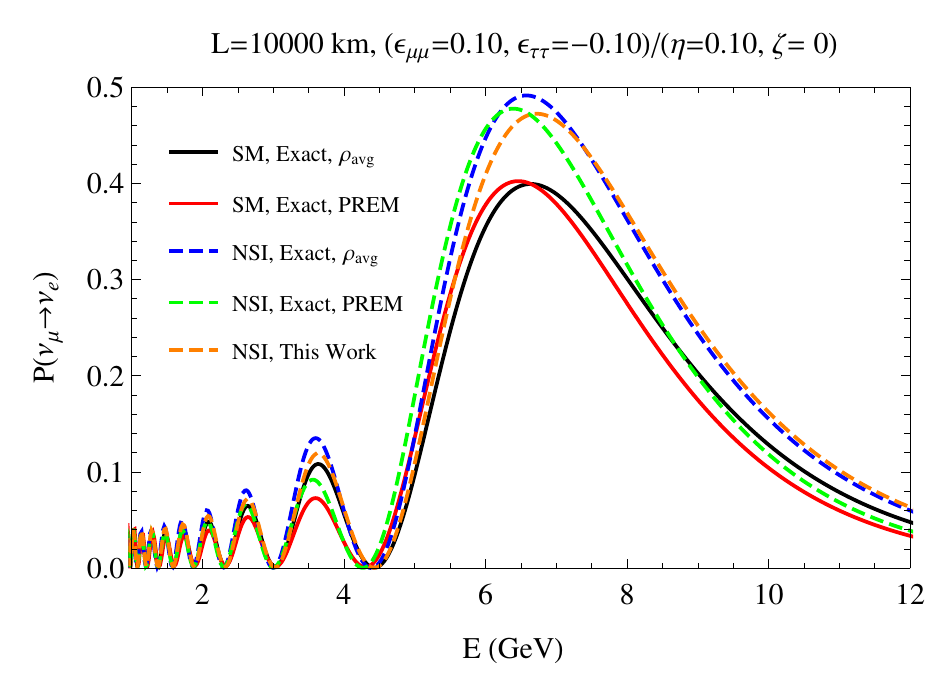}
\includegraphics[width=7.5cm,height=5.5cm]{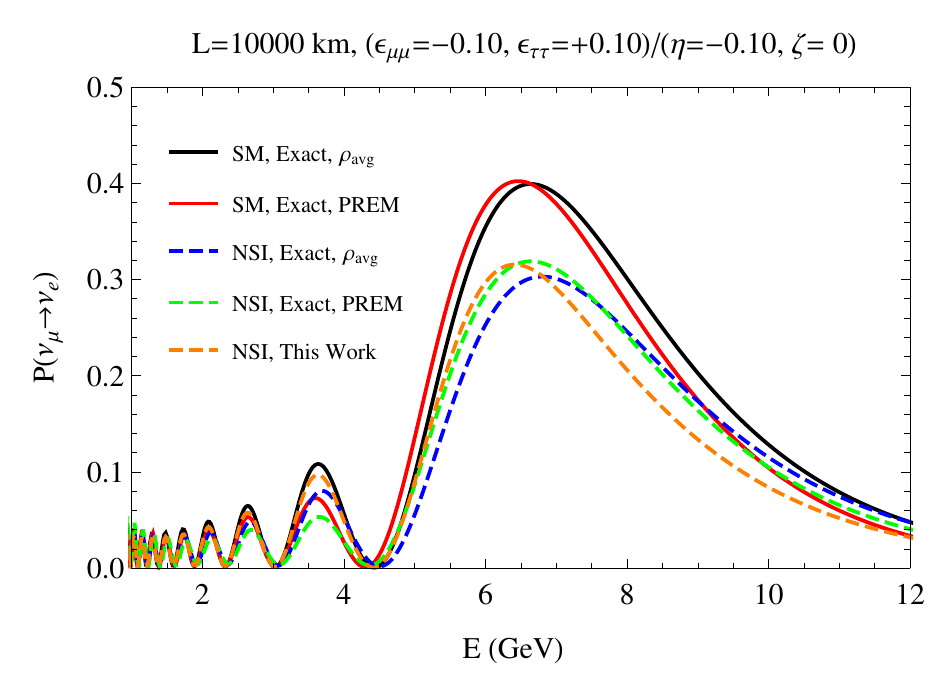}
\caption{$\nu_\mu\rightarrow\nu_e$ transition probability as a function of neutrino energy $E$ 
in GeV for 8770 km (10000 km) baseline in upper (lower) panels. We compare the exact 
numerical probabilities considering the constant (denoted by `$\rho_{\rm avg}$'), and
varying (labelled by `PREM') Earth density profiles for both the SM and NSI cases.
We also plot our approximate analytical expressions in the presence of NSI. 
For the NSI's, we take $\eta=0.1, \zeta=0$ (left panels), and $\eta=-0.1, \zeta=0$ (right panels).
In all the panels, we consider $\theta_{23}=40^{\circ}$, $\delta=0^{\circ}$, and normal mass hierarchy.}
\label{fig:constant-vs-varying-matter-density}
\end{figure}
 
So far, we considered the line-averaged constant Earth matter density for a given 
baseline which has been estimated using the PREM profile to present our results.
Now, it would be quite interesting to study how the exact numerical probabilities 
would be affected if we consider the more realistic varying Earth density profile
instead of the line-averaged constant matter density for the baselines as large as 
8770 km and 10000 km. In Fig.~\ref{fig:constant-vs-varying-matter-density}, 
we show the exact numerical probabilities considering the constant and varying 
Earth density profiles for 8700 km (upper panels) and 10000 km (lower panels) 
baselines. In the figure legends, the line-averaged constant matter density cases
are denoted by `$\rho_{\rm avg}$' and the varying Earth density cases are 
labelled by `PREM'. We perform these comparisons for both the SM and
NSI scenarios. For the NSI's, we take $\eta=0.1, \zeta=0$ (left panels), and 
$\eta=-0.1, \zeta=0$ (right panels). For the sake of completeness, we also plot the 
approximate probability expressions in the presence of NSI that we derived
in this paper assuming the line-averaged constant Earth matter density based
on the PREM profile. Fig.~\ref{fig:constant-vs-varying-matter-density} 
clearly shows that though the line-averaged constant matter density probability 
does not completely overlap with the PREM-based varying matter density 
profile probability, it is still fairly accurate. Moreover, the effect due to the 
inclusion of flavor-diagonal NSI's is correctly captured in our approximate 
analytical expressions.

\end{appendix}

\newpage

\bibliographystyle{JHEP}
\bibliography{osc-references-v2}

\end{document}